\documentclass{MYbook}
\usepackage{epsfig}

\title{Channeling, Radiation and Reactions in Crystals at High Energy}              % set your book title here

\makeindex

\begin{document}
\thispagestyle{empty}

\begin{center}

\Large{Vladimir Baryshevsky}

\vspace{1cm}

{\bf{\Large{Channeling, Radiation and Reactions \\in Crystals at
High Energy}}}

\vspace{1cm}

\begin{figure}[htb]
  \centering
  \includegraphics[width=4 cm]{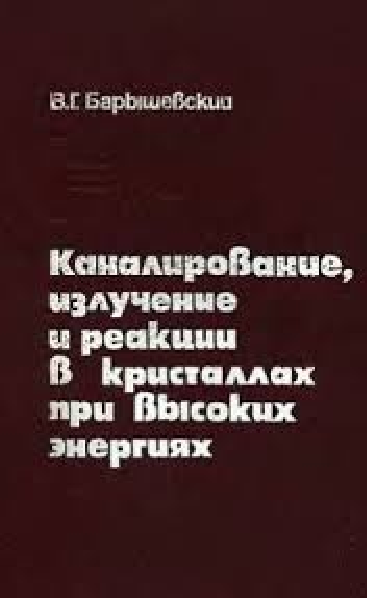}\\
  %\caption*{}
  %\label{fig1}
  \nonumber
\end{figure}

Author-made translation of the book \\ published in Russian in
1982

\end{center}

\newpage
\setcounter{page}{1}
\tableofcontents

%\setcounter{page}{1}

%\input Introduction.tex

%%%%%%%%%%%%%%%%%%%%%%%%%%%%%%%%%%%  Channelling Chapter 1 %%%%%%%%%%%%%%%%%%%%%%

\chapter{Channeling of High-Energy Particles in Crystals}
\label{ch:1}

%%%%%%%%%%%%%%%%%%%%%%%%%%%%%%%%%%%  Section 1 %%%%%%%%%%%%%%%%%%%%%%

\section{Channeling and Diffraction of Particles}
\label{sec:1.1}

A fast particle passing through a single crystal undergoes elastic
and inelastic scattering due to the interaction with electrons and
nuclei and causes various reactions. From a quantum mechanical
viewpoint, scattering processes and reactions excite secondary
(scattered) waves in a crystal. One should bear in mind that
secondary waves, which describe elastic scattering, interfere with
one another and with the incident wave. This leads to the
formation of a sum coherent wave in a crystal. Since the formation
of a coherent wave  is caused by the processes of elastic
scattering, its transmission through the crystal can be described
by introducing the effective periodic potential $V(\vec{r})$
averaged over temperature oscillations of the atoms (nuclei). The
expansion of $V(\vec{r})$ into the Fourier series has the form
\cite{13,14,15}
\begin{equation}
\label{1.1} V(\vec{r})=\sum_{\vec{\tau}}V(\vec{\tau})e^{i
\vec{\tau }\vec{r}}\,,
\end{equation}
where $\vec{\tau}$ is the reciprocal  lattice vector of the
crystal;
\[
V(\vec{\tau})=\frac{1}{\Omega}\sum_{j}V_{j0}(\vec{\tau})e^{-w_{\vec{j}}(\vec{\tau})}e^{-i\vec{\tau}\vec{r}_{j}}
\]

is the Fourier component of the potential; here $\Omega$ is the
crystal unit cell volume; $\vec{r}_{j}$ is the coordinate of the
$j$-type atom (nucleus) in the unit cell; the square of
$e^{-w_{j}(\vec{\tau})}$ is equal to the thermal factor, or the
Debye-Waller factor, known from X-ray and neutron scattering;
$V_{j0}(\vec{\tau})$ is the Fourier component of the interaction
potential between the particle and the atom whose center of
gravity rests in the origin of coordinates.

When a particle of charge $\pm e$ ($e$ is the value of the
electron charge) passes through a crystal, $V(\vec{r})$ represents
the ordinary Coulomb interaction, while $V_{j0}$ is determined by
the expression
\[
V_{j0}(\vec{\tau})=\pm\frac{4\pi
e^{2}}{(\vec{\tau})^{2}}[z_{j}-F_{j}(\vec{\tau})]\,,
\]
where $z_{j}$ is the charge of the nucleus located at point
$\vec{r}_{j}$ in the unit cell; $F_{j}(\vec{\tau})$ is the form
factor of the atom located at point $\vec{r}_{j}$ \cite{16}.

\begin{figure}[htbp]
\centering \epsfxsize = 8 cm
\centerline{\epsfbox{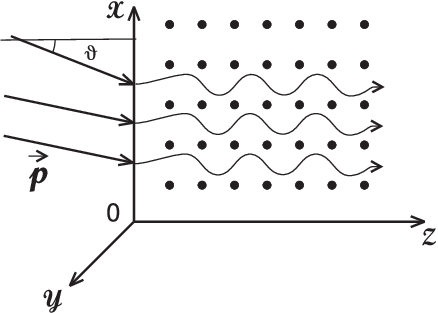}} \caption{Particle
channeling in a crystal} \label{Channeling Figure 1}
\end{figure}

%%%%%%%%%%%%%%%%%

Consider in more detail the case when a particle enters a single
crystal at  a certain small angle $\vartheta$  with respect to the
crystallographic planes (axes) of the crystal (Fig. 1). If this
angle is smaller than the so-called Lindhard angle, the particle
in the crystal moves in the channeling regime \cite{1,2,3}.

Theoretical analysis of the channeling effect should take into
account that when a high-energy particle, for which the wavelength
$\lambda$ is much smaller than the interatomic distance, is
incident on a crystal at a small angle $\vartheta$, the
periodicity of chains and  planes of the crystal along the
direction of particle motion has almost no influence on the nature
of particle motion \cite{15,17,18}. As a result, the particle
behavior is determined by the averaged potential of the crystal
axes (planes), which is constant along the direction of particle
incidence and periodic in the transverse plane.

Direct the $z$-axis of the coordinate system along the crystal
axes (planes), relative to which the particle moves at a small
angle. In this case the periodic along the $x$-axis potential of
planes,  which describes  planar channeling, can be written as
follows \cite{15,17,18}:
%%%%%
\begin{equation}
\label{1.2}
V(x)=\sum_{\tau_{x}}V(2\pi\tau_{x})e^{-i2\pi\tau_{x}x}(\tau_{y}=\tau_{z}=0).
\end{equation}
Axial channeling is described in terms of the two-dimensional
periodic in a transverse plane potential

\begin{equation}
\label{1.3}
V(\vec{\rho})=\sum_{\vec{\tau}_{\perp}}V(2\pi\vec{\tau}_{\perp})e^{-i2\pi\vec{\tau}_{\perp}\vec{\rho}},
\end{equation}
where $\vec{\rho}=(x,y)$;
$\vec{\tau}_{\perp}=(\tau_{x},\tau_{y})$; $\tau_{z}=0$.

 To determine the influence of a
single crystal on a passing relativistic particle in the general
case, it is necessary to study the solution of the Dirac equation.
With this aim in view, it is convenient to convert it into a
second--order equation, very much similar in form to the
Schr\"odinger equation \cite{19}:
\begin{eqnarray}
\label{1.4} \left[-\hbar^{2}\Delta_{r}-p^{2}+\frac{2}{c^{2}}E
V(\vec{r})-\frac{1}{c^{2}}V^{2}(\vec{r})
-i\frac{\hbar}{c}\vec{\alpha}\vec{\nabla}
V(\vec{r})\right]\psi(\vec{r})=0\,,
\end{eqnarray}
where $p^{2}=(E^{2}-m^{2}c^{4})/c^{2}$ is the  momentum of the
particle entering the crystal; $E$ is its energy; $\vec{\alpha}$
are the Dirac matrices; $\psi$ is the bispinor.

According to (\ref{1.4}), the effective potential acting on a
relativistic particle is the sum of three terms, one of which
increases with the growth of particle energy $E$. For these
reason, the terms including $V^{2}$ and $\vec{\alpha}$  can be
dropped when analyzing spatial and angular distribution of the
particles which have interacted with the crystal. However, when
analyzing the polarization properties of particles transmitted
through a crystal, it is crucial that the term containing the
matrices  $\vec{\alpha}$ should be taken into account \cite{20}.

Dropping the terms proportional to $V^{2}$ and $\vec{\alpha}$,
from  (\ref{1.4}) we obtain the following equation
\begin{equation}
\label{1.5} \left[-\hbar^{2}\Delta_{r}-p^{2}+\frac{2}{c^{2}}E
V(\vec{r})\right]\psi(\vec{r})=0\,.
\end{equation}
Upon dividing  (\ref{1.5}) first  by $2m$ and then by $2m\gamma$
($\gamma=E/mc^{2}$ is the particle Lorentz factor), we can recast
it in two forms:
\begin{equation}
\label{1.6} \left[-\frac{\hbar^{2}}{2m}\Delta_{r}+\gamma
V(\vec{r})\right]\psi(\vec{r})=\varepsilon\psi(\vec{r})\,,
\end{equation}
and
\begin{equation}
\label{1.7} \left[-\frac{\hbar^{2}}{2m\gamma}\Delta_{r}+
V(\vec{r})\right]\psi(\vec{r})=\varepsilon^{\prime}\psi(\vec{r})\,,
\end{equation}
where $\varepsilon=p^{2}/2m$;
$\varepsilon^{\prime}=p^{2}/2m\gamma=\varepsilon\gamma^{-1}$.
Recall that $p^{2}=(E^{2}-m^{2}c^{4})/c^{2}$.

Equation (\ref{1.6}) coincides with the nonrelativistic
Schrodinger equation  for a particle moving in a potential growing
with the increase in the particle energy. Equation (\ref{1.7})
coincides with the Schr\"{o}dinger equation  for a particle with a
relativistic mass $m\gamma$.

The eigenfunctions of the Dirac (Schrodinger) equations with a
periodic potential are known to be the Bloch functions \cite{21}.
Hence, an arbitrary solution of equations (\ref{1.4})-(\ref{1.7})
is described by the superpositions of the Bloch functions. This
fact makes it possible to draw some general conclusions about the
nature  of the particle-crystal interaction. Further we follow the
line of reasoning given in \cite{22} for the case of the
interaction between non-relativistic electrons and a crystal,
which is also suitable for our case  due to the mathematical
equivalence of equations (\ref{1.4}) (\ref{1.7}) and the
non-relativistic Schrodinger equation.

%%%%%%%%%%%%%%%%%%%%%%%%%%

Let a beam of particles with the momentum  $\vec{p}$ fall on a
plane--parallel crystal plate bounded by the planes $z=0$ and
$z=L$ (see \fref{Channeling Figure 1}). The corresponding plane
wave that describes the incident particle is determined by the
expression \footnote{Unless otherwise stated, assume that
$\hbar=c=1$.}
\begin{equation}
\label{1.8}
\psi_{0}(\vec{r})=\exp(i\vec{p}\vec{r})=\exp(i\vec{p}_{\perp}\vec{\rho}+p_{z}z)\,,
\end{equation}
where $\vec{\rho}$ is the vector with the components $x$ and $y$;
the $z$-axis is directed into the interior of the crystal
perpendicular to its entrance surface. For simplicity, we shall
further assume that the crystal lattice is rectangular and has the
lattice constants $a$, $b$, $c$ in the directions $x$, $y$, $z$,
respectively.

%%%
The interaction between the wave $\psi_{0}$ and a crystal gives
rise to secondary waves.  The potential $V$ equals zero outside
the crystal, and the secondary waves can be represented as a
superposition of the eigenfunctions of (\ref{1.6}), (\ref{1.7})
when $V =0$, i.e., as a superposition of plane waves. Therefore
outside the crystal on the side of incidence, i.e. at $z>0$, there
are  the waves reflected from the crystal, which have the  form
\begin{equation}
\label{1.9} \psi_{\mathrm{ref}}=\sum_{n}A_{n}\exp[i(\vec{p}_{\perp
n}\vec{\rho}-p_{zn}z)]\,.
\end{equation}

%%%%%
The transversal components of the momentum $\vec{p}_{\perp n}$ are
still arbitrary. As (\ref{1.9}) should describe the particle flow
moving to the left of the crystal, the values of $p_{zn}$ are
always positive. Moreover, since the energy of a scattered
particle equals the energy of an incident particle, the momentum
is
\[
p_{zn}=\sqrt{E^{2}-p^{2}_{\perp n}-m^{2}}=\sqrt{p^{2}-p^{2}_{\perp
n}}\,.
\]
So, within the range $z<0$, the wave function
\begin{equation}
\label{1.10} \psi=\psi_{0}+\psi_{\mathrm{ref}}\,.
\end{equation}
On the other side of the crystal, at $z>L$,  there is  a
transmitted wave alone
\begin{equation}
\label{1.11}
\psi=\sum_{n}A^{\prime}_{n}\exp[i(\vec{p}^{\prime}_{\perp
n}\vec{\rho}+p^{\prime}_{zn}z)]\,,
\end{equation}
where $p^{\prime}_{zn}=\sqrt{p^{2}-p^{\prime 2}_{\perp n}}$.

%%%%%%%%%%%%%%%5
Inside the plate, the potential $V(\vec{r})$ differs from zero.
The eigenfunctions are the Bloch waves, and the general solution
inside the crystal is described by the superposition of the Bloch
waves. It is known that the Bloch wave for band  $n$ can be
written accurate to the normalization factor in the form:
\begin{equation}
\label{1.12} \psi_{\kappa
n}(\vec{r})=e^{i\vec{\kappa}\vec{r}}u_{\kappa n}(\vec{r})\,,
\end{equation}
where $\kappa$ is the reduced quasimomentum; $u_{\kappa
n}(\vec{r})$ is the periodic function with the period of the
crystal. The function $u_{\kappa n}(\vec{r})$ is likely to be
expanded into a Fourier series. As a result, (\ref{1.12}) can be
represented as
\begin{equation}
\label{1.13} \psi_{\kappa
n}(\vec{r})=\sum_{\tau}a_{n\kappa}(\vec{\tau})\exp[i(\vec{\kappa}+\vec{\tau})\vec{r}]\,,
\end{equation}
where $\vec{\tau}$ is the reciprocal lattice vector with the
components $\tau_{x}=\lambda/a$, $\tau_{y}=\mu/b$,
$\tau_{z}=\nu/c$ ($\lambda$, $\mu$, $\nu$ are integral numbers,
running over the integers from $-\infty$ to $+\infty$).

%%%%%%

Near the plate surface the wave function and its first derivative
in the $z$ direction should be  continuous.   The continuity
condition implies that the superposition of the functions
(\ref{1.3}), which describes the wave in a crystal should only
contain such Bloch functions for which the sum of vector
$\kappa_{\perp}$ and a certain reciprocal lattice vector
$2\pi\vec{\tau}_{0\perp}$ equals $\vec{p}_{\perp}$, i.e.,
$\vec{\kappa}_{\perp}+2\pi\vec{\tau}_{0\perp}=\vec{p}_{\perp}$.

%%%%%%%%%%
Thus, the arbitrary solution of (\ref{1.6}), (\ref{1.7}) inside a
crystal can also be represented as the superposition of plane
waves.

Since inside the plate  the wave function contains plane waves
with transversal momenta
$\vec{\kappa}_{\perp}+2\pi\vec{\tau}_{\perp}$, due to the boundary
conditions at the $z=L$ surface, the waves having the same
transversal momenta should propagate from a crystal into vacuum.
As a consequence behind the plate ($z>L)$)
\begin{equation}
\label{1.14}
\psi=\sum_{\vec{\tau}_{\perp}}A^{\prime}(\vec{\tau}_{\perp})\exp[i(\vec{p}_{\perp}+
2\pi\vec{\tau}_{\perp})\vec{\rho}]\exp[ip_{z}(\vec{\tau}_{\perp})z],
\end{equation}
where
$p_{z}(\vec{\tau}_{\perp})=\sqrt{p^{2}-(\vec{p}_{\perp}+2\pi\vec{\tau}_{\perp})^{2}}$;
before the plate $(z<0)$
\begin{eqnarray}
\label{1.15}
\psi=e^{i\vec{p}\vec{r}}+\sum_{\vec{\tau}_{\perp}}A(\vec{\tau}_{\perp})\exp[i(\vec{p}_{\perp}+2\pi\vec{\tau}_{\perp})\vec{\rho}]\times\nonumber\\
\exp[-ip_{z}(\vec{\tau}_{\perp})z].
\end{eqnarray}

 An important result (which was already emphasized in \cite{22}) follows from equalities (\ref{1.14}), (\ref{1.15}):
 the direction of scattered waves leaving a plane--parallel plate is uniquely determined by the incident direction and the
 magnitude of the momentum (energy, wavelength) of the incident particles  in the same way as in the elementary kinematic
 Laue theory of interference developed for thin plates, when the  effects of wave refraction may be neglected, namely the
 projection of the momentum of each scattered wave onto the crystal surface  differs from the corresponding value for the
 incident wave by a reciprocal lattice vector $2\pi\vec{\tau}_{\perp}$. The possible refraction only leads to the redistribution
 of intensity among the scattered waves.

 This conclusion is valid for any particles interacting with a single--crystal plate and any angles of particle entrance
 into the crystal (even for those smaller than the Lindhard angle). It means that the channeling phenomenon is just a
 particular case of diffraction by a periodically arranged set of scatterers  (see also \cite{1}).

%%%%%%%%%%%%%%%%%%%%%%%%%%%%%%%%%%%  Section 2 %%%%%%%%%%%%%%%%%%%%%%
\section{Principles of the Quantum Theory of Channeling}
\label{sec:1.2}

The possibility to describe the interaction of fast particles with
a crystal in terms of an averaged potential (\ref{1.2}),
(\ref{1.3}) enables carrying out a more detailed analysis of the
peculiarities of their transmission through a crystal. A thorough
quantum mechanical study of this problem on the basis of a
time-dependent density matrix (temporal) is given by Kagan and
Kononetz in \cite{15,23,24}. A similar problem in a stationary
representation of wave scattering by a crystal was examined by
Kalashnikov and Strikhanov in  \cite{18,25} who scrutinized the
extreme case of particle scattering by a one plane (axis). Further
we will follow the analysis performed by the author together with
Dubovskaya \cite{5,17}.

First of all, we shall make use of the fact that in the case of
axial channeling, due to the constant character of the potential
(\ref{1.2}), (\ref{1.3}) along the $z$-axis (in the case of planar
channeling, along the $y$- and $z$-axes), the particle motion in
these directions is free and can be characterized by a
well--defined  momentum.  As a consequence, it is possible to
separate  variables in  (\ref{1.6}) and (\ref{1.7}) and then
analyze the equations, which depend only on the coordinates
relative to which the potential $V$ is periodic. Thus, in the
axial case  from  (\ref{1.6}) we obtain
\begin{equation}
\label{2.1} \left[-\frac{1}{2m}\Delta_{\rho}+\gamma
V(\vec{\rho})\right]\psi_{n\kappa}(\vec{\rho})=\varepsilon_{n\kappa}\psi_{n\kappa}(\vec{\rho})\,.
\end{equation}
The two--dimensional Bloch functions $\psi_{n\kappa}(\rho)$
(one--dimensional in the planar case) are the eigenfunctions of
 (\ref{2.1}). The corresponding eigenvalues are
$\varepsilon_{n\kappa}$.

%%%%%%

Expand the function $\psi(\vec{r})$ into the eigenfunctions which
are determined by equation (\ref{2.1}):
\begin{equation}
\label{2.2} \psi(\vec{r})=\sum_{n^{\prime}}\int
d^{2}\kappa^{\prime}b_{n^{\prime}\kappa^{\prime}}(z)\psi_{n^{\prime}\kappa^{\prime}}(\vec{\rho}).
\end{equation}
Substitution of (\ref{2.1}) into (\ref{1.6}), further
multiplication of (\ref{1.6}) by $\psi_{n\kappa}^{*}(\vec{\rho})$
and its integration  with respect to  $\vec{\rho}$ with due
account of the orthogonality condition
\begin{equation}
\label{2.3}
\int\psi^{*}_{n\kappa}(\vec{\rho})\psi_{n^{\prime}\kappa^{\prime}}(\vec{\rho})d^{2}\rho=\delta_{nn^{\prime}}\delta(\vec{\kappa}-\vec{\kappa}^{\prime}),
\end{equation}
gives the equation determining the quantities $b_{n\kappa}(z)$:
\begin{equation}
\label{2.4} -\frac{1}{2m}\frac{\partial^{2}}{\partial
z^{2}}b_{n\vec{\kappa}}(z)=(\varepsilon-\varepsilon_{n\vec{\kappa}})b_{n\vec{\kappa}}(z).
\end{equation}
The solutions of (\ref{2.4}) are plane waves
\begin{equation}
\label{2.5} b_{n\vec{\kappa}}(z)\sim\exp[\pm
ip_{zn}(\vec{\kappa})z],
\end{equation}
where the momentum
$p_{zn}(\vec{\kappa})=\sqrt{2m(\varepsilon-\varepsilon_{n\vec{\kappa}})}$,
i.e.,
$p_{zn}(\vec{\kappa})=\sqrt{p^{2}-2m\varepsilon_{n\vec{\kappa}}}$.
(Recall that $p$ is the momentum of a particle incident on a
crystal.)
%%%%%%%%%%%%%%%%%%%%%%%%%%%%55
So, the general solutions of (\ref{1.6}) in a crystal can be
written as the superposition:
\begin{equation}
\label{2.6} \psi(\vec{r})=\sum_{n}\int
\tilde{c}_{n\kappa}\psi_{n\kappa}(\vec{\rho})\exp[ip_{zn}(\vec{\kappa})z]d^{2}\kappa\,.
\end{equation}
The waves of the form  $\exp[-ip_{zn}(\vec{\kappa})z]$ are not
included into the superposition because they describe the mirror
reflected waves whose amplitudes for particles incident at not a
very small angle relative to the crystal surface are negligible.
At the entrance surface of the crystal ($z=0$), it is necessary to
join the superposition of (\ref{2.4}) and the solution of
(\ref{1.10}), where the mirror reflected waves are also neglected.
Thus, at $z=0$, we have the equality
\begin{equation}
\label{2.7} \exp(i\vec{p}_{\perp}\vec{\rho})=\sum_{n}\int
d^{2}\kappa\tilde{c}_{n\vec{\kappa}}\psi_{n\vec{\kappa}}(\vec{\rho})\,.
\end{equation}
%%%%%%%5
Multiplying (\ref{2.7}) by
$\psi^{*}_{n^{\prime}\kappa^{\prime}}(\vec{\rho})$ and integrating
it with respect to $d^{2}p$, we directly find the expansion
coefficients
\begin{equation}
\label{2.8}
\tilde{c}_{n\vec{\kappa}}=\int\exp(i\vec{p}_{\perp}\vec{\rho})\psi^{*}_{n\vec{\kappa}}(\vec{\rho})d^{2}\rho\,.
\end{equation}
Now make use of the fact that the Bloch function can be written in
the form
\[
\psi_{n\vec{\kappa}}(\vec{\rho})=\exp(i\vec{\kappa}\vec{\rho})u_{n\vec{\kappa}}(\vec{\rho})\,,
\]
where $u_{n\vec{\kappa}}(\vec{\rho})$ is the function periodic in
a transverse plane. The integration with respect to $\vec{\rho}$
in (\ref{2.8}) is split into the sum of integrals over the unit
cells and then
 (\ref{2.8}) can be written as follows
\begin{equation}
\label{2.9}
\tilde{c}_{n\vec{\kappa}}=\sum_{\vec{\tau}_{\perp}}\delta(\vec{p}_{\perp}-\vec{\kappa}-2\pi\vec{\tau}_{\perp})c_{n\vec{\kappa}};
\end{equation}
\begin{equation}
\label{2.10}
c_{n\vec{\kappa}}=\frac{(2\pi)^{2}}{S}\int_{s}e^{i\vec{p}_{\perp}\vec{\rho}}\psi^{*}_{n\kappa}(\rho)d^{2}\rho\,.
\end{equation}

%%%
As $\vec{\kappa}$ is the reduced momentum, (\ref{2.9}) means the
requirement of the equality of vector $\vec{\kappa}$ and the
reduced part of the transversal momentum of the incident particle
$\vec{p}_{\perp}-2\pi\vec{\tau}_{\perp}$. It follows from
(\ref{2.9}) and (\ref{2.10}) that the wave function of a particle
inside a crystal (\ref{2.6}) can be written in the form
\begin{equation}
\label{2.11}
\psi(\vec{r})\sum_{n}c_{n\vec{\kappa}}\psi_{n\vec{\kappa}}(\vec{\rho})\exp(ip_{\vec{z}n}z),
\end{equation}
where $\vec{\kappa}=\vec{p}_{\perp}-2\pi\vec{\tau}_{\perp}$
($\vec{\tau}_{\perp}$ is chosen from the condition of  reduction
of $p_{\perp}$ to the first Brillouin zone);
$p_{zn}=\sqrt{p^{2}-2m\varepsilon_{n\vec{\kappa}}}$.

%%%%%%%%%%%%%55
If a particle moves in a regime of planar channeling, then the
motion along the $y$-axis is also free (the $x$-axis is directed
perpendicular to the family of planes, along which the particle is
channeled). In this case
\begin{equation}
\label{2.12}
\psi(\vec{r})=\sum_{n}c_{n\kappa}\psi_{n\kappa}(x)e^{ip_{y}y}e^{ip_{zn}z};
\end{equation}
\begin{equation}
\label{2.13}
c_{n\kappa}=\frac{2\pi}{a}\int_{0}^{a}e^{ip_{x}x}\psi^{*}_{n\kappa}(x)dx,
\end{equation}
where $a$ is the lattice spacing along the $x$-axis;
$\kappa=p_{x}-2\pi\vec{\tau}_{x}$.

%%%%%%%%%55
Let us present the expressions relating the Bloch functions to the
localized Wannier functions, which  come in handy when analyzing
the behavior of a channeled particle. A detailed treatment of
their properties for a three--dimensional case  is given in
\cite{21}.  In the  one--and two--dimensional cases, which are of
interest for us, the Wannier function centered in a well with the
coordinate of its center $\vec{\rho}_{m}(x_{m})$ is determined as
follows:
\begin{equation}
\label{2.14}
W_{n}(\vec{\rho}-\vec{\rho}_{m})=\frac{\sqrt{S}}{2\pi}\int
e^{-i\vec{\kappa}\vec{\rho}_{m}}\psi_{n\vec{\kappa}}(\vec{\rho})d^{2}\kappa
\end{equation}
or
\[
W_{n}(x-x_{m})=\sqrt{\frac{a}{2\pi}}\int e^{-i\kappa
x_{m}}\psi_{n\kappa}(x)d\kappa\,.
\]
Integration with respect to $\kappa$ is made within the first
Brillouin zone.

%%%%%%%%%%%%%55
The Bloch functions expressed in terms of the Wannier functions
have the form
\begin{eqnarray}
\label{2.15}
\psi_{n\vec{\kappa}}(\vec{\rho})=\frac{\sqrt{S}}{2\pi}\sum_{m}e^{i\vec{\kappa}\vec{\rho}_{m}}W_{n}(\vec{\rho}-\vec{\rho}_{m});\nonumber\\
\psi_{n\vec{\kappa}}(x)=\sqrt{\frac{a}{2\pi}}\sum_{m}e^{i\kappa
x_{m}}W_{n}(x-x_{m})\,.
\end{eqnarray}
In normalizing in a finite volume, the factor $\sqrt{S}/2\pi$
$(\sqrt{a}/2\pi)$ should be replaced by
$1/\sqrt{N}(1/\sqrt{N_{x}})$; $N(N_{x})$ is the number of unit
cells (the number of crystal spacings) in the ($x,\,y$) plane
(along the $x$-axis). Equalities (\ref{2.11})--(\ref{2.13}) enable
analyzing  the features of behavior of  fast particles in a
crystal in the general case.

%%%%%

Now consider a wave produced by a particle behind a crystal.
According to (\ref{1.14}) it is necessary to find the explicit
form of the coefficients $A^{\prime}(\vec{\tau}_{\perp})$. For
this purpose join the solutions of (\ref{2.11}) and (\ref{1.14})
in the $z=L$ plane:
\begin{equation}
\label{2.16}
\sum_{n}c_{n\vec{\kappa}}\psi_{n\vec{\kappa}}(\vec{\rho})e^{ip_{zn}L}=
\sum_{\vec{\tau}_{\perp}}A^{\prime}(\vec{\tau}_{\perp})e^{i(\vec{p}_{\perp}+
2\pi\vec{\tau}_{\perp})\rho}e^{ip_{z}(\tau_{\perp})L}.
\end{equation}
Substitute the expansion of the Bloch function
$\psi_{n\vec{\kappa}}(\vec{\rho})$ into Fourier series into
(\ref{2.16}) (see (\ref{1.13})):
$$
\psi_{n\vec{\kappa}}(\vec{\rho})=\sum_{\vec{\tau}_{\perp}}a_{n\vec{\kappa}}(\vec{\tau}_{\perp})e^{i(\kappa+2\pi\vec{\tau}_{\perp})\vec{\rho}};
$$
$$
a_{n\vec{\kappa}}(\vec{\tau}_{\perp})=\frac{1}{S}\int_{S}u_{n\vec{\kappa}}(\vec{\rho})e^{-i\pi\vec{\tau}_{\perp}\vec{\rho}}.
$$

%%%%%%

Since (\ref{2.16}) should be fulfilled at an arbitrary point
$\rho$, it will hold if the coefficients of the identical
exponents on the right and left sides of (\ref{2.16}) are equal.
As a result, we have
\begin{equation}
\label{2.17}
A^{\prime}(\vec{\tau}_{\perp})=\sum_{n}c_{n\vec{\kappa}}a_{n\vec{\kappa}}(\vec{\tau}_{0\perp}
+\vec{\tau}_{\perp})e^{i(p_{zn}-p_{z}(\tau_{\perp}))L}\,,
\end{equation}
where $\vec{\tau}_{0\perp}=\vec{p}_{\perp}-\vec{\kappa}$.

The coefficients $A^{\prime}(\vec{\tau}_{\perp})$ have the meaning
of the probability amplitudes to find a particle with the
transversal momentum $\vec{p}_{\perp}+\vec{\tau}_{\perp}$ in the
wave which has passed through a crystal, and, hence, they actually
determine the angular distribution of particles behind the
crystal.

The expressions obtained above make it possible to determine in
the general case  all the required characteristics of particles in
a crystal and outside it and study the features of the reactions
they initiate.

%%%%%%%%%%%%%%%%%%%%%%%%%%%%%%%%%%%  Section 3 %%%%%%%%%%%%%%%%%%%%%%

\section{The Energy--Band Spectrum of Electrons and Positrons
Channeled in a Single Crystal} \label{sec:1.3}

Give a more detailed treatment of the structure of a particle wave
function $\psi(\vec{r})$ in a crystal, which is described by
equality (\ref{2.11}). According to (\ref{2.11}) $\psi(\vec{r})$
is represented as the superposition of the Bloch functions
corresponding to a potential periodic in a transverse plane. The
contribution of each wave is determined by the coefficient
$c_{n\kappa}$, whose squared absolute value defines the
probability to find a particle in the state of the band energy
spectrum $n\vec{\kappa}$. The formation of the superposition
(\ref{2.11}) at the particle entrance into the crystal is due to
the fact that there is a quasi-momentum, not a momentum remaining
in a crystal. As a consequence the state with the defined
momentum, which describes the particle falling upon the crystal is
not stationary inside the crystal. As a result, upon entering into
the crystal a particle (for instance, a muon) does not appear to
be in some specified band state, but populates the whole set of
such states.

To understand the character of the band energy spectrum of
channeled particles and the features of its population, it is
important to have quantitative results for the stated quantities,
which have been  obtained by using real interplanar potentials. As
demonstrated in \cite{26}, such quite exact calculations for
channeled particles can be carried out, using a quasi-classical
WKB (Wentzel-Kramer-Brillouin) method. A detailed treatment of the
band spectrum theory in the quasi-classical approximation for
planar channeling was given by I.D. Feranchuk and B.A. Chevganov.

%%%%%%%%%%%%%5
First, pay attention to the fact that the general dispersion
equation, which defines the band spectrum of a particle in a
one-dimensional periodic potential can be obtained without any
approximations \cite{27}. Indeed, in the range where
$\varepsilon^{\prime}>V(x)$, two linearly independent solutions of
equation (\ref{2.1}) correspond to every value of
$\varepsilon^{\prime}$. Let $f$ and $f^{*}$ denote these
solutions, respectively. Then the general solution of equation
(\ref{2.1}) at $ld<x<(l+1)d$, $l=0, 1$ may be represented as
\begin{equation}
\label{3.1} \varphi_{l}(x)=c_{1}f(x)+c_{2}f^{*}(x).
\end{equation}
%%%%%%%%%%%%%%%%%%%%5
Translation to the range $(l+1)d\leq x\leq (l+2)d$ transforms both
function $f$ and $f^{*}$
 into a linear superposition  of the same functions, i.e.,
\begin{eqnarray}
\label{3.2}
f(x+d)=Df(x)+Rf^{*}(x)\,,\nonumber\\
f^{*}(x+d)=D^{*}f^{*}(x)+R^{*}f(x)\,.
\end{eqnarray}
From the the periodicity condition follows  $|D|^{2}=1+|R|^{2}$.
Then
\begin{equation}
\label{3.3}
\varphi_{l+1}(x)=\varphi_{l}(x+d)=c_{1}(Df+Rf^{*})+c_{2}(D^{*}f^{*}+R^{*}f)\,,
\end{equation}
and in accordance with the Bloch theorem,
$\varphi_{l+1}=e^{i\kappa d}\varphi_{l}(x)$, where $\kappa$ is the
quasimomentum.
%%%%%%%%%%%%%%%%5

As the functions $f$ and $f^{*}$ are independent, (\ref{3.3})
yields the system of equations
\begin{eqnarray}
\label{3.4}
c_{1}D+c_{2}R^{*}=c_{1}e^{i\kappa d}\,,\nonumber\\
c_{1}R+c_{2}D^{*}=c_{2}e^{i\kappa d}\,.
\end{eqnarray}
The condition of the existence of a nontrivial solution of the
system (\eref{3.4}) leads to the desired dispersion equation.
\begin{equation}
\label{3.5} \cos \kappa d=|D|\cos\varphi(\varepsilon)\,, \qquad
\mbox{where}\qquad
D(\varepsilon)=|D(\varepsilon)|e^{i\varphi(\varepsilon)}\,.
\end{equation}
For convenience sake, instead of the quantities $R$ and $D$, we
shall further use the coefficients of reflection $R_{1}$ and
transmission $D_{1}$ related to them,  which can be determined in
a conventional manner. For example, for a wave passing through a
potential barrier from the right to the left, we obtain
\[
f(x)+R_{1}f^{*}(x)=D_{1}f(x+d)\,,
\]
i.e.,
\begin{eqnarray}
\label{3.6}
& &D  =  \frac{1}{D_{1}},\quad R=\frac{R_{1}}{D_{1}},\quad D_{1}=|D_{1}|e^{i\varphi_{1}(\varepsilon)}\,,\nonumber\\
& &\cos \kappa d  =  \frac{1}{|D_{1}|}\cos
\varphi_{1}(\varepsilon)\,.
\end{eqnarray}

%%%%%%%%%%%%%%%%%%%%%%%%%%%%%5
To determine the explicit form of the coefficient $D_{1}$, it is
necessary to turn to a certain approximation and find the
functions $f$ and $f^{*}$. Here  the quasiclassical approximation
is applied with the accuracy for the given equation determined by
the parameter  \cite{28} $\xi\simeq 1/n^{2}$, where $n$ is the
number of bound levels in an isolated potential well coinciding
with the channel potential.  In the stated approximation, the
functions $f$ and $f^{*}$ have the form
\begin{eqnarray}
\label{3.7}
f(x) &=& \frac{e^{i\int p(x)dx}}{\sqrt{p(x)}}, \quad f^{*}(x)=\frac{e^{-i\int p(x)dx}}{\sqrt{p(x)}}\,,\nonumber\\
p(x) &=& \sqrt{2E(\varepsilon^{\prime}-V(x))}\,.
\end{eqnarray}

\begin{figure}[htbp]
\centering \epsfxsize =8 cm
\centerline{\epsfbox{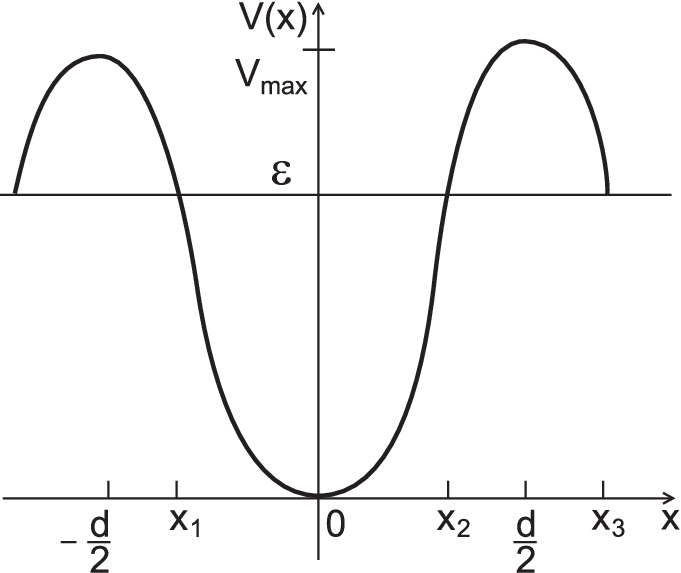}} \caption{Location of
the turning points} \label{Channeling Figure 2}
\end{figure}

Within the range $\varepsilon^{\prime}<V_{\mathrm{max}}$, the
coefficient $D_{1}$ is determined by a well--known expression
which is true when $|D_{1}|\ll 1$:
\[
|D_{1}|\approx e^{-\tau_{1}}, \quad R_{1}\approx  1, \quad
\varphi(\varepsilon^{\prime})=\sigma_{1}(\varepsilon^{\prime})\,.
\]
Here
\[
\sigma_{1}=\int_{x_{1}}^{x_{0}}\sqrt{2E(\varepsilon^{\prime}-V(x))}dx\,,\quad
\tau_{1}=\int_{x_{2}}^{x_{3}}\sqrt{2E(V(x)-\varepsilon^{\prime})}dx\,.
\]
Location of the turning points $x_{1}$, $x_{2}$, $x_{3}$  is
presented in \fref{Channeling Figure 2}, the energy is counted off
from the minimum of the potential.

%%%%%%%%%%%%%%%%%%%%%%%%555

%%%%%%%%%%%%%%%%%%%%%%%%%%%5
In contrast to the classical case, the reflection coefficient is
nonzero in the range $\varepsilon^{\prime}>V_{\mathrm{max}}$ too.
The method for calculating $D_{1}$ in the quasiclassical
approximation at $\varepsilon^{\prime}>V_{\mathrm{max}}$ developed
in \cite{27,28} is based on  application  of the path of
integration, which passes through the complex turning points
defined by the following equalities
\begin{equation}
\label{3.8} V(z_{0}^{l})=\varepsilon^{\prime}\,,\quad
z_{0}^{l}=ld+iz_{0}\,, \quad l=0, 1, 2, \ldots\,.
\end{equation}
%%%%%%%%%%%555
Without reproducing the calculations carried out in \cite{28} (see
also \cite{27}), we only present the expression for the reflection
coefficient
\begin{equation}
\label{3.9}
D_{1}=\exp\left\{i\int_{z_{0}^{l}}^{z_{0}^{l+1}}\sqrt{2E(\varepsilon^{\prime}-V(x))}dz\right\}\,.
\end{equation}
%%%%%%%%%%%%%%%%%%%%%%%%%%%%55

%%%%%%%%%%%%%%%%%%
If the function $V(x)$ is symmetrical with respect to  the  line
$x=d/2$ as it usually occurs for real potentials, then the path of
integration can be chosen so that the quantities $|D_{1}|$ and
$\varphi_{1}(\varepsilon^{\prime})$, which we are concerned with,
would be represented as real integrals:
\begin{equation}
\label{3.10}
\varphi_{1}(\varepsilon^{\prime})=\sigma_{0}(\varepsilon^{\prime})=
\int_{-d/2}^{d/2}\sqrt{2E(\varepsilon^{\prime}-V(x))}dx\,, \quad
|D_{1}|=e^{-\tau_{2}}\,,
\end{equation}
where
\[
\tau_{2}=2\int^{y_{0}}_{0}\sqrt{2E\left(\varepsilon^{\prime}-V\left(\frac{d}{2}+iy\right)\right)}dy\,,
\quad z_{0}=\frac{d}{2}+iy_{0}\,.
\]

%%%%%%%%%%%%%%%%%%
When the energy of  transverse motion is close to the top of the
potential barrier,  (\ref{3.9}) and (\ref{3.10}) for $D_{1}$ are
not applicable. We intend to obtain the formulas, which are valid
at $\varepsilon^{\prime}\approx V_{\mathrm{max}}$ and go over to
(\ref{3.9}) or (\ref{3.10}) in the corresponding limiting
cases.\footnote{A similar investigation was performed in \cite{29}
for a different problem and only for the case when
$\varepsilon^{\prime}>V_{\mathrm{max}}$.}

At the top of the barrier,
\[
V(x)\approx V_{\mathrm{max}}-kx_{1}^{2}\,,\quad
k=2V^{\prime\prime}(\frac{d}{2})\,,\quad x_{1}\equiv
x-\frac{d}{2}\,,
 \]
and the Schr\"{o}dinger equation in this case has the analytical
solution
\begin{equation}
\label{3.11}
\varphi(x)=A_{1}{\cal{D}}_{m}(x_{1}e^{-i\frac{\pi}{4}}\sqrt[4]{4\lambda})+
B_{1}{\cal{D}}_{-m-1}(x_{1}e^{i\frac{\pi}{4}}\sqrt[4]{4\lambda})\,,
\end{equation}
where ${\cal{D}}_{m}(z)$ is the parabolic cylinder function;
\[
\lambda=\sqrt{2kE};\quad
m=-\frac{1}{2}-\frac{i}{2}\sqrt{\lambda}x_{0}^{2}\,,\quad
x_{0}^{2}= -\frac{\varepsilon^{\prime}-V_{\mathrm{max}}}{k}\,.
\]
%%%%%%%%%%%%%%%%%%%%
 On the right of the
potential barrier, i.e., at $x_{1}>0$, the functions
${\cal{D}}_{m}$ and ${\cal{D}}_{-m-1}$  asymptotically go into
 functions $f$ and $f^{*}$, correspondingly, which are  determined by
(\ref{3.7}). As we are concerned with the coefficients of
reflection and transmission at the singular barrier alone, let us
assume that $B_{1}=0$. Then at $x_{1}\sqrt[4]{4\lambda}\gg 1$,
\begin{eqnarray}
\label{3.12} & &\varphi(x)\simeq\frac{A_{1}}{\sqrt[4]{4\lambda
x_{1}^{2}}}\exp i\left(\frac{\sqrt{\lambda}}{2}x_{1}^{2}
-\frac{\sqrt{\lambda}x_{0}^{2}}{2}\ln\sqrt[4]{4\lambda}x_{1}\right.\nonumber\\
&
&\left.+i\frac{\pi}{8}-\frac{\pi}{8}\sqrt{\lambda}x_{0}^{2}\right)\simeq\frac{A^{\prime}}{\sqrt{p}}
\exp\left(i\int_{0}^{x_{1}}pdx\right)\,.
\end{eqnarray}
Upon translation to the region $x_{1}<0$, the function
$\varphi(x)$ is transformed as follows \cite{30}:
%%%%%%%%%%%%%%
\begin{eqnarray}
\label{3.13} \varphi(x)&=&A_{1}{\cal{D}}_{m}(\sqrt[4]{4\lambda}
x_{1}e^{i3\pi/4})
=A_{1}\left[{\cal{D}}_{m}(\sqrt[4]{4\lambda}x_{1}e^{-i\pi/4})\right.\nonumber\\
& &\left.-\frac{\sqrt{2\pi}}{\Gamma(-m)}e^{i\pi
m}{\cal{D}}_{-m-1}(\sqrt[4]{4\lambda} x_{1}e^{i\pi/4})\right]
\end{eqnarray}
and (at $|x_{1}\sqrt[4]{4\lambda}|\gg 1$) it  should
asymptotically go into a linear combination
\[
\varphi(x)\approx
A\frac{\exp(i\sigma_{0}-i\int^{x_{1}}_{0}pdx)}{\sqrt{p}}+B\frac{\exp(-i\sigma_{0}+i\int^{x_{1}}_{0}pdx)}{\sqrt{p}}\,,
\]
and $A_{1}A^{-1}=D_{1}$,  $BA^{-1}=R_{1}$.

Using the known asymptotics of the functions ${\cal{D}}_{m}$ and
${\cal{D}}_{-m-1}$ \cite{30}, we find
\[
{{D}}_{1}=(\sqrt{2\pi})^{-1}\Gamma\left(\frac{1}{2}+\frac{i}{2}\sqrt{\lambda}x_{0}^{2}\right)
\exp\left(i\sigma_{0}-\frac{\pi}{4}\sqrt{\lambda}x_{0}^{2}\right)\,,
\]
\[
R_{1}=(\sqrt{2\pi})^{-1}\Gamma\left(\frac{1}{2}+\frac{i}{2}\sqrt{\lambda}x_{0}^{2}\right)\times
\exp\left(2i\sigma_{0}-i\frac{\pi}{2}+\frac{\pi}{4}\sqrt{\lambda}x_{0}^{2}\right)\,.
\]
where $\Gamma(z)$ is the gamma function.

%%%%%%%%%%%%%%
Thus, near the top of the barrier
\begin{eqnarray}
\label{3.14} |{{D}}_{1}|=&
&(\sqrt{2\pi})^{-1}\left|\Gamma\left(\frac{1}{2}+\frac{i}{2}\sqrt{\lambda}x_{0}^{2}\right)\right|
\exp\left(-\frac{\pi}{4}\sqrt{\lambda}x_{0}^{2}\right)\nonumber\\
& &=\frac{\exp\left(-\frac{\pi}{2}\sqrt{\lambda}x_{0}^{2}\right)}{2\cosh\frac{\pi}{2}\sqrt{\lambda}x_{0}^{2}}\,,\nonumber\\
\varphi(\varepsilon^{\prime})& &=\arg D_{1}=\sigma_{0}+\delta;
\delta=\arg\Gamma\left(\frac{1}{2}+\frac{i}{2}\sqrt{\lambda}x_{0}^{2}\right)\,.
\end{eqnarray}
Now take into account  that in the domain  of applicability of the
solution of (\ref{3.11}), the following equalities hold
\begin{eqnarray}
 \label{3.15}
&
&\pi\sqrt{\lambda}x_{0}^{2}=\int_{x_{2}}^{x_{3}}\sqrt{2E[V(x)-\varepsilon^{\prime}]}dx=\tau_{1}\,,\quad
\mbox{at}
\quad\varepsilon^{\prime}<V_{\mathrm{max}}\,,\\
&
&\pi\sqrt{\lambda}x_{0}^{2}=-2\int_{0}^{y_{0}}\sqrt{2E\left[\varepsilon^{\prime}
-V\left(\frac{d}{2}+iy\right)\right]}dy=-\tau_{2}\,,\quad\mbox{at}\quad
\varepsilon^{\prime}>V_{\mathrm{max}}\,.\nonumber
\end{eqnarray}
As a result, with the considered accuracy, the dispersion equation
takes the form
\begin{eqnarray}
\label{3.16}
\cos\kappa d=2\cosh\frac{\tau_{1}}{2}e^{\tau_{1}/2}\cos\sigma\,, \quad\mbox{at}\quad \varepsilon^{\prime}<V_{\mathrm{max}}\,,\nonumber\\
\cos\kappa d=2\cosh\left(\frac{1}{2}\tau_{2}\right)e^{-\tau_{2}/2}
\cos\sigma_{0}\quad\mbox{at}\quad
\varepsilon^{\prime}>V_{\mathrm{max}}
\end{eqnarray}
and enables plotting a band spectrum within the whole energy range
of a particle.

%%%%%%%%%%%%%%

Consider the limiting cases for which the analytical solution of
the
equation can be constructed:\\
 (1)
$\varepsilon^{\prime}<V_{\mathrm{max}}, \tau_{1}\gg 1$. In this
case \cite{31}
\begin{eqnarray}
\label{3.17}
\sigma=& & pi\left(n+\frac{1}{2}\right)+e^{-\tau_{1}}\cos\kappa d, n=0, 1, 2, \ldots\, ,\nonumber\\
\varepsilon^{\prime}_{n\kappa}&
&=\varepsilon_{n}^{(0)}+\Delta\varepsilon_{n\kappa}\,,
\end{eqnarray}
where $\varepsilon_{n}^{(0)}$ coincide with the energy levels of
the discrete spectrum in the isolated potential well;
\[
\int^{x_{2}}_{x_{1}}\sqrt{2E[\varepsilon_{n}^{(0)}-V(x)]}dx=\pi\left(n+\frac{1}{2}\right)\,,
\]
and the quantities
\[
\Delta\varepsilon_{n\kappa}\approx
e^{-\tau_{1}(\varepsilon_{n}^{(0)})}\cos\kappa
d\ll\varepsilon_{n+1}^{(0)}-\varepsilon_{n}^{(0)}
\]
determine the energy levels of the allowed band with the width
considerably smaller than the distance between the bands;\\
(2)  $\varepsilon^{\prime}>V_{\mathrm{max}}, \tau_{2}\gg 1$:
\begin{eqnarray}
\label{3.18}
& &\cos\kappa d=(1+e^{-\tau_{2}})\cos\sigma,\nonumber\\
& &\int_{-d/2}^{d/2}\sqrt{2E(\varepsilon^{\prime}_{n\kappa}-V(x))}dx=\pi n+\kappa d+\delta_{n}(\kappa)\,,\nonumber\\
& &\delta_{n}\ll 1, 0\leq\kappa d\leq \pi \,,\\
&
&\delta_{n}(\kappa)=(-1)^{n}c\pm\sqrt{\kappa^{2}+2\exp[-\tau_{2}(\varepsilon^{\prime}_{n\kappa})]}\,,
\quad c=\pi-\kappa d\,,\nonumber
\end{eqnarray}
in this case the energy spectrum consists of wide allowed bands
and narrow forbidden bands with the width
\begin{equation}
\label{3.19}
\Delta\varepsilon_{\mathrm{forb}}=2\sqrt{2\exp[-\tau_{2}(\varepsilon^{\prime}_{n\kappa})]}\,,
\end{equation}
and for $\varepsilon^{\prime}\gg V_{\mathrm{max}}$,
$\varepsilon^{\prime}_{n\kappa}=(\kappa d+\pi n)^{2}/2E$.

%%%%
The exact solutions of equation (\ref{3.16}) were numerically
found for a silicon crystal. The potential obtained in the Moliere
approximation and averaged over temperature oscillations \cite{3}
was used as an interplanar potential.The potential obtained in the
Moliere approximation and averaged over temperature oscillations
\cite{3} was used for computation. This potential has the form:

for positrons
\begin{eqnarray}
V_{e^{+}}=&
&\psi_{a}^{2}\sum_{i=1}^{3}\frac{1}{2}\gamma_{i}e^{\tau_{i}}
\left\{e^{-\beta_{i}x/a}erfc\left[\frac{1}{\sqrt{2}}\left(\frac{\beta_{i}u_{1}}{a}-\frac{x}{u_{1}}\right)\right]\right.\nonumber\\
&
&\left.+e^{\beta_{i}x/a}erfc\left[\frac{1}{\sqrt{2}}\left(\frac{\beta_{i}u_{1}}{a}+\frac{x}{u_{1}}\right)\right]\right\}\,,\nonumber
\end{eqnarray}

for electrons
\[
V_{e^{-}}(x)=-V_{e^{+}}\left(x-\frac{d}{2}\right)+V_{\mathrm{max}}\,.
\]
Here $\psi_{a}$, $\gamma_{i}$, $\beta_{i}$, $\tau_{i}$, $u_{1}$
are the potential parameters determined in \cite{3}.

%%%%%%%%%%%%

It should be noted that when using the programme of numerical
solution of equation (\ref{3.16}) arranged in the optimum way,
plotting of the whole energy spectrum for electrons and
antielectrons (Fig. 3) with 1\% accuracy  takes a few minutes on
the EC 1030 computers.

\begin{figure}[htbp]
\centering
\epsfxsize = 8 cm \centerline{\epsfbox{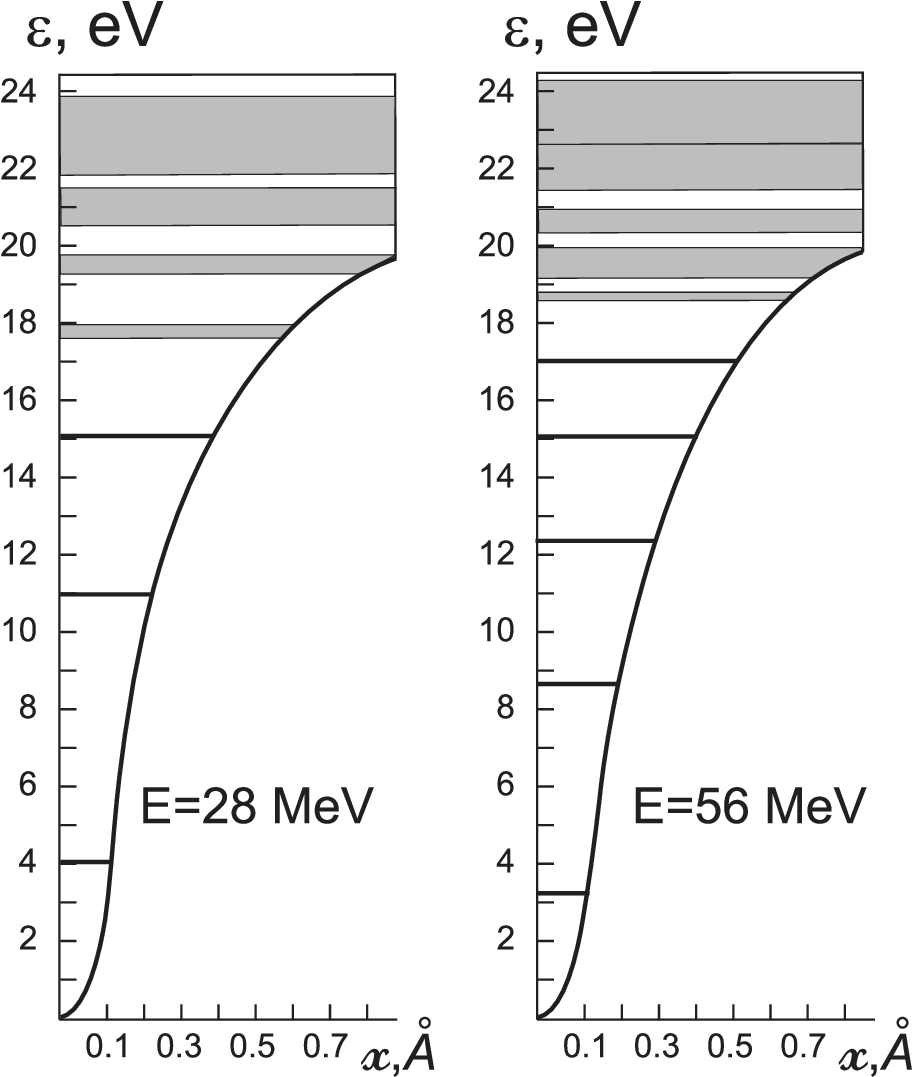}}
\caption{Energy bands calculated for electrons with the energies
$E=28$ and 56\,MeV. Channeling along the plane $(110)$ in a
silicon crystal. Solid horizontal lines - the positions of the
energy levels in the Moliere potential} \label{Channeling Figure
3}
\end{figure}

\begin{figure}[htbp]
\centering
\epsfxsize = 6 cm \centerline{\epsfbox{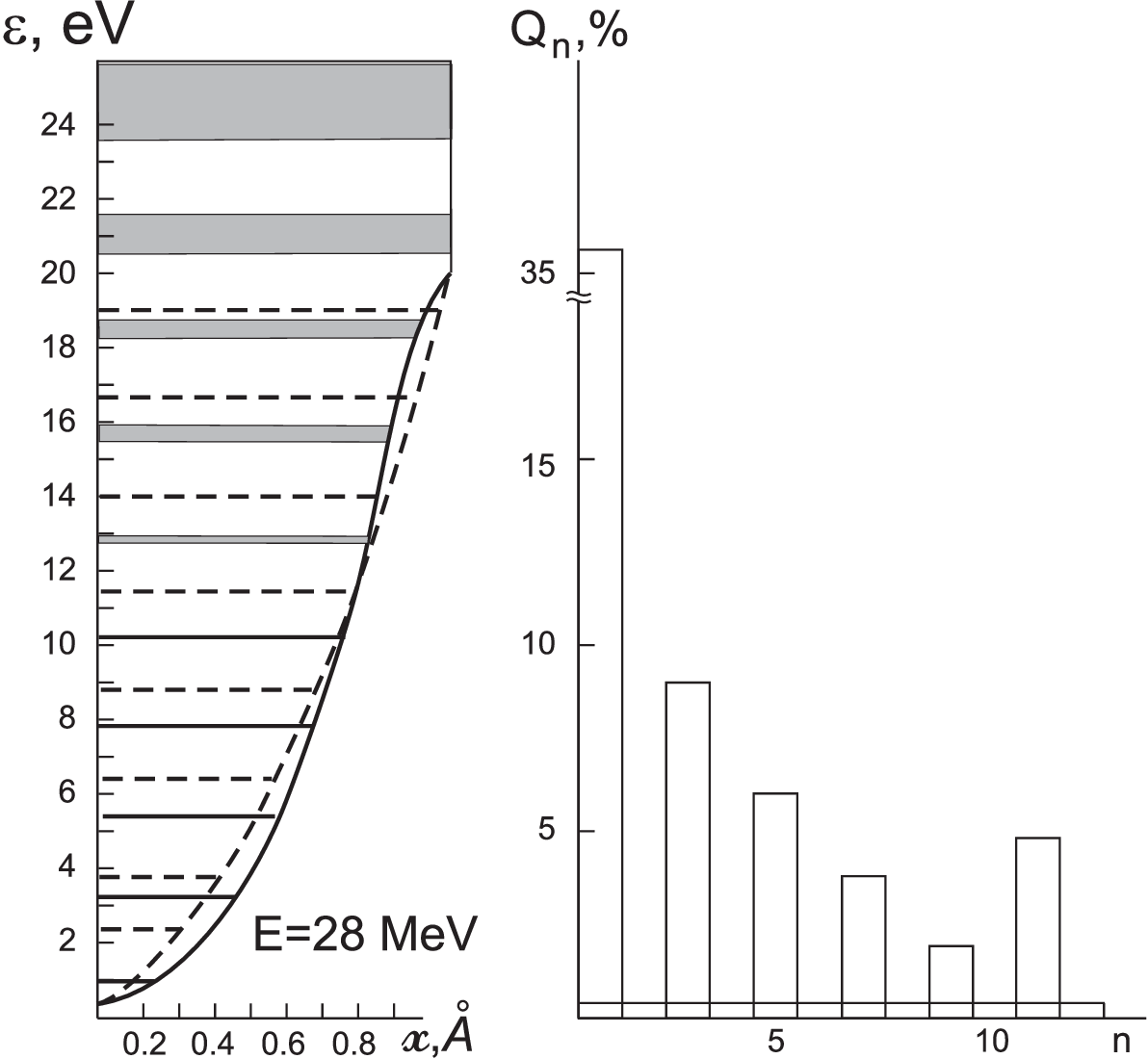}}
\caption{Energy bands  (a) and occupancy coefficients of the
energy levels (b) for positrons at  zero entrance angle. Dashed
curve - the parabolic potential, dashed horizontal line - the
position of levels in it} \label{Channeling Figure 4}
\end{figure}

%%%%%%%%%%%%%%%%

Now proceed to  considering  normalization of the wave functions
and the occupancy coefficients for the energy levels.

According to (\ref{3.1}) and (\ref{3.4}), the stationary particle
wave function in a channel $\varphi(x)=c_{1}f(x)+c_{2}f^{*}(x)$,
and $c_{2}=q(\varepsilon)c_{1}$,
$q(\varepsilon)=R_{1}^{-1}[D_{1}(\varepsilon)e^{i\kappa d}-1]$,
the coefficient $c_{1}$ is  determined from the normalizing
condition
\[
|c_{1}|^{2}(1+|q|^{2})J=1,\, J=\int_{-d/2}^{d/2}|f|^{2}dx\,,
\]
where the integrals of rapidly oscillating function $f^{2}$ and
$f^{*2}$ are dropped.

For the energy levels lying far from the top of the potential
barrier:
\begin{eqnarray}
\label{3.20}
J=\frac{1}{E}T(\varepsilon^{\prime})=\left\{\begin{array}{c}
                                              \int_{x_{1}}^{x_{2}}\frac{dx}{p(x)} \qquad \mbox{at}\qquad \varepsilon^{\prime}<V_{\mathrm{max}}\\
                                              \int_{-d/2}^{d/2}\frac{dx}{p(x)} \qquad\mbox{at}\qquad \varepsilon^{\prime}>V_{\mathrm{max}}
                                            \end{array}\right.\,,
\end{eqnarray}
where $T(\varepsilon^{\prime})$ is the classical flight time of
particles between the planes. From this  the normalizing constant
is
\begin{equation}
\label{3.21} c_{1}=\frac{1}{\sqrt{1+|q|^{2}}}\sqrt{\frac{E}{T}}\,.
\end{equation}

%%%%%%%%%%%%%%%%

If the quantum effects of tunneling and over-barrier reflection
are neglected, then $|q|^{2}$ takes on only two values:
$|q|^{2}=0$ at $\varepsilon^{\prime}>V_{\mathrm{max}}$ and
$|q|^{2}=1$ at $\varepsilon^{\prime}<V_{\mathrm{max}}$. The
normalizing constant abruptly changes by a factor of $\sqrt{2}$
when $\varepsilon^{\prime}$ goes from the subbarrier to the
over-barrier range, whereas in reality the quantity $|q|^{2}$
smoothly changes from 0 to 1.

This fact also manifests itself within the classical approach to
the problem: in this case when a particle passes through a
barrier, the cycle of particle motion  changes abruptly. The false
opposition of particles executing infinite motion and channeled
particles appears.

Formula (\ref{3.21}) becomes unsuitable in the vicinity of the top
of the barrier when a classical cycle $T$ becomes infinite.
However, using the analytical solution of (\ref{3.11}) enables one
to regularize the expressions for the normalization integral
(\ref{3.21}). Indeed, let us introduce a certain passing point
with the coordinate $a$ satisfying the conditions compatible in
the case of  quasi-classical motion:
\[
\frac{d}{2}-a\ll \frac{d}{2}\,,
\]
 but
 \[
 \sqrt[4]{4\lambda}\left(\frac{d}{2}-a\right)\gg 1\,.
 \]
  Then
we obtain for a potential  symmetrical with respect to the
$d/2$-axis
%%%%%%%%%%5
\begin{eqnarray}
\label{3.22}
J\approx 2\int_{x_{2}-a}^{x_{2}}|f|^{2}dx+2\int^{x_{2}-a}_{0}\frac{dx}{p(x)}\nonumber\\
\simeq
2\left[c(\varepsilon^{\prime})\int^{x_{2}}_{0}|{\cal{D}}_{m}
(x\sqrt[4]{4\lambda}e^{-i\frac{\pi}{4}})|^{2}dx\right.\nonumber\\
+\left.\int^{x_{2}}_{0}\left(\frac{1}{p(x)}-\frac{1}{p_{0}(x)}\right)dx\right]\,,
\end{eqnarray}
where
\[
p_{0}(x)=\sqrt{2E(\varepsilon^{\prime}-V_{\mathrm{max}}+kx^{2})}\,,
\]
\[
c(\varepsilon^{\prime})=\left\{\begin{array}{cc}
                               e^{-\frac{1}{4}\tau_{1}}\,, \quad\mbox{at}\quad \varepsilon^{\prime}<V_{\mathrm{max}} \\
                               e^{-\frac{1}{4}\tau_{2}}\,, \quad\mbox{at}\quad \varepsilon^{\prime}>V_{\mathrm{max}}\,,
                             \end{array}\right.
\]
and the asymptotics of (\ref{3.12}) is used for the function
${\cal{D}}_{m}(z).$

Transform the formula for $|q|^{2}$ allowing for the dispersion
equation:
\begin{eqnarray}
\label{3.23} |q|^{2}=\left\{\begin{array}{c}
          \displaystyle\frac{A^{2}+1-2\cos^{2}\sigma-2\sin\sigma\sqrt{A^{2}-\cos^{2}\sigma}}{1+A^{2}}\,,
           \quad \varepsilon^{\prime}<V_{\mathrm{max}}  \\
          \\
          \displaystyle\frac{B^{2}+1-2\cos^{2}\sigma_{0}-2\sin\sigma_{0}\sqrt{B^{2}-\cos^{2}\sigma_{0}}}{1+B^{2}}\,,
           \quad \varepsilon^{\prime}>V_{\mathrm{max}}\,,
        \end{array}\right.
\end{eqnarray}
and
\[
A=\left(2\cosh\frac{\tau_{1}}{2}e^{\frac{\tau_{1}}{2}}\right)^{-1}\,,\quad
B=\left(2\cosh\frac{\tau_{2}}{2}e^{-\frac{\tau_{2}}{2}}\right)^{-1}\,.
\]

%%%%%%%%%%5
Formulas (\ref{3.22}) and (\ref{3.23}) enable one to calculate the
normalization constant $c_{1}$ at all possible values of
$\varepsilon^{\prime}$. It should be pointed out that for the
integral of $|{\cal{D}}_{m}|^{2}$ in (\ref{3.22}), it is possible
to obtain the analytical expression in terms of the $G$-function
of Meyer \cite{30}. But it is more reasonable to find this
integral numerically, using the integral representation of
$\cal{D}$-functions.

%%%%%%%%%%%%%%%5

Now go over  to calculating the occupancy coefficients
$Q_{n\kappa}=|c_{n\kappa}|^{2}$ of the energy levels. Consider
first  those values of $E$, at which the quantum mechanical
effects are insignificant, as we did when calculating the
normalization integral. Then
\begin{eqnarray}
\label{3.24}
c_{n\kappa} &=& c_{1}\int_{-d/2}^{d/2}\exp\left[-i\int_{x_{1}}^{x}p(x^{\prime})dx^{\prime}+ip_{0x}x\right]\frac{dx}{\sqrt{p(x)}}\,,\\
p_{0x} &=& p_{0z}\theta\,,
\end{eqnarray}

and to calculate the integral (\ref{3.24}) in the given
approximation, one can use the saddle--point method. As a result,
at
\[
0<\varepsilon^{\prime}_{n\kappa}-\frac{p^{2}_{0x}}{2E}<V_{\mathrm{max}}\,,
\]
 we
find the expression derived in \cite{32}:
\begin{equation}
\label{3.25}
c_{n\kappa}=\frac{\sqrt{\pi}c_{1}}{\sqrt{E|V^{\prime}(x_{0})|}}\exp\left[ip_{0x}x-i\int_{x_{1}}^{x}p(x^{\prime})dx^{\prime}\right]\,,
\end{equation}
and the saddle point $x_{0}$ is determined by the condition
\begin{equation}
\label{3.26} p_{0x}=p(x_{0})\,, \quad\mbox{i.e.}\,,\quad
V(x_{0})=\varepsilon^{\prime}_{n\kappa}-\frac{p^{2}_{0x}}{2E}\,.
\end{equation}
%%%%%%%%%5

In the case when
\[
\varepsilon^{\prime}_{n\kappa}-\frac{p^{2}_{0x}}{2E}>V_{\mathrm{max}}
\qquad\mbox{or}\qquad
\varepsilon^{\prime}_{n\kappa}-\frac{p^{2}_{0x}}{2E}<0\,,
\]
(\ref{3.26}) does not have real roots (recall that for electrons,
as well as for positrons, the energy is counted off from the
potential minimum), but the solution $z_{0}=x_{0}+iy_{0}$ always
exists in a complex plane. Appearance of the imaginary part of the
saddle point coordinate means, in fact, the exponential
attenuation of the occupancy coefficients in these energy bands.
Analytical continuation  of quasiclassical wave functions makes it
possible to obtain the following expression for the occupancy
coefficients:
\begin{eqnarray}
\label{3.27} Q_{n\kappa}=\left\{
\begin{array}{c}
   \displaystyle\frac{\pi|c_{1}|^{2}}{E|V^{\prime}(x_{0})|}\,,\quad \mbox{at}\quad 0< \varepsilon^{\prime}_{n\kappa}-
   \frac{p^{2}_{0x}}{2E}<V_{\mathrm{max}}\,,\\
  \\
    \displaystyle\frac{\pi|c_{1}|^{2}}{E|V^{\prime}(z_{0})|}e^{-2\int_{0}^{y_{0}}\sqrt{2E[\varepsilon^{\prime}_{n\kappa}
  -V(\frac{d}{2}+i y)]}dy+2p_{0x}y_{0}}\,,\\
  \\
   \mbox{at}\quad\varepsilon^{\prime}_{n\kappa}-\displaystyle\frac{p^{2}_{0x}}{2E}>V_{\mathrm{max}}\,,\\
  \\
   \displaystyle\frac{\pi|c_{1}|^{2}}{E|V^{\prime}(z_{0})|}e^{2\int_{0}^{y_{0}}\sqrt{2E[\varepsilon^{\prime}_{n\kappa}-
   V(0+i y)]}dy-2p_{0x}y_{0}}\,,\\
   \\
  \mbox{at}\quad\varepsilon^{\prime}_{n\kappa}-\displaystyle\frac{p^{2}_{0x}}{2E}<0\,,
\end{array}\right.
\end{eqnarray}
and at any $p_{0x}$, the  complex turning point $z_{0}$ is defined
by the equality
\[
\varepsilon^{\prime}_{n\kappa}-\frac{p^{2}_{0x}}{2E}=V(z_{0})\,.
\]

%%%%%%%%%%5
Expression (\ref{3.27}) becomes inapplicable at
$|V^{\prime}(z_{0}|\approx 0$, i.e., when the saddle point is
located either near the top of the barrier or near the bottom of
the potential well. In the vicinity of these points, the real
potential is approximated by a parabola, and the values of
$Q_{n\kappa}$ in this energy band can be calculated, using
analytical solutions of the Schr\"{o}dinger equation:
\begin{eqnarray}
\label{3.28}
\varphi_{1}&=&A_{1}[{\cal{D}}_{m_{1}}(x\sqrt[4]{4\lambda_{1}})+{\cal{D}}_{-m_{1}-1}(ix\sqrt[4]{4\lambda_{1}})]\,,\nonumber\\
\varepsilon^{\prime}_{n\kappa}&-&\frac{p^{2}_{0x}}{2E}\simeq 0\,,\nonumber\\
\varphi_{2}&=&A_{2}{\cal{D}}_{m_{2}}(x_{1}e^{-i\frac{\pi}{4}}\sqrt[4]{4\lambda_{2}})\,,
\quad \varepsilon^{\prime}_{n\kappa}-\frac{p^{2}_{0x}}{2E}\simeq
V_{\mathrm{max}}\,,
\end{eqnarray}
where
\[
m_{1}=-\frac{1}{2}+i\frac{\sqrt{\lambda_{1}}}{2}\frac{(\varepsilon^{\prime}-V_{max})}{k_{1}}\,,
\]
\[
m_{2}=-\frac{1}{2}+\frac{\sqrt{\lambda_{2}}}{2}\frac{\varepsilon^{\prime}}{k_{2}}\,,\quad
\lambda_{1,2}=\sqrt{2Ek_{1,2}}\,,
\]
\[
k_{1}=2V^{\prime\prime}(0)\,,\quad
k_{2}=2V^{\prime\prime}\left(\frac{d}{2}\right )\,,\quad
x_{1}=x-\frac{d}{2}\,.
\]
The coefficients in linear combinations of the parabolic cylinder
functions in (\ref{3.28}) are chosen on condition that at
$|m_{1,2}|\gg 1$, the functions $\varphi_{1,2}$ should go over to
a quasi-classical solution
\[
\frac{1}{\sqrt{p(x)}}\exp[\pm i\int p(x)dx]\,.
\]
Equation (\ref{3.27}) enables one to calculate the occupancy
coefficients of the energy levels within the whole energy range
(\fref{Channeling Figure 4}).

%%%%%%%%%%%%%%%%%%%%%%%%%%%%%%%%%%%  Cannelling Chapter 2 %%%%%%%%%%%%%%%%%%%%%%

\chapter[A Channeled
Fast Particle as a 2D (1D) Relativistic Atom ]{A Channeled Fast
Particle as a Two-Dimensional (One-Dimensional) Relativistic Atom}
\label{ch:2}

%%%%%%%%%%%%%%%%%%%%%%%%%%%%%%%%%%%  Section 4 %%%%%%%%%%%%%%%%%%%%%%

\section[Spontaneous Photon Radiation in Radiation Transitions Between the Bands of Transverse Energy of Channeled Particles]{Spontaneous Photon Radiation in Radiation Transitions Between the Bands of Transverse Energy of Channeled Particles}
\label{sec:2.4}

Transverse motion of a channeled particle is characterized by a
distinct band energy spectrum (see Fig. 3). The bands deep in the
wells are very narrow. In this case it is possible to speak of
discrete levels in a well. Kalashnikov, Koptelov and Ryazanov in
\cite{4,33} put forward the  idea  that the emission of X-ray and
$\gamma$-radiation may occur through radiative capture of the
electron entering a crystal at the levels of transverse motion  in
a well formed by the axis (plane). According to \cite{6} at the
transition of a channeled electron with the energy of the order of
a few mega  electron-volts between the levels in a well, one
should expect the emission of optical radiation. A detailed
treatment carried out by the author  together with Dubovskaya in
\cite{5,17,34} demonstrated that the stated above effects of
photon formation are a particular case of the general mechanism of
$\gamma$-quantum emission at radiative transitions between the
energy bands of the transverse motion of particles passing through
a crystal, which occurs for both electrons and positrons.

Within the framework of the quantum mechanical correspondence
principle every radiative transition may be described as the
radiation of a certain classical oscillator. Since a particle has
a transversal momentum, we shall deal with a moving one- or
two-dimensional "atom" whose radiation spectrum is considerably
influenced by the Doppler effect \cite{34}. From the viewpoint of
the classical theory the possibility of the  induction of
$\gamma$-radiation by channeled electrons and positrons and the
importance of the Doppler effect in this process were pointed out
by Kumakhov \cite{7}. Note, however, that the idea of the
induction of X-ray and $\gamma$-radiation at radiative transitions
between the energy bands of transverse motion of relativistic
particles in crystals still was not articulated in this work.

Interestingly enough, that  the concept of the possibility of
optical and soft X-ray radiation of diffracted particles in
crystals at  interband transition was expressed in \cite{13}. But
only in \cite{5,7,34}the authors came to clear awareness of the
crucial role of the Doppler effect causing the transformation of
relatively low-frequency particle oscillations in a crystal
(characteristic frequencies are in the optical and soft X-ray
spectral regions) into hard X-ray and $\gamma$-radiation, whose
frequency increases with the growth of particle energy.

%%%%%%%%%%%%%%%%%%%%%%%%%%%%%5

The major characteristics of the radiation produced by channeled
particles may be deduced by the simple reasoning given below
\cite{17,34}.

Let a particle with the momentum $\vec{p}$ and the energy $E$ fall
upon a plane-parallel crystal plate. Its collision with the
crystal results in the emission of a  photon with the energy
$\omega$ and momentum $k$.  In the final state the particle energy
and momentum take on the values $E_{1}$ and $\vec{p}_{1}$. It is
important to remember that  if the reaction proceeds  in an
arbitrary constant field, the energy (not the momentum) of the
system is conserved. Thus, for  particle energies we have the
equality
\begin{equation}
\label{4.1} E=E_{1}+\omega.
\end{equation}
Due to the periodicity in a transverse plane of the crystal
potential responsible for channeling,  the transversal component
of the momentum retains accurate to the reciprocal lattice vector
of the crystal (see Chapter I),
\begin{equation}
\label{4.2}
\vec{p}_{\perp}=\vec{p}_{1\perp}+\vec{k}_{\perp}+\vec{\tau}_{\perp}\,.
\end{equation}
In the longitudinal  direction, the potential responsible for
channeling  is constant, and the particle has a certain
longitudinal momentum $p_{zn}$ (see Chapter I), so
\begin{equation}
\label{4.3} p_{zn}=p_{1zn}+k_{z}n(k_{z})\,,
\end{equation}
where $n(k_{z})$ is the  refractive index  of the crystal, still
considered to be real.

%%%%%%%%%%%%%%%%%555
According to the analysis made in \cite{35} the photon momentum in
a medium is $kn$. In the representation of (\ref{4.2}),
(\ref{4.3}) we have taken into account the fact that at  radiation
in a finite plate the transversal component of the momentum  does
not change through refraction at the boundary, but the
longitudinal component of the photon momentum undergoes an abrupt
change. Equalities (\ref{4.2}), (\ref{4.3}) follow from the
rigorous theory of radiation in a plate of thickness $L$ (see
Chapter I).

%%%%%%%%%%%%%%%%%%%%%%
Consider thoroughly equality (\ref{4.3}) determining the change in
the particle longitudinal  momentum through photon emission. Write
the explicit form of (\ref{4.3}) in terms of the particle energy.
According to Chapter I, section 2
\[
p_{zn}=\sqrt{p^{2}-2m\varepsilon_{n\kappa}(E)}; \qquad
p_{1zf}=\sqrt{p^{2}_{1}-2m\varepsilon_{f\kappa_{1}}(E_{1})}\,,
\]
$\kappa$ is the reduced quasi-momentum corresponding to the
transversal momentum of the particle in the initial state
$\vec{p}_{\perp}$; $\kappa_{1}$ is the quasi-momentum of the
particle in the final state, which is obtained from (\ref{4.2}) by
reduction of $p_{1\perp}$ to the first Brillouin zone.
%%%%%%%%%%%%%%%%%%%%%%%%%%%55
Using the equalities for  $p_{zn}$ and $p_{1zn}$, equation
(\ref{4.3}) can be written in the form
\begin{equation}
\label{4.4}
\sqrt{E^{2}-m^{2}-2m\varepsilon_{n\kappa}(E)}=\sqrt{E^{2}_{1}-m^{2}-2m\varepsilon_{f\kappa_{1}}(E_{1})}+k_{z}n(k_{z}).
\end{equation}
As the total particle energy is much greater than the energy
associated with the transversal motion of a particle in a crystal,
it is possible to expand the square roots in equality (\ref{4.4}).

%%%%%%%%%%%%%%%%%%%%%5
In the most interesting case in consideration of radiation under
channeling of particles with the energy less than a few
gigaelectronvolts $\omega\ll E,E_{1}$. As a result (\ref{4.4}) can
be recast as
\begin{equation}
\label{4.5} \omega[1-\beta
n(\omega)\cos\vartheta]-\frac{m}{E}(\varepsilon_{n\kappa}-\varepsilon_{f\kappa_{1}})=0.
\end{equation}
In writing (\ref{4.5}), it is assumed that $\cos\vartheta$ in the
expression for $n(k_{z})=n(\omega\cos\vartheta)\simeq n(\omega)$
is equal to unity due to the fact that for relativistic particles
the effective angle of photon radiation is
\[
\vartheta\sim\frac{m}{E}=\frac{1}{\gamma}\ll 1\,, \quad
\beta=v_{z}\,,
\]
 From
(\ref{4.5}) follows
\begin{equation}
\label{4.6}
\omega=\frac{(\varepsilon_{n\kappa}-\varepsilon_{f\kappa_{1}})\gamma^{-1}}{1-\beta
n(\omega)\cos\vartheta}\,.
\end{equation}

%%%%%%%%%%%%%%%%%%%%5
To clarify the meaning of equality (\ref{4.6}), let us compare it
with the expression determining the frequency of photons emitted
by an oscillator moving in a medium:
\begin{equation}
\label{4.7} \omega=\frac{\Omega}{1-\beta n(\omega)\cos\vartheta},
\end{equation}
where $\Omega$ is the oscillator frequency in the laboratory frame
of reference;
$\Omega=\Omega_{0}\sqrt{1-\beta^{2}}=\Omega_{0}\gamma^{-1}$;
$\Omega_{0}$ is the oscillator frequency in its rest frame.
Comparing (\ref{4.6}) and (\ref{4.7}), one can notice that a
particle under channeling conditions can be considered as a moving
in a medium oscillator having the following frequency in its rest
frame (i.e., the frame with a zero longitudinal particle velocity)
\begin{equation}
\label{4.8}
\Omega_{0nf}=\varepsilon_{n\kappa}-\varepsilon_{f\kappa_{1}}\,.
\end{equation}
Thus, the frequency $\Omega_{0nf}$ is determined by the difference
of energies between  the discrete zones (levels) of  particle
transverse motion \cite{34}. In the laboratory frame the frequency
of such an oscillator is
\begin{equation}
\label{4.9}
\Omega_{nf}=(\varepsilon_{n\kappa}-\varepsilon_{f\kappa_{1}})\gamma^{-1}=
\varepsilon^{\prime}_{n\kappa}-\varepsilon^{\prime}_{f\kappa_{1}}\,.
\end{equation}
%%%%%%%%%%%%%%%%%%%%%%55

It should be pointed out that unlike a conventional oscillator,
the frequency of the oscillator correlated with a channeled
particle in the rest frame depends on the particle energy owing to
the fact that the value of the potential $u_{c}(\vec{\rho})$,
produced by crystal axes (planes) depends on the particle energy
$u_{c}(\vec{\rho})=\gamma u(\vec{\rho})$ ($u_{c}(\vec{\rho})$ is
the potential of axes (planes) in the laboratory frame). In this
regard it is interesting that equation (\ref{2.1}) can be treated
as the equation describing the spectrum of a particle transverse
motion in the coordinate system where its longitudinal momentum is
equal to zero.

%%%%%%%%%%%%%%%%%%%%%%5555
To be more specific, suppose that a particle undergoes transitions
between the zones of transverse motion located inside the well
(see\textbf{ Figure }{Channeling Figure 3}). In this case the
energy zones may be treated as discrete levels. Their dependence
on the particle energy can be found explicitly for the simplest
cases. Let, for example, a potential well be rectangular. Then
$\varepsilon_{n}=\pi^{2}n^{2}/2md^{2}$ ($n=1, 2, 3, ...:$ $d$ is
the well width). At the transition between the levels with
specified values of ($n, f$), the frequency $\Omega_{nf}\sim
\gamma^{-1}$ and the frequency of a forward-emitted photon
(without regard to the refraction effect) is
\begin{equation}
\label{4.10}
\omega=2(\varepsilon_{n}-\varepsilon_{f})\gamma=\frac{\pi^{2}}{m^{2}d^{2}}(n^{2}-f^{2})E\,.
\end{equation}
Thus, the radiation frequency increases linearly with the increase
in the particle energy, and for
$\varepsilon_{n}-\varepsilon_{f}\ll m$ it is always $\omega\ll E$.
%%%%%%%%%%%%%%%%%%%%%%5555

If we consider the transition between the level located at the
well edge ($\varepsilon_{n}\sim\gamma u$, $u$ is the well depth)
and the lower state (for example,
$\varepsilon_{f}=\pi^{2}/2md^{2}$), then $\omega=2u\gamma^{2}$,
and the maximum photon frequency in this case increases
quadratically with increasing energy. At the energies close to
$m^{2}/u$, the frequency $\omega$ is comparable to $E$, and in
(\ref{4.4}) it is important that a significant change in $E_{1}$
should be taken into account. At the transitions between the
levels located at the well edge $n\sim f\sim\sqrt{\gamma}$,
$\varepsilon_{n}-\varepsilon_{f}\sim\sqrt{\gamma}$. As a
consequence, $\omega\sim\gamma^{3/2}$. The frequency of a
forward-emitted photon exhibits the same energy-dependent behavior
pattern when moving in an oscillatory well \cite{36}, as well as
in a quasi-classical approximation for the transition between
neighboring levels \cite{37}.

%%%%%%%%%%%55555

However, the stated energy-dependent behavior of the frequency
$\omega$ holds true only in the absence of refraction and
absorption of photons (the refractive index is $n=1$). If recall
that $n$ is different from unity, equality (\ref{4.6}) in fact
turns into the equation determining the value of the frequency
$\omega$. As a result, it is possible that additional frequencies
determined by the dependence of the refractive index of a medium
on the frequency of a produced photon appear in the radiation
spectrum of a channeled particle, i.e., the complex and anomalous
Doppler effect may arise \cite{5,34}.
In the case under study, due to the above mentioned similarity of
the laws governing the process of photon emission by a channeled
particle and those concerning the process of the photon emission
by a moving atom, the theory of the complex and anomalous Doppler
effects is formed in a perfect analogy with the that given by
Frank in \cite{38,39,40,41} for the case of moving atoms.
According to \cite{38}, the region of the complex photon spectrum
existence is determined by the condition
\[
\frac{v\cos\vartheta}{W(\omega)}\geq 1,
\]
where $W(\omega)=d\omega/dk$ is  the photon group velocity.

%%%%%%%%%%%%%%%%%%%%55
In the X-ray and harder spectral ranges $n-1<10^{-5}$. Hence, $W$
is close to the velocity of light in a vacuum. In order to observe
the manifestation of a few frequencies within the stated spectral
range, the oscillator in a medium  is to be started  up to achieve
very high energies. For instance, if we are concerned about the
emission of photon  with the energy $\omega\geq 1$\,keV, then at
$\vartheta=10^{-3}$ rad the particle velocity should satisfy the
condition $v\geq 1-10^{5}$, which corresponds to the energies
$E\geq 3\cdot 10^{2}\, m$.  Such energies are really difficult to
achieve for atoms and nuclei, but at the same time they are
attainable for a channeled electron (positron). Thus, the study of
radiation of channeled particles enables us to investigate the
complex and anomalous Doppler effects even within the X-ray
spectrum \cite{5,34}.

%%%%%%%%%%%%%%%%%%%%%%%%%%5
Using (\ref{4.6}) and the  following explicit expression for the
refractive index in the X-ray spectrum far from the characteristic
atomic frequencies
\[
n(\omega)=1-\frac{\omega^{2}_{L}}{2\omega^{2}}
\]
($\omega^{2}_{L}=4\pi zN_{A}e^{2}/m$ is the plasma frequency of
the medium; $N_{A}$ is the number of atoms per 1 cm$^{3}$), we get
the explicit expression for the possible frequencies of the
emitted photons at the transitions $n\rightarrow f$ inside a well,
when the zone width may be neglected, being considered as a
discrete level in the form
\begin{equation}
\label{4.11}
\omega_{nf}^{(1,2)}=\frac{2m(\varepsilon_{n}-\varepsilon_{f})\pm[4m^{2}(\varepsilon_{n}-\varepsilon_{f})^{2}
-8E^{2}\omega^{2}_{L}(1-\beta\cos\vartheta)]^{1/2}}{4E(1-\beta\cos\vartheta)}\,.
\end{equation}
%%%%%%%%%%%%%%%%%%555

According to (\ref{4.11}) in the spectral range in question, two
frequencies of the emitted photons the difference between which
depends on the energy of the incident particle and the observation
angle $\vartheta$ correspond to the given transition. If the
difference $(\varepsilon_{n}-\varepsilon_{f})$ changes with the
energy growth slower than $E^{2}$, then for the given nonzero
value of the angle $\vartheta$, the difference between the
frequencies $\omega^{(1)}$ and $\omega^{(2)}$ decreases, vanishing
at a certain value of $E=E_{nf}$.
%%%%%%%%%%%%%%%555

%%%%%%%%%%%%%%%555
At $E>E_{nf}$ the frequencies (\ref{4.11}) become complex.
 This means that the radiation of hard photons at a selected angle $\vartheta$
is impossible. As there is a one--to--one correspondence between
the frequencies of the emitted photons and the angle of radiation
of $\gamma$-quanta, it is obvious that for a given value of
$\omega_{nf}$, we obtain the constraints for the possible angles
of observation  of this frequency. At $\vartheta\rightarrow 0$ the
threshold energy value grows, and at $\vartheta=0$ the upper limit
(threshold) disappears. In this case (compare with \cite{40})
\begin{equation}
\label{4.12}
\omega_{nf}^{(1,2)}=\left[(\varepsilon_{n}-\varepsilon_{f})\pm\sqrt{(\varepsilon_{n}-\varepsilon_{f})^{2}-
\omega^{2}_{L}}\right] \frac{E}{m}\,.
\end{equation}

%%%%%%%%%%%%%%%%5555

It follows from (\ref{4.12}) that certain restrictions are also
imposed on the possible values of the difference of  the energies
of transitions  $(\varepsilon_{n}-\varepsilon_{f})$. Namely,  it
is necessary that
${|\varepsilon_{n}-\varepsilon_{f}|>\omega_{L}}$. At
${|(\varepsilon_{n}-\varepsilon_{f})|< \omega_{L}}$, the
frequencies characterizing the transitions between the discrete
levels $n$ and $f$ are not observed in the radiation spectrum. The
restrictions obtained agree well with the criterion of the
appearance of the Doppler effect for an oscillator moving in a
medium \cite{39}. If
$\varepsilon_{n}-\varepsilon_{f}\gg\omega_{L}$, then
\begin{equation}
\label{4.13} \omega_{nf}^{(1)}\simeq
2(\varepsilon_{n}-\varepsilon_{f})\frac{E}{m};\qquad
\omega_{nf}^{(2)}\simeq\frac{\omega^{2}_{L}}
{2(\varepsilon_{n}-\varepsilon_{f})}\frac{E}{m}\,.
\end{equation}
%%%%%%%%%%%%%%%5

This makes it clear that the medium has practically no influence
on hard radiation. Soft radiation is totally dependent on the
refractive properties of the medium. If
$(\varepsilon_{n}-\varepsilon_{f})\sim\sqrt{\gamma}$ (the
oscillatory well, the transitions between neighboring levels in
the quasiclassical case), the frequency
$\omega_{nf}^{(2)}\sim\sqrt{\gamma}$, i.e., the frequency goes up
slowly with the growth of energy. If the difference
$(\varepsilon_{n}-\varepsilon_{f})\sim\gamma$, then
$\omega_{nf}^{(2)}=\mbox{const}$. It also follows from the
apparent requirement $\omega_{nf}^{(1,2)}\geq 0$  that in view of
(\ref{4.11}), the transitions  to lower energy levels
$\varepsilon_{f}<\varepsilon_{n}$ are only possible.

%%%%%%%%%%%%%%%%%%%%%%%%%%%%%%%%%%%  Section 5 %%%%%%%%%%%%%%%%%%%%%%

\section{Complex and Anomalous Doppler Effects in an Absorption Medium}
\label{sec:2.5} Now consider the how the radiation spectrum
changes of in an absorbing medium \cite{42}. In this case the
refractive index is complex, and equality (\ref{4.3}), which in
fact shows that the momentum transmitted to the medium
$q_{znf}=p_{zn}-p_{1zf}-k_{zn}$ is zero, does not hold. However,
the smaller $q_{znf}$, the greater the probability of photon
emission is. The radiation  probability  will have its peak value
at the $\gamma$-quantum frequencies  $\omega_{nf}$, for which the
longitudinal transmitted momentum has a minimum value.

With the presence of the imaginary part of $n$ and the fulfillment
of the condition $\omega\ll E$, the longitudinal transmitted
momentum may be written as follows:
\begin{equation}
\label{5.1} q_{znf}=\omega-\omega\beta
n^{\prime}(\omega)\cos\vartheta-\frac{m}{E}(\varepsilon_{n\vec{\kappa}}-\varepsilon_{f\vec{\kappa}_{1}})-i\omega\beta
n^{\prime\prime}(\omega)\cos\vartheta,
\end{equation}
where $n=n^{\prime}+in^{\prime\prime}$; $n^{\prime}$ is the real
part of $n$; $n^{\prime\prime}$ is the imaginary part of $n$.
According to (\ref{5.1}), the minimum value of $q_{znf}$ is
limited by its imaginary part.
\begin{equation}
\label{5.2} \texttt{Im}q_{znf}=\omega\beta
n^{\prime\prime}(\omega)\cos\vartheta\equiv \delta(\omega).
\end{equation}
The corresponding photon frequencies for which the quantity
$q_{znf}$ is minimum, and, hence, the radiation probability is
maximum, are determined from the condition
\begin{equation}
\label{5.3} \texttt{Re}q_{znf}=\omega-\omega\beta
n^{\prime}\cos\vartheta-\frac{m}{E}(\varepsilon_{n\vec{\kappa}}-\varepsilon_{f\vec{\kappa}_{1}})=|\varepsilon|\delta(\omega),
\end{equation}
where $|\varepsilon|\leq 1$.

At $\varepsilon=0$, the condition (\ref{5.3}) determines the
frequencies corresponding to the frequencies in the center of the
given intensity maximum. All other frequencies at $\varepsilon\neq
0$ are located in some vicinity on either side of the central
frequency. The radiation intensity corresponding to them is
comparable with the radiation intensity of the central frequency.
Therefore equation (\ref{5.3}) in fact determines the radiation
spectrum and may be recast as follows
\begin{equation}
\label{5.4}
\omega_{nf}=\frac{\varepsilon^{\prime}_{n\vec{\kappa}}-\varepsilon^{\prime}_{f\vec{\kappa}_{1}}+|\varepsilon|\delta(\omega_{nf})}{1-\beta
n^{\prime}(\omega_{nf})\cos\vartheta}.
\end{equation}

Note that solving equation (\ref{5.4}), one should bear in mind
that the reduced quasi-momentum $\vec{\kappa}_{1}$ depends on the
frequency $\omega_{\vec{n}f}$. If
$|\vec{k}_{\perp}|\ll\frac{\pi}{a}$,
$\varepsilon_{f\vec{\kappa}_{1}}$ can be expanded into a series:
$\varepsilon_{f\vec{\kappa}_{1}}=\varepsilon_{f\vec{\kappa}}+\vec{k}_{\perp}\vec{\nabla}_{\vec{\kappa}}\varepsilon_{f\vec{\kappa}}+\ldots$
($\vec{k}_{\perp}=\omega_{nf}\vec{n}_{\perp}$;
$\vec{n}_{\perp}=\vec{k}_{\perp}/|\vec{k}_{\perp}|$;
$\vec{\nabla}_{\vec{\kappa}}\varepsilon_{f\vec{\kappa}}$) is the
particle velocity in the state $f_{\vec{\kappa}}$). Near the
extremums of the bands  the first expansion term is zero, and it
is  necessary to allow for the following terms of the series.
Taking account of the stated dependence is crucial when analyzing
the formation of photons through intraband transitions, when the
difference
$\varepsilon_{n\vec{\kappa}}-\varepsilon_{f\vec{\kappa}_{1}}$ is
determined just by the correction terms
$\vec{k}_{\perp}\vec{\nabla}_{\vec{\kappa}}\varepsilon_{f\vec{\kappa}}+\ldots$
.

Expressions (\ref{5.3}), (\ref{5.4}) are obtained without using
the explicit form of the refractive index $n(\omega)$, so they
are also applicable for the analysis of the radiation spectrum of
$\gamma$-quanta in the frequency range, where a large contribution
to $n(\omega)$ comes from the crystal nuclei with, for example,
low resonances. As in this case the medium under consideration is
strongly absorbing, the entire  frequency spectrum is given by
(\ref{5.4}) which allows for the imaginary part of the refractive
index. Though, as it has already been pointed out, in order to
find central frequencies in the intensity maxima, it is sufficient
to make use of equation (\ref{4.5}).

The refractive index within the X-ray frequency range for
Mossbauer crystals can be represented in the form
\begin{equation}
\label{5.5}
n^{\prime}=1-\frac{\omega_{L}^{2}}{2\omega^{2}}-\mu\frac{(\omega-\omega_{0})}{(\omega-\omega_{0})+\Gamma^{2}/4},
\end{equation}
where $\mu\equiv\frac{\pi
N}{2\omega_{0}^{2}}\frac{2I+1}{2I_{0}+1}\frac{\Gamma}{1+\alpha_{\gamma}}f_{M}$;
$f_{M}$ is the Lamb-Mossbauer factor; $\alpha_{\gamma}$ is the
internal conversion coefficient; $I$ and $I_{0}$ are the spins of
the initial and final states of the nucleus, respectively;
$\omega_{0}$ is the resonant frequency of the nuclear
$\gamma$-transition; $\Gamma$ is the nuclear level width.

It was stated in \cite{43} that the refractive index in a
Mossbauer crystal (\ref{5.5}) may become greater than unity. This
enables observation of the Vavilov-Cherenkov effect, and hence,
the anomalous Doppler effect for short-wavelength photons at which
the emitting particle moves to a higher energy level. Substituting
(\ref{5.5}) into (\ref{4.5}), we obtain the following expression
for central frequencies in the maximum;
\begin{equation}
\label{5.6}
\omega-\omega\beta\cos\vartheta+\frac{\beta\omega^{2}_{L}}{2\omega}+\frac{\mu\beta(\omega-\omega_{0})\omega}{(\omega-\omega_{0})^{2}+\Gamma^{2}/4}-\Omega_{nf}=0.
\end{equation}

According to (\ref{5.5}) $n^{\prime}(\omega)$ can become greater
than unity only in a narrow range near the resonant frequency
$\omega_{0}$ (for instance, for $^{57}Fe
\Delta\omega\equiv\omega-\omega_{0}=10\Gamma$ \cite{43}. Using
this fact, the frequencies determined by the anomalous Doppler
effect can be sought in the form $\omega=\omega_{0}-\Delta$, where
$\Delta\ll\omega_{0}$. It is clear from (\ref{5.6}) that in this
range the equation is solvable, when $\Omega_{nf}$ is less than
zero, which corresponds to the system transition to a higher
energy level through radiation. Consequently, from (\ref{5.6}) we
may obtain the following expression for anomalous Doppler
frequencies corresponding to the central frequencies in the
intensity maximum
\begin{equation}
\label{5.7}
\omega_{nf}^{(1,2)}\equiv\omega_{0}-\Delta_{nf}^{(1,2)}=\omega_{0}-\frac{\mu}{2A}\mp\left[\left(\frac{\mu}{2A}\right)^{2}-\frac{\Gamma^{2}}{4}\right]^{1/2},
\end{equation}
where
$$
A\equiv\frac{\vartheta^{2}}{2}+\frac{m^{2}}{2E^{2}}+\frac{\omega^{2}_{L}}{2\omega^{2}_{0}}+\frac{|\Omega_{nf}|}{\omega_{0}}.
$$

Far from the frequency $\omega_{0}$ the contribution of the
resonance term in the refractive index may be neglected. Finally
we turn back to the case of radiation considered above, which is
described by formula (\ref{4.11}), from which we obtain the other
two solutions of equation (\ref{5.6}) corresponding to the normal
Doppler frequencies. In view of (\ref{5.7}) at
$$
E<E^{\prime}=m|\Omega_{nf}|\left[\sqrt{2}\omega_{0}\left(\frac{\mu}{\Gamma}-\frac{\vartheta^{2}}{2}-
\frac{\omega^{2}_{L}}{2\omega_{0}^{2}}\right)\right]^{-1}
$$
the frequencies $\omega_{nf}^{(1,2)}$ become complex. As a result,
at such energies the anomalous Doppler effect  is impossible.

Interestingly enough, the phenomenon of photon emission
accompanied by the excitation of the emitting system itself does
not only arise as a result of the anomalous Doppler effect, or
when the  velocity of the source is higher than the velocity of
light in a vacuum. This process also occurs when  the oscillator
moves at subluminal velocity in a medium with $n<1$, if the
coherent radiation length is limited (for example, due to the
photon absorption in the medium, the presence of the crystal
boundaries, multiple scattering \cite{42}). Indeed, in the case of
absorbing medium there is a whole set of frequencies  for which
$q_{znf}\sim\omega \texttt{Im}\, n$. As a result the radiation
intensities for these frequencies are comparable with one another,
so from (\ref{5.8}) we  may obtain the following expression for a
photon spectrum
\begin{equation}
\label{5.8}
\omega_{nf}^{(1,2)}=\frac{[\Omega_{nf}+|\varepsilon|\delta(\omega_{nf})]\pm\left\{[\Omega_{nf}+|\varepsilon|\delta(\omega_{nf})]^{2}-2\omega^{2}_{L}(1-\beta\cos\vartheta)\right\}^{1/2}}{2(1-\beta\cos\vartheta)}
\end{equation}
It follows from (\ref{5.8}) that in the case of absorbing medium
the photon radiation accompanied by  the excitation of the
emitting system itself becomes possible. Indeed, in view of
(\ref{5.8}) the following conditions should be fulfilled to make
this process possible:
\begin{eqnarray}
\label{5.9}
\omega_{nf}>0; |\varepsilon|\delta(\omega_{nf})>|\Omega_{nf}|;\nonumber\\
\left[\Omega_{nf}+|\varepsilon|
\delta(\omega_{nf})\right]^{2}-2\omega^{2}_{L}(1-\beta\cos\vartheta)\geq
0.
\end{eqnarray}
The conditions (\ref{5.9}) may be reduced to one
\begin{equation}
\label{5.10}
|\varepsilon|\delta(\omega_{nf})>|\Omega_{nf}|+\sqrt{2}\omega_{L}(1-\beta\cos\vartheta)^{1/2}.
\end{equation}
It is seen from the expression for frequency $\Omega_{nf}$ that
with the increase in the energy of the channeled particle the
requirement  (\ref{5.10}) becomes less strict  and proves to be
feasible for a larger number of levels $n$ and $f$ of the discrete
spectrum of particle transverse motion. If the condition
(\ref{5.10}) is not satisfied, the radiation  corresponding to the
transition between the given energy levels $n$ and $f$ of the
transverse motion will only occur when the system moves to a lower
energy level.

As it has already been mentioned, the presence of the target
boundaries and  multiple scattering of a channeled particle  along
with absorption,  lead to limitation of the minimum value of the
longitudinal component of the momentum transmitted to the medium,
and hence, to limitation of the coherent length. Thus, for
instance, for thin crystal plates with $L<(\omega
n^{\prime\prime}\cos\vartheta)^{-1}$ the maximum coherent length
$l\sim 1/q_{znf}$ determining the process of radiation cannot
exceed $L$. The frequency spectrum in this case is described by
formula (\ref{5.8}) with $\delta(\omega)$ replaced by $L^{-1}$.

Note also that with the presence of boundaries, the momentum
transmitted along the normal to the crystal surface, should no
longer be zero (or $2\pi\vec{\tau}_{z}$) even for a thick
nonabsorbing medium. In this case the frequency spectrum can be
written in the form
\begin{equation}
\label{5.11}
\omega_{nf}^{(1,2)}=\frac{(l_{nf}^{-1}(\omega)-\Omega_{nf})\pm[(l_{nf}^{-1}(\omega)-\Omega_{nf})^{2}-2\omega^{2}_{L}(1-\beta\cos\vartheta)]^{1/2}}{2(1-\beta\cos\vartheta)}
\end{equation}
where $l_{nf}(\omega)=(p_{zn}-p_{1zf}-k_{z}n)^{-1}$.

So, radiation of a channeled particle accompanied by the
excitation of the emitting system is possible  in a medium with
$n<1$ not only for a source moving at the velocity greater than
the velocity of light in vacuum but also for an oscillator moving
with  subluminal velocity. From the viewpoint of physics the
phenomenon in question  can be understood, taking into account the
fact that the limitation of the coherent length, and hence, the
magnitude of the longitudinal momentum transmitted to the medium
gives rise to uncertainty in the real part of such a momentum.
From the conservation laws follows that this is equivalent to the
appearance of uncertainty in the value of the energy of the
particle transverse motion. If the uncertainty in the energy which
results from the limitation of the coherent length exceeds the
distance between the discrete levels of transverse motion in a
laboratory system, it will cause virtual elimination of the
distinction between the levels in the given interval of changes in
the transverse momentum. Consequently, the system through
radiation can move to both lower and higher levels of transverse
motion.

Until now we considered a crystal as an optically isotropic medium
for photons. Note, however, that  a crystal can exhibit optical
anisotropy in both optical and X-ray (and shorter wavelength)
spectral ranges \cite{14}. When analyzing the radiation process,
the refractive index in conservation laws means one of the major
target refractive indices \cite{34}.  In a short wavelength range
the optical anisotropy of crystals is manifest in the case
diffraction of  $\gamma$-quanta in them. Then both real and
imaginary parts of the crystal refractive index strongly depend on
the direction of photon propagation, which results in a
significant change in spectral, angular and polarization
characteristics of all types of radiation excited by a charged
particle in a crystal [\cite{5,34,44,45,46,47}] (see
(\ref{ch:4})).

It should be pointed out that radiation associated with the
transitions between the levels of discrete spectrum  of the
particle transverse motion in a crystal may be treated as
spontaneous radiation of a channeled particle. When a crystal is
put in the area occupied by an electromagnetic field (for,
example, light radiation), one can stimulate induced transitions
between the stated levels, which will give rise to induced
radiation \cite{17,48}.

%%

%%%%%%%%%%%%%%%%%%%%%%%%%%%%%%%%%%%  Chapter 3 %%%%%%%%%%%%%%%%%%%%%%

\chapter[The Foundations of the Theory of $\gamma$-quanta Emission in Crystals under Channeling ...]{The
Foundations of the Theory of $\gamma$-quanta Emission in Crystals
under Channeling Conditions} \label{ch:3}

%%%%%%%%%%%%%%%%%%%%%%%%%%%%%%%%%%%  Section 6 %%%%%%%%%%%%%%%%%%%%%%

\section[The Cross Section of Photon Generation by Particles in an External Field]{The Cross Section of Photon Generation by Particles in an External Field}
\label{sec:3.6}

Theoretical study of the process of photon production by channeled
particles has been carried out from various viewpoints. The
emission of $\gamma$-quanta in crystals with the thicknesses
smaller than the length of transformation of the wave function of
an incident particle from a plane wave to a superposition of the
Bloch waves was examined in  \cite{49,50,51}. According to
\cite{49,50,51} the process of electron emission for such
thicknesses can be analyzed in terms of the concept of radiative
capture of a particle incident on a crystal into the channeling
regime. In \cite{7}, there  considered radiation  in an infinite
crystal within the framework of the classical model of a particle
motion in a parabolic potential. Within the framework of this
model there is only one radiation frequency corresponding to the
Doppler shifted frequency of particle oscillation in a harmonic
well involved in the formation of the radiation spectrum. Further
analysis of the problem given in \cite{36,37,52,53} was also
performed for an infinite crystal.
%%%%

At the same time as far back as in our early works \cite{5,17,34},
devoted to the problem of photon radiation under channeling
conditions, it was shown that in a real potential the radiation
spectrum is produced by frequencies corresponding to a wide range
of the particle transitions between the levels of transverse
motion. Such transitions result in the fact that when exploring
the radiation spectrum at a given angle to the direction of a
particle motion, a discrete set of spectral lines is to be
observed. That spectrum was experimentally revealed in
\cite{54,55,56}. Moreover, according to \cite{5,17,23,24}, when a
particle enters the crystal, the whole set of transverse motion
levels is necessarily populated. As a result, not only sub-barrier
transitions (inside a well) but also the over-barrier transitions,
as well as the transitions from over-barrier to sub-barrier states
take part in the spectrum formation. Over-barrier states located
near the barrier edge are characterized by a wide regions of
transverse motion, which was completely ignored in
\cite{36,37,52,53}, and it was only in \cite{57a,57b} where this
fact was taken into consideration.

The population of all the above-mentioned states  depends on the
type of a particle, the angle at which it enters the crystal, and
the shape of a well. This fact has a considerable impact on the
shape of the spectrum formed by particles during radiative
transitions between the levels of transverse motion, which was
convincingly demonstrated by Bayer, Katkov and Strakhovenko by
using  numerical calculations \cite{58}. In \cite{59,60} the
important role of the radiation produced through over-barrier
transitions has also been emphasized recently.

According to \cite{5,17,34}, refraction, absorption, and
diffraction of photons in crystals  also considerably affect the
radiative spectrum. Below we  presented the results obtained in
our investigations \cite{5,17,34,44,45}.

%%%%%%%%%%%%%%%%%%%%%%

Let a beam of charged particles with the momentum $\vec{p}$ and
energy $E$ fall on a crystal of volume $V$. As a result of
collision with the crystal, the particle momentum changes and the
particle, undergoing  acceleration, emits radiation. Theoretical
analysis of the process under study implies that each particle
corresponds to a wave packet, produced in a generator (particle
accelerator) at a certain moment $t_{0}$. Due to the particle
interaction with a medium, at long distances from the crystal in
addition to a primary wave packet diverging spherical waves which
describe the scattered and newly produced particles (in this case
- photons) also appear (Fig. 5). To calculate the cross-section,
it is necessary to know the transition probability per unit time
for the process, when one particle scattered in a constant field
produces in its final state a certain number of other particles.
In view of the  quantum mechanical theory of reactions it may be
represented by the general formula of the form (see, for example,
\cite{19}) ($\hbar=c=1$):
%%%%%%%%%%%%%%%
\begin{equation}
\label{6.1}
dW=2\pi\delta(E_{f}-E)\overline{|M_{fi}|^{2}}\frac{1}{2E{\cal{L}}^{3}}\prod_{a}\frac{d^{3}p_{a}}{(2\pi)^{3}2E_{a}}\,,
\end{equation}
where $E$ is the energy of the initial particle; $E_{f}$ is the
energy of the final state; $\vec{p}_{a}$ and $E_{a}$ are the
momenta and energies of the final particles; ${\cal{L}}^{3}$ is
the normalization volume; $M_{fi}$ is the amplitude of scattering
from the initial $i$ to the final $f$ state; the overline means
averaging over the spin states of the particles involved in the
reaction.
%%%%%%%%%%%%%%%
The scattering cross section $d\sigma$ is obtained by dividing
$dW$ by the incident particle flux density $j=v/{\cal{L}}^{3}$,
where $v=|\vec{p}|/E$ is the velocity of the primary particle. as
a result, we get
\begin{equation}
\label{6.2}
d\sigma=2\pi\delta(E_{f}-E)\overline{|M_{fi}|^{2}}\frac{1}{2|\vec{p}|}\prod_{a}\frac{d^{3}p_{a}}{(2\pi)^{3}2E_{a}}\,.
\end{equation}

In the case of interest there is a particle and a photon in the
final state. Write the radiation cross-section as
\begin{equation}
\label{6.3}
d\sigma=2\pi\delta(E_{1}+\omega-E)\overline{|M(\vec{p}_{1},\vec{k};\vec{p})|^{2}}\frac{d^{3}p_{1}d^{3}k}{8(2\pi)^{6}pE_{1}\omega}\,,
\end{equation}
where $\vec{p}$ is the  momentum of the primary particle
(electron, positron); $\vec{k}$ is the photon momentum; $\omega$
is the photon frequency; $E_{1}$ and $\vec{p}_{1}$ are the energy
and momentum of the particle in the final state.

%%%%%%%%%%%%%%%%%%%%%%%%%%%%%

The matrix element $M$ describing the process of photon emission
in an arbitrary external field can be represented in the form
\begin{equation}
\label{6.4} M(\vec{p}_{1},\vec{k};\vec{p})=
e\int\Psi_{p_{1}}^{(-)*}(\vec{r})\vec{\alpha}\vec{A}_{\vec{k}}^{(-)*}(\vec{r})\Psi_{p}^{(+)}(\vec{r})d^{3}r\,,
\end{equation}
where $\Psi_{p}^{+}(\vec{r})$, $\Psi_{p_{1}}^{-}(\vec{r})$ are the
exact solutions of the Dirac equation for particle scattering in
the external field, having different asymptotics far from the
crystal: $\Psi_{p}^{+}(\vec{r})$, for the primary particle
(asymptotics type --- an incident plane wave plus diverging
spherical waves), $\Psi_{p_{1}}^{-}(\vec{r})$, for the final
particle (the asymptotics type --- an incident plane wave plus
converging spherical waves); $\vec{A}_{\vec{k}}^{(-)}(\vec{r})$ is
the vector potential of the emitted photon, being the exact
solution of Maxwell equations and  describing photon scattering by
a crystal (the asymptotic type --- an incident plane wave plus
converging spherical wave) \cite{17,61}.
%%%%%%%%%%%%%%%%%%%%%%%%%%%%%%%%%

The wave functions of all the particles are normalized to one
particle  within the volume ${\cal{L}}^{3}$. T he terms
$1/\sqrt{2E{\cal{L}}^{3}}$  $1/\sqrt{2E_{1}{\cal{L}}^{3}}$ and
$1/\sqrt{2\omega{\cal{L}}^{3}}$ appearing in them are shown
explicitly,and they are included in the definitions of $dW$ and
$d\sigma$ (see(\ref{6.3})). Thus, if the photon-crystal
interaction is ignored, the vector potential
$\vec{A}_{\vec{k}}^{(-)}(\vec{r})$ has the form \cite{19}
\begin{equation}
\label{6.5}
\vec{A}_{\vec{k}}^{(-)}(\vec{r})=\sqrt{4\pi}\vec{e}_{s}e^{i\vec{k}\vec{r}},
\end{equation}
where $\vec{e}_{s}$ is the photon polarization vector. Then the
matrix element in (\ref{6.4}) is written as follows:
\begin{equation}
\label{6.6} M(\vec{p}_{1},\vec{k};\vec{p})\equiv
e\sqrt{4\pi}{\cal{M}}_{fi}=
e\sqrt{4\pi}\int\Psi_{p_{1}}^{(-)*}(\vec{r})\vec{\alpha}\vec{e}_{s}^{*}e^{-i\vec{k}\vec{r}}\Psi_{\vec{p}}^{(+)}(\vec{r})d^{3}r.
\end{equation}
To find the explicit form for $d\sigma$, one should know the wave
functions $\psi^{(\pm)}$. (The general analysis of the
characteristics of the functions describing particle scattering by
a crystal was given in (\ref{ch:1}). Considering the photon
radiation in a crystal, when solving the Dirac equation it is
necessary (as well as  in bremsstrahlung by a screened Coulomb
potential \cite{62}) to take into account the terms proportional
to $\alpha$. This occurs through the fact that for fast particles
the matrix element  $\vec{\alpha}$ involved in (\ref{6.6}) is the
vector, whose direction is close to $\vec{k}$. Therefore the major
term $\vec{\alpha}\vec{e}$  proves to be small, and the correction
terms have the same order of magnitude \cite{19}.

%%%%%%%%%%%%%%%%%%

Write the Dirac equation (\ref{1.4}) in the form:
\begin{equation}
\label{6.7}
[\Delta_{r}+p^{2}-2EV(\vec{r})]\Psi(\vec{r})=-i\vec{\alpha}\nabla
V(\vec{r})\Psi(\vec{r})\,.
\end{equation}
The term proportional to $V^{2}$  may still be ignored. Dividing
both sides of equation (\ref{6.7}) by $2m\gamma$, we obtain
\begin{equation}
\label{6.8}
\left[\frac{1}{2m\gamma}\Delta_{r}+\varepsilon^{\prime}-V(\vec{r})\right]\Psi(\vec{r})=-\frac{i}{2m\gamma}(\vec{\alpha}\nabla
V(\vec{r}))\Psi(\vec{r})\,.
\end{equation}
Remember that (see (\ref{1.7}))
\[
\varepsilon^{\prime}=\frac{p^{2}}{2m\gamma}
\]
and  $\Psi(\vec{r})$ is sought as
\begin{equation}
\label{6.9}
\Psi(\vec{r})=\Psi^{(0)}(\vec{r})+\Psi^{(1)}(\vec{r})\,,
\end{equation}
where $\Psi^{(0)}(\vec{r})$ satisfies equation (\ref{6.8}) with a
zero right--hand side, and $\Psi^{(1)}(\vec{r})$ is the desired
correction. The equation for $\Psi^{(0)}(\vec{r})$ does not
include spin matrices. Therefore the spin state of a particle
passing through a crystal cannot change in this approximation: it
coincides with the spin state of a particle in a plane wave
incident on a crystal. It is convenient to extract a bispinor
amplitude describing the particle spin state in its explicit form,
and represent $\Psi(\vec{r})$ as follows \cite{19}:
%%%%%%%%%%%%%%%%%%%%%%%%

%%%%%%%%%%%%%%%%%%%%%
\begin{equation}
\label{6.10}
\Psi(\vec{r})=e^{i\vec{p}\vec{r}}[u(\vec{p})\varphi_{p}(\vec{r})+\varphi^{(1)}_{p}(\vec{r})]\,,
\end{equation}
where $u(\vec{p})$ is the constant bispinor amplitude of the plane
wave incident on the crystal normalized by the condition
\begin{equation}
\label{6.11} \bar{u}(\vec{p})u(\vec{p})=2m\,.
\end{equation}
Substituting (\ref{6.10}) into (\ref{6.8}) and retaining the
first--order terms over $\vec{\alpha}\nabla V$, we come to the
equation
\begin{eqnarray}
\label{6.12}
\left[\frac{1}{2m\gamma}\Delta+\frac{i}{m\gamma}\vec{p}\nabla-V(\vec{r})\right]\varphi_{p}^{(1)}(\vec{r})\nonumber\\
=-\frac{i}{2m\gamma}u(\vec{p})(\vec{\alpha}\nabla
V(\vec{r}))\varphi_{p}(\vec{r})\,.
\end{eqnarray}
To solve (\ref{6.12}), make use of the fact that the function
$\varphi_{p}(\vec{r})$ satisfies the equation
\begin{equation}
\label{6.13}
\left[\frac{1}{2m\gamma}\Delta+\frac{i}{m\gamma}\vec{p}\nabla-V(\vec{r})\right]\varphi_{p}(\vec{r})=0\,.
\end{equation}
Application of the operation $\nabla$ to equation (\ref{6.13})
gives
\begin{equation}
\label{6.14}
\left[\frac{1}{2m\gamma}\Delta+\frac{i}{m\gamma}\vec{p}\nabla-V(\vec{r})\right]\nabla\varphi_{p}(\vec{r})=\varphi_{p}(\vec{r})\nabla
V(\vec{r})\,.
\end{equation}
Upon multiplying (\ref{6.14}) by
\[
-\frac{i}{2m\gamma}u(\vec{p})\vec{\alpha}
\]
 and comparing the result with (\ref{6.12}), we obtain immediately
\begin{equation}
\label{6.15}
\varphi_{p}^{(1)}(\vec{r})=-\frac{i}{2m\gamma}\vec{\alpha}\nabla\varphi_{p}(\vec{r})u(\vec{p})\,.
\end{equation}
Thus, finally
\begin{equation}
\label{6.16}
\Psi(\vec{r})=e^{i\vec{p}\vec{r}}\left(1-\frac{i}{2m\gamma}\vec{\alpha}\nabla\right)\varphi_{p}(\vec{r})u(\vec{p})\,.
\end{equation}

%%%%%%%%%%%%%%%%%%%%%%%
It should be emphasized that, as demonstrated by the direct
comparison between the expansion of (\ref{6.16}) and the exact
solution of the Dirac equation for the Kronig-Penney model
\cite{17}, relation (\ref{6.16}) in the case of not very thick
crystals is always suitable (applicable) (for example, for $E=1$
GeV the thickness is $l\leq 1$ cm, for  $E=500$ keV $\div 1 $ MeV
$l\sim 10^{-3}\div 10^{-2}$ cm). In fact, the stated expansion
holds true, when the parameter $\Omega l/v$ is small ($\Omega$ is
the characteristic energy of spin-orbit interaction between the
incident particle spin and the crystal axis; $v$ is the particle
velocity).

Substitute the wave functions (\ref{6.16}) into the expression for
the matrix element ${\cal{M}}_{fi}$ (see (\ref{6.6})):
\begin{eqnarray}
\label{6.17} {\cal{M}}_{fi}&=&\int
d^{3}re^{-i(\vec{p}_{1}+\vec{k}-\vec{p})\vec{r}}u^{+}(\vec{p}_{1})\left(1+\frac{i}{2E_{1}}\vec{\alpha}\nabla\right)
\varphi_{p_{1}}^{(-)*}(\vec{r})\nonumber\\
&\times&\vec{\alpha}\vec{e}_{s}^{*}\left(1-\frac{i}{2E_{1}}\vec{\alpha}\nabla\right)\varphi_{p}^{(+)}(\vec{r})u(\vec{p})\,,
\end{eqnarray}
%%%%%%%%%%%%%%%%%%%%%%%
i.e.,
\begin{equation}
\label{6.18}
{\cal{M}}_{fi}=u^{+}(\vec{p}_{1})[\vec{\alpha}\vec{e}^{\,*}_{s}I_{1}+
(\vec{\alpha}\vec{e}^{\,*}_{s})(\vec{\alpha}\vec{I}_{2})+(\vec{\alpha}\vec{I}_{3})(\vec{\alpha}\vec{e}^{\,*}_{s})]u\vec{p}\,,
\end{equation}
where, by analogy with \cite{62}, the following quantities are
introduced
\begin{eqnarray}
\label{6.19}
I_{1}&=&\int e^{-i\vec{q}\vec{r}}\varphi_{p_{1}}^{(-)*}(\vec{r})\varphi_{p}^{(+)}(\vec{r})d^{3}r\,,\nonumber\\
\vec{I}_{2}&=&-\frac{i}{2E}\int e^{-i\vec{q}\vec{r}}\varphi_{p_{1}}^{(-)*}(\vec{r})\nabla_{r}\varphi_{p}^{(+)}(\vec{r})d^{3}r\,,\nonumber\\
\vec{I}_{3}&=&\frac{i}{2E_{1}}\int
e^{-i\vec{q}\vec{r}}(\nabla_{r}\varphi_{p_{1}}^{(-)*}(\vec{r}))\varphi_{p}^{(+)}(\vec{r})d^{3}r\,,
\end{eqnarray}
$\vec{q}=\vec{p}_{1}+\vec{k}-\vec{p}$ is the transmitted momentum.

For further consideration it should be noted that the integrals in
(\ref{6.19}) are related to each other \cite{62}.

%%%%%%%%%%%%%%%

Integration by parts in the equality for $\vec{I}_{3}$ gives:
\begin{eqnarray}
\label{6.20}
\vec{I}_{3}&=&-\frac{i}{2E_{1}}\int\varphi_{p_{1}}^{(-)*}(\vec{r})\nabla_{r}[e^{-i\vec{q}\vec{r}}\varphi_{p}^{(+)}(\vec{r})]d^{3}r\nonumber\\
&=&-\frac{i}{2E_{1}}\int\varphi_{p_{1}}^{(-)*}(\vec{r})e^{-i\vec{q}\vec{r}}\nabla_{r}\varphi_{p}^{(+)}(\vec{r})d^{3}r\nonumber\\
&-&\frac{\vec{q}}{2E_{1}}\int
e^{-iqr}\varphi_{p_{1}}^{(-)*}(\vec{r})\varphi_{p}^{(+)}(\vec{r})d^{3}r\,.
\end{eqnarray}
Comparison of (\ref{6.20}) and (\ref{6.19}) gives
\begin{equation}
\label{6.21}
\vec{I}_{3}=\frac{E}{E_{1}}\vec{I}_{2}-\frac{\vec{q}}{2E_{1}}I_{1}\,.
\end{equation}

%%%%%%%%%%%%%%%%%%%%%%%%%%%%%%%%%%
To further simplify the matrix element in (\ref{6.18}), it is
convenient to cast (\ref{6.18}), using two-component spinor
functions:
\begin{eqnarray}
\label{6.22} u(\vec{p})=\sqrt{E+m}\left(
                       \begin{array}{c}
                         1 \\
                         \sqrt{\frac{E-m}{E+m}}(\vec{\sigma}\vec{n}_{p})  \\
                       \end{array}
                     \right)w;\nonumber\\
u(\vec{p}_{1})=\sqrt{E_{1+m}}\left(
                              \begin{array}{c}
                                1 \\
                               \sqrt{\frac{E_{1}-m}{E_{1}+m}}(\vec{\sigma}\vec{n}_{p_{1}})   \\
                              \end{array}
                            \right)w_{1}; \,\vec{\alpha}=\left(
                                                              \begin{array}{cc}
                                                                0 & \vec{\sigma} \\
                                                                 \vec{\sigma} & 0 \\
                                                              \end{array}
                                                            \right)\,,
\end{eqnarray}
%%%%%%%%%%%%%%%%
where $\vec{\sigma}$ are the Pauli matrices; $w(w_{1})$ is the
two-component spinor \cite{19}; $\vec{n}_{p}$ is the unit vector
in the $\vec{p}$ direction; $\vec{n}_{p_{1}}$ is the same for
$\vec{p}_{1}$. Substitution of (\ref{6.22}) into (\ref{6.18}),
gives quite an awkward expression for ${\cal{M}}_{ji}$, which,
however, simplifies at $E$, $E_{1}\gg m$. The calculations in this
approximation are perfectly analogous to those performed by Olsen
and Maximon in \cite{62}, making it possible to write
${\cal{M}}_{fi}$ as follows:
\begin{equation}
\label{6.23} {\cal{M}}_{fi}=2\sqrt{\frac{E}{E_{1}}}w_{1}^{+}
\left\{(E+E_{1})(\vec{g}\vec{e}_{s}^{\,*})+i\omega\vec{\sigma}[\vec{g}\times\vec{e}_{s}^{\,*}]\right\}w\,.
\end{equation}
%%%%%%%%%%%%%%%%%%
The vector
\begin{equation}
\label{6.24} \vec{g}=\vec{g}_{\perp\vec{k}}+\vec{g}_{\parallel}\,,
\end{equation}
where
\[
\vec{g}_{\perp\vec{k}}=\vec{I}_{2\perp\vec{k}}+\frac{1}{2}\vec{n}_{\vec{p}\perp\vec{k}}I_{1}\,,
\quad \vec{g}_{\parallel}=-\frac{m}{2E}\vec{n}_{\parallel}I_{1}\,,
 \]
 the symbol ($\perp\vec{k}$) means the projection of the corresponding vector
 onto
 the plane perpendicular to the direction of the photon momentum  $\vec{k}$; the symbol
 $\parallel$ is for the projection of the vector onto the $\vec{k}$ direction; $\vec{n}_{\parallel}$  is the unit vector in the $\vec{k}$ direction.

%%%%%%%%%%%%%%%%%%%%%
Upon introducing the polarization density matrices of the initial
$\rho$ and the final $\rho_{1}$ electrons, we obtain for the
average square of the matrix element involved in the cross-section
of (\ref{6.3}),
\[
\overline{|{\cal{M}}_{fi}|^{2}}= \textrm{Tr} \,\rho
{\cal{M}}^{+}\rho_{1}{\cal{M}}\,,
\]
where
\begin{eqnarray}
\label{6.25}
{\cal{M}}&=&2\sqrt{\frac{E}{E_{1}}}\left\{(E+E_{1})(\vec{g}\vec{e}_{s}^{\,*})
+i\omega\vec{\sigma}[\vec{g}\times\vec{e}_{s}^{\,*}]\right\};\nonumber\\
\rho&=&\frac{1}{2}(1+\vec{\xi}\vec{\sigma})\,,\qquad
\rho_{1}=\frac{1}{2}(1+\vec{\xi}_{1}\vec{\sigma})\,,
\end{eqnarray}
$\vec{\xi}(\vec{\xi}_{1})$ is the polarization vector of the
particle in the initial (final) state; $0\leq\xi$; $\xi_{1}\leq
1$.
%%%%%%%%%%%%%%%%%%%%%%%%%%%
After taking the trace appearing in (\ref{6.25}), we get
\begin{eqnarray}
\label{6.26} &
&\overline{|{\cal{M}}_{fi}|^{2}}=4\frac{E}{E_{1}}\left\{\frac{1}{2}\omega^{2}|\vec{g}|^{2}+2EE_{1}(1+\vec{\xi}\vec{\xi}_{1})
|\vec{g}\vec{e}^{\,*}|^{2}\right.\nonumber\\
&
&+\frac{1}{2}\omega^{2}\texttt{Re}\left\{|\vec{g}|^{2}\vec{\xi}_{1}\vec{\xi}-2(\vec{g}\vec{\xi})(\vec{g}^{\,*}\vec{\xi}_{1})\right\}
+\omega E_{1}\texttt{Re}\left\{\left[|\vec{g}|^{2}(\vec{\xi}\vec{e})\right.\right.\nonumber\\
&
&\left.\left.-2(\vec{g}\vec{e})(\vec{g}^{\,*}\vec{\xi})\right]\vec{\xi}_{1}\vec{e}^{\,*}\right\}-\omega
E\,
\texttt{Re}\left\{\left[|\vec{g}|^{2}(\vec{\xi}_{1}\vec{e})-2(\vec{g}\vec{e})(\vec{g}^{\,*}\vec{\xi}_{1})\right]
(\vec{\xi}\vec{e}^{\,*})\right\}\nonumber\\
&
&+\frac{1}{2}\omega|\vec{g}|^{2}(E\vec{\xi}+E_{1}\vec{\xi}_{1})(i\vec{e}\times\vec{e}^{\,*})+\frac{1}{2}\omega
\,
\texttt{Re}\left\{|\vec{g}|^{2}\left(E_{1}\vec{\xi}\right.\right.\nonumber\\
&
&\left.\left.+E\vec{\xi}_{1}\right)(i\vec{e}\times\vec{e}^{\,*})-2(g(E_{1}\vec{\xi}+E\vec{\xi}_{1}))(\vec{g}^{\,*}
(i\vec{e}\times\vec{e}^{\,*}))\right\}\nonumber\\
&
&+\frac{1}{2}\omega\left[\omega(1+\vec{\xi}\vec{\xi}_{1})(i\vec{e}\times\vec{e}^{\,*})+(E+E_{1})(\vec{\xi}\times\vec{\xi}_{1})
\times(i\vec{e}\times\vec{e}^{\,*})\right.\nonumber\\
&
&\left.\left.+(E+E_{1})(\vec{\xi}+\vec{\xi}_{1})-2\texttt{Re}\left\{\vec{e}^{\,*}((E\vec{\xi}+E_{1}\vec{\xi}_{1})\vec{e})\right\}\right]
(i\vec{g}\times\vec{g}^{\,*})\right\}\,.
\end{eqnarray}
%%%%%%%%%%%%%%%%%%%%
Using (\ref{6.25}) and (\ref{6.26}), we get the required
expression  for the  cross-section of the photon radiation in a
crystal allowing for polarization of all the particles involved in
the reaction.
\begin{equation}
\label{6.27} d\sigma=e^{2}\delta(E_{1}+\omega-E) sp\,\rho
{\cal{M}}^{+}\rho_{1}{\cal{M}}\frac{d^{3}p_{1}d^{3}k}{4(2\pi)^{4}pE_{1}\omega}\,.
\end{equation}
The relationships (\ref{6.26}) and (\ref{6.27}) solve in the
general form the problem of finding $d\sigma$.

If we are not concerned about the polarization of the final
particle, then with $\xi_{1}$ in (\ref{6.27}) assumed to be zero,
the entire expression (\ref{6.27}) should to be multiplied by  2.

%%%%%%%%%%%%%%%%%%%%%%%%%%%%%%%%%%%%%%%%

%%%%%%%%%%%%%%%%%%%%%%%%%%%%%%%%%%%  Section 7 %%%%%%%%%%%%%%%%%%%%%%

\section[Photon Generation in Crystals under Channeling Conditions]{Photon Generation in Crystals under Channeling Conditions}
\label{sec:3.7} We now turn to a more detailed treatment of the
cross-section of (\ref{6.27}). Let us take into account that in
(\ref{6.19}) the linear dimensions of the domain of integration
only exceed the linear dimensions of the crystal by the magnitude
of the vacuum coherence length
\begin{equation}
\label{chan_7.1}
l_{\mathrm{coh}}\sim\frac{1}{q_{z}}=\frac{2}{\omega}\frac{E(E-\omega)}{m^{2}}\,.
\end{equation}
(A thorough treatment of the properties of $l_{coh}$ see in
\cite{63,64}). For this reason, analyzing the radiation process in
a crystal target with lateral dimensions  much larger than its
thickness, we can apply the expressions describing scattering of a
plane wave by a crystal plate with infinite lateral dimensions,
i.e., the functions considered in (\ref{ch:1}). Substituting these
functions into (\ref{6.19}) with due account of the relation
\[
\varphi_{p}^{(+)}(\vec{r})=\varphi_{-p}^{(-)*}(\vec{r})
\]
%%%%%%%%%%%%%%%%%%%%
and  integrating it with respect to the momentum $\vec{p}_{1}$, we
get the below expression for the spectral--angular distribution of
the number of photons emitted by a channeled particle
$dN=\frac{1}{S}d\sigma$ ($S$ is the area of the target surface):
\begin{eqnarray}
\label{7.2} & &\frac{d^{2}N_{s}}{d\omega
d\Omega}=\frac{e^{2}\omega}{4\pi^{2}}\texttt{Re}\sum_{nfj}Q_{nj}e^{\tilde{i\Omega}_{nj}L}
\left[ \frac{1-\exp(iq^{*}_{zjf}L)}{q^{*}_{zjf}} \right]
\nonumber\\
& &\times \left[\frac{1-\exp(-iq_{znf}L)}{q_{znf}} \right] \left\{
\frac{\omega^{2}}{2E_{1}^{2}}\vec{g}_{nf}\vec{g}^{\,*}_{jf}
+2\frac{E}{E_{1}}(1+\vec{\xi}\vec{\xi}_{1})
\right. \nonumber\\
& & \left.
\times(\vec{g}_{nf}\vec{e}^{\,*}_{s})(\vec{g}_{jf}^{\,*}\vec{e}_{s})+\frac{\omega^{2}}{2E^{2}_{1}}\texttt{Re}
[\vec{g}_{nf}\vec{g}_{jf}^{\,*}(\vec{\xi}\vec{\xi}_{1})
-2(\vec{g}_{nf}\vec{\xi})(\vec{g}_{jf}^{\,*}\vec{\xi}_{1})]
\right. \nonumber\\
& & \left. +\frac{\omega}{E_{1}}\texttt{Re}
\left\{[\vec{g}_{nf}\vec{g}_{jf}^{\,*}(\vec{\xi}\vec{e}_{s})
-2(\vec{g}_{nf}\vec{e}_{s})(\vec{g}^{\,*}_{jf}\vec{\xi})](\vec{\xi}_{1}\vec{e}_{s}^{\,*})
\right\} \right. \\
& & \left. -\frac{\omega E}{E_{1}^{2}}\texttt{Re}
\left\{[\vec{g}_{nf}\vec{g}^{*}_{jf}(\vec{\xi}_{1}\vec{e}_{s})
-2(\vec{g}_{nf}\vec{e}_{s})(\vec{g}_{jf}^{\,*}\vec{\xi}_{1})](\vec{\xi}\vec{e}^{\,*}_{s})
\right\} \right. \nonumber\\
& & \left. \left.
+\frac{\omega}{2E_{1}^{2}}(\vec{g}_{nf}\vec{g}^{\,*}_{jf})(E\vec{\xi}+E_{1}\vec{\xi}_{1})
[i\vec{e}_{s}\times\vec{e}^{\,*}_{s}]+\frac{\omega}{2E_{1}^{2}}\texttt{Re}
\left\{(\vec{g}_{nf}\vec{g}^{\,*}_{jf})
(E_{1}\vec{\xi}+E\vec{\xi}_{1})[i\vec{e}_{s}\times\vec{e}^{\,*}_{s}]\right. \right. \right. \nonumber\\
& & \left. \left.
-2(\vec{g}_{nf}(E_{1}\vec{\xi}+E\vec{\xi}_{1}))(\vec{g}^{\,*}_{jf}[i\vec{e}_{s}\times\vec{e}^{\,*}_{s}])\right\} \right. \nonumber\\
& & \left.
+\frac{\omega}{2E_{1}^{2}}\left[\omega(1+\vec{\xi}\vec{\xi}_{1})[i\vec{e}_{s}\times\vec{e}^{\,*}_{s}]
+(E+E_{1})\left[[\vec{\xi}\times\vec{\xi}_{1}][i\vec{e}_{s}\times\vec{e}^{\,*}_{s}]\right]
\right. \right.
\nonumber\\
& & \left. \left.
+(E+E_{1})(\vec{\xi}+\vec{\xi}_{1})-2\texttt{Re}\left\{\vec{e}^{\,*}_{s}
\left(\left(E\vec{\xi}+E_{1}\vec{\xi}_{1}\right)
\vec{e}_{s}\right)\right\}\right][i\vec{g}_{nf}\times\vec{g}^{\,*}_{jf}]
\right\}\,,\nonumber
\end{eqnarray}
where $q_{znf}=p_{zn}-p_{1zf}-k_{z}$ is the longitudinal momentum
transmitted through radiation;
$\tilde{\Omega}_{nj}=\varepsilon^{\prime}_{n\kappa}(E)-\varepsilon^{\prime}_{j\kappa}(E)=
\frac{1}{\gamma}(\varepsilon_{n\kappa}(E)-\varepsilon_{j\kappa}(E))$;
the argument $E$ in the notation for the transverse energy of the
initial state emphasizes that the particle in the initial state
has the energy $E$.

In the general two--dimensional case (axial channeling), the
following relations are valid
\begin{equation}
\label{7.3}
Q_{nj}=c_{n}(\vec{p}_{\perp})c^{*}_{j}(\vec{p}_{\perp})\,, \quad
c_{n}(\vec{p}_{\perp})=\sqrt{\frac{N_{\perp}}{S}}\int_{S}e^{i\vec{p}_{\perp}\vec{\rho}}\psi_{n\kappa}^{*}(\vec{\rho})d^{2}\rho\,,
\end{equation}
where $N_{\perp}$ is the number of two--dimensional unit cells in
a transverse plane of the crystal; $s$ is the area of the unit
cell. When a particle is channeled along the planes located
periodically along the $x$-axis:
\begin{equation}
\label{7.4} Q_{nj}=c_{n}(p_{x})c_{j}^{*}(p_{x}); \quad
c_{n}(p_{x})=\sqrt{\frac{N_{x}}{a}}\int_{0}^{a}e^{ip_x
x}\psi^{*}_{n\kappa}(x)dx\,,
\end{equation}
where $N_{x}$ is the number of the crystal periods along the
$x$-axis; $a$ is the lattice spacing along the $x$-axis;
\[
\psi_{n\kappa}(x)=\frac{1}{\sqrt{N_{x}}}e^{i\kappa
x}u_{n\kappa}(x)
 \]
is the Bloch function describing the transverse motion in zone $n$
of a particle  with the reduced quasi-momentum
\[
\kappa=p_{x}-\frac{2\pi l}{a};
\]
the integral number $l$ is found from the condition
\[
\left|p_{x}-\frac{2\pi l}{a}\right|<\frac{\pi}{a}.
\]
In the two--dimensional case
\[
\psi_{n\vec{\kappa}}(\vec{\rho})=\frac{1}{N_{\perp}}e^{i\vec{\kappa}\vec{\rho}}u_{n\kappa}(\vec{\rho})
\]
is the Bloch function with
$\vec{\kappa}=\vec{p}_{\perp}-\vec{\tau}_{\perp}$;
$\vec{\tau}_{\perp}$ is obtained from the condition of the
reduction of $\vec{p}_{\perp}$ to the first Brillouin zone.

%%%%%%%%%%%%%%%%%%%%%%%%%%%%%%%%
Vector $\vec{g}_{nf}$ in a two-dimensional (axial) case has the
form
\begin{eqnarray}
\label{7.5} \vec{g}_{nf}&=&\vec{g}_{\perp nf}+\vec{g}_{\parallel
nf}=\frac{1}{2E}\vec{W}_{nf}=
\frac{1}{2E}(\vec{I}_{2nf}+\vec{p}_{z\perp\vec{k}}I_{1nf}-m\vec{n}_{\parallel}I_{1nf})\,,\nonumber\\
\vec{I}_{2nf}&=&-iN_{\perp}\int_{s}e^{-i\vec{k}_{\perp}\vec{\rho}}\psi^{*}_{f\vec{\kappa}_{1}}
(\vec{\rho})\vec{\nabla}_{\rho}\psi_{n\vec{\kappa}}(\vec{\rho})d^{2}\rho\,,\nonumber\\
I_{1nf}&=&N_{\perp}\int_{s}e^{-i\vec{k}_{\perp}\vec{\rho}}\psi^{*}_{f\vec{\kappa}}(\vec{\rho})\psi_{n\vec{\kappa}}(\vec{\rho})d^{2}\rho;\,\,
\vec{p}_{\vec{z}\perp\vec{k}}=p\vec{n}_{\vec{z}\perp\vec{k}},
\end{eqnarray}
where $\vec{n}_{z}$ is the unit vector along the $z$-axis
direction; recall that the symbol ($\perp\vec{k}$ ) stands for the
projection of the corresponding vector onto the plane
perpendicular to the direction of the photon momentum $\vec{k}$.
In the one--dimensional (planar) case, vector
%%%%%%%%%%%%%%%%%%%%%
\begin{eqnarray}
\label{7.6} \vec{g}_{nf}&=&\frac{1}{2E}\vec{W}_{nf}=
\frac{1}{2E}(I_{2nf}\vec{n}_{x}+(\vec{p}-\vec{p}_{x})_{\perp k}I_{1nf}-m\vec{n}_{\parallel}I_{1nf})\,,\nonumber\\
\vec{I}_{2nf}&=&-iN_{x}\int_{0}^{a}e^{-ik_{x}x}\psi^{*}_{f\kappa_{1}}(x)\frac{\partial}{\partial x}\psi_{n\kappa}(x)dx\,,\nonumber\\
I_{1nf}&=&N_{x}\int_{0}^{a}e^{-ik_{x}x}\psi^{*}_{f\kappa_{1}}(x)\psi_{n\kappa}(x)dx,
\end{eqnarray}
where $\vec{p}_{x}=p_{x}\vec{n}_{x}$; $\vec{n}_{x}$ is the unit
vector along the $x$-axis;
\[
\kappa_{1}=p_{x}-k_{x}-\frac{n_{0}}{a}\,,
 \]
$n_{0}$ is found from the condition of the reduction of
$p_{x}-k_{x}$ to the first Brillouin zone, i.e.,
\[
|p_{x}-k_{x}-\frac{n_{0}}{a}|<\frac{\pi}{a}\,,
 \]
$\vec{\kappa}_{1}=\vec{p}_{\perp}-\vec{k}_{\perp}-\vec{\tau}_{0}$;
$\vec{\tau}_{0}$ is found from the condition of the reduction of
$\vec{p}_{\perp}-\vec{k}_{\perp}$ to the first Brillouin zone.

%%%%%%%%%%%%%%%%%%%%%
The formulas obtained above enable one to describe angular,
spectral and polarization properties of radiation formed in a
crystal in detail.

Let particles incident on a crystal be nonpolarized ($\xi=0$), and
the polarization of final particles be of no interest to us. As
has already been mentioned, in this case it should be assumed that
$\xi_{1}=0$ and the expression for the cross section should be
multiplied by two. As a result, we obtain
\begin{eqnarray}
\label{7.7} \frac{d^{2}N_{s}}{d\omega d
\Omega}&=&\frac{e^{2}\omega}{2\pi^{2}}\texttt{Re}\sum_{nfj}Q_{nj}e^{i\tilde{\Omega}_{nj}L}
\left[\frac{1-\exp(iq_{zjf}^{*}L)}{q_{zjf}^{*}}\right]
\left[\frac{1-\exp(iq_{znf}L)}{q_{znf}^{*}}\right]\\
& \times &
\left\{\frac{\omega^{2}}{2E^{2}_{1}}\vec{g}_{nf}\vec{g}^{\,*}_{jf}
+2\frac{E}{E_{1}}(\vec{g}_{nf}\vec{e}^{\,*}_{s})(\vec{g}_{jf}^{\,*}\vec{e}_{s})+\frac{\omega^{2}}{2E^{2}_{1}}[i\vec{e}_{s}\times\vec{e}^{\,*}_{s}]
[i\vec{g}_{nf}\times\vec{g}_{jf}^{\,*}]\right\}\,.\nonumber
\end{eqnarray}
%%%%%%%%%

According to (\ref{7.7}), the spectral angular distribution of
photons oscillates with the change in the crystal thickness $L$
at frequencies $\tilde{Q}_{nj}$ determined by the differences
between the energies of the transverse motion levels which are
populated when a particle enters the crystal. These oscillations
of the radiation intensity are quite similar to those observed at
radiation of atoms at the given angle under pulse-excitation into
the superposition of states. If the characteristic frequencies
$\tilde{\Omega}_{nj}$ and the crystal thickness $L$ are such that
$\tilde{\Omega}_{nj}L\gg 1$, the averaging of (\ref{7.7}) over the
thickness spread leads to the averaging of oscillations, and it
should be assumed that in (\ref{7.7}) $j=0$ (integration of
(\ref{7.7}) with respect to $d\omega$ or $d\Omega$ also leads to
vanishing of the oscillations). The characteristic oscillation
frequencies in the transverse plane
\[
\tilde{\Omega}\sim\frac{1}{T}=\frac{v_{\perp}}{d}\simeq\frac{\vartheta_{L}}{d}
\]
where $T$ is the oscillation period in the transverse plane;
$v_{\perp}$ is the velocity of transverse motion; $d$ is the
channel width (cm); $\vartheta_{L}$ is the Lindhard angle;
$v_{\perp}=\vartheta_{L}c$ ($c$ is the velocity of light), at
$c=1$ $v_{\perp}=\vartheta_{L}$. Consequently, the inequality
$\tilde{\Omega}L\gg 1$ can be cast as follows \cite{15}
\begin{equation}
\label{7.8} \frac{L\vartheta_{L}}{d}\gg 1
\end{equation}
%%%%%%%%%%%%%%%%%%%%%%%%

At $\vartheta_{L}\simeq 10^{-4}$ for positrons with the energy 1
GeV and $d=10^{-8}$ cm the inequality holds true for the
thicknesses $L\gg 10^{-4}$ cm (in the absence of degeneracy of
energy levels).

Thus, if $L\gg 1/\tilde{\Omega}_{nj}$, the sum in (\ref{7.7})
should only contain the terms with $n=j$, which leads to the
following relation
\begin{eqnarray}
\label{7.9}
\frac{d^{2}N_{s}}{d\omega d \Omega}& &=\frac{e^{2}\omega}{2\pi^{2}}\sum_{nf}Q_{nn}\left|\frac{1-\exp(-iq_{znf}L)}{q_{znf}}\right|^2\\
&
&\times\left\{\frac{\omega^{2}}{2E^{2}_{1}}|\vec{g}_{nf}|^2+2\frac{E}{E_{1}}|\vec{g}_{nf}\vec{e}^{\,*}_{s}|^{2}
+\frac{\omega^{2}}{2E^{2}_{1}}[i\vec{e}_{s}\times\vec{e}^{\,*}_{s}][i\vec{g}_{nf}\times\vec{g}_{nf}^{\,*}]\right\}\,.\nonumber
\end{eqnarray}
If we do not concern ourselves with the polarization of an emitted
photon,  (\ref{7.9}) is to be summed over the polarization states:
\begin{eqnarray}
\label{7.10}
\frac{d^{2}N}{d\omega d\Omega}=\frac{e^{2}\omega}{2\pi^{2}}\sum_{nf}Q_{nn}\left|\frac{1-\exp(-iq_{znf}L)}{q_{znf}}\right|^{2}\nonumber\\
\times\left[\left(2\frac{E}{E_{1}}+\frac{\omega^{2}}{E^{2}_{1}}\right)|\vec{g}_{\perp
nf}|^{2}+ \frac{\omega^2}{E_1^2}|\vec{q}_{\parallel
nf}|^{2}\right]\,.
\end{eqnarray}
%%%%%%%%%%%%%%%%%%%%%%%%%%%%%%%%%%%55
Recall that
\begin{eqnarray}
\label{7.11}
\vec{g}_{\perp nf}&=&\frac{1}{2E}\vec{W}_{\perp nf}=\frac{1}{2E}(\vec{I}_{2nf}+\vec{p}_{z\perp k}I_{1nf})\,,\nonumber\\
\vec{g}_{\parallel nf}&=&\frac{1}{2E}\vec{W}_{\parallel
nf}=-\frac{m}{2E}\vec{n}_{\parallel}I_{1nf}\,.
\end{eqnarray}
The transferred momentum $q_{znf}$ in the planar case can be
written as follows
\begin{eqnarray}
\label{7.12}
 q_{znf}& &\simeq
\frac{\omega}{2(E-\omega)}\left[\vartheta^{2}(E-\omega\cos^{2}\varphi)+\frac{m^{2}}{E}-2\Omega_{nf}
+2\varepsilon^{\prime}_{n\kappa}(E)\right]-\Omega_{nf}\nonumber\\
& &\equiv\frac{\omega}{2(E-\omega)}
\left[\vartheta^{2}(E-\omega\cos^{2}\varphi)
+\frac{m^{2}}{E}\right]-(\varepsilon^{\prime}_{n\kappa}(E)-\varepsilon^{\prime}_{n\kappa_{1}}(E_{1}))\,,\nonumber\\
\Omega_{nf} &
&=\frac{m}{E}(\varepsilon_{n\kappa}(E)-\varepsilon_{n\kappa_{1}}(E_{1}))\,.
\end{eqnarray}
%%%%%%%%%%%%%%%%%%%%%%%%%%%%%%%%%5
As far as we still analyze the process of photon radiation within
the range of frequencies $\omega$ and crystal thicknesses, where
the absorption and refraction of emitted quanta may be neglected,
the expressions involved in (\ref{7.9}) and (\ref{7.10}) with high
accuracy can be recast in the form
\[
\left|\frac{1-\exp(-iq_{znf}L)}{q_{znf}}\right|^{2}\simeq 2\pi
L\delta(q_{znf})\,.
\]
As a consequence,
\begin{eqnarray}
\label{7.13} & &\frac{d^{2}N_{s}}{d\omega
d\Omega}=\frac{e^{2}\omega L}{\pi}\sum_{nf}Q_{nn}
\left\{\frac{\omega^{2}}{2E^{2}_{1}}|\vec{g}_{nf}|^{2}+2\frac{E}{E_{1}}
|\vec{g}_{nf}\vec{e}_{s}^{\,*}|^{2}\right.\nonumber\\
& &
\left.+\frac{\omega^{2}}{2E_{1}^{2}}[i\vec{e}_{s}\times\vec{e}^{\,*}_{s}][i\vec{g}_{nf}\times\vec{g}^{\,*}_{nf}]\right\}\delta(q_{znf})\,,
\end{eqnarray}
\begin{eqnarray}
\label{7.14} & &\frac{d^{2}N}{d\omega
d\Omega}=\sum_{s}\frac{d^{2}N_{s}}{d\omega
d\Omega}=\frac{e^{2}\omega L}{\pi}\sum_{nf}Q_{nn}
\left\{\left(2\frac{E}{E_{1}}+\frac{\omega^{2}}{E^{2}_{1}}\right)|\vec{g}_{\perp nf}|^{2}\right.\nonumber\\
& &\left.+\frac{\omega^{2}}{E_{1}^{2}}|\vec{g}_{\parallel
nf}|^{2}\right\}\delta(q_{znf})\,.
\end{eqnarray}

%%%%%%%%%%%%%%%%%%%%%%%%%%%%5555
If expression (\ref{7.14}) only includes the sub-barrier
transitions, then it coincides with that derived in \cite{37}.
Such a restriction, however, as we have pointed out repeatedly
\cite{5,17,44}, does not fit the real experimental conditions,
when at particle entering at a certain angle to the axis (plane),
the above-barrier states (regions) are also necessarily populated.

In a most typical case, photons of frequency $\omega\ll E$ are
emitted through channeling. If in this case the energies
$\varepsilon_{n\kappa}$ and $\varepsilon_{f\kappa_{1}}\ll m$
(i.e., the transverse motion in the system with zero longitudinal
velocity of a particle is nonrelativistic, which occurs for
particles, whose energy is less than a few gigaelectronvolts),
then expressions (\ref{7.13}), (\ref{7.14}) simplify considerably:
%%%%%%%%%%%%%%5
\begin{equation}
\label{7.15} \frac{d^{2}N_{s}}{d\omega d\Omega}=\frac{2e^{2}\omega
L}{\pi}\sum_{nf}Q_{nn}|\vec{g}_{\perp nf}\vec{e}^{\,*}_{s}|^{2}
\delta(\omega(1-\beta\cos\vartheta)-\Omega_{nf})\,,
\end{equation}
\begin{equation}
\label{7.16} \frac{d^{2}N}{d\omega d\Omega}=\frac{2e^{2}\omega
L}{\pi}\sum_{nf}Q_{nn}|\vec{g}_{\perp
nf}|^{2}\delta(\omega(1-\beta\cos\vartheta)-\Omega_{nf})\,,
\end{equation}
where $\beta=v_{z}/c$ and at $c=1$, the value of  $\beta=v_{z}$;
$v_{z}$ is the longitudinal particle velocity; the component
$\vec{g}_{\parallel}$ in this approximation does not contribute to
(\ref{7.15}), (\ref{7.16}).

%%%%%%%%%%%%%%%%%%%%

It is worthy of mention that the quantity $\vec{W}_{\perp
nf}=\vec{I}_{2nf}+\vec{p}_{z\perp k}I_{1nf}$ appearing in the
expression for vector $\vec{g}_{\perp nf}$ can be represented in
several equivalent forms. Using the notations agreed in \cite{44},
we have the following expression for $\vec{W}_{\perp}$
\begin{equation}
\label{7.17}
\vec{W}_{\gamma\eta}=\vec{p}_{n}J_{\gamma\eta}-\vec{n}_{x}I_{\gamma\eta},\,
\vec{p}_{n}=\vec{p}-\frac{2\pi n_{0}}{a}\vec{n}_{x};
\end{equation}
$n_{0}$ is obtained by reduction of vector $p_{x}-k_{x}$ to the
first Brillouin zone, i.e., $|p_{x}-k_{x}-\frac{2\pi
n_{0}}{a}|<\frac{\pi}{a}$;
\begin{eqnarray}
\label{7.18}
J_{\gamma\eta}=\int_{0}^{a}e^{-i\frac{2\pi(l-n_{0})}{a}x}u_{\gamma p_{x}}(x)u^{*}_{\eta p_{1x}}(x)dx;\nonumber\\
I_{\gamma\eta}=\frac{1}{i}\int_{0}^{a}e^{-i\frac{2\pi(l-n_{0})}{a}x}u_{\gamma
p_{x}}(x)\frac{d}{dx}u^{*}_{\eta p_{1x}}(x)dx;
\end{eqnarray}
$l$ is found from the condition $|p_{x}-\frac{2\pi
l}{a}|<\frac{\pi}{a}$; Integration of the expression for
$I_{\gamma\eta}$ by parts gives
\begin{eqnarray}
\label{7.19}
I_{\gamma\eta}=-\frac{1}{i}\int_{0}^{a}u^{*}_{\eta p_{1x}}(x)\frac{d}{dx}(e^{-i\frac{2\pi(l-n_{0})}{a}x}u_{\gamma p_{x}}(x))dx=\nonumber\\
\frac{2\pi(l-n_{0})}{a}J_{\gamma\eta}-I_{\gamma\eta}^{\prime},
\end{eqnarray}
where
$$
I_{\gamma\eta}^{\prime}=-i\int_{0}^{a}e^{-i\frac{2\pi(l-n_{0})}{a}x}u^{*}_{\eta
p_{1x}}(x)\frac{d}{dx}u_{\gamma p_{x}}(x)dx.
$$
From the definitions of $n_{0}$ and $l$ follows that $\frac{2\pi
l}{a}=p_{x}-\kappa$, and $\frac{2\pi
n_{0}}{a}=p_{x}-k_{x}-\kappa_{1}$. Hence, we can write:
\begin{eqnarray}
\label{7.20}
J_{\gamma\eta}=\int_{0}^{a}e^{-i(k_{x}+\kappa_{1}-\kappa)x}u_{\gamma p_{x}}(x)u^{*}_{\eta p_{1x}}(x)dx;\nonumber\\
I_{\gamma\eta}=(k_{x}+\kappa_{1}-\kappa)J_{\gamma\eta}-I^{\prime}_{\gamma\eta};\nonumber\\
I^{\prime}_{\gamma\eta}=-i\int_{0}^{a}e^{-i(k_{x}+\kappa_{1}-\kappa)x}u^{*}_{\eta
p_{1x}}(x)\frac{d}{dx}u_{\gamma p_{x}}(x)dx.
\end{eqnarray}
Recall that the Bloch function is
\begin{eqnarray}
\label{7.21}
\psi_{\gamma p_{x}}(x)\equiv\psi_{\gamma\kappa}(x)=\frac{1}{\sqrt{N_{x}}}e^{i\kappa x}u_{\gamma p_{x}}(x);\, u_{\gamma p_{x}}(x)\equiv u_{\gamma\kappa}(x);\nonumber\\
\psi_{\eta p_{1x}}(x)\equiv
\psi_{\eta\kappa_{1}}(x)=\frac{1}{\sqrt{N_{x}}}e^{i\kappa_{1}x}u_{\eta
p_{1x}}(x).
\end{eqnarray}
Consequently,
\begin{equation}
\label{7.22} u_{\gamma p_{x}}=\sqrt{N_{x}}e^{-i\kappa
x}\psi_{\gamma p_{x}}(x);\, u_{\eta
p_{1x}}(x)=\sqrt{N_{x}}e^{-i\kappa_{1} x}\psi_{\eta p_{1x}}(x).
\end{equation}
Substituting (\ref{7.22}) into (\ref{7.20}), we obtain the
following equality from (\ref{7.17}):
\begin{equation}
\label{7.23}
\vec{W}_{\gamma\eta}=(\vec{p}-\vec{p}_{x})J_{\gamma\eta}+I_{\gamma\eta}\vec{n}_{x}.
\end{equation}
As  in (\ref{7.15}) (see also (9) in \cite{44}) vector $\vec{W}$
is multiplied by the photon polarization vector $\vec{e}_{s}$,
$(\vec{p}-\vec{p}_{x})$ in (\ref{7.23}) can be replaced by
$(\vec{p}-\vec{p}_{x})_{\perp k}$. \footnote{We obtained formula
(\ref{7.15}) in \cite{44} in a more general form (with  the term
$\frac{1-\exp(-iq_{znf})L}{q_{znf}}$ instead of
$\delta$-function). Two years after the work was published, the
coincident formula was derived in \cite{57a,57b}. The authors of
\cite{57a,57b} first did not notice that their relations coincide
with those we had obtained before and declared our theory invalid.
In \cite{20,65} we proved that the  criticism from the authors of
\cite{57a,57b} is unfounded. Now compare (\ref{7.15}) and the
coincident formula (9) in \cite{44} with formula (7) derived in
\cite{57a,57b}. According to \cite{57a,57b} the formula for
spectral-angular distribution of radiation at spontaneous
transitions in the planar case has the form
\begin{eqnarray}
\label{7.24} \frac{d^{2}W}{d\omega
d\Omega}=\frac{e^{2}\omega}{2\pi}\sum_{f}\left\{e_{\sigma}|j_{if}^{(x)}(k_{x})|^{2}\sin^{2}\varphi+e_{\pi}
|j^{(z)}_{if}(k_{x})\theta\right.\nonumber\\
\left.-j^{(x)}_{if}(k_{x})\cos\varphi|^{2}\right\}\delta\left[\omega(\frac{\theta^{2}+E^{-2}}{2}-
\frac{\partial\varepsilon_{f}(E^{\prime\prime}_{i})}{\partial
E_{i}})-\tilde{\omega}_{if}\right].
\end{eqnarray}
The notations in (\ref{7.24}) are the same as in \cite{57a,57b}.
If a particle populates only one level, (\ref{7.15}) could differ
from (\ref{7.24}) by the expression between the braces in
(\ref{7.24}), and by the
$\frac{1}{E^{2}}|\vec{W}_{nf}\vec{e}_{s}|^{2}$. We will
demonstrate that there is no difference. Consider
$\pi$-polarization. In this case the polarization vector
$\vec{e}_{s}=\vec{e}_{\pi}$ is in the plane formed by the particle
and photon momenta. As a consequence,
$\vec{e}_{\pi}\vec{n}_{z\perp k}=-\vartheta$, $\vartheta$ is the
photon radiation angle;
$\vec{e}_{\pi}\vec{n}_{x}\simeq\cos\varphi$. Then
\begin{equation}
\label{7.25}
\left|\frac{\vec{e}_{\pi}\vec{W}_{nf}}{E}\right|^{2}=\left|J_{1nf}\vartheta-\frac{1}{E}I_{2nf}\cos\varphi\right|^{2}.
\end{equation}
It is clear from the definition of $J_{1nf}$ and $I_{2nf}$,
$j^{(x)}$ and $j^{(z)}$ that
$j_{if}^{(x)}=\frac{1}{E}I^{*}_{2nf}$,
$j^{(z)}_{if}(k_{x})=J^{*}_{if}$ and, hence, the formulae for
spectral-angular distribution of radiation coincide. Consider
$\sigma$-polarization. Now $\vec{e}_{s}=\vec{e}_{\sigma}$ is
perpendicular to the plane made up by the momenta of a photon and
a particle. As a result, $\vec{e}_{\sigma}\vec{n}_{z\perp k}=0$,
$\vec{e}_{\sigma}\vec{n}_{x}\simeq\sin\varphi$ and
\begin{equation}
\label{7.26}
\left|\frac{\vec{e}_{\sigma}\vec{W}_{nf}}{E}\right|^{2}=\frac{1}{E^{2}}|I_{2nf}|^{2}\sin^{2}\varphi,
\end{equation}
so the formulae coincide completely.}

The presence of $\delta$-functions in the derived expressions
enables one to easily find spectral or angular distribution of
emitted photons. It should be emphasized that the finite width of
the bands for transverse motion leads to the fact that the
radiation in question appears  not only at transitions between
different levels but also at the transitions within a given band.
In the case of narrow bands the corresponding radiation for
high-energy particles lies within the optical spectrum. For wide
over-barrier bands these transitions cause radiation in the X-ray
and shorter wavelength spectra.
 As follows from the presence of the $\delta$-function in expressions (\ref{7.15}), (\ref{7.16}),
the corresponding equation defining the photon frequency at the
intraband transition has the form
\begin{equation}
\label{7.27}
(1-\beta\cos\vartheta)\omega-(\varepsilon^{\prime}_{n\kappa}-\varepsilon^{\prime}_{n\kappa_{1}})=0\,.
\end{equation}
At fixed frequency, this equation determines the radiation angle
of a quantum. Note that in solving (\ref{7.27}) in the case of
over-barrier states it is vital to remember that
$\varepsilon^{\prime}_{n\kappa_{1}}$ depends on $\omega$ and
$\vartheta$.

%%%%%%%%%%%%%%%%%%%%%%%%%%%%%%%%%%%  Section 8 %%%%%%%%%%%%%%%%%%%%%%

\section[Spectral and Angular Distributions of Photons in the Dipole Approximation]{Spectral and Angular Distributions of Photons in the Dipole Approximation}
\label{sec:3.8}

Though simple at first sight, expressions (\ref{7.15}),
(\ref{7.16}) are rather complicated. Matrix elements $I_{2nf}$ and
$I_{1nf}$ defining vector $\vec{g}_{nf}$ are quite analogous to
matrix elements used in the theory of atomic radiation (see, for
example, \cite{19}). Investigating the properties of radiation in
the range where photon frequencies and exit angles are such that
$k_{\perp}a\ll 1$, the  exponentials in $I_{2nf}$ and $I_{1nf}$
may be expanded, and the reduced vectors $\kappa$ and $\kappa_{1}$
in wave functions may be equated. At the same time, when solving
(\ref{7.27}), the distinction between $\kappa$ and $\kappa_{1}$
should be taken into account especially for intrabad transitions.
Under the condition $k_{\perp}a\ll 1$
$\varepsilon^{\prime}_{n\kappa_{1}}$ can be expanded in terms of
$k_{\perp}$. As a result,
$\varepsilon^{\prime}_{n\kappa}-\varepsilon^{\prime}_{n\kappa_{1}}\simeq\frac{d\varepsilon^{\prime}}{dk_{\perp}}$
$k_{\perp}=vk_{\perp}$. Velocity $v$ has the order of magnitude
$\vartheta_{L}c$, i.e., $v\sim 10^{6}$ cm/s for $\vartheta_{L}\sim
10^{-4}$. From this $vk_{\perp}\sim 10^{12}-10^{13}$ sec$^{-1}$
for $k_{\perp}\sim 10^{7}$ cm$^{-1}$. The frequency $vk_{\perp}$
in this case is much smaller than the characteristic frequency of
of interband transitions, so the corresponding radiation lies in a
substantially softer spectra (in this case it lies in the optical
region even for particles with energies of the order of 1 GeV ).
For this reason, when analyzing the radiation spectrum in the
X-ray and shorter wavelength spectral regions, we shall not take
into account intraband transitions, assuming that
$\kappa=\kappa_{1}$ in the interband transition frequencies. As a
result, equation (\ref{7.17}) is easily solvable, and integration
of (\ref{7.15}), (\ref{7.17}) with respect to the photon exit
angles with the maximum collimation angle $\vartheta_{k}\ll m/E$,
gives in the dipole approximation the following expressions for
the spectrum \cite{45,66,67}:
%%%%%%%%%%%%%%%%%%%%%%%555
\begin{eqnarray}
\label{8.1}
\frac{dN_{s}}{d\omega}=e^{2}L\sum_{nf}Q_{nn}|\vec{\rho}_{nf}\vec{e}^{*}_{s}|^{2}\Omega^{2}_{nf}
\left[1-\frac{\omega}{\Omega_{nf}}(1-\beta^{2})\right.\nonumber\\
\left.+\frac{\omega^{2}}{2\Omega^{2}_{nf}}(1-\beta^{2})^{2}\right]
\theta\left(\frac{\vartheta^{2}_{k}}{2}-\alpha_{nf}(\omega)\right)\theta(\alpha_{nf}(\omega))\,,
\end{eqnarray}
\begin{eqnarray}
\label{8.2} \frac{d
N}{d\omega}=e^{2}L\sum_{nf}Q_{nn}|\vec{\rho}_{nf}|^{2}\Omega^{2}_{nf}
\left[1-\frac{\omega}{\Omega_{nf}}(1-\beta^{2})\right.\nonumber\\
\left.+\frac{\omega^{2}}{2\Omega^{2}_{nf}}(1-\beta^{2})^{2}\right]
\theta\left(\frac{\vartheta^{2}_{k}}{2}-\alpha_{nf}(\omega)\right)\theta(\alpha_{nf}(\omega))\,,
\end{eqnarray}
where
$\vec{\rho}_{nf}=N_{\perp}\int_{s}\psi_{n\kappa}(\vec{\rho})\vec{\rho}\psi_{f\kappa}(\vec{\rho})d^{2}\rho$;
$\theta(z)=1$ at $z>0$ and $\theta(z)=0$ at $z<0$;
$\alpha_{nf}(\omega)=(1-\beta)(\frac{\tilde{\omega}_{nf}}{\omega}-1)$;
$\tilde{\omega}_{nf}=\Omega_{nf}/(1-\beta)$ is the maximum
radiation frequency at the $n\rightarrow f$ transition. If the
collimation angle $\vartheta_{k}=\pi$,
\begin{eqnarray}
\label{8.3} \frac{d
N}{d\omega}=e^{2}L\sum_{nf}Q_{nn}|\vec{\rho}_{nf}|^{2}\Omega^{2}_{nf}
\left\{1-\frac{\omega}{\Omega_{nf}}(1-\beta^{2})\right.\nonumber\\
\left.+\frac{\omega^{2}}{2\Omega^{2}_{nf}}(1-\beta^{2})^{2}\right\}\theta(2-\alpha_{nf}(\omega))\theta(\alpha_{nf}(\omega))\,.
\end{eqnarray}

In the particular case when only sub-barrier transitions remain in
the sum over $n$, $f$, expression (\ref{8.3}) turns into the one
analyzed in \cite{37}.

%%%%%%%%%%%%%%%%%%%%%%%%%%%
Now consider the angular distribution. With this aim in view,
integrate (\ref{7.15}), (\ref{7.16}) over the frequencies. Under
real conditions, the detector registers the photons within a
certain spectral interval ${\omega_{1}\leq\omega\leq\omega_{2}}$.
Integration within this interval gives
\begin{eqnarray}
\label{8.4}  \frac{dN_{s}}{d\Omega}& &
=A_{s}\frac{(1-\beta\cos\vartheta)^{2}-(1-\beta^{2})
\sin^{2}\vartheta\cos^{2}\varphi}{(1-\beta\cos\vartheta)^{4}}\nonumber\\
& & \times\theta[\cos\vartheta-b_{nf}(\omega_{1})]\theta[b_{nf}(\omega_{2})-\cos\vartheta]\,,\\
& &\vartheta\leq\frac{m}{E}\,,\quad
b_{nf}(\omega)=\frac{1}{\beta}\left(1-\frac{\Omega_{nf}}{\omega}\right)\,,\quad
A_{s}=\frac{e^{2}L}{2\pi}\sum_{nf}Q_{nn}|\vec{x}_{nf}\vec{e}^{\,*}_{s}|^{2}\Omega^{3}_{nf}\,,\nonumber
\end{eqnarray}
\begin{eqnarray}
\label{8.5}
\frac{d N}{d\Omega}& &=A\frac{(1-\beta\cos\vartheta)^{2}-(1-\beta^{2})\sin^{2}\vartheta\cos^{2}\varphi}{(1-\beta\cos\vartheta)^{4}}\nonumber\\
& &\times\theta[\cos\vartheta-b_{nf}(\omega_{1})]\theta[b_{nf}(\omega_{2})-\cos\vartheta]\,,\nonumber\\
A & &
=\frac{e^{2}L}{2\pi}\sum_{nf}Q_{nn}|\vec{x}_{nf}|^{2}\Omega^{3}_{nf}\,.
\end{eqnarray}
%%%%%%%%%%%%%%
From formulas (\ref{8.4}), (\ref{8.5}) follows the well-known
result that the angular distribution of the radiation from a
relativistic particle whose velocity and acceleration are mutually
perpendicular is a universal function independent of the shape of
the potential in which the particle moves \cite{68}.

The number of $\gamma$-quanta $\Delta N_{\omega}$ emitted by a
channeled particle in the frequency interval $\Delta\omega$ in the
dipole approximation can be estimated as follows:
$$
\Delta N_{\omega}\simeq e^{2}x^{2}_{0}\Omega^{2}L\Delta\omega,
$$
where $x_{0}$ is the  amplitude of  particle oscillations in the
channel. From this follows, for example, that the particle with
the energy $E\sim 1$ GeV ($x_{0}\sim 10^{-8}$ cm, $\Omega\sim
10^{16}$ s$^{-1}$) passing through a silicon plate of length
$L\sim 10^{-2}$ cm  in the spectral interval
$\frac{\Delta\omega}{\omega}=10^{-3}$ emits the number of quanta
$\Delta N_{\omega}\sim 10^{-3}\div 10^{-4}$ in the vicinity of the
maximum frequency, and $\Delta N_{\omega}\sim 10^{-7}\div 10^{-6}$
in the range of X-ray photons with the frequency of the order of
tens of kiloelectron-volts (according to our estimations
\cite{5,17,34}).

%%%%%%%%%%%%%%%%

In the case of excitation of resonance nuclear levels the number
of quanta formed in the interval of the order of the level width
is important, which leads for example, for the number of quanta
produced in  $^{57}Fe$ Mossbauer target to the estimated value of
$\Delta N_{\omega}\sim 10^{-14}$ quanta \cite{34}. The stated
values follow from the formulae given in \cite{36}, if taking into
account that the estimate is given per unit length and the entire
spectral interval.

It should be emphasized that for numerous applications in solid
state physics and other fields, it is necessary to know the number
of photons in a certain narrow frequency interval, rather than in
the entire spectral interval. As a result, in narrow spectral
intervals within the ranges of tens and hundreds of
kiloelectronvolts the so-called parametric radiation often appears
to be much more intense (see Section (\ref{sec:4.15}).

%%%%%%%%%%%%%%%%%%%%%%%%%%%%%%%%%%%  Chapter 4 %%%%%%%%%%%%%%%%%%%%%%

\chapter[The Influence of $\gamma$-Quanta Refraction and Diffraction on ....]{The Influence of $\gamma$-Quanta Refraction and
Diffraction on Angular and Spectral Characteristics of Radiation
Produced by Particles in Crystals} \label{ch:4}

%%%%%%%%%%%%%%%%%%%%%%%%%%%%%%%%%%%  Section 9 %%%%%%%%%%%%%%%%%%%%%%

\section[Radiation in a Refractive Medium]{Radiation in a Refractive Medium}
\label{sec:4.9}

Consider the theory of photon radiation in crystals when the
effects caused by refraction and diffraction are of importance.
The results obtained also describe radiation of diffracted
electrons \cite{69,70,71,72,73}.

Refraction and diffraction are significant when the crystal
thickness is $L>1/k|n-1|$. As shown in Chapter (\ref{ch:2}), in
this case spectral and angular distributions change drastically.
In particular, the effects caused by diffraction lead to the
appearance of radiation at large angles with the spectrum
depending on the effects of anomalous transmission of
$\gamma$-quanta through a crystal \cite{74}. Moreover, diffraction
gives rise to a new, quite a vigorous radiation mechanism, the
so-called parametric mechanism for generating $\gamma$-quanta
\cite{75,76,77} (see also \cite{78,79,80,81}).

Theoretical description of such phenomena requires (see Chapter
(\ref{ch:3}) finding  the transition matrix element $M$ determined
by the photon wave function being the exact solution of
homogeneous Maxwell equations describing propagation of an
electromagnetic wave in a medium. It should be emphasized that the
photon wave function of the type $A^{(-)}$ satisfies Maxwell
equations with the complex conjugate dielectric permittivity
\cite{14}, and ignoring the asymptotic requirements may lead to
the formation of misbehaving wave functions in an absorbing
medium.

%%%%%%%%%%%%%%%%%%%%%%%%%%%%%%55
As before, consider photon emission in a plane-parallel crystal
plate. If the photon exit angle of is not equal to the Wulff-Bragg
angle, then in the X-ray and the frequency ranges with shorter
wavelengths, where $|n-1|\ll 1$, the expression for $A^{-}_{ks}$
has the form \cite{17}
\begin{eqnarray}
\label{9.1}
A^{(-)}_{ks}(\vec{r})& &= \sqrt{4\pi}\left\{\vec{e}_{s}e^{i\vec{k}\vec{r}}e^{-ik_{z}n^{*}L}\theta(-z)\right.\nonumber\\
& &\left.+\vec{e}_{s}e^{i\vec{k}\vec{r}}e^{-ik_{z}n^{*}L}e^{ik_{z}(n^{*}-1)z}\theta(z)\theta(L-z)\right.\nonumber\\
&
&\left.+\vec{e}_{s}e^{i\vec{k}\vec{r}}e^{-ik_{z}L}\theta(z-L)\right\}\,.
\end{eqnarray}
According to (\ref{9.1}) inside the plate with  boundaries $0\leq
z\leq L$
\begin{equation}
\label{9.2} A^{(-)}_{ks}(\vec{r})=
\sqrt{4\pi}\,\vec{e}_{s}e^{i\vec{k}_{1}^{*}\vec{r}}e^{-ik_{z}n^{*}L}\,,
\end{equation}
where $\vec{k}_{1}=(\vec{k}_{\perp}, k_{z}n)$. Comparison of
(\ref{9.2}) and the photon wave function (\ref{6.5}) shows that
taking into account the refractive effects in matrix elements is
reduced to the substitution of vector $\vec{k}^{*}_{1}$ for vector
$\vec{k}$. In other words, all general formulas written out in
Chapter (\ref{ch:3}) preserve their form (for this purpose, we
retained the complex conjugation symbol in $q_{znf}$). As a
result, for example, at $\omega\gg E$ the spectral-angular
distribution of photons has the form\footnote{Formula (\ref{9.3})
is obtained if, when integrating matrix elements only the
integrals over the path inside the crystal are retained, and the
integrals over the  path in a vacuum are discarded. The total
radiation cross-section including vacuum terms is given in
\cite{17}. The vacuum terms are important for a soft spectral
range, when the vacuum coherent radiation length appears to be
comparable with the plate thickness or the quantum absorption
depth.}
\begin{eqnarray}
\label{9.3} \frac{d^{2}N_{s}}{d\omega
d\Omega}=\frac{e^{2}\omega}{\pi^{2}}\texttt{Re}\sum_{nfj}Q_{nj}e^{i\tilde{\Omega}_{nj}L}
\left[\frac{1-\exp(iq^{*}_{zjf}L)}{q^{*}_{zjf}}\right]\nonumber\\
\times\left[\frac{1-\exp(-iq_{znf}L)}{q_{znf}}\right](\vec{g}_{nf}\vec{e}_{s}^{\,*})(\vec{g}_{jf}^{\,*}\vec{e}_{\,s})\,.
\end{eqnarray}
When the crystal thickness $L$ is much greater than the photon
absorption depth $l_{\mathrm{abs}}$ in a crystal, (\ref{9.3})
simplifies
\begin{equation}
\label{9.4} \frac{d^{2}N_{s}}{d\omega
d\Omega}=\frac{e^{2}\omega}{\pi^{2}}\texttt{Re}\sum_{nfj}Q_{nj}e^{i\tilde{\Omega}_{nj}L}\frac{1}{q^{*}_{zjf}q_{znf}}
(\vec{g}_{nf}\vec{e}^{\,*}_{s})(\vec{g}_{jf}^{\,*}\vec{e}_{s})\,.
\end{equation}
%%%%%%%%%%%%%%%%%%%%%%%%%%%%%%
Integration of expression (\ref{9.3}) for the double-differential
radiation spectrum over the angles with maximum opening
$\vartheta_{k}$ equal the collimation angle of the photon beam
that exits the crystal, we obtain the radiation spectrum in the in
the dipole approximation as
\begin{eqnarray}
\label{9.5}
\frac{d N}{d\omega} =\frac{e^{2}}{4\pi}\sum_{nf}Q_{nn}|x_{nf}|^{2}\left\{\frac{[\Omega_{nf}-\omega(1-\beta^{2}n^{\prime 2})]^{2}+\Omega_{nf}^{2}}{\omega^{2}n^{\prime}n^{\prime\prime}}\right.\nonumber\\
\left.\times\left[(1+e^{-2\omega
n^{\prime\prime}L})\left(\arctan\frac{\alpha_{nf}}{n^{\prime\prime}}+
\arctan\frac{\frac{\vartheta^{2}_{k}}{2}-\alpha_{nf}}{n^{\prime\prime}}\right)\right.\right.\nonumber\\
\left.\left.-2\pi e^{-2\omega n^{\prime\prime}L}\theta\left(\frac{\vartheta^{2}_{k}}{2}-\alpha_{nf}\right)\theta(\alpha_{nf})+\frac{1}{2i}\left(\xi_{nf}^{+}\right.\right.\right.\nonumber\\
\left.\left.\left.+\delta_{nf}^{-}-\xi_{nf}^{-}\delta_{nf}^{+}\right)\right]+(1+e^{-2\omega n^{\prime\prime}L})\frac{3\vartheta_{k}^{2}}{2}\right.\nonumber\\
\left.\times\theta\left(\frac{\vartheta_{k}^{2}}{2}-\alpha_{nf}\right)\theta(\alpha_{nf})+\left[2(1-\beta^{2}n^{\prime 2})\right.\right.\nonumber\\
\left.\left.-2\frac{\Omega_{nf}}{\omega}+\frac{\Omega_{nf}^{2}}{\omega^{2}}\right]\left[(1+e^{-2\omega n^{\prime\prime}L})\right.\right.\nonumber\\
\left.\left.\times\ln\frac{\alpha^{2}_{nf}+\left(\frac{n^{\prime\prime}}{n^{\prime}}\right)^{2}}{\left(\frac{\vartheta^{2}_{k}}{2}
-\alpha_{nf}\right)^{2}+\left(\frac{n^{\prime\prime}}{n^{\prime}}\right)^{2}}-(\xi_{nf}^{+}+\delta_{nf}^{-}-\delta_{nf}^{+}+\xi_{nf}^{-})\right]\right\},
\end{eqnarray}
where
\[
\alpha_{nf}=(1-\beta
n^{\prime})\left(\frac{\tilde{\omega}_{nf}}{\omega}-1\right);\,
\tilde{\omega}_{nf}=\frac{\Omega_{nf}}{1-\beta n^{\prime}};
\]
\[
n^{\prime}=\texttt{Re}\,n(\omega);\,
n^{\prime\prime}=\texttt{Im}\,n(\omega);
\]
\begin{eqnarray}
\label{9.6}
\xi^{\pm}_{nf}=E i(-\omega n^{\prime\prime}L\pm iL\omega\alpha_{nf})\nonumber\\
-E i\left[-\omega n^{\prime\prime}L\pm iL\omega\left(\alpha_{nf}-\frac{\vartheta^{2}_{k}}{2}\right)\right];\nonumber\\
\delta^{\pm}_{nf}=e^{-2\omega n^{\prime\prime}L}\left\{Ei\left[\omega n^{\prime\prime}L\pm i L\omega\left(\alpha_{nf}-\frac{\vartheta^{2}_{k}}{2}\right)\right]\right.\nonumber\\
\left.-E i(\omega n^{\prime\prime}L\pm
iL\omega\alpha_{nf})\right\}
\end{eqnarray}
($Ei(z)$ is the integral exponential function ).

As the explicit form of the refractive index was not used in
(\ref{9.5}), this equation also holds true for crystals containing
resonant nuclei. In this case it is essential to take account of
the photon absorption in a crystal (the absorption length in such
crystals can be of the order of $10^{-5}\div 10^{-4}$ cm).

Allowing for absorption is also  necessary when considering
radiation in a relatively soft X-ray spectrum (the absorption
length $l_{abs}$ of the photons with $\omega<10$ keV in a $Si$
crystal proves to be less than $10^{-2}$ cm (Figure 6).

\bigskip

\textit{Figure 6. Spectral distribution of photons in relative
units ($f=\frac{dN\omega}{d\omega}A$ ). Photons are emitted by a
particle  ($E=1$  GeV) in $Si$ (110) crystals (crystal thicknesses
in cm). Dashed curves - photon distribution without absorption;
solid curves - photon distribution with account of absorption.}

\bigskip

In the limiting case ($L\gg l_{abs}$ the expression (\ref{9.5})
for the radiation spectrum simplifies and takes the form
\begin{eqnarray}
\label{9.7}
\frac{d N}{d\omega}=\frac{e^{2}\omega}{2\pi}\sum_{nf}Q_{nn}|x_{nf}|^{2}\left\{\frac{l_{abs}}{\omega}\left[\Omega_{nf}^{2}+(\Omega_{nf}\right.\right.\nonumber\\
\left.\left.-(1-\beta^{2}n^{\prime 2})\omega)^{2}\right]\left(\arctan\frac{\alpha_{nf}}{n^{\prime\prime}}+\arctan\frac{\frac{\vartheta_{k}^{2}}{2}-\alpha_{nf}}{n^{\prime\prime}}\right)\right.\nonumber\\
\left.+\frac{3}{4}\vartheta_{k}^{2}\theta\left(\frac{\vartheta_{k}^{2}}{2}-\alpha_{nf}\right)\theta(\alpha_{nf})+\left[(1-\beta^{2}n^{\prime 2})\right.\right.\nonumber\\
\left.\left.-\frac{\Omega_{nf}}{\omega}+\frac{\Omega_{nf}^{2}}{2\omega^{2}}\right]\ln\frac{\alpha^{2}_{nf}+\left(\frac{n^{\prime\prime}}{n^{\prime}}\right)^{2}}{\left(\frac{\vartheta_{k}^{2}}{2}-\alpha_{nf}\right)^{2}+\left(\frac{n^{\prime\prime}}{n^{\prime}}\right)^{2}}\right\},
\end{eqnarray}
where $l_{abs}=\frac{1}{2\omega n^{\prime\prime}(\omega)}$.

For radiation at a small angle with respect to the direction of
particle motion the last two terms in (\ref{9.7}) are small, and
the spectral intensity of radiation is practically proportional to
the photon absorption length in a crystal:
\begin{eqnarray}
\label{9.8}
\frac{dN}{d\omega}=\frac{e^{2}\omega}{2\pi}\sum_{nf}Q_{nn}|x_{nf}|^{2}\frac{l_{abs}}{\omega}\left[\Omega_{nf}^{2}+\left(\Omega_{nf}\right.\right.\nonumber\\
\left.\left.-(1-\beta^{2}n^{\prime
2})\omega\right)^{2}\right]\left(\arctan\frac{\alpha_{nf}}{n^{\prime\prime}}+\arctan\frac{\frac{\vartheta_{k}^{2}}{2}-\alpha_{nf}}{n^{\prime\prime}}\right).
\end{eqnarray}

The analysis shows that at an arbitrary ratio of the crystal
thickness to the photon absorption depth, quite a simple formula
may express the radiation spectrum in an absorbing crystal with
high accuracy \cite{67}:
\begin{equation}
\label{9.9}
\frac{dN}{d\omega}=l_{abs}(\omega)\left(1-\exp\left(-\frac{L}{l_{abs}(\omega)}\right)\right)\frac{d\tilde{N}}{d\omega},
\end{equation}
where $\frac{d\tilde{N}}{d\omega}$ is the radiation spectrum in
the absence of absorption, i.e., at $n^{\prime\prime}=0$.

%%%%%%%%%%%%%%%%%%%%%%%%%%%%%%%%%%%  Section 10 %%%%%%%%%%%%%%%%%%%%%%

\section[Optical Radiation Produced by Channeled Particles]{Optical Radiation Produced by Channeled Particles}
\label{sec:4.10}

The effects associated with photon refraction in a medium in the
optical spectrum, where the refractive index $n$ (the dielectric
permittivity $\varepsilon(\omega)$) can be appreciably different
from unity, appear to be of particular importance. Formulae
describing radiation of a moving oscillator in a refractive
infinite medium were derived by Frank in \cite{38}. For channeled
(diffracting) particles the presence of boundaries in a crystal is
essential. Classical theory generalizing \cite{38} for the case of
photon emission by an oscillating particle traversing a plate of
finite thickness is given by the author together with I.M. Frank.
In this case the formula describing spectral-angular
characteristics of radiation of the oscillator moving along the
z-axis and oscillating along the x-axis with the amplitude $x_{0}$
has the form
\begin{eqnarray}
\label{10.1} \frac{d^{2}N}{d\omega
d\Omega}=\frac{e^{2}\omega^{3}x_{0}^{2}}{4\pi^{2}\hbar c^{3}}
\left\{|G_{1}|^{2}\cos^{2}\varphi|\beta\varepsilon(\omega)-\sqrt{\varepsilon(\omega)-\sin^{2}\vartheta}|^{2}\right.\nonumber\\
\left.+|G_{2}|^{2}\sin^{2}\varphi|1-\beta\sqrt{\varepsilon(\omega)-\sin^{2}\vartheta}|^{2}\right\}|S(L)|^{2}\,,
\end{eqnarray}
where
\[
S(L)=\frac{\exp\left\{i\left[\omega\left(1-\beta\sqrt{\varepsilon(\omega)-\sin^{2}\vartheta}\right)-\omega_{0}\right]
\frac{L}{v_{z}}\right\}-1}
{i\left[\omega\left(1-\beta\sqrt{\varepsilon(\omega)-\sin^{2}\vartheta}\right)-\omega_{0}\right]}\,,
\]
$\omega_{0}$ is the oscillation frequency of the oscillator in the
laboratory system of coordinates; $\beta=v_{z}/c$; $v_{z}$ is the
velocity along the $z$-axis;
\begin{eqnarray}
G_{1} & = &g_{1}\left(\sqrt{\varepsilon(\omega)-\sin^{2}\vartheta}+\varepsilon(\omega)\cos\vartheta\right)\cos\vartheta;\nonumber\\
G_{2} &=&  g_{2}\left(\sqrt{\varepsilon(\omega)-\sin^{2}\vartheta}+\cos\vartheta\right)\cos\vartheta\,,\nonumber\\
g_{1}
&=&\left[\left(\sqrt{\varepsilon(\omega)-\sin^{2}\vartheta}+\varepsilon(\omega)\cos\vartheta\right)^{2}
\exp\left(-i\frac{L\omega}{c}\sqrt{\varepsilon(\omega)-\sin^{2}\vartheta}\right)\right.\nonumber\\
&
&\left.-\left(\sqrt{\varepsilon(\omega)-\sin^{2}\vartheta}-\varepsilon(\omega)\cos\vartheta\right)^{2}
\exp\left(i\frac{L\omega}{c}\sqrt{\varepsilon(\omega)-\sin^{2}\vartheta}\right)\right]^{-1}\,,\nonumber\\
g_{2}&=&\left[\left(\sqrt{\varepsilon(\omega)-\sin^{2}\vartheta}+\cos\vartheta\right)^{2}
\exp\left(-i\frac{L\omega}{c}\sqrt{\varepsilon(\omega)-\sin^{2}\vartheta}\right)\right.\nonumber\\
&
&\left.-\left(\sqrt{\varepsilon(\omega)-\sin^{2}\vartheta}-\cos\vartheta\right)^{2}
\exp\left(i\frac{L\omega}{c}\sqrt{\varepsilon(\omega)-\sin^{2}\vartheta}\right)\right]^{-1}\,.\nonumber
\end{eqnarray}

To derive the expression
\[\frac{d^{2}N}{d\omega d\Omega}
 \]
describing the distribution of photons produced by a quantum
emitter, suffice it to replace $x_{0}$ by a doubled  matrix
element of the transition from the emitter's coordinate:
$x_{0}^{2}\rightarrow 4|x_{nf}|^{2}$; and the oscillation
frequency $\omega_{0}$ by the transition frequency $\Omega_{nf}$.

Without absorption
\begin{equation}
\label{10.2}
|S(L)|^{2}=4\frac{\sin^{2}[\omega(1-\beta\sqrt{\varepsilon(\omega)-\sin^{2}\vartheta})-
\omega_{0}]\frac{L}{2v_{z}}}{[\omega(1-\beta\sqrt{\varepsilon(\omega)-\sin^{2}\vartheta})-\omega_{0}]^{2}}\,.
\end{equation}
The contribution due to the waves reflected from the vacuum--plate
entrance boundary should also be added to the intensity in
(\ref{10.1}). It is obtained by replacing  $\beta\rightarrow
-\beta$ and the sign "+" between the brackets of the multiplier
appearing in $G_{1(2)}$ after $g_{1(2)}$ with "--".  In the
vicinity of the frequencies and angles for which
$\sqrt{\varepsilon(\omega)-\sin^{2}\vartheta}$ vanishes, the
interference terms may also gain in importance.

%%%%%%%%%%%%%%%%%%%%%%%
In view of the fact that the path $L$ is finite, every angle
$\vartheta$ has a corresponding frequency spectrum, covering the
range $\Delta\omega$, which are close to the Doppler frequency.
Photons with such frequency are emitted  within the finite range
of angles $\Delta\vartheta$ \cite{38}. With increasing $L$ the
range of angles $\Delta\vartheta$ reduces. Thus, following
\cite{38}, in our case we have the below equality for the range
$\Delta\omega$ in the absence of absorption
\begin{eqnarray}
\label{10.3} & &\Delta\omega=\frac{\pm\pi}{|1-\beta
n(\omega_{\vartheta},\vartheta)\cos\vartheta|-\omega_{\vartheta}\frac{dn(\omega_{\vartheta},\,\vartheta)}
{d\omega_{\vartheta}}\cos\vartheta}\frac{2v}{L}\,,\nonumber\\
& &\omega_{\vartheta}(1-\beta
n(\omega_{\vartheta},\,\vartheta)\cos\vartheta)=\omega_{0}\,,
\end{eqnarray}
where we introduce the refractive index
\begin{equation}
\label{10.4}
n(\omega,\vartheta)=\frac{\sqrt{\varepsilon-\sin^{2}\vartheta}}{|\cos\vartheta|}=
\sqrt{1+\frac{\varepsilon(\omega)-1}{\cos^{2}\vartheta}}\,.
\end{equation}
If we are not concerned with the width of the peak, then with high
accuracy
\begin{eqnarray}
\frac{\sin^{2}[\omega(1-\beta\sqrt{\varepsilon(\omega)-\sin^{2}\vartheta})-\omega_{0}]
\frac{L}{2v}}{[\omega(1-\beta\sqrt{\varepsilon(\omega)-\sin^{2}\vartheta})-\omega_{0}]^{2}}\nonumber\\
\simeq\frac{L}{2v}\pi\delta(\omega(1-\beta\sqrt{\varepsilon(\omega)-\sin^{2}\vartheta})-\omega_{0})
\end{eqnarray}
and the number of photons emitted by a linear oscillator is
defined by formula
\begin{eqnarray}
\label{10.5} & &\frac{d^{2}N}{d\omega
d\Omega}=\frac{e^{2}x_{0}^{2}\omega^{3}L}{2\pi\hbar c^{3}v}
\left\{|G_{1}|^{2}\cos^{2}\varphi(\beta\varepsilon(\omega)-\sqrt{\varepsilon(\omega)-\sin^{2}\vartheta})^{2}\right.\nonumber\\
& &\left.+|G_{2}|^{2}\sin^{2}\varphi(1-\beta\sqrt{\varepsilon(\omega)-\sin^{2}\vartheta})^{2}\right\}\nonumber\\
&
&\times\delta(\omega(1-\beta\sqrt{\varepsilon(\omega)-\sin^{2}\vartheta})-\omega_{0})\,.
\end{eqnarray}

%%%%%%%%%%%%%%%%%%%%%%%%%%%

Find the angular distribution of photons with frequencies $\omega$
lying in the range $\omega_{1}\leq \omega \leq\omega_{2}$:
\begin{eqnarray}
\label{10.6} \frac{d N}{d\Omega}=\frac{e^{2}x_{0}^{2}L}{2\pi\hbar
c^{3}v}\sum_{\alpha}\omega_{\vartheta_{\alpha}}^{3}\cal{M}
(\omega_{\vartheta_{\alpha}},\, \vartheta, \varphi)\eta_{\alpha}\nonumber\\
\times\left|1-\beta n(\omega_{\vartheta_{\alpha}},
\vartheta)\cos\vartheta-\beta\omega_{\vartheta_{\alpha}} \frac{d
n(\omega_{\vartheta_{\alpha}},
\,\vartheta)}{d\omega_{\vartheta_{\alpha}}}\cos\vartheta\right|^{-1}\,,
\end{eqnarray}
where $\cal{M}$ denotes the curly bracket, appearing in
(\ref{10.5}), taken at the frequency value
$\omega=\omega_{\vartheta_{\alpha}}$; the sign $\eta_{\alpha}$
reminds that (\ref{10.6}) is nonzero in the range of polar angles
$\vartheta$, which is determined by the direction of the photon
escape with the maximum $\omega_{2}$ and minimum $\omega_{1}$
frequencies;
\[
\omega_{\vartheta_{\alpha}}(1-\beta n(\omega_{\vartheta_{\alpha}},
\vartheta)\cos\vartheta)=\omega_{0}\,.
\]

%%%%%%%%%%%%%%

Now consider spectral distribution. Integration of expression
(\ref{10.5}) with respect to the angle $\varphi$ is reduced to
replacing $\cos^{2}\varphi$ and $\sin^{2}\varphi$ by $\pi$. As a
consequence,
\begin{eqnarray}
\label{10.7} \frac{d
N}{d\omega}=\frac{e^{2}x_{0}^{2}\omega^{3}L}{2\hbar
c^{3}v}\int\limits_{\vartheta_{\mathrm{min}}}^{\vartheta_{\mathrm{max}}}
\left\{|G_{1}|^{2}(\beta\varepsilon-\sqrt{\varepsilon-\sin^{2}\vartheta})^{2}+|G_{2}|^{2}\right.\nonumber\\
\left.\times(1-\beta\sqrt{\varepsilon-\sin^{2}\vartheta})^{2}\right\}\delta(\omega(1-\beta\sqrt{\varepsilon-\sin^{2}\vartheta})
-\omega_{0})\sin\vartheta d\vartheta\,,
\end{eqnarray}
where $\vartheta_{\mathrm{min}}$, $\vartheta_{\mathrm{max}}$ are
the minimum and maximum angles defining the boundaries of the
angular range within which the radiation is detected.

To determine $dN/d\omega$, it is necessary to find the roots of
equation
\begin{equation}
\label{10.8}
\omega(1-\beta\sqrt{\varepsilon-\sin^{2}\vartheta})-\omega_{0}=0\,.
\end{equation}
%%%%%%%%%%%%%%%%%%%%%%%555
From (\ref{10.8}) follows that
\begin{equation}
\label{10.9}
\cos\vartheta_{1,2}=\pm\sqrt{\left(\frac{\omega-\omega_{0}}{\beta\omega}\right)^{2}-(\varepsilon-1)}\,.
\end{equation}
If the radiation propagating at an acute angle relative to the
particle velocity is registered, the contribution to (\ref{10.7})
comes from only one positive root of (\ref{10.9}). As a result,
\begin{eqnarray}
\label{10.10}
\frac{dN}{d\omega}=\frac{e^{2}\omega^{2}x_{0}^{2}L}{2cv^{3}}\left\{|G_{1}|^{2}\left(\beta\varepsilon-
\frac{1}{\beta}+\frac{\omega_{0}}{\omega}\right)^{2}\right.\nonumber\\
\left.+|G_{2}|^{2}\left(\frac{\omega_{0}}{\omega}\right)^{2}\right\}
\left|1-\frac{\omega_{0}}{\omega}\right|\frac{\xi_{\vartheta}}{\cos\vartheta_{1}}\,.
\end{eqnarray}
%%%%%%%%%%%%%%%%%%
Here the functions $G_{1}$ and $G_{2}$, which include the
quantities $\cos\vartheta$ and
$\sqrt{\varepsilon-\sin^{2}\vartheta}$ are expressed in terms of
frequency according to (\ref{10.8}), (\ref{10.9}), e.g.,
$\sqrt{\varepsilon-\sin^{2}\vartheta}=\frac{1}{\beta}(1-\frac{\omega_{0}}{\omega})$.
 The symbol $\xi_{\vartheta}$ reminds that (\ref{10.10}) is nonzero within the frequency range determined by the
 frequency values of the photons escaping at the angle $\vartheta_{\mathrm{max}}$ and $\vartheta_{\mathrm{min}}$.

Note that to describe the phenomena occurring under the anomalous
Doppler effect, it takes only to replace $\omega_{0}$ by
$-\omega_{0}$ (in the quantum case $\Omega_{nf}>0$ under the
normal Doppler effect, and $\Omega_{nf}<0$ under the anomalous
one) in all the above formulas.

The relations derived  simplify appreciably if mirror-reflected
waves can be neglected, i.e., for example, in the case when
$n(\omega,\,\vartheta)$ slightly differs from unity. Under such
conditions with good accuracy
\[
|G_{1}|^{2}=\frac{\cos^{2}\vartheta}{(\varepsilon\cos\vartheta+\sqrt{\varepsilon-\sin^{2}\vartheta})^{2}}\,;\quad
|G_{2}|^{2}=\frac{\cos^{2}\vartheta}{(\cos\vartheta+\sqrt{\varepsilon-\sin^{2}\vartheta})}\,.
\]

%%%%%%%%%%%%%%%%%%%%%%%%%%%%%%%%55
Generalization of formulae (\ref{10.1}) to the two-dimensional
case  (axial channeling) was given by the author together with
I.Ya. Dubovskaya. Spectral-angular distribution of radiation has
the form:

for $\sigma$ -polarization ($\hbar=c=1$)
\begin{eqnarray}
\label{10.11}
\frac{d^{2}N_{\sigma}}{d\omega d\Omega}=\frac{e^{2}\omega}{4\pi^{2}}\left\{\frac{|D_{\sigma}|^{2}}{q_{z}^{2}}\left(\frac{\vec{\beta}[\vec{n}\times\vec{n}_{z}]}{\sin\vartheta}\right)^{2}+\frac{2}{q_{z}}\frac{\vec{\beta}[\vec{n}\times\vec{n}_{z}]}{\sin\vartheta}\right.\nonumber\\
\left.\times \texttt{Im}\sum_{nf}Q_{nf}\Omega_{nf}D_{\sigma}^{*}\left[\left(\frac{1}{q_{z}}+\frac{A^{r}_{\sigma}}{\tilde{q}_{z}}\right)e^{i(p_{zn}-p_{1zf})L}\right.\right.\nonumber\\
\left.\left.+q^{-1}_{znf}B_{\sigma}(e^{iq_{znf}L}-1)+B^{r}_{\sigma}\tilde{q}_{znf}^{-1}(e^{i\tilde{q}_{znf}L}-1)\right]\right.\nonumber\\
\left.\times(\rho_{xnf}\sin\varphi-\rho_{ynf}\cos\varphi)+\texttt{Re}\sum_{nfj}Q_{nj}\Omega_{nf}\Omega_{jf}\right.\nonumber\\
\left.\times[B_{\sigma}^{*}q_{zjf}^{_1*}(e^{-iq^{*}_{zjf}L}-1)+B_{\sigma}^{*r}\tilde{q}_{zjf}^{-1*}(e^{-i\tilde{q}^{*}_{zjf}L}-1)]\right.\nonumber\\
\left.\times\left[B_{\sigma}q_{znf}^{-1}(e^{iq_{znf}L}-1)+B^{r}_{\sigma}\tilde{q}_{znf}^{-1}(e^{i\tilde{q}_{znf}L}-1)\right.\right.\nonumber\\
\left.\left.-2\left(\frac{1}{q_{z}}+\frac{A^{r}_{\sigma}}{\tilde{q}_{z}}\right)e^{i(p_{zn}-p_{1zf})L}\right](\rho_{xnf}\sin\varphi-\rho_{ynf}\cos\varphi)\right.\nonumber\\
\left.\times(\rho_{xjf}^{*}\sin\varphi-\rho^{*}_{yjf}\cos\varphi)-(1-\sin 2\varphi)\right.\nonumber\\
\left.\times
\texttt{Re}\sum_{nj}Q_{nj}e^{i(p_{zn}-p_{zj})L}F_{nj}\right\},
\end{eqnarray}
where
$$
q_{z}=p_{z}-p_{1z}-k_{z}\simeq(1-\beta_{z}\cos\vartheta)\omega;
$$
$$
\tilde{q}_{z}=p_{z}-p_{1z}+k_{z}\simeq(1+\beta_{z}\cos\vartheta)\omega;
$$
$$
q_{znf}=p_{zn}-p_{1zf}-k_{1z}\simeq
\omega(1-\beta_{z}\sqrt{\varepsilon(\omega)-\sin^{2}\vartheta})-\Omega_{nf};
$$
$$
k_{1z}=\omega\sqrt{\varepsilon-\sin^{2}\vartheta};\,\,\tilde{q}_{znf}=p_{zn}-p_{1zf}+k_{1z}
$$
$$
\simeq\omega(1+\beta_{z}\sqrt{\varepsilon(\omega)-\sin^{2}\vartheta})-\Omega_{nf};
$$
$$
A^{r}_{\sigma}=(1-\varepsilon(\omega))(e^{-i\omega
a_{0}L}-e^{i\omega a_{0}L})g_{2};
$$
$$
a_{0}=\sqrt{\varepsilon(\omega)-\sin^{2}\vartheta};\,\,
B_{\sigma}=2G_{2};
$$
$$
B_{\sigma}^{r}=2\cos\vartheta(cos\vartheta-\sqrt{\varepsilon(\omega)-\sin^{2}\vartheta})g_{2};
$$
$$
D_{\sigma}=4a_{0}\cos\vartheta g_{2}; \,\,
Q_{nf}=c_{n}(\vec{p}_{\perp})c_{f}^{*}(\vec{p}_{1\perp});
$$
$\vec{n}=\frac{\vec{k}}{\omega}$: $\vec{n}_{z}$ is the unit vector
along the z-axis.
$$
F_{n\vec{j}}=\frac{1}{E^{2}}N_{\perp}\int
\varphi_{n\vec{\kappa}}(\vec{\rho})\Delta_{\vec{\rho}}\varphi_{j\vec{\kappa}}^{*}(\vec{\rho})d^{2}\rho;
$$

for $\pi$-polarization
\begin{eqnarray}
\label{10.12}
\frac{d^{2}N_{\sigma}}{d\omega d\Omega}=\frac{e^{2}\omega}{4\pi^{2}}\left\{\frac{|D_{\pi}|^{2}|\beta_{z}-(\vec{\beta}\vec{n})\cos\vartheta|^{2}}{q_{z}^{2}\sin^{2}\vartheta}+\beta_{z}^{2}\sin^{2}\vartheta\right.\nonumber\\
\left.\times\left|\frac{1}{q_{z}}+\frac{A^{r}_{\pi}}{\tilde{q}_{z}}\right|^{2}+2 \texttt{Im}\sum_{nf}Q_{nf}\frac{D_{\pi}^{*}(\beta_{z}-(\vec{\beta}\vec{n})\cos\vartheta)}{q_{z}\sin\vartheta}\right.\nonumber\\
\left.\times\left[(\beta_{z}\omega\sin^{2}\vartheta-\Omega_{nf}a_{0})B_{\pi}q^{-1}_{znf}(e^{iq_{znf}L}-1)\right.\right.\nonumber\\
\left.\left.+(\beta_{z}\omega\sin^{2}\vartheta+\Omega_{nf}a_{0})B_{\pi}^{r}\tilde{q}^{-1}_{znf}(e^{i\tilde{q}_{znf}L}-1)\right.\right.\nonumber\\
\left.\left.+(\beta_{z}\omega\sin^{2}\vartheta+\Omega_{nf}\cos\vartheta)q^{-1}_{z}e^{i(p_{zn}-p_{1zf})L}\right.\right.\nonumber\\
\left.\left.+(\beta_{z}\omega\sin^{2}\vartheta+\Omega_{nf}\cos\vartheta)A^{r}_{\pi}\tilde{q}^{-1}_{z}e^{i(p_{zn}-p_{1zf})L}\right]\right.\nonumber\\
\left.\times(\rho_{xnf}\cos\varphi+\rho_{ynf}\sin\varphi)+ \texttt{Re}\sum_{nfj}Q_{nj}\left[\left(\beta_{z}\omega\sin^{2}\vartheta\right.\right.\right.\nonumber\\
\left.\left.\left.-\Omega_{jf}a_{0}^{*}\right)B_{\pi}^{*}q_{zjf}^{-1*}(e^{-iq^{*}_{zjf}L}-1)+\left(\beta_{z}\omega\sin^{2}\vartheta\right.\right.\right.\nonumber\\
\left.\left.\left.+\Omega_{jf}a_{0}^{*}\right)B_{\pi}^{r*}\tilde{q}_{zjf}^{-1*}(e^{-i\tilde{q}^{*}_{zjf}L}-1)\right]\left[(\beta_{z}\omega\sin^{2}\vartheta-\Omega_{nf}a_{0})B_{\pi}q^{-1}_{znf}\right.\right.\nonumber\\
\left.\left.\times(e^{iq_{znf}L}-1)+(\beta_{z}\omega\sin^{2}\vartheta+\Omega_{nf}a_{0})B_{\pi}^{r}\tilde{q}^{-1}_{znf}(e^{i\tilde{q}_{znf}L}-1)\right.\right.\nonumber\\
\left.\left.-2(\beta_{z}\omega\sin^{2}\vartheta-\Omega_{nf}\cos\vartheta)q^{-1}_{z}e^{i(p_{zi}-p_{1zf})L}\right.\right.\nonumber\\
\left.\left.-2(\beta_{z}\omega\sin^{2}\vartheta+\Omega_{nf}\cos\vartheta)A^{r}_{\pi}\tilde{q}_{z}^{-1}e^{i(p_{zn}-p_{1zf})L}\right]\right.\nonumber\\
\left.\times(\rho_{xnf}\cos\varphi+\rho_{ynf}\sin\varphi)(\rho^{*}_{xjf}\cos\varphi+\rho_{yjf}^{*}\sin\varphi)\right.\nonumber\\
\left.-2\beta_{z}\sin\vartheta\cos\vartheta \texttt{Im}\left[\left(\frac{1}{q_{z}}+\frac{A_{\pi}^{r}}{\tilde{q}_{z}}\right)\left(\frac{1}{q_{z}}-\frac{A_{\pi}^{r*}}{\tilde{q}_{z}}\right)\sum_{nj}Q_{nj}\right.\right.\nonumber\\
\left.\left.\times e^{-i\tilde{\Omega}_{nj}L}\tilde{\Omega}_{nj}(\rho_{xnj}\cos\varphi+p_{ynj}\sin\varphi)\right]-\cos^{2}\vartheta(1+\sin2\varphi)\right.\nonumber\\
\left.\times\left|\frac{1}{q_{z}}-\frac{A_{\pi}^{r}}{\tilde{q}_{z}}\right|^{2}
\texttt{Re}\sum_{nj}Q_{nj}e^{-i\tilde{\Omega}_{nj}L}F_{nj}\right\};
\end{eqnarray}
\[
A_{\pi}^{r}=(\varepsilon^{2}(\omega)\cos^{2}\vartheta-a_{0}^{2})(e^{-i\omega
a_{0}L}-e^{i\omega a_{0}L})g_{1};
\]
\[
B_{\pi}=2\cos\vartheta(\varepsilon(\omega)\cos\vartheta+a_{0})g_{1}=2G_{1};
\]
\[
B_{\pi}^{r}=2\cos\vartheta(a_{0}-\varepsilon\cos\vartheta)g_{1};\,\,
D_{\pi}=4a_{0}\cos\vartheta\varepsilon g_{1}.
\]

%%%%%%%%%%%%%%%%%%%%%%%%%%%%%%%%%%%  Section 11 %%%%%%%%%%%%%%%%%%%%%%

\section[Angular Distribution of Radiation Produced by Particles in a Crystal under Refraction]{Angular Distribution of Radiation Produced by Particles in a Crystal under Refraction}
\label{sec:4.11}

Let us give a more detailed treatment of $dN/d \Omega$ in the case
of radiation in a crystal whose thickness exceeds the photon
absorption length in a medium, i.e., assume that the condition
$\omega \texttt{Im}\,n(\omega)L\gg 1$ is satisfied. In this case
in order to integrate the cross-section (\ref{9.4}) with respect
to frequencies, we shall make use of the fact that the function
$1/|q_{znf}|^{2}$ has a sharp maximum in the vicinity of the point
$\omega_{v}$, where $\texttt{Re}\, q_{znf}=0$. At the same time
other terms appearing in (\ref{9.4}) change smoothly with the
change in the photon frequency. Therefore the function before
$|q_{znf}|^{-2}$ may be factored outside the integral sign at the
maximum point $\omega=\omega_{\vartheta}$.

Expand $\texttt{Re}\,q_{z}$ into a series in the vicinity of the
point  $\omega=\omega_{\vartheta}$:
\begin{equation}
\label{11.1} \texttt{Re}\,
q_{z}\simeq\frac{d(\texttt{Re}\,q_{z})}{d\omega}(\omega-\omega_{\vartheta})+\frac{1}{2}\frac{d^{2}
(\texttt{Re}\,q_{z})}{d\omega^{2}}(\omega-\omega_{\vartheta})^{2}
\end{equation}
and expand the limits of integration to the infinite interval. As
a result, the angular distribution of photons emitted by a
channeled particle is written as follows:
\begin{eqnarray}
\label{11.2}
& &\frac{dN}{d\Omega}=\sum_{nfa}Q_{nn}|x_{nf}|^{2}\frac{e^{2}l_{abs}(\omega^{(a)}_{\vartheta})\Omega_{nf}^{3}}{2\pi}\nonumber\\
& &\times\frac{[1-\beta n^{\prime}(\omega^{(a)}_{\vartheta})\cos\vartheta]^{2}-[1-\beta^{2}n^{\prime 2}(\omega^{(a)}_{\vartheta})]\sin^{2}\vartheta\cos^{2}\varphi}{[1-\beta n^{\prime}(\omega^{(a)}_{\vartheta})\cos\vartheta]^{4}}\nonumber\\
&
&\times\Delta_{nf}(\omega^{(a)}_{\vartheta})\eta_{a}(\omega_{1},\omega_{2})\,,
\end{eqnarray}
where $l_{\mathrm{abs}}(\omega^{(a)}_{\vartheta})$ is the
absorption length of the photon with the frequency
$\omega^{(a)}_{\vartheta}$.

%%%%%%%%%%%%%%%%%%%%%55555
Summation over $(a)$ indicates summation over all possible
solutions of equation $\texttt{Re}\, q_{z}(\omega)=0$ in the
spectral range of the detector $[\omega_{1},\omega_{2}]$. The term
$\Delta_{nf}$ is due to the dispersion of the medium, and it has
the form
\begin{eqnarray}
\label{11.3} \Delta_{nf}(\omega^{(a)}_{\vartheta})&=&\frac{1-\beta
n(\omega^{(a)}_{\vartheta})\cos\vartheta} {\left|1-\beta
n(\omega^{(a)}_{\vartheta})\cos\vartheta-\omega^{(a)}_{\vartheta}\beta\cos\vartheta\left(\frac{\partial
n(\omega)}
{\partial\omega}\right)_{\omega=\omega^{(a)}_{\vartheta}}\right|}\nonumber\\
& =
&\left|1-\frac{(\omega^{(a)}_{\vartheta})^{2}\beta\cos\vartheta}{\Omega_{nf}}\left(\frac{\partial
n(\omega)}
{\partial\omega}\right)_{\omega=\omega^{(a)}_{\vartheta}}\right|^{-1}\,.
\end{eqnarray}
%%%%%%%%%%%%%%%%%%%%%55

The angular distribution of photons for which
$\omega(n^{\prime}-1)L\gg 1$, $l_{\mathrm{abs}}<L$ is obtained by
replacing $l_{\mathrm{abs}}$ in (\ref{11.2})  with the crystal
thickness. Due to a particular relationship between the observed
frequency  and the photon emission angle, the shape of angular
distribution  depends significantly on the frequency value within
the detection range $(\omega_{1},\omega_{2})$.  The function
$\eta_{a}(\omega_{1},\omega_{2})$ takes account of this
circumstance. For example, if the frequencies $\omega_{1}$ and
$\omega_{2}$ lie in the X-ray spectrum, where
\[
n^{\prime}(\omega)=1-\frac{\omega^{2}_{L}}{2\omega^{2}}\,,
 \]
the function   $\eta_{a}(\omega_{1},\omega_{2})$ may be
represented as
\begin{eqnarray}
\label{11.4}
\eta_{a}(\omega_{1},\omega_{2})=\left\{\begin{array}{l}
                                  \theta(b_{nf}(\omega_{a})-\cos\vartheta)\theta(\cos\vartheta-\cos\vartheta_{m})
                                  \,\quad\mbox{at}\quad  \omega_{1}<\omega_{0}<\omega_{2}\,, \\
                                  \theta(b_{nf}(\omega_{1})-\cos\vartheta)\theta(\cos\vartheta-b_{nf}(\omega_{2}))
                                  \quad\mbox{at}\quad\omega_{1},\quad \omega_{2}>\omega_{0}\,, \\
                                  \theta(b_{nf}(\omega_{2})-\cos\vartheta)\theta(\cos\vartheta-b_{nf}(\omega_{1}))
                                 \quad\mbox{at}\quad\omega_{0}>\omega_{1},\,\,\omega_{2}\,,
                              \end{array}\right.\nonumber\\
\end{eqnarray}
here $\omega_{0}=\omega^{2}_{L}/\Omega_{nf}$.
%%%%%%%%%%%%%%%%%5
The multiplier
\begin{eqnarray}
\label{11.5}
\Delta_{nf}(\omega_{\vartheta}^{(a)})\simeq\left|1-\frac{\omega^{2}_{L}}{\omega_{\vartheta}^{(a)}\Omega_{nf}}\right|^{-1}\,,\nonumber\\
b_{nf}(\omega)=\frac{1}{\beta}\left(1-\frac{\Omega_{nf}}{\omega}\right)\,,
\end{eqnarray}
where $\omega_{\vartheta}^{(a)}$ is defined by the formula
\begin{equation}
\label{11.6}
\omega_{\vartheta}^{(a)}=\frac{\Omega_{nf}\pm\sqrt{\Omega_{nf}^{2}-2\omega^{2}_{L}(1-\beta\cos\vartheta)}}{2(1-\beta\cos\vartheta)}\,.
\end{equation}
According to (\ref{11.5}), for the radiation angles $\vartheta$,
at which
\[
\omega_{\vartheta}^{(a)}\gg\frac{\omega^{2}_{L}}{\Omega_{nf}}\,,
\]
the multiplier $\Delta_{nf}(\omega_{\vartheta}^{(a)})$ may be
assumed equal to unity and consequently, the effect of the
frequency dispersion of the medium on the angular distribution of
quanta may be neglected.
%%%%%%%%%%%%%%%%5

To define the frequency range of the spectrum, where $\Delta_{nf}$
being different from unity is of importance, rewrite (\ref{11.3})
as follows
\begin{equation}
\label{11.7}
\Delta_{nf}=\left(1-\frac{\beta\cos\vartheta}{v_{\mathrm{ph}}(\omega_{\vartheta}^{(a)})}\right)
\left(1-\frac{\beta\cos\vartheta}{W(\omega_{\vartheta}^{(a)})}\right)^{-1}\,,
\end{equation}
where $v_{\mathrm{ph}}(\omega_{\vartheta}^{(a)})$ and
$W(\omega_{\vartheta}^{(a)})$ are the phase and group velocities
of light in the medium at the frequency
$\omega=\omega_{\vartheta}^{(a)}$.

According to (\ref{11.7}), the multiplier $\Delta_{nf}$ becomes
essential for photons with the frequencies, at which the group and
phase velocities in a medium are different. This occurs, for
example, in the vicinity of the resonances.

%%%%%%%%%%%%%%%%%%%%%%%%%%5

The expression for angular distribution (\ref{11.2}) simplifies if
the condition
\[
\delta=\frac{2\omega^{2}_{L}}{\Omega^{2}_{nf}}(1-\beta\cos\vartheta)\ll
1
 \]
 is fulfilled (e.g. for an electron with the energy $E\geq 100$\,MeV  at $\vartheta=0$ in a silicon crystal $\delta<10^{-5}$).
 Then the root of (\ref{11.6}) may be decomposed, which gives
\begin{equation}
\label{11.8}
\omega_{\vartheta}^{(1)}\simeq\frac{\Omega_{nf}}{1-\beta\cos\vartheta}
\end{equation}
for the upper frequency radiation mode and
\begin{equation}
\label{11.9}
\omega_{\vartheta}^{(2)}\simeq\frac{\omega^{2}_{L}}{2\Omega_{nf}}\left(1+\frac{\omega^{2}_{L}}
{\Omega_{nf}^{2}}\left(\frac{1}{\gamma^{2}}+\vartheta^{2}\right)\right)
\end{equation}
for the lower one. In view of (\ref{11.9}), at the lower mode, the
observed photon frequency is practically independent of the
radiation angle $\vartheta$. As a result, for
$\omega_{1}<\omega_{0}<\omega_{2}$, the expression for angular
distribution takes the form
%%%%%%%%%%%%%%%%%555
\begin{eqnarray}
\label{11.10} \frac{dN}{d\Omega}& &
=\frac{e^{2}L}{2\pi}\sum_{nf}Q_{nn}|x_{nf}|^{2}\left\{\Omega_{nf}^{3}\Phi(\vartheta,\,\varphi)\theta
\left[\cos\vartheta -
b(\omega_{0})\right]\theta\left[b(\omega_{2})-\cos\vartheta\right]\right.\nonumber\\
& &\left.+\frac{\omega^{6}_{L}\beta^{2}}
{8\Omega_{nf}^{3}}\sin^{4}\vartheta\cos^{2}\varphi-\frac{\omega^{4}_{L}\beta}{2\Omega_{nf}}
\sin^{2}\vartheta\cos\vartheta\cos^{2}\varphi\right.\nonumber\\
&
&\left.+\Omega_{nf}\omega^{2}_{L}(1-\sin^{2}\vartheta\cos^{2}\varphi)\right\}\,,
\end{eqnarray}
where $\omega_{0}=\omega^{2}_{L}/\Omega_{nf}$. If
$\omega_{1,2}>\omega_{0}$, the angular distribution of radiation
is described by (\ref{8.4}).

At a certain energy $E_{\mathrm{cr}} $ of a channeled particle (or
at a fixed particle energy for a limiting radiation angle
$\vartheta_{\mathrm{max}}$), being such that the condition
\begin{equation}
\label{11.11}
\Omega_{nf}^{2}-2\omega^{2}_{L}(1-\beta\cos\vartheta)=0
\end{equation}
is fulfilled, the difference between the frequencies
$\omega_{\vartheta}^{(1)}$  and  $\omega_{\vartheta}^{(2)}$
disappears, and at an angle $\vartheta_{\mathrm{max}}$  a
$\gamma$-quantum with the frequency
$\omega_{0}=\omega_{L}^{2}/\Omega_{nf}$ is emitted (at the angles
$\Omega>\Omega_{\mathrm{max}}$, photon emission by channeled
particles is impossible).

%%%%%%%%%%%%%%%%%%%%%%%%55

Under the conditions of one frequency observation the photon group
velocity $W(\omega_{0})$ is equal to the projection of the
particle velocity along the direction of $\gamma$-quantum emission
$v \cos\vartheta$ \cite{38}. In this case expression (\ref{11.2})
is not applicable for describing angular distribution, as the
first derivative in the expansion of (\ref{11.1}) to which we
confined ourselves when calculating (\ref{11.2}) vanishes.
Therefore, the quadratic expansion terms in (\ref{11.1}) should be
taken into account when integrating the differential cross-section
in (\ref{9.4}) over the frequencies:
\begin{equation}
\label{11.12}
\texttt{Re}q_{z}\simeq-\frac{\beta\cos\vartheta}{2}\left(2\frac{dn(\omega)}{d\omega}+\omega\frac{d^{2}n(\omega)}
{d\omega^{2}}\right)_{\omega=\omega_{\vartheta}}(\omega-\omega_{\vartheta})^{2}\,.
\end{equation}

As a result, when the frequency
$\omega_{0}=\omega_{L}^{2}/\Omega_{nf}$ is within the range
$(\omega_{1},\omega_{2})$, which is the domain of integration of
the detector, the number of $\gamma$-quanta emitted by a channeled
particle over the angular range $\Delta\vartheta$ near the angle
$\vartheta=\vartheta_{max}$ is given by the following expression:
\begin{eqnarray}
\label{11.13} &
&\frac{dN}{d\varphi}\simeq\frac{e^{2}\omega^{3}_{L}}{\sqrt{2}}\sum_{nf}Q_{nn}|x_{nf}|^{2}
\sqrt{\Omega_{nf}}l_{\mathrm{abs}}^{3/2}(\omega_{0})\\
&
&\times\left\{1-\cos\varphi\left(1-\frac{\omega^{4}_{L}}{\gamma^{4}\Omega^{4}_{nf}}\right)\right\}
\theta_{\mathrm{max}}\Delta\theta\,,\quad
\theta_{\mathrm{max}}\simeq\left(\frac{\Omega^{2}_{nf}}
{\omega^{2}_{L}}-\frac{1}{\gamma^{2}}\right)^{1/2}\nonumber\,,
\end{eqnarray}
where for a lower mode $\Delta\theta\sim\theta_{\mathrm{max}}$\\
 and for an upper mode
\[
\theta_{\mathrm{max}}\Delta\theta\sim\frac{\Omega_{nf}}{\omega_{L}}\sqrt{n^{\prime\prime}}\,.
\]

The characteristic feature of angular distribution near the
critical confluence point of the two frequencies is the following
dependence of $dN/d\varphi$ on the absorption length:
$l_{\mathrm{abs}}^{3/2}$ (compare \cite{37}).

%%%%%%%%%%%%%%%%%%%%%%%%%%%%%%%%%%%  Section 12 %%%%%%%%%%%%%%%%%%%%%%

\section[Influence of Diffraction on the Process of Photon Emission in Crystals]
{Influence of Diffraction on the Process of Photon Emission in
Crystals} \label{sec:4.12}

Diffraction of produced photons in a crystal gives rise to a new
phenomenon: emission of $\gamma$-quanta at large angles with
respect to the direction of the fast particle motion, and
formation of a characteristic diffraction pattern. Two
fundamentally different mechanisms contribute to the latter
\cite{5,17}: one caused by the deceleration of electrons in a
single crystal, being most pronounced in the process of photon
emission through radiative transitions between the bands (levels)
of transverse energy; the other, occurring even for a particle
moving at a constant velocity, is due to scattering of
pseudo-photons associated with a particle by atoms and crystal
nuclei (the so-called parametric radiation
\cite{74,75,76,77,78,79,80,81}). The number of photons in the
diffraction peak appears to be quite large, which enables
obtaining information about the crystal structure directly from
the analysis of the frequency and angular photon spectra. At the
particle energy exceeding several tens of megaelectron volts the
spectral density of radiation caused by a parametric mechanism in
the frequency range  up to several hundreds of kiloelectron volts,
proves to be one or two orders of magnitude higher than density of
radiation emerging at radiative transitions between the levels of
the particle transverse energy \cite{82}.

As mentioned above, to determine the radiation intensity, one
should first find the photon wave function $A_{ks}^{(-)}$ under
diffraction conditions. If the photon wave length is comparable
with a lattice spacing, $A_{ks}^{(-)}$  may be found using the
two-wave approximation of the dynamical theory of diffraction. If
it is much less than the lattice spacing, the theory developed for
the case of  electron channeling is applicable (see Section
(\ref{sec:1.1}, \ref{sec:1.2}) \cite{83,84}.

In the two-wave approximation of the dynamical theory of
diffraction (see, for example, \cite{85}) the wave function
$A_{ks}^{(-)}$ may be represented in the general form as follows:
\begin{equation}
\label{12.1}
A_{ks}^{(-)}(\vec{r})=\vec{e}_{s}\Phi(z)e^{i\vec{k}\vec{r}}+\vec{e}_{1s}\Phi_{1}(z)e^{k_{1}r},
\end{equation}
where $\vec{e}_{s}$ and $\vec{e}_{1s}$ are the polarization
vectors of the direct and diffracted waves satisfying the the
transversality condition:
$(\vec{e}_{s}\vec{k})=(\vec{e}_{1s}\vec{k}_{1})=0$;
$\vec{k}_{1}=\vec{k}+2\pi\vec{\tau}$; $s=1, 2$;
$\vec{e}_{1}\parallel\vec{e}_{11}\parallel[\vec{k},\,
2\pi\vec{\tau}]$; $e_{2}\parallel[\vec{k}[\vec{k}, \,2\pi\tau]]$;
$\vec{e}_{12}\parallel[\vec{k}_{1}[\vec{k}, \,2\pi\vec{\tau}]]$;
$2\pi\vec{\tau}$ is the reciprocal lattice vector characterizing
the family of planes, where the photon diffraction occurs. Note
here that in the general case of diffraction in polarized and
magnetically ordered crystals equations (\ref{12.1}) turns out to
be more complicated. Methods of constructing solutions describing
such a diffraction see in \cite{14,86}.

The photon wave functions corresponding to various cases of the
Laue  and Bragg diffraction only differ by the shape of the
amplitudes $\Phi(z)$ and $\Phi_{1}(z)$:

a. The Bragg case ($k_{z}>0$, $k_{z}+2\pi\tau_{z}<0$):
\begin{eqnarray}
\label{12.2}
A_{ks}^{(-)}(\vec{r})=\vec{e}_{s}[\gamma^{0*}_{1s}+\gamma_{2s}^{0*}]e^{i\vec{k}\vec{r}}e^{-ik_{z}L}\theta(-z)\nonumber\\
+\left\{\vec{e}_{s}e^{i\vec{k}\vec{r}}e^{-ik_{z}L}\left[\gamma^{0*}_{1s}e^{i\frac{\omega}{\gamma_{0}}\varepsilon^{*}_{1s}z}+\gamma_{2s}^{0*}e^{i\frac{\omega}{\gamma_{0}}\varepsilon^{*}_{2s}z}\right]\right.\nonumber\\
\left.+\vec{e}_{1s}\beta_{1}\gamma_{s}^{\tau *}e^{i\vec{k}_{1}\vec{r}}e^{-ik_{z}L}\left[e^{i\frac{\omega}{\gamma_{0}}\varepsilon^{*}_{1s}z}-e^{i\frac{\omega}{\gamma_{0}}\varepsilon^{*}_{2s}z}\right]\right\}\nonumber\\
\times\theta(z)\theta(L-z)+\left\{e_{s}e^{i\vec{k}\vec{r}}e^{-ik_{z}L}+\vec{e}_{1s}\beta_{1}\gamma_{s}^{\tau *}\right.\nonumber\\
\left.\times\left[e^{i\frac{\omega}{\gamma_{0}}\varepsilon^{*}_{1s}L}+e^{i\frac{\omega}{\gamma_{0}}\varepsilon^{*}_{2s}L}\right]e^{i\vec{k}_{1}\vec{r}}e^{-ik_{z}L}\right\}\theta(z-L),
\end{eqnarray}
where
$$
\gamma_{0}=\cos\theta;\,\,
\theta=\frac{\vec{k}\vec{n}_{z}}{\omega};\,\,
\vec{k}_{1}=\vec{k}+2\pi\vec{\tau};
$$
$$
\gamma^{0}_{1,2s}=\pm\frac{2\varepsilon_{2,1s}-g_{00}}{\Delta_{s}^{*}};\,\,
\gamma_{s}^{\tau}=\frac{g_{10}^{s}}{\Delta_{s}^{*}};
$$
$$
\Delta_{s}=(2\varepsilon_{2s}^{*}-g_{00}^{*})e^{i\frac{\omega}{\gamma_{0}}\varepsilon^{*}_{1s}L}-(2\varepsilon_{1s}^{*}-g_{00}^{*})e^{i\frac{\omega}{\gamma}\varepsilon^{*}_{2s}L};
$$
$$
\varepsilon_{1,2s}=\frac{1}{4}\left\{g_{00}+\beta_{1}g_{11}-\beta_{1}\alpha
\pm
    \sqrt{\left(g_{00}+\beta_{1}g_{11}-\alpha\beta_{1}\right)^{2}-4\beta_{1}(\alpha
    g_{00}-g_{00}g_{11}+g^{s}_{10}g^{s}_{01})
    }
\right\};
$$
$$
\beta_{1}=\frac{k_{z}}{k_{z}+2\pi\tau_{z}}=\frac{k_{z}}{k_{1z}};\,\,
\alpha=\frac{2(\vec{k}2\pi\vec{\tau})+(2\pi\vec{\tau})^{2}}{\omega^{2}};
$$
the quantities $g^{s}_{ik}$ are determined by the expansion of the
dielectric permittivity of a crystal into a series in terms of
reciprocal lattice vectors. The crystal dielectric permittivity is
a periodic function of the position of nuclei and atoms;

b. the Bragg case ($k_{z}<0, \, k_{z}+2\pi\tau_{z}>0$):
\begin{eqnarray}
\label{12.3}
A_{ks}^{(-)}(\vec{r})=\left\{\vec{e}_{s}e^{i\vec{k}\vec{r}}-\vec{e}_{1s}\beta_{1}\gamma_{s}^{\tau *}e^{i\vec{k}_{1}\vec{r}}\right.\nonumber\\
\left.\times\left[e^{-i\frac{\omega}{|\gamma_{0}|}\varepsilon^{*}_{2s}L}
-e^{-i\frac{\omega}{|\gamma_{0}|}\varepsilon^{*}_{1s}L}\right]\right\}\theta(-z)\nonumber\\
+\left\{\vec{e}_{s}e^{i\vec{k}\vec{r}}\left[\gamma_{1s}^{0*}e^{-i\frac{\omega}{|\gamma_{0}|}(\varepsilon^{*}_{2s}L+\varepsilon^{*}_{1s}z)}+\gamma_{2s}^{0*}e^{-i\frac{\omega}{|\gamma_{0}|}(\varepsilon^{*}_{1s}L+\varepsilon^{*}_{2s}z)}\right]\right.\nonumber\\
\left.-\vec{e}_{1s}\beta_{1}\gamma_{s}^{\tau *}e^{i\vec{k}_{1}{r}}\left[e^{-i\frac{\omega}{|\gamma_{0}|}(\varepsilon^{*}_{2s}L+\varepsilon^{*}_{1s}z)}-e^{-i\frac{\omega}{|\gamma_{0}|}(\varepsilon^{*}_{1s}L+\varepsilon^{*}_{2s}z)}\right]\right\}\nonumber\\
\times\theta(z)\theta(L-z)+\vec{e}_{s}e^{i\vec{k}\vec{r}}\left[\gamma^{0*}_{1s}e^{-i\frac{\omega}{|\gamma_{0}|}(\varepsilon^{*}_{2s}+\varepsilon^{*}_{1s})L}\right.\nonumber\\
\left.+\gamma^{0*}_{2s}e^{-i\frac{\omega}{|\gamma_{0}|}(\varepsilon^{*}_{1s}+\varepsilon^{*}_{2s})L}\right]e^{-ik_{z}L}\theta(z-L);
\end{eqnarray}

c. the Laue case ($k_{z}>0,\, k_{z}+2\pi\tau_{z}>0$):
\begin{eqnarray}
\label{12.4}
A_{ks}^{(-)}(\vec{r})=\left\{\vec{e}_{s}e^{i\vec{k}\vec{r}}\left[-\xi_{1s}^{0*}e^{-i\frac{\omega}{\gamma_{0}}\varepsilon^{*}_{1s}L}-\xi_{2s}^{0*}e^{-i\frac{\omega}{\gamma_{0}}\varepsilon^{*}_{2s}L}\right]\right.\nonumber\\
\left.+\vec{e}_{1s}e^{i\vec{k}_{1}\vec{r}}\beta_{1}\left[\xi_{1s}^{\tau *}e^{-i\frac{\omega}{\gamma_{0}}\varepsilon^{*}_{1s}L}+\xi_{2s}^{\tau *}e^{-i\frac{\omega}{\gamma_{0}}\varepsilon^{*}_{2s}L}\right]\right\}\theta(-z)\nonumber\\
+\left\{\vec{e}_{s}e^{i\vec{k}\vec{r}}\left[-\xi_{1s}^{0*}e^{-i\frac{\omega}{\gamma_{0}}\varepsilon^{*}_{1s}(L-z)}-\xi_{2s}^{0*}e^{-i\frac{\omega}{\gamma_{0}}\varepsilon^{*}_{2s}(L-z)}\right]\right.\nonumber\\
\left.+\vec{e}_{1s}\beta_{1}e^{i\vec{k}_{1}\vec{r}}\left[\xi_{1s}^{\tau *}e^{-i\frac{\omega}{\gamma_{0}}\varepsilon^{*}_{1s}(L-z)}+\xi_{2s}^{\tau *}e^{-i\frac{\omega}{\gamma_{0}}\varepsilon^{*}_{2s}(L-z)}\right]\right\}\nonumber\\
\times\theta(z)\theta(L-z)+\vec{e}_{s}e^{i\vec{k}\vec{r}}e^{-ik_{z}L}\theta(z-L),
\end{eqnarray}
where
$$
\xi_{1,2s}^{0}=\mp\frac{2\varepsilon_{2,
1s}-g_{00}}{2(\varepsilon_{2s}-\varepsilon_{1s})};\,\,
\xi_{1,2s}^{\tau}=\mp\frac{g_{01}^{s}}{2(\varepsilon_{2s}-\varepsilon_{1s})};
$$

d. the Laue case ($k_{z}<0,\, k_{z}+2\pi\tau_{z}<0$):
\begin{eqnarray}
\label{12.5}
A_{ks}^{(-)}(\vec{r})=\vec{e}_{s}e^{i\vec{k}\vec{r}}\theta(-z)+\left\{\vec{e}_{s}e^{i\vec{k}\vec{r}}\left[-\xi_{1s}^{0*}e^{-i\frac{\omega}{|\gamma_{0}|}\varepsilon^{*}_{1s}z}\right.\right.\nonumber\\
\left.\left.-\xi_{2s}^{0*}e^{-i\frac{\omega}{|\gamma_{0}|}\varepsilon^{*}_{2s}z}\right]+\vec{e}_{1s}\beta_{1}e^{i\vec{k}_{1}\vec{r}}\left[\xi_{1s}^{\tau *}e^{-i\frac{\omega}{|\gamma_{0}|}\varepsilon^{*}_{1s}z}\right.\right.\nonumber\\
\left.\left.-\xi_{2s}^{\tau *}e^{-i\frac{\omega}{|\gamma_{0}|}\varepsilon^{*}_{2s}z}\right]\right\}\theta(z)\theta(L-z)\nonumber\\
+\left\{\vec{e}_{s}^{i\vec{k}\vec{r}}\left[-\xi_{1s}^{0*}e^{-i\frac{\omega}{|\gamma_{0}|}\varepsilon^{*}_{1s}L}-\xi_{2s}^{0*}e^{-i\frac{\omega}{|\gamma_{0}|}\varepsilon^{*}_{2s}L}\right]\right.\nonumber\\
\left.+\vec{e}_{1s}\beta_{1}e^{i\vec{k}_{1}\vec{r}}e^{i2\pi\tau_{z}L}\left[\xi_{1s}^{\tau *}e^{-i\frac{\omega}{|\gamma_{0}|}\varepsilon^{*}_{1s}L}\right.\right.\nonumber\\
\left.\left.+\xi_{1s}^{\tau
*}e^{-i\frac{\omega}{|\gamma_{0}|}\varepsilon^{*}_{2s}L}\right]\right\}\theta(z-L).
\end{eqnarray}

%%%%%%%%%%%%%%%%%%%%%%%%%%%%%%%%%%%  Section 13 %%%%%%%%%%%%%%%%%%%%%%

\section[Spectral-Angular Distribution in the Bragg and Laue Cases]{Spectral-Angular Distribution in the Bragg and Laue Cases}
\label{sec:4.13}

Consider the influence of diffraction on spectral-angular
distribution of photons emitted by a particle passing through a
crystal. The particle enters the crystal at a certain small angle
with respect to the z-axis directed perpendicular to the crystal
surface.

Let, for example, a photon be diffracted by a family of
crystallographic planes described by the reciprocal lattice vector
$2\pi\vec{\tau}$, which is directed  anti-parallel to the z-axis,
i.e., $2\pi\tau_{x}=2\pi\tau_{y}=0$, $2\pi\tau_{z}<0$. In this
case under diffraction conditions $k_{z}+2\pi\tau_{z}<0$;
according the Wulff-Bragg condition, the anomalies in the photon
spectrum should be expected at the frequencies $\omega=\frac{\pi
n}{a}$ ($a$ is the lattice spacing along the z-axis; $n=1, 2...$).

In the case in question (\textit{the Bragg case a.} ) the photon
wave function has the form (\ref{12.2}). Its substitution into the
general expression for the radiation cross-section enables us to
find the explicit form of the cross-section (see \cite{44},
formula (8)) (the photon is emitted at a small angle with respect
to the particle momentum.):
\begin{eqnarray}
\label{13.1}
\frac{d^{2}N_{s}}{d\omega d\Omega}=\frac{e^{2}\omega}{\pi^{2}}\sum_{nf}Q_{nn}|\vec{g}_{nf}\vec{e}_{s}^{*}|^{2}\nonumber\\
\times\left|\sum_{\mu=1,2}\gamma_{\mu s}^{(0)}l_{nf}^{\mu
s}(1-e^{-iL/l_{nf}^{\mu s}})\right|^{2},
\end{eqnarray}
where $l_{nf}^{\mu s}=(q_{znf}^{\mu s})^{-1}$ is the coherent
length; $q_{znf}^{\mu
s}=p_{zn}-p_{1zf}-k_{z}-\omega\varepsilon_{\mu s}$. The
cross-section will take on its maximum value for all the
frequencies satisfying the inequality
\begin{eqnarray}
\label{13.2}
\texttt{Re}\,q_{znf}^{\mu s}\equiv\frac{\omega}{2}\left(\frac{m^{2}}{E^{2}}+\theta^{2}\right)-(\varepsilon^{\prime}_{n\kappa}-\varepsilon^{\prime}_{f\kappa_{1}})\nonumber\\
-\omega \texttt{Re}\varepsilon_{\mu s}\leq\omega
\texttt{Im}\,\varepsilon_{\mu s}.
\end{eqnarray}
Using (\ref{13.2}), the spectrum can be written as
\begin{equation}
\label{13.3} \omega_{nf}^{\mu
s}=\frac{(\varepsilon^{\prime}_{n\kappa}-\varepsilon^{\prime}_{f\kappa_{1}})+|\varepsilon|\delta_{\mu
s}}{\frac{1}{2}\left(\frac{m^{2}}{E^{2}}+\vartheta^{2}\right)-\texttt{Re}\,\varepsilon_{\mu
s}},
\end{equation}
where $\delta_{\mu s}=\omega_{nf}^{\mu
s}\texttt{Im}\,\varepsilon_{\mu s}$; $|\varepsilon|\leq 1$.

Consider \textit{the Bragg case b.}. Now the emitted photons can
fly into the  left half-plane from the crystal target. The
diffraction pattern obtained coincides with that produced by a
polychromatic beam of  photons incident along the z-axis with the
opening angle  $\vartheta\sim m/E$.

\textit{The Laue case d.} The analysis shows that the radiation
intensity is sharply suppressed, as none of the coherent lengths
can become large.

\textit{The Laue case c.} Spectral-angular distribution for the
number of photons escaping (outcoming) at a large angle with
respect to the direction of particle motion has the form
\cite{45,47}
\begin{eqnarray}
\label{13.4}
\frac{d^{2}N_{s}}{d\omega d\Omega}=\frac{e^{2}\beta_{1}^{2}\omega}{\pi^{2}}\sum_{nf}Q_{nn}|\vec{e}_{1s}^{*}\vec{g}_{\vec{n}f}|^{2}\nonumber\\
\times\left|\sum_{\mu=1,2}\xi_{\mu
s}^{\tau}\frac{1-e^{-iq_{znf}^{\mu s}L}}{q_{znf}^{\mu
s}}\right|^{2}.
\end{eqnarray}

The quantity (\ref{13.4}) attains the maximum value at the minimum
$q_{znf}^{\mu s}$. As $q_{znf}^{\mu s}$ is the complex value the
minimum value is limited by the imaginary part. The inequality
\begin{eqnarray}
\label{13.5}
\texttt{Re}\,q_{znf}^{\mu s}=p_{zn}-p_{1zf}-(k_{z}+2\pi\tau_{z})\nonumber\\
-\frac{\omega}{\gamma_{1}}\texttt{Re}\,\varepsilon_{\mu
s}\leq\frac{\omega}{\gamma_{1}}\texttt{Im}\,\varepsilon_{\mu s}
\end{eqnarray}
leads to the relation between $\omega$ and the emission angle of
the photons, and, thus determining the photon spectrum. It should
be emphasized that, due to the effect of anomalous transmission,
the imaginary part of $q_{znf}$ in the case of the Laue
diffraction may become anomalously small, which results in an
appreciable increase in the radiation intensity of $\gamma$-quanta
as compared to the case of the absence of diffraction.

Pay attention to the fact that upon introducing the notation
$k_{1z}=k_{z}+2\pi\tau_{z}$, inequality (\ref{13.5}) takes the
form analogous to that of the longitudinal momentum at the
emission of a photon with the wave vector $\vec{k}_{1}$.

It is common knowledge that at photon emission by fast particles,
the photon emission angle is small. Hence, the angle that vector
$\vec{k}_{1}$ makes with the direction of particle motion is small
too. From this follows that a large emission angle is exhibited by
a photon whose wave vector $\vec{k}$ is such that together with
vector $2\pi\vec{\tau}$ it sums up into vector $\vec{k}_{1}$,
which makes a small angle with the direction of the particle
momentum. As a result, the analysis of kinematics is perfectly
analogous to the case of emission at a small angle $\vartheta$.

Expression (\ref{13.4}) can be simplified considerably at
$\texttt{Re}\,\varepsilon_{\mu s}\gg\texttt{Im}\,\varepsilon_{\mu
s}$ and the crystal thickness small as compared to the absorption
depth, or much greater than absorption depth for a
$\gamma$-quantum. Using in the former case the relation
$\left|\frac{1-e^{-iqL}}{q}\right|^{2}\simeq 2\pi L\delta(q)$, we
may integrate (\ref{13.4}) with respect to, for example,
frequencies and obtain the photon angular distribution. As
diffraction is most pronounced within the range of photon wave
lengths $\lambda\sim 10^{-8}-10^{-9}$ cm, and the angle of
$\vec{k}+2\pi\vec{\tau}$ with $\vec{p}$ is small, then
$(\vec{k}+2\pi\vec{\tau})_{x}a\ll 1$, and in $\vec{I}_{nf}$ and
$I_{1nf}$ we may expand the exponents into a series \cite{17,44}.
Confining ourselves to the first nonzero terms, we can obtain, for
example for planar channeling, the following expression for the
angular distribution of $\gamma$-quanta emitted at a large angle
to the polarization plane $\vec{e}_{11}$ perpendicular to the
diffraction plane \cite{45,47}:
\begin{eqnarray}
\label{13.6}
\frac{dN_{11}^{\tau}}{d\Omega}=\frac{e^{2}L}{2\pi}\sum_{nf}Q_{nn}|x_{nf}|^{2}\sum_{\mu}\frac{(\omega_{nf}^{\mu1})^{2}}{\Omega_{nf}}
\beta_{1}^{2}|\xi_{\mu 1}^{\tau}(\omega_{nf}^{\mu1})|^{2}\nonumber\\
\times\left[1-\frac{(\omega_{nf}^{\mu1})^{2}}{\gamma_{1}\Omega_{nf}}\texttt{Re}
\left(\frac{\partial\varepsilon\mu_{1}}{\partial\omega}\right)_{\omega=\omega_{nf}^{\mu1}}\right]^{-1}\nonumber\\
\times\left[\beta_{1}\omega_{nf}^{\mu1}\sin^{2}\theta\cos\varphi\frac{\tau_{y}\cos\varphi-\tau_{x}\sin\varphi}{|\tau_{\perp}|}\right.\nonumber\\
\left.+\Omega_{nf}\frac{\tau_{z}\sin\theta\sin\varphi-\tau_{y}\cos\theta}{|\tau_{\perp}|}\right]^{2},
\end{eqnarray}
where
$\omega_{nf}^{\mu1}=\Omega_{nf}(1-\beta\cos\theta-\gamma_{1}^{-1}\texttt{Re}\,\varepsilon_{\mu1}(\omega_{nf}^{\mu1}))^{-1}$;
$\theta$ is the angle of $\vec{k}+2\pi\vec{\tau}$ with the z-axis.

The angular distribution of $\gamma$-quanta emitted with the same
polarization at a small angle with respect to the particle
momentum is described by the same expression (\ref{13.6}), where
$\xi^{\tau}\rightarrow\xi^{0}$, $\gamma_{1}\rightarrow\gamma_{0}$,
$\beta_{1}=1$, $\theta=\vartheta$ is the photon emission angle. If
the polarization of $\gamma$-quanta is $\vec{e}_{2}$, i.e., it
lies in the diffraction plane, then their angular distribution is
obtained by additional replacement of
$\tau_{y}\rightarrow\tau_{x}$, $\tau_{x}\rightarrow-\tau_{y}$ in
the augend of (\ref{13.6}), and
$tau_{y}\cos\theta\rightarrow\tau_{x}$ in the addend.

Angular distribution of $\gamma$-quanta emitted at a large angle
with the polarization $\vec{e}_{12}$ lying in the diffraction
plane differs from (\ref{13.6}) by  lengthy terms of the order of
unity and has the form
\begin{eqnarray}
\label{13.7}
\frac{dN_{12}^{\tau}}{d\Omega}=\frac{e^{2}L\beta_{1}^{2}}{2\pi}\sum_{nf}Q_{nn}|x_{nf}|^{2}\sum_{\mu}\frac{(\omega_{nf}^{\mu 2})^{2}}{\Omega_{nf}}|\xi_{\mu 2}^{\tau}(\omega_{nf}^{\mu 2})|^{2}\nonumber\\
\times\left[1-\frac{(\omega_{nf}^{\mu 2})^{2}}{\gamma_{1}\Omega_{nf}}\texttt{Re}\left(\frac{\partial\varepsilon\mu_{2}}{\partial\omega}\right)_{\omega=\omega_{nf}^{\mu 2}}\right]^{-1}\nonumber\\
\left\{\frac{\beta_{1}\sin\theta\cos\varphi[\cos\theta(\vec{n}_{1}\vec{\tau})-\tau_{z}]\omega_{nf}^{\mu
2}}{|\tau_{\perp}|}
+\Omega_{nf}\frac{[\sin^{2}\theta\cos\varphi(\vec{n}_{1}\vec{\tau})-\tau_{x}]}{|\tau_{\perp}|}\right\}^{2},
\end{eqnarray}
where
$\vec{n}_{1}=\frac{\vec{k}+2\pi\vec{\tau}}{|\vec{k}+2\pi\vec{\tau}|}$.

The derived expressions for angular distribution of radiation
simplify considerably, if the particle energy is such, that
$1-\beta\gg \frac{1}{\gamma_{1}}\texttt{Re}\varepsilon_{\mu s}$.
In this case we may assume that the frequency corresponding to the
radiation maximum is very likely to be independent of dielectric
properties of a crystal, being determined only by the radiation
angle and the frequency of the corresponding transition, i.e.,
\begin{equation}
\label{13.8} \omega_{nf}^{\mu
s}\simeq\omega_{nf}=\frac{\Omega_{nf}}{1-\beta\cos\vartheta}.
\end{equation}
As a result, for example, angular distributions under diffraction
conditions in the Laue case will be recast as follows

1. For radiation at a large angle with respect to the direction of
the particle motion
\begin{equation}
\label{13.9}
\frac{dN_{s}^{\tau}}{d\Omega}=\frac{e^{2}L}{2\pi}\beta_{1}^{2}\sum_{nf}Q_{nn}|x_{nf}|^{2}\Omega_{nf}^{3}R^{\tau}_{s}(\theta,\varphi)B^{\tau}_{s}(\omega_{nf})
\end{equation}
(recall that $s$ means photon polarization which may be of two
types: $\sigma$-polarization
$\vec{e}_{\sigma}^{\tau}\parallel[\vec{k}, 2\pi\vec{\tau}]$ and
$\pi$-polarization
$\vec{e}_{\pi}^{\tau}\parallel[\vec{k}_{1}[\vec{k},
2\pi\vec{\tau}]]$), where
\begin{eqnarray}
\label{13.10}
R_{\sigma}^{\tau}(\theta,\varphi)=\left[\frac{\beta\sin^{2}\theta\cos\varphi(\tau_{y}\cos\varphi-\tau_{x}\sin\varphi)}{(1-\beta\cos\vartheta)^{2}\sqrt{\tau^{2}-(\vec{n}_{1}\vec{\tau})^{2}}}\right.\nonumber\\
\left.+\frac{(\tau_{z}\sin\theta\sin\varphi-\tau_{y}\cos\theta)}{(1-\beta\cos\theta)\sqrt{\tau^{2}-(\vec{n}_{1}\vec{\tau})^{2}}}\right]^{2};\nonumber\\
R_{\sigma}^{\tau}(\theta,\varphi)=\left[\frac{\beta\sin\theta\cos\varphi[\cos\theta(\vec{n}_{1}\vec{\tau})-\tau_{z}]}{(1-\beta\cos\theta)^{2}\sqrt{\tau^{2}-(\vec{n}_{1}\vec{\tau})^{2}}}\right.\nonumber\\
\left.+\frac{[\sin\theta\cos\varphi(\vec{n}_{1}\vec{\tau})-\tau_{x}]}{(1-\beta\cos\theta)^{2}\sqrt{\tau^{2}-(\vec{n}_{1}\vec{\tau})^{2}}}\right]^{2};\nonumber\\
\vec{n}_{1}=\frac{\vec{k}+2\pi\vec{\tau}}{|\vec{k}+2\pi\vec{\tau}|};\,\,
B_{s}^{\tau}(\omega_{nf})\equiv\sum_{\mu}|\xi_{\mu
s}^{\tau}(\omega_{nf})|^{2};
\end{eqnarray}
$\theta$ is the angle of vector $\vec{k}+2\pi\vec{\tau}$ with the
z-axis, $\varphi$ is the polar angle in the $xy$ plane.

2. For radiation  along the direction of particle motion
\begin{equation}
\label{13.11}
\frac{dN_{s}^{0}}{d\Omega}=\frac{e^{2}L}{2\pi}\sum_{nf}Q_{nn}|x_{nf}|^{2}\Omega_{nf}^{3}R^{0}_{s}(\theta,\varphi)B^{0}_{s}(\omega_{nf})
\end{equation}
where
$$
R_{\sigma}^{0}(\vartheta,\varphi)=\left[\frac{\beta\sin^{2}\vartheta\cos\varphi(\tau_{y}\cos\varphi-\tau_{x}\sin\varphi)}{(1-\beta\cos\vartheta)^{2}\sqrt{\tau^{2}-(\vec{n}\vec{\tau})^{2}}}\right.
$$
$$
\left.+\frac{\tau_{z}\sin\vartheta\sin\varphi-\tau_{y}\cos\vartheta}{(1-\beta\cos\vartheta)\sqrt{\tau^{2}-(\vec{n}\vec{\tau})^{2}}}\right]^{2};
$$
$$
R_{\sigma}^{0}(\vartheta,\varphi)=\left[\frac{\sin^{2}\vartheta\cos\varphi(\tau_{x}\cos\varphi+\tau_{y}\sin\varphi)}{(1-\beta\cos\vartheta)^{2}\sqrt{\tau^{2}-(\vec{n}\vec{\tau})^{2}}}\right.
$$
$$
\left.+\frac{\tau_{z}\sin\vartheta\cos\varphi-\tau_{x}}{(1-\beta\cos\vartheta)\sqrt{\tau^{2}-(\vec{n}\vec{\tau})^{2}}}\right]^{2};
$$
$$
B_{s}^{0}(\omega_{nf})\equiv\sum_{\mu}|\xi_{\mu
s}^{0}(\omega_{nf})|^{2};\, \, \vec{n}=\frac{\vec{k}}{|\vec{k}|}
$$
($\vartheta$ is the angle of vector $\vec{k}$  with the z-axis;
$\varphi$ is the polar angle in the $xy$ plane).

Approximate integral expressions for the number of $\gamma$-quanta
emitted at a large angle with respect to the direction of particle
motion within the diffraction peak may be found, using the fact
that the frequency of the photon produced and the position of the
diffraction peak are determined, on the one hand, by the Bragg
condition, and, on the other hand, by the  laws of conservation in
emission for a corresponding transition $\Omega_{nf}$. As a
result,  we obtain the following expressions for the number of
$\gamma$-quanta emitted within the diffraction peak at a large
angle with respect to the direction of particle motion:

a. for $\pi$-polarization
\begin{eqnarray}
\label{13.12}
\Delta N^{\tau}_{\pi}\approx\frac{\pi e^{2}L\beta_{1}^{2}\tau^{4}|g^{\prime}_{00}|}{8|\tau_{z}|^{3}}\sum_{nf}Q_{nn}|x_{nf}|^{2}\Omega_{nf}^{2}\nonumber\\
\times\left\{\frac{\tau_{x}^{2}}{\tau_{\perp}^{2}}+\frac{\pi\tau^{2}|\tau_{z}|\Omega_{nf}}{\tau_{\perp}^{2}\tilde{\omega}_{nf}^{2}}\left(1-\frac{\pi\tau^{2}}{|\tau_{z}|\tilde{\omega}_{nf}}\right)\right\},\,
\tilde{\omega}_{nf}=2\Omega_{nf}\gamma^{2};
\end{eqnarray}

b. for $\sigma$-polarization
\begin{eqnarray}
\label{13.13}
\Delta N^{\tau}_{\sigma}\approx\frac{\pi e^{2}\beta_{1}^{2}L\tau^{4}|g^{\prime}_{00}|}{8|\tau_{z}|^{3}}\sum_{nf}Q_{nn}|x_{nf}|^{2}\Omega_{nf}^{2}\nonumber\\
\times\left\{1-\frac{\pi\tau^{2}(\tau_{x}^{2}-\tau_{y}^{2})}{|\tau_{z}|\tau_{\perp}^{2}\tilde{\omega}_{nf}}\right\}.
\end{eqnarray}

To estimate the number of quanta emitted within the diffraction
peak, note that in the order of magnitude expression (\ref{13.9})
can be represented as the product of the spectrum of photon
emission by a channeled particle without regard to diffraction
into the function $|\xi|^{2}$ characterizing reflection of photons
by a crystal in the presence of diffraction. The value of the
stated function is close to unity under the fulfillment of the
Bragg conditions in the range of angles $\delta\vartheta\sim
\varepsilon_{\mu s}$, close to the Bragg ones, i.e., in the range
$\delta\vartheta\sim 10^{-6}$ rad, vanishing rapidly at great
deviation from the diffraction condition. Hence, the number of
quanta emitted within the diffraction peak is of the same order of
magnitude as that emitted without diffraction in the range of
angles $\delta\vartheta\sim 10^{-6}$ rad near the intensity
maximum. As follows from the estimations \cite{5,17,44,45,46,47}
(see Section (\ref{sec:3.8}), depending on the energy and the type
of matter, $10^{-8} E/m$  quanta (it is assumed that
$m/E>\delta\vartheta$) will be emitted in the range
$\delta\vartheta\sim 10^{-6}$ rad over the crystal thickness
$L\sim 10^{-2}$ cm. From this follows that at the current of
$10^{-6}$ A and the energy of, for example, 50 MeV one should
expect emission of about $10^{7}$ quanta/sec, which appreciably
exceeds the intensity of conventional X-ray sources for the same
angular and spectral ranges.

The above formulae also hold true in the case when the crystal
thickness is much greater than the absorption depth of quanta, if
in expressions (\ref{13.6})-(\ref{13.11}) the thickness $L$ is
understood as the quantum absorption depth
$$
L(\omega_{nf}^{\mu s})=\gamma_{1(0)}[2\omega_{nf}^{\mu
s}\texttt{Im}\varepsilon_{\mu s}(\omega_{nf}^{\mu s})]^{-1},
$$
where the subscripts $1(0)$ refer to radiation at a large (small)
angle with respect to the direction of particle motion,
respectively.

Thus, radiation produced through radiative transitions between the
levels of transverse motion of a channeled particle, form behind a
crystal a diffraction pattern which can be decoded by means of the
methods applied in X-ray structural analysis.

%%%%%%%%%%%%%%%%%%%%%%%%%%%%%%%%%%%  Section 14 %%%%%%%%%%%%%%%%%%%%%%

\section[Radiation Spectrum in the Quasi-classical Approximation]{Radiation Spectrum in the Quasi-classical Approximation}
\label{sec:4.14}

Due to a large relativistic mass, transverse motion of
ultra-relativistic channeled particles is quasi-classical for a
vast majority of levels. This makes it possible to apply particle
wave functions in the quasi-classical approximation to calculate
the matrix elements $x_{nf}$ appearing in
(\ref{13.6})-(\ref{13.11}) \cite{31,45,47,66,67}.  The sum over
$a$ appearing in (\ref{13.6})-(\ref{13.10}) is split into two sums
relating to: 1. over-barrier states and 2. sub-barrier states
\cite{5,17,45}.

%%%%%%%%%%%%%%%%%%%%%%%%%%

The Bloch functions of sub-barrier states which are not located
near the the barrier top may be taken in the tight binding
approximation with quasiclassical functions in a well. For
example, for planar channeling we may write
\begin{equation}
\label{14.1}
\psi_{n\kappa}(x)=c_{n}\frac{1}{\sqrt{p_{n}(x)}}\cos\left(\int_{x_{n}}^{x}p_{n}(x^{\prime})dx^{\prime}-\frac{\pi}{4}\right)
\end{equation}
with the quantization condition
${\int_{x_{n}}^{a-x_{n}}p_{n}(x^{\prime})dx^{\prime}=\pi(n+\frac{1}{2})}$,
where
${p_{n}(x)=\left\{2E[\varepsilon^{\prime}_{n}-V(x)]\right\}^{1/2}}$;
$V(x)=V(x+a)$ is the one--dimensional periodic in $x$ potential of
crystal planes; $x_{n}$ is the turning point in the well for  $n$
level; $c_{n}^{2}=4E(T^{n}_{\mathrm{sub}})^{-1}$;
$T^{n}_{\mathrm{sub}}$ is the period of particle motion in the
well:
\[
T^{n}_{\mathrm{sub}}=2E\int_{x_{n}}^{a-x_{n}}p_{n}^{-1}(x^{\prime})dx^{\prime}.
\]
The Bloch functions of the over-barrier states in this
approximation may be written as follows
\begin{equation}
\label{14.2}
\psi_{n\kappa}(x)=\tilde{c}_{n}\frac{1}{\sqrt{p_{n}(x)}}e^{i\int^{x}_{0}p_{n}(x^{\prime})dx^{\prime}}
\end{equation}
with the quantization condition
%%%%%%%%%%%%%%%%%%%%%%
$\int^{a}_{0}p_{n}(x^{\prime})dx^{\prime}=\kappa a+2\pi n$, where
${\tilde{c}_{n}=E(N_{x}T^{n}_{\mathrm{over}})^{-1}}$;
$T^{n}_{\mathrm{over}}$ is the period of the over-barrier motion:
\[
T^{n}_{\mathrm{over}}=E\int_{0}^{a}p_{n}^{-1}(x^{\prime})dx^{\prime}.
\]
Using the stated wave functions, one may obtain the following
expressions for the occupation coefficients:
\begin{equation}
\label{14.3}
Q_{nn}=2\pi(aT^{n}_{\mathrm{sub}}|V^{\prime}(x_{0})|)^{-1}
\quad\mbox{for sub-barrier states}\,,
\end{equation}
\begin{equation}
\label{14.4}
Q_{nn}=2\pi(aT^{n}_{\mathrm{over}}|V^{\prime}(x_{0})|)^{-1}
\quad\mbox{for over-barrier states}\,.
\end{equation}
The point $x_{0}$ is found from the condition
$\varepsilon^{\prime}_{n}=\frac{p_{x}^{2}}{2E}+V(x_{0})$\,.

%%%%%%%%%%%%%%%%%%%%%%%%%%
As a result we have, for example, in the case when
$\frac{n-f}{n}\ll 1$:\\
 for sub-barrier transitions
\begin{eqnarray}
\label{14.5}
x_{nf}&=&\frac{2E}{T^{n}_{\mathrm{sub}}}\int_{x_{n}}^{a-x_{n}}xp_{n}^{-1}(x)\cos
\left(\Omega_{nf}^{\mathrm{sub}}\int_{x_{n}}^{x}p_{n}^{-1}(x^{\prime})Edx^{\prime}\right)dx,\nonumber\\
\Omega_{nf}^{\mathrm{sub}}&=&\frac{2\pi(n-f)}{T^{n}_{\mathrm{sub}}}\,,
\end{eqnarray}

for over-barrier transitions
\begin{eqnarray}
\label{14.6}
x_{nf}&=&\frac{E}{T^{n}_{\mathrm{sub}}}\int_{0}^{a}p_{n}^{-1}(x)x
\exp\left\{i\Omega_{nf}^{\mathrm{over}}\int_{0}^{x}p_{n}^{-1}(x^{\prime})Edx^{\prime}\right\}dx,\nonumber\\
\Omega_{nf}^{\mathrm{over}}&=&\frac{2\pi(n-f)}{T^{n}_{\mathrm{over}}}\,.
\end{eqnarray}
%%%%%%%%%%%%%%%%%%%%%%%%%%

Using the expressions derived for $x_{nf}$, it is possible to
demonstrate directly that formulas of the type (\ref{7.16})
calculated in the quasiclassical approximation for the transitions
\[
\frac{n-f}{n}\ll 1,
\]
coincide with  similar expressions obtained by means of classical
electrodynamics calculations.

Consider in more detail the spectrum of forward radiation in
(\ref{8.2}) in the case when refraction and absorption can be
neglected. From (\ref{8.2}) follows that in the dipole
approximation the spectral distribution of radiation intensity in
the absence of absorption and refraction may be written as follows
\begin{eqnarray}
\label{14.7}
\frac{dw_{\omega}}{d\omega}& &=L e^{2}\omega\sum_{nf}Q_{nn}|x_{nf}|^{2}\\
& &\times
\Omega_{nf}^{2}\left[1-\frac{\omega}{\Omega_{nf}}(1-\beta^{2})+\frac{\omega^{2}}{2\Omega_{nf}^{2}}(1-\beta^{2})^{2}\right]
\theta\left[\frac{\vartheta_{k}^{2}}{2}-\alpha_{nf}(\omega)\right]\theta[\alpha_{nf}(\omega)]\,.\nonumber
\end{eqnarray}
%%%%%%%%%%%%%%%%%%%%%%%%%%%%%%%%%%

If $\vartheta_{k}=\pi$, then in view of (\ref{8.3}) we have
\begin{eqnarray}
\label{14.8}
\frac{dw_{\omega}}{d\omega}& &= L e^{2}\omega\sum_{nf}Q_{nn}|x_{nf}|^{2}\\
& &\times
\Omega_{nf}^{2}\left\{1-\frac{\omega}{\Omega_{nf}}(1-\beta^{2})+\frac{\omega^{2}}{2\Omega_{nf}^{2}}(1-\beta^{2})^{2}\right\}
 \theta[2-\alpha_{nf}(\omega)]\theta[\alpha_{nf}(\omega)]\,.\nonumber
\end{eqnarray}
Note that in the particular case $Q_{nn}=\delta_{n\bar{n}}$, where
$\bar{n}$ belongs to the states lying inside the well, expression
(\ref{14.9}) converts into the expression discussed in
\cite{36,37}.

%%%%%%%%%%%%%%%%%%%%%%%%%%%%%%

On the other hand, according to [\cite{68}, formula to problem 2
on p. 278] in the case when the particle deviation angle in the
field is small in comparison with the radiation angle, we have
\begin{equation}
\label{14.9}
\frac{dw_{\omega}}{d\omega}=\frac{e^{2}\omega}{2\pi}\int\limits_{\frac{\omega}{2}(1-\beta^{2})}^{\infty}
\frac{|w_{\omega^{\prime}}|^{2}}{\omega^{\prime
2}}\left[1-\frac{\omega}{\omega^{\prime}}(1-\beta^{2})+\frac{\omega^{2}}{2\omega^{\prime
2}}(1-\beta^{2})^{2}\right]d\omega^{\prime}\,,
\end{equation}
where $w_{\omega^{\prime}}$ is the Fourier transform of the
particle acceleration, which, due to the periodic character of
motion in a transverse plane, is the set of harmonics multiple of
$2\pi/T$, where $T$ is the period of classical motion for the
given initial conditions.
 Substitution of $w_{\omega^{\prime}}$
for the periodic motion in (\ref{14.9}) gives formula (\ref{14.9})
at $Q_{nn}=\delta_{nn^{\prime}}$. Averaging of (\ref{14.9}) over
various initial conditions of motion, which in (\ref{14.8})
corresponds to summation over $n$ with the weights $Q_{nn}$ leads
to the complete coincidence of these formulas. The theory of
radiation of channeled particles based on (\ref{14.9}) is given in
\cite{87}.

%%%%%%%%%%%%%%%%%%%%%%

Now consider another extreme case, when at particle motion in the
field produced by crystallographic axes (planes), the particle
deviation angle (which is of the order of the Lindhard angle $
\vartheta_{\pi}=\sqrt{\frac{2V}{E}})$ is much larger than the
characteristic angle of the photon emission
$\vartheta_{\gamma}\sim\frac{m}{E}$.  Coherent radiation length
$l=\frac{2}{\omega}\left(1-\frac{\omega}{E}\right)\gamma^{2}$  is
small as compared to the spatial period of particle oscillation in
a channel. In view of \cite{68}, radiation in the given direction
occurs mainly from that part of the classical trajectory of the
particle, where the particle velocity is almost parallel to this
direction.
Along this part, the field acting on the particle  may be
considered constant, and the part of the trajectory contributing
to radiation may be considered a circle. This enables application
of the theory of photon emission in uniform circular motion for
analyzing the problem. As a result, in view of the problem 1 $\S
77$ in \cite{68}, the spectral distribution of radiation intensity
has the form
\begin{equation}
\label{14.10}
\frac{dw_{\omega}}{d\omega}=-\frac{2e^{2}\omega}{\sqrt{\pi}\gamma^{2}}
\int\limits_{-\infty}^{+\infty}\left[\frac{\Phi^{\prime}(u)}{u}+\frac{1}{2}\int\limits_{u}^{\infty}\Phi(u^{\prime})du^{\prime}\right]dt\,,
\end{equation}
where $\Phi(u)$ is the Airy function of argument
\[
u=\left[\frac{m\omega}{e{\cal{E}}(\vec{r}(t))\gamma^{2}}\right]^{2/3}\,,
 \]
${\cal{E}}(\vec{r}(t))$ in our case is the magnitude of the
electric field strength at the particle location point.

%%%%%%%%%%%%%%%%%%%%%%%%

Next consider planar channeling
${\cal{E}}(\vec{r}(t))={\cal{E}}(x(t))$. Change the variables
\[
dt=\frac{dx}{v(x,x_{0})}\,,
\]
where
\[
v(x,x_{0})=\sqrt{\frac{2}{E}(E_{\perp}(x_{0})-V(x))}
 \]
is the velocity in the transverse plane of the particle entering
the channel at point $x_{0}$;
\[
E_{\perp}(x_{0})=E\frac{\vartheta^{2}}{2}+V(x_{0})\,
 \]
where $\vartheta$ is the particle angle of incidence with respect
to the chosen family of crystallographic planes; $E=m\gamma$ is
the energy of the particle entering the crystal.

Take into account that the particle motion in a periodic potential
is periodic. The time of particle motion from the left turning
point to the right one (see Figure (\ref{Channeling Figure 2}))
\[
\tau(x_{0})=\int\limits^{a-x_{1}(x_{0})}_{x_{1}(x_{0})}\frac{dx}{v(x,\,x_{0})}\,,
 \]
$x_{1}(x_{0})$ is determined from the equation
$E_{\perp}(x_{0})=V(x_{1}(x_{0}))$.

 If $E_{\perp}(x_{0})$ is greater than the maximum value of $V$, then $x_{1}(x_{0})=0$.
 Hence, the entire integral over $t$ may be represented as a sum
 of  $L/\tau(x_{0})$  identical integrals, i.e., (\ref{14.10}) can be written as follows:
\begin{eqnarray}
\label{14.11}
\frac{dw_{\omega}(\vartheta, x_{0})}{d\omega}&=&-\frac{2e^{2}\omega}{\sqrt{\pi}\gamma^{2}}\frac{L}{\tau(x_{0})}\\
&\times &
\int\limits_{x_{1}(x_{0})}^{a-x_{1}(x_{0})}\left[\frac{\Phi^{\prime}(u)}{u}+\frac{1}{2}\int_{u}^{\infty}
\Phi(u^{\prime})du^{\prime}\right]
\frac{dx}{v(x,\,x_{0})}\,.\nonumber
\end{eqnarray}
%%%%%%%%%%%%%%%%%%

Upon averaging (\ref{14.11}) over the points of entrance and
initial angular distribution of the incident particle, we obtain
\begin{eqnarray}
\label{14.12} \frac{dw_{\omega}}{d\omega}=\frac{1}{a}\int\limits
f(\vartheta)d(\vartheta)\int_{0}^{a}dx_{0}\frac{dw_{\omega}(\vartheta,
x_{0})}{d\omega}\,.
\end{eqnarray}

 Equation (\ref{14.10}) is derived using the methods of classical electrodynamics,
 so it is valid for describing the spectrum of soft photons with the energy $\omega\ll E$
 (but one should bear in mind that the coherence length $l$ should be less than the characteristic
 spatial period of the trajectory). To analyze the spectrum in a short--wave range  $\omega\sim E$,
 make use of the fact that, as shown by Nikishov and Ritus \cite{88a,88b}, with due account of the
 quantum recoil effects the spectral distribution of radiation produced by a particle moving along a circular trajectory has the form:
\begin{equation}
\label{14.13}
\frac{dI}{d\omega}=-\frac{e^{2}m^{2}}{\sqrt{\pi}E}\frac{\eta}{1+\eta}
\left\{\int\limits^{\infty}_{\xi}\Phi(\xi^{\prime})d\xi^{\prime}+\frac{2}{\xi}\left(1+\frac{\eta^{2}}{2(1+\eta)}\right)
\Phi^{\prime}(\xi)\right\}\,,
\end{equation}
where $\eta=\omega/E-\omega$; $\xi=(\eta/\chi)^{2/3}$;
$\chi=eH\gamma/m^{2}$, $H$ is the strength of the external
magnetic field.

In a similar manner as has been done above, replacing  the
strength of the external magnetic field  by  the strength of the
electric field which acts on a particle moving at a certain small
angle with the crystallographic axis (plane) and integrating
(\ref{14.13}) over the flight time, we obtain  the following
expression for the spectral distribution of radiation energy:
\begin{equation}
\label{14.14}
\frac{dw}{d\omega}=-\frac{e^{2}m^{2}}{\sqrt{\pi}E}\frac{\eta}{1+\eta}\int\limits_{-\infty}^{+\infty}
\left\{\int\limits_{\xi}^{\infty}\Phi(\xi^{\prime})d\xi^{\prime}
+\frac{2}{\xi}\left(1+\frac{\eta^{2}}{2(1+\eta)}\right)\Phi^{\prime}(\eta)\right\}dt\,.
\end{equation}
From this we obtain for planar channeling
\begin{eqnarray}
\label{14.15} & &\frac{dw(\vartheta,
x_{0})}{d\omega}=-\frac{e^{2}m^{2}}{\sqrt{\pi}E}\frac{\eta}{1+\eta}\frac{L}{\tau(x_{0})}\\
& &
\times\int\limits_{x_{1}(x_{0})}^{a-x_{1}(x_{0})}\left\{\int\limits_{\xi}^{\infty}\Phi(\xi^{\prime})
d\xi^{\prime}+\frac{2}{\xi}\left(1+\frac{\eta^{2}}{2(1+\eta)}\right)\Phi^{\prime}(\eta)\right\}\frac{dx}{v(x_{1}x_{0})}\,.\nonumber
\end{eqnarray}

Averaged spectral distribution is given by  (\ref{14.12}).

%%%%%%%%%%%%%%%%%%%%%%%%%%%%%%%%%%%  Section 15 %%%%%%%%%%%%%%%%%%%%%%

\section[Parametric Radiation]{Parametric Radiation}
\label{sec:4.15}

As mentioned above, the contribution to radiation intensity under
diffraction conditions comes from radiation through transition
between the levels along with radiation which is due to scattering
of pseudo-photons associated with a particle by crystal atoms and
nuclei (parametric radiation). This mechanism manifests itself in
its purest form in particle motion in a crystal beyond the
channeling regime. Recall that parametric radiation is the photon
production in the transmission of a uniformly moving charged
particle through a periodically inhomogeneous medium.

Parametric optical radiation in a one-dimensional medium with
dielectric permittivity of one-dimensional periodicity was first
studied by Fainberg and Khizhnyak \cite{89}. The phenomenon of
photon production when a particle passes through a medium with
space-periodic dielectric permittivity was reviewed by
Ter-Mikaelyan \cite{63,64}. In \cite{74} attention was focused on
the fact that the effect of anomalous transmission can drastically
change the spectral properties of radiation produced by a particle
in a thick crystal. Classical theory of parametric radiation in a
thick crystal, when the effects caused by anomalous transmission
are of importance was developed by Feranchuk and the author
\cite{75,76,77}, Garibyan and Yan Shi \cite{78,79}. Thorough
analysis carried out in \cite{75,76,77,81} made it possible to not
only find general expressions for spectral-angular distributions
of emitted photons but also to obtain explicit expressions for the
number of quanta emitted by a particle within the diffraction peak
as well as to analyze the process of radiation in crystals
containing Mossbauer nuclei. Formulae for the number of quanta
produced by a particle analogous to those in \cite{75,76,77} were
later derived in \cite{80}.

To obtain the formulae describing parametric radiation in its pure
form  sufficient it to assume that the angle of a particle
entrance  into the crystal is much larger than the Lindhard angle.
In this case the particle wave functions  in a crystal are plane
waves. As a result, we have the following expression for the
differential number of quanta emitted by a particle forward into
the narrow cone along the direction of its velocity $\vec{v}$ in
the Laue case (c.): From this we obtain for planar channeling
\begin{eqnarray}
\label{15.1}
dN_{s}^{(0)}=\frac{e^{2}}{\pi^{2}}(\vec{e}_{s}\vec{v})^{2}\left|\sum_{\mu=1,2}\frac{2\varepsilon_{\mu s}-g_{00}}{2(\varepsilon_{2s}-\varepsilon_{1s})}(l_{0}^{(0)}-l_{\mu s}^{(0)})\right.\nonumber\\
\left.\times(e^{-iL/l_{\mu
s}^{(0)}}-1)\right|^{2}\vartheta_{0}d\vartheta_{0}d\varphi_{0}\omega
d\omega,
\end{eqnarray}
where $\vartheta_{0}$ and $\varphi_{0}$ are the polar and
azimuthal angles of the photon.

Moreover, there appears radiation concentrated in the narrow cone
with the axis along the direction
$\omega^{\tau}_{B}\vec{v}+2\pi\vec{\tau}$,
$\omega^{\tau}_{B}=\frac{\pi\tau^{2}}{|(\vec{\tau}\vec{v})|}$.
\footnote{Note that here and below, unlike \cite{75,76,77,82}, for
the sake of uniformity of symbols, we use the notation $2\pi\tau$
to denote the reciprocal lattice vector. In \cite{75,76,77,81} the
reciprocal lattice vector is denoted by $\vec{\tau}$.} The
differential number of quanta emitted in the direction of
diffraction is given by
\begin{eqnarray}
\label{15.2}
dN_{s}^{(\tau)}=\frac{e^{2}}{\pi^{2}}(\vec{e}_{1s}\vec{v})^{2}\left|\sum_{\mu=1,2}\frac{(-1)^{\mu} g_{\tau}^{(s)}}{2(\varepsilon_{2s}-\varepsilon_{1s})}(l_{0}^{(\tau)}-l_{\mu s}^{(\tau)})\right.\nonumber\\
\left.\times(e^{-iL/L_{\mu
s}^{\tau}}-1)\right|^{2}\vartheta_{\tau}d\vartheta_{\tau}d\varphi_{\tau}\omega
d\omega,
\end{eqnarray}
where
$\cos\vartheta_{\tau}=(\vec{k},\omega^{\tau}_{B}\vec{v}+2\pi\vec{\tau})/\omega^{2}$.
Coherent radiation lengths $l_{\mu s}^{(\nu)}$ are determined as
follows:
\begin{eqnarray}
\label{15.3}
l_{\mu s}^{(\nu)}=\frac{1}{q_{z\mu s}^{(\nu)}}=\frac{2}{\omega}\left(\frac{m^{2}}{E^{2}}+\vartheta_{\nu}^{2}-2\varepsilon_{\mu s}\right)^{-1};\nonumber\\
l_{\mu
s}^{(\nu)}=\frac{2}{\omega}\left(\frac{m^{2}}{E^{2}}+\vartheta_{\nu}^{2}\right)^{-1}.
\end{eqnarray}

From the analysis of (\ref{15.1}) and (\ref{15.2}) follows  that
the radiation cross-section is maximum when the real part of the
longitudinal momentum transmitted to the medium vanishes. From the
requirement $\texttt{Re}q^{(\nu)}_{z\mu s}=0$ we find the
dispersion equation defining the condition for emergence of
parametric radiation in a crystal \cite{90}.
\begin{equation}
\label{15.4}
\cos\vartheta_{\nu}=\frac{1}{v}-\texttt{Re}2\varepsilon_{1,2s}.
\end{equation}
Equation (\ref{15.4}) differs from the equation defining the
condition for emergence of Vavilov-Cherenkov radiation in a
homogeneous medium by the dielectric permittivity
$\varepsilon(\omega)$, substituted for the corresponding
expression for a crystal $1+2\varepsilon_{1,2s}$.

In two limiting cases  of  thin ($\omega L|g_{00}|\ll 1$) and
thick ($\omega L\texttt{Im}g_{00}\gg 1$) crystals, it is possible
to obtain analytical expressions for the total number of quanta
produced by one particle, which are valid at $\ln\frac{E}{m}\gg
1$:

a. $\omega_{B}L|g_{00}|\ll 1$, with the results for the Laue and
Bragg cases coinciding:
\begin{equation}
\label{15.5}
N_{s}^{\tau}=e^{2}\frac{(2\pi\vec{\tau})^{2}|(2\pi\vec{\tau}_{\perp})^{2}-(2\pi\tau_{z})^{2}|}{8|2\pi\tau_{z}|^{3}}L|g_{10}^{s}(\omega_{B})|^{2}\ln\frac{E}{m},
\end{equation}
where $|2\pi\tau_{z}|\gg\omega_{B}\frac{m}{E}$;
$\frac{m^{2}}{E^{2}}\geq |g_{00}|$;

b. $\omega_{B}L\texttt{Im}g_{00}\gg 1$. In the Laue case:
\begin{eqnarray}
\label{15.6}
N_{s}^{\tau}=e^{2}\frac{|(2\pi\vec{\tau}_{\perp})^{2}-(2\pi\tau_{z})^{2}|}{8(2\pi\vec{\tau}_{z})^{2}}\frac{|g_{10}^{s}(\omega_{B}^{\tau})|^{2}}{|g_{00}^{\prime\prime}(\omega_{B}^{\tau})}\nonumber\\
\times\left|\ln\left\{\left(\frac{m^{2}}{E^{2}}+\delta_{s}-g_{00}^{\prime}\right)^{2}+|g^{s}_{10}|^{2}-\delta_{s}^{2}\right\}\right|,
\end{eqnarray}
where $g_{00}=g_{00}^{\prime}+ig_{00}^{\prime\prime}$;
$\delta_{s}=\frac{\texttt{Re}\sqrt{g_{10}^{s}g_{01}^{s}}\texttt{Im}\sqrt{g_{10}^{s}g_{01}^{s}}}{g_{00}^{\prime\prime}}$;
$g_{00}^{\prime}$ is the real part of $g_{00}$;
$g_{00}^{\prime\prime}$ is the imaginary part of $g_{00}$. And the
angular divergence of quanta
$\Delta\vartheta=\sqrt{\frac{m^{2}}{E^{2}}+g_{00}^{\prime}}$, the
order of magnitude of the frequency spread near
$\omega=\omega_{B}^{\tau}$ is defined by the formula
\begin{equation}
\label{15.7}
\frac{\Delta\omega}{\omega_{B}^{\tau}}\simeq\sqrt{\frac{m^{2}}{E^{2}}+g_{00}^{\prime}}.
\end{equation}

In the Bragg case when the below condition is satisfied
\begin{equation}
\label{15.8}
\frac{m^{2}}{E^{2}}-2g_{10}^{s\prime}-2\sqrt{|\beta_{1}|\texttt{Re}(g_{10}^{s}g_{01}^{s})}\gg
g_{00}^{\prime\prime}.
\end{equation}
the intensity is defined by formula (\ref{15.6}). If the condition
(\ref{15.8}) is violated,
\begin{eqnarray}
\label{15.9}
N_{s}^{\tau}\simeq e^{2}\frac{|(2\pi\vec{\tau}_{\perp})^{2}-(2\pi\tau_{z})^{2}|}{8(2\pi\tau_{z})^{2}}\nonumber\\
\times\frac{|g_{10}^{s}(\omega_{B})|^{2}\ln|\frac{m^{2}}{E^{2}}+g_{00}^{\prime}|}{[\sqrt{|\beta_{1}|\texttt{Re}(g_{10}^{s}g_{01}^{s})}(g_{00}^{\prime\prime}+\sqrt{|\beta_{1}|g_{10}^{\prime\prime}g_{01}^{\prime\prime}})]^{1/2}}.
\end{eqnarray}
Numerical analysis showed that the values of $N_{s}^{\tau}$  found
from formulae (\ref{15.6})-(\ref{15.7}) coincide with the results
of calculations by the exact formulae with the accuracy of
$5-10\%$

Now go over to considering the frequency spectrum of parametric
radiation concentrated along the direction of particle motion.
Assume a crystal to be quite thick ($\omega L\texttt{Im}g_{00}\gg
1$). Expression (\ref{15.1}) can be represented in the form

$$
dN_{s}^{(0)}=dN_{s}^{n}+d\tilde{N}_{s},
$$
where
\begin{equation}
\label{15.10}
dN_{s}^{n}=\frac{4e^{2}}{\pi^{2}}p^{2}_{s}\frac{|g_{00}|^{2}}{(\gamma^{-2}+\vartheta^{2})^{2}|\gamma^{-2}+\vartheta^{2}-g_{00}|^{2}}\vartheta^{3}d\vartheta
d\varphi\frac{d\omega}{\omega}
\end{equation}
\begin{eqnarray}
\label{15.11}
d\tilde{N}_{s}&=&\frac{4e^{2}}{\pi^{2}}p^{2}_{s}\nonumber\\
&
&\times\frac{(2\varepsilon_{2s}-g_{00})(2\varepsilon_{1s}-g_{00})}
{(\gamma^{-2}+\vartheta^{2})^{2}|\gamma^{-2}+\vartheta^{2}-g_{00}|^{2}|\gamma^{-2}+\vartheta^{2}
-2\varepsilon_{1s}|^{2}|\gamma^{-2}+\vartheta^{2}-2\varepsilon_{2s}|^{2}}\nonumber\\
& &\times\texttt{Re}\left[4\varepsilon_{2s}\varepsilon_{1s}(\gamma^{-2}+\vartheta^{2}-g_{00})\right.\nonumber\\
& &\left.+2g_{00}(\gamma^{-2}+\vartheta^{2})(\varepsilon_{2s}-\varepsilon_{1s})-g_{00}(\gamma^{-2}+\vartheta^{2})\right.\nonumber\\
&
&\left.\times(2\gamma^{-2}+2\vartheta^{2}+g_{00})\right]\vartheta^{3}d\vartheta
d\varphi\frac{d\omega}{\omega},
\end{eqnarray}
and $p_{1}=\sin\varphi$; $p_{2}=\cos\varphi; \gamma=E/m$. Formula
(\ref{15.10}) coincides with the expression for the cross-section
of transient radiation in a homogeneous medium with dielectric
permittivity $\varepsilon(\omega)=1+2g_{00}$. The addend is
associated with parametric radiation, it contains information
about the crystal structure.

Analysis of expression (\ref{15.11}) shows that $d\tilde{N}_{s}$
has a pronounced resonance character: when the conditions
(\ref{15.4}) hold, its value exceeds $dN_{s}^{n}$ by a factor of
$(g_{0}^{\prime}/g_{0}^{\prime\prime})^{2}$ (i.e., by a factor of
$10^{4}\div10^{5}$). The width of the peak formed by parametric
radiation is very small (see (\ref{15.7})), so the contribution to
the integral intensity of forward radiation due to parametric
effect is insignificant as compared to the intensity of transient
radiation.

As seen from a through analysis carried out by Feranchuk
\cite{81}, the study of the energy spectrum of the forward-emitted
photons enables one to simultaneously measure a larger number of
structure amplitudes, which may appreciably reduce the duration of
physical experiments in X-ray diffraction analysis. Below we
follow the same line of reasoning as in \cite{81}.

The most direct method to measure $d\tilde{N}_{s}$ is to use X-ray
detectors with high angular and frequency resolution. But good
reliability of the parametric effect study  against transient
radiation is possible when the relative angular and energy
resolution of a detector is not poorer than $10^{-2}\%$. Though
the investigation of radiation spectrum with such a resolution is
feasible, using another single crystal with known parameters as a
detector, such an experiment seems to be tedious, and above all,
it leads to a considerable loss of radiation intensity.

Another opportunity is to use detectors, which enable detecting
X-ray radiation with given (preset) polarization. In this case
suffice it to register photons polarized perpendicular to the
radiation plane, i.e., the plane formed by vectors $\vec{k}$ and
$\vec{v}$. The radiation registered by such a detector will be
completely associated with the parametric effect. Nevertheless,
this method also exhibits the shortcomings mentioned above.

Therefore we only give a more detailed analysis of one
experimental method which seems to provide the simplest way of
measuring $d\tilde{N}_{s}$ making the most out of the advantages
of the parametric Vavilov-Cherenkov effect: high intensity and the
possibility of simultaneous study of numerous structure
amplitudes.

Thus, suppose that a detector registers the total radiation
propagating in the cone with the apex angle
$\Delta\vartheta=\sqrt{\gamma^{-2}+g_{00}^{\prime}}$ along the
direction of particle motion  and has a relative energy resolution
$\Delta\omega/\omega=\omega\sim 3-5\%$, typical of semiconductor
detectors. Assume also that the electron beam does not get into
the detector after leaving the crystal. For this purpose one may
use a holed detector, or change the beam direction after the
crystal by means of a magnetic field.

The number of photons with the frequency $\omega_{0}$ registered
by the detector per unit time, which are formed in the
transmission of a beam of monochromatic electrons with the energy
$E$ and current $J$ through a thick perfect single crystal, is
defined by the expression derived from (\ref{15.10}),
(\ref{15.11}) by integration with respect to the exit angles and
frequency and summation over polarizations of quanta:
\begin{eqnarray}
\label{15.12}
n_{0}(\omega_{0})=J\frac{4e^{2}}{\pi}\left[\frac{2\gamma^{-2}-g_{00}^{\prime}(\omega_{0})}{|g_{00}^{\prime}(\omega_{0})|}\right.\nonumber\\
\left.\times\ln\frac{\gamma^{-2}-g_{00}^{\prime}(\omega_{0})}{\gamma^{-2}}-2\right]\frac{\Delta\omega}{\omega},
\end{eqnarray}
if  $\omega_{0}<\omega_{B}^{\tau}-\Delta\omega$ or
$\omega_{0}>\omega_{B}^{\tau}+\Delta\omega$, and
\begin{eqnarray}
\label{15.13}
n(\omega_{0})=n(\omega_{0})+n_{\tau}(\omega_{0})=n_{0}(\omega_{0})+Je^{2}\frac{|\tau_{\perp}^{2}-\tau_{z}^{2}|}{\tau^{2}}\nonumber\\
\times\frac{|g_{\tau}(\omega_{0})|^{2}}{g_{00}^{\prime\prime}(\omega_{0})}|\ln
B|f\left(\frac{\Delta\omega}{\omega^{\tau}_{B}}\right),
\end{eqnarray}
if
$\omega^{\tau}_{B}-\Delta\omega<\omega_{0}<\omega^{\tau}_{B}+\Delta\omega$.
Here
$$
\omega^{\tau}_{B}=\frac{(2\pi\tau)^{2}}{2|2\pi\tau_{z}|};\,\,B=\ln[(\gamma^{-2}+\delta-g_{00}^{\prime})^{2}+|g_{\tau}|^{2}-\delta^{2}];
$$
$$
\delta=g_{\tau}^{\prime}\frac{g_{\tau}^{\prime\prime}(\omega_{0})}{g_{00}^{\prime\prime}(\omega_{0})};
$$
$$
f(x)=\left\{\begin{array}{cc}
       1 & x>\sqrt{\gamma^{-2}+g_{00}^{\prime}}, \\
       x/\sqrt{\gamma^{-2}+g_{00}^{\prime}} & x<\sqrt{\gamma^{-2}+g_{00}^{\prime}};
     \end{array}\right.
$$

$\Delta\omega$ is the energy resolution of the detector, we shall
assume to be appreciably less than the distance between the
nearest resonance frequencies, i.e.,
$$
\Delta\omega\ll
\min_{\tau_{1}\tau_{2}}[\omega^{\tau_{1}}_{B}-\omega^{\tau_{2}}_{B}]\simeq\pi\tau_{\min}.
$$

Complete information about the crystal structure is contained in
the quantities $n_{\tau}$, which, according to (\ref{15.13}) are
determined by structure amplitudes. The relative value of
$n_{\tau}$ as compared to the background counting rate $n_{0}$
associated with the transient radiation depends on $\Delta\omega$:
\begin{eqnarray}
\label{15.14}
\xi\equiv\frac{n_{\tau}}{n_{0}}\approx\frac{|\tau_{\perp}^{2}-\tau_{z}^{2}}{\tau^{2}|}\frac{|g_{\tau}|^{2}}{g_{00}^{\prime\prime}}\frac{\omega^{\tau}_{B}}{\Delta\omega};\nonumber\\
\frac{\Delta\omega}{\omega^{\tau}_{B}}\geq\sqrt{\gamma^{-2}+g_{00}^{\prime}}.
\end{eqnarray}
From (\ref{15.14}) follows that at a relative resolution of the
detector $\Delta\omega/\omega\sim 0.03$ the quantity $\xi$ for
real crystals varies within the limits from 0.01 to 0.1.

The method enabling one to select a weak signal with the intensity
$n_{s}$ against the noise of intensity $n_{n}\gg n_{s}$ ia
applicable to measure $n_{\tau}$. This method is widely used in
the problems dealing with the measurement of weak luminous fluxes
\cite{91}. It is based on splitting of the total measurement time
$t$ into two equal parts $t_{1}$ and $t_{2}$, with all the photons
associated with both mainstream and noise flows being registered
during time $t_{1}$. During the time period $t_{2}$ only noise
pulses are taken into account. Then the difference of the number
of photons $N_{1}$, gathered in time $t_{1}$ and the number of
photons registered by the detector in time $t_{2}$ determines the
signal intensity:
\begin{equation}
\label{15.15} n_{s}=2(N_{2}-N_{1})/t\pm\Delta n_{s},
\end{equation}
the relative accuracy $\Delta n/n_{s}$ is obviously dependent on
the measurement time $t$. The time necessary to attain the the
given accuracy $\beta$ can be easily found
\begin{equation}
\label{15.16} t_{\beta}=2(2n_{b}+n_{s})/n_{s}^{2}\beta^{2}.
\end{equation}

In the problem in question this method may be used as follows.
Suppose that a multichannel analyzer with the channel width
$\Delta\omega$ corresponding to the energy resolution of the
detector is used to study the pulses from the x-ray detector. Let
during time $t/2$ the pulses be summed up in each analyzer
channel, which appear at the detector output when an X-ray quantum
with the energy corresponding to the given channel gets into the
detector. Then the crystal should be turned through the angle
$\psi$ satisfying the condition
\begin{equation}
\label{15.17} 2\pi\tau_{\min}\sin\psi\gg\Delta\omega
\end{equation}
about the direction of the velocity of the electrons, with
$\tau_{\min}$ being the smallest of the vectors $\vec{\tau}$.

If the condition (\ref{15.17}) is fulfilled, the photon frequency
$\omega_{1}$, which was close to the resonance one for a certain
reciprocal lattice vector $\vec{\tau}(\omega_{1}\approx
\omega_{B}^{\tau})$, after the crystal rotation will appreciably
differ from it, so that the intensity of quanta with the frequency
$\omega_{1}$ will only be determined by the quantity $n_{0}$. If
now in each channel of the analyzer we subtract the number of
quanta registered by the detector during time $t/2$ after the
crystal rotation, then in the analyzer channels corresponding to
the resonance frequencies in the first time period, the number of
pulses $N_{\tau}$  will be  defined by formula
\begin{equation}
\label{15.18} N_{\tau}=n_{\tau}T/2\pm\Delta N,
\end{equation}
while in the rest of the channels the number of pulses is equal in
magnitude to $\Delta N$ - the number of pulses due to statistic
fluctuations of photons, and
\begin{equation}
\label{15.19} \Delta N\simeq\sqrt{(n_{\vec{\tau}}+n_{0})t/2}.
\end{equation}
Using (\ref{15.18}), we may find $n_{\tau}$ along with the
structure amplitude $F(\vec{\tau})$ with the absolute error
determined by the quantity $2\Delta N/t$. The time necessary to
measure $F(\tau)$ with the specified relative accuracy $\beta$ is
found, using (\ref{15.16}), if assume that $n_{s}=n_{\tau}$,
$n_{b}=n_{0}$:
\begin{equation}
\label{15.20} t_{\beta}=\frac{1}{\beta^{2}J|\ln
B|e^{2}}\frac{g_{00}^{\prime\prime
2}}{|g_{\tau}|^{4}}\frac{\Delta\omega}{\omega^{\tau}_{B}}.
\end{equation}

To estimate $t_{\beta}$ choose the electron current $J=10^{-6}$ A,
$E=50$  MeV, $\Delta\omega/\omega^{\tau}_{B}=0.03$,
$|g_{\tau}|=10^{-6}$, $g_{00}^{\prime\prime}=10^{-8}$ are the
typical values for real crystals. Then, to measure $F(\tau)$ with
the relative accuracy 0.01, we need the time $t_{\beta}\simeq
10^{-2}$ s.

Mention also a simpler way of selecting transient radiation
suitable for investigating crystals containing atoms with the
small number of electrons, when the frequencies
$\omega^{\tau}_{B}$ of photon emitted in parametric effect are
greater than characteristic atomic frequencies. In this case
$n_{0}$ has a universal dependence on the frequency $\omega$. That
is why it is not necessary to rotate a crystal to determine
$n_{\tau}$: suffice it to subtract
$N_{k}=N_{0}\omega_{0}^{2}/\omega_{k}^{2}$, where $N_{0}$ is the
number of pulses in the channel corresponding to the frequency
$\omega_{0}$ ($\omega_{0}$ satisfies $\omega_{0}<\pi\tau_{\min}$)
from the total number of photons registered in the channel of the
analyzer corresponding to the frequency $\omega_{k}$. It may be
demonstrated that in this case the time of accumulation is also
determined by (\ref{15.20}).

%%%%%%%%%%%%%%%%%%%%%%%%%%%%%%%%%%%  Chapter 5 %%%%%%%%%%%%%%%%%%%%%%

\chapter[Classical Theory of Radiation Formation by Particles in a Medium]{Classical Theory of Radiation Formation by Particles in a Medium}
\label{ch:5}
%\chaptermark{}

%%%%%%%%%%%%%%%%%%%%%%%%%%%%%%%%%%%  Section 16 %%%%%%%%%%%%%%%%%%%%%%

\section{Particle Radiation in a Medium in the Presence of Scattering and Energy Losses}
\label{sec:5.16}

Classical theory of production of electromagnetic radiation by
particles passing through a single crystal without regard to
refraction, absorption and diffraction was developed by M.A.
Kumakhov \cite{7,36}, M.I. Podgoretsky \cite{92,93}, A.I.
Akhiezer, V.F. Boldyshev and N.F. Shulga \cite{87}, D.A. Alferov,
Yu. A. Bashmakov, E. G. Bessonov \cite{94,95a,95b}, V.N. Baier,
V.M. Katkov, V.M. Strakhovenko \cite{58}.

Presented below is the classical theory of photon formation by
particles in a medium with due account of the effects caused by
refraction, absorption and diffraction, which also enables one
directly to allow for possible multiple scattering of particles
\cite{14,97,98,99,100}.

%%%%%%%%%%%%%%%
So, let a charge move in a medium (e.g., in a crystal) in an
arbitrary manner. The spectral density of radiation energy per
unit solid angle $W_{\vec{n}\omega}$($\vec{n}=\vec{k}/k$; the
differential number of quanta
$dN_{\vec{n}\omega}=W{\vec{n}\omega}/\hbar\omega$) as well as the
polarization characteristics of radiation may be easily obtained
if the field $\vec{E}(\vec{r},\,\omega)$ produced by a charge at
large distances from the crystal is known. For instance,
\begin{equation}
\label{16.1}
W_{\vec{n}\omega}=\frac{cr^{2}}{4\pi^{2}}\overline{\left|\vec{E}(\vec{r},\omega)\right|^{2}}\,,
\end{equation}
where $c$ is the speed of light; the vinculum means averaging over
all possible states of the system under consideration. To find the
field $\vec{E}(\vec{r},\,\omega)$, one should solve Maxwell's
equations which for an arbitrary medium have the form
\begin{equation}
\label{16.2} \left[-\textrm{rot}\, \textrm{rot}\,
\vec{E}(\vec{r},\,\omega)+\frac{\omega^{2}}{c^{2}}\vec{E}(\vec{r},\,\omega)\right]_{i}
+\frac{4\pi i\omega}{c^{2}}\hat{\sigma}_{ij}E_{j}=-\frac{4\pi
i\omega}{c^{2}}j_{0i}(\vec{r},\,\omega)\,,
\end{equation}
where $\hat{\sigma}_{ij}$ is the conductivity tensor of matter;
$j_{0i}(\vec{r},\,\omega)$ is the Fourier transform of the $i$-th
component of the current induced by a moving charge. In the
quantum mechanical case by "$j_{0i}(\vec{r},\,\omega)$" one should
understand the unaveraged over the crystal states current of
transition from one quantum mechanical state to another.

%%%%%%%%%%%%%%%%%%%%%%%%%%%%%%%%5
The transverse solution of (\ref{16.2}) can be found, using the
Green function $G$ of this equation satisfying the relation of the
form
\begin{equation}
\label{16.3} G=G+G_{0}\frac{i\omega}{c^{2}}\hat{\sigma}G,
\end{equation}
where $G_{0}$ is the transverse Green function of  equation
(\ref{16.2}) at $\hat{\sigma}=0$ (its explicit form see, for
example, in \cite{101}). Using $G$, it is easy to find the field
we are concerned with:
\begin{equation}
\label{16.4} \vec{E}(\vec{r},\omega)=\int
G_{il}(\vec{r},\vec{r}^{\,\prime},
\omega)\frac{i\omega}{c^{2}}j_{0l}(\vec{r}^{\,\prime})d^{3}\vec{r}^{\,\prime}\,.
\end{equation}
According to \cite{14} at $r\rightarrow\infty$ the Green function
is expressed via the solution of homogeneous Maxwell's equations
$E_{i}^{(-)}(\vec{r},\omega)$ containing a converging spherical
wave at infinity:
\begin{equation}
\label{16.5}
\lim_{r\rightarrow\infty}G_{il}(\vec{r},\vec{r}^{\,\prime},
\omega)=\frac{e^{ikr}}{r}\sum_{s}e_{i}^{s}E_{\vec{k}l}^{s(-)^{*}}(\vec{r}^{\,\prime},\omega)\,,
\end{equation}
%%%%%%%%%%%%%%%%%%%%%%%%%%%5
\begin{equation}
\label{16.6} \left[-\textrm{rot}\, \textrm{rot}\,
\vec{E}^{(-)}(\vec{r},\,\omega)+\frac{\omega^{2}}{c^{2}}\vec{E}^{(-)}(\vec{r},\,\omega)\right]_{i}
-\frac{4\pi i\omega}{c^{2}}\hat{\sigma}_{ij}^{*}E_{j}^{(-)}=0\,,
\end{equation}
where $\vec{e}^{\,s}$ is the transverse unit vector of
polarization; $s=1,\,2$.

If the wave is incident onto the object of finite dimensions,
then, at $r\rightarrow\infty$,
\begin{equation}
\label{16.7}
\vec{E}_{\vec{k}l}^{(-)}(\vec{r},\omega)=\vec{e}^{\,s}e^{i\vec{k}\vec{r}}+
\quad\mbox{const}\quad \frac{e^{-ikr}}{r}\,.
\end{equation}
Using (\ref{16.4}) and (\ref{16.5}), we find
\begin{equation}
\label{16.8}
E_{i}(\vec{r},\omega)=\frac{e^{-ikr}}{r}\frac{i\omega}{c^{2}}\,
\sum_{s}e^{s}_{i}\int\vec{E}_{\vec{k}}^{(-)s^{*}}(\vec{r}^{\,\prime},\omega)\vec{\jmath}(\vec{r}^{\,\prime},\omega)d^{3}r^{\prime}\,.
\end{equation}
%%%%%%%%%%%%%%%
In view of (\ref{16.1}) and (\ref{16.8}), the spectral density of
radiation is
\begin{equation}
\label{16.9}
W_{\vec{n}\omega}=\sum_{s}W_{s\vec{n}\omega}=\frac{\omega^{2}}{4\pi^{2}c^{3}}
\,\sum_{s}
\left|\overline{\int\vec{E}_{\vec{k}}^{s(-)^{*}}(\vec{r},\omega)\vec{\jmath}(\vec{r},\omega)d^{3}r}\right|^{2}\,,
\end{equation}
where $W_{s\vec{n}\omega}$ is the spectral density of radiation
per unit solid angle for photons, characterized by the
polarization vector $\vec{e}^{\,s}$. To explicitly find
$W_{\vec{n}\omega}$, it is necessary to know the field
$\vec{E}_{\vec{k}}^{s(-)}$ and the current $\vec{\jmath}$. With
known solution $\vec{E}_{\vec{k}}^{s(+)}$ of the homogenous
Maxwell equations describing the process of photon scattering by
the target, field $\vec{E}_{\vec{k}}^{s(-)}$ can be found  using
the below relation:
\begin{equation}
\label{16.10}
\vec{E}_{\vec{k}}^{s(-)^{*}}=\vec{E}_{-\vec{k}}^{s(+)}\,.
\end{equation}
Introduce the following explicit expression for the Fourier
transform of the current into (\ref{16.9}):
\begin{eqnarray}
\label{16.11}
\vec{\jmath}(\vec{r},\omega)=\int e^{i\omega t}\vec{\jmath}(\vec{r},t)dt\,,\nonumber\\
\vec{\jmath}(\vec{r},t)=e\vec{v}(t)\delta(\vec{r}-\vec{r}(t))\,.
\end{eqnarray}
%%%%%%%%%%%%%%%%%

Substitution of (\ref{16.11}) into (\ref{16.9}) gives
\begin{eqnarray}
\label{16.12}
W_{s\vec{n}\omega}=\frac{e^{2}\omega^{2}}{4\pi^{2}c^{3}}\int_{t_{1}}^{t_{2}}\int(\vec{E}_{\vec{k}}^{(-)s^{*}}(\vec{r}(t))\vec{v}(t))e^{i\omega t}\nonumber\\
\times(\vec{E}_{\vec{k}}^{(-)s}(\vec{r}(t^{\prime}))\vec{v}(t^{\prime}))e^{-i\omega
t^{\prime}}dtdt^{\prime},
\end{eqnarray}
where $t_{1}$ and $t_{2}$ are the starting and finishing moments
of the charge motion, respectively.

In (\ref{16.2}) perform averaging over the possible particle
trajectories in a medium. Such averaging is usually performed with
the combined probability density $w(\vec{r},\vec{v},t;
\vec{r}^{\,\prime},\vec{v}^{\,\prime},t^{\prime})$ of finding  the
coordinate $\vec{r}$  and the velocity $\vec{v}$ of a particle at
moment $t$,  the coordinate $\vec{r}^{\,\prime}$ and the velocity
$\vec{v}^{\,\prime}$ at moment $t^{\prime}$. However, when
investigating the effects of the energy losses, it is more
convenient to perform averaging with a similar function, which
depends on variables $\vec{r}$ and $\vec{p}$, where $\vec{p}$ is
the particle momentum. As a result ($c=1$),
\begin{eqnarray}
\label{16.13} W_{s
\omega}=\frac{e^{2}\omega^{2}}{4\pi^{2}}\int\int\limits_{t_{1}}^{t_{2}}\int
\left(\vec{E}_{\vec{k}}^{(-)s^{*}}(\vec{r})\frac{\vec{p}}{E(p)}\right)
\left(\vec{E}_{\vec{k}}^{(-)s}(\vec{r}^{\,\prime})\frac{\vec{p}^{\,\prime}}{E^{\prime}(p^{\,\prime})}\right)\nonumber\\
\times w(\vec{r},\vec{p}\,,
 t\,,\, \vec{r}^{\,\prime}, \vec{p}^{\,\prime}, t^{\prime})
 e^{i\omega(t-t^{\prime})}d^{3}rd^{3}r^{\prime}d^{3}pd^{3}p^{\prime}dtdt^{\prime}\,.
\end{eqnarray}

%%%%%%%%%%%%%%%%%%%%%%%%%

Choose the coordinate system so that the $x\,y$ plane  coincides
with the  matter--vacuum boundary. Direct the $z$-axis from the
medium to vacuum. Suppose that a particle with  momentum
$\vec{p}_{0}$ directed along the $z$-axis starts moving at time
$(-T)$ at point $(0,\,0,\,-z_{0})$ inside the medium. Let it cross
the matter--vacuum boundary at  moment $t=0$.

In the case of high energies of $\gamma$-quanta we are concerned
with, we may neglect mirror reflected waves in the expressions for
the fields $\vec{E}_{\vec{k}}^{(-)s}$. As a result,
\begin{eqnarray}
\label{16.14} \vec{E}_{\vec{k}}^{(-)s}=\left\{\begin{array}{cc}
                           \vec{e}^{\,s}e^{i\vec{k}\vec{r}} &  \mbox{at}\, z>0\,, \\
                          \vec{e}^{\,s}e^{i\vec{k}^{\prime *}\vec{r}} & \mbox{at} \,z<0\,,
                         \end{array}\right.
\end{eqnarray}
where $\vec{k}^{\prime}$ is the photon wave vector in the medium
with the components
$\vec{k}^{\prime}_{\perp}=\omega\vec{n}_{\perp}$,
$k^{\prime}_{z}=\omega\sqrt{\varepsilon}\vec{n}_{z}$.

%%%%%%%%%%%%%%%%%
Using (\ref{16.14}) and going from variables ($p$, $\theta$,
$\varphi$) to  variables ($E$, $\vec{\theta}$), where $E$ is the
energy and
$\vec{\theta}=\theta_{x}\vec{\imath}+\theta_{y}\vec{\jmath}$ is
the transverse angular vector, one can obtain the following
expression for the intensity distribution
$W_{\parallel\vec{n}\omega}$ of photons polarized in the  plane of
exit from matter:
\begin{eqnarray}
\label{16.15} &
&W_{\parallel\vec{n}\omega}=\frac{e^{2}\omega^{2}\vartheta^{2}}{4\pi^{2}}
\left\{\int\limits_{0}^{\infty}dt\int\limits_{0}^{\infty}dt^{\prime}\int\limits d\xi d\xi^{\prime}F(\theta)F(\theta^{\prime})\right.\nonumber\\
& &\left.\times\exp[-i\vec{k}(\vec{r}-\vec{r}^{\,\prime})+i\omega(t-t^{\prime})]\right.\nonumber\\
& &\left.\times w_{1}(\vec{r},\vec{\theta},\, E,\,
t+\tau)w_{2}(\vec{r},\vec{\theta},\,
E,\, t|\vec{r}^{\,\prime},\vec{\theta}^{\prime}, E^{\prime}, t^{\prime})\right.\nonumber\\
& &\left.+2 \texttt{Re}\int\limits_{-T}^{0}dt\int\limits_{0}^{\infty}dt^{\prime}\int\limits d\xi d\xi^{\prime}F(\theta)F(\theta^{\prime})\right.\nonumber\\
& &\left.\times\exp[-i(\vec{k}^{\prime}\vec{r}-\omega t)+i(\vec{k}\vec{r}^{\,\prime}-\omega t^{\prime})]\right.\nonumber\\
& &\left.\times w_{1}(\vec{r},\vec{\theta}, E, t+T)w_{2}(\vec{r},
\vec{\theta},\, E,\, t|\vec{r}^{\,\prime},\vec{\theta}^{\prime}, E^{\prime}, t^{\prime})\right.\nonumber\\
& &\left.+2 \texttt{Re}\int\limits_{-T}^{0}dt\int\limits_{0}^{-t}d\tau\int\limits d\xi d\xi^{\prime}F(\theta)F(\theta^{\prime})\right.\nonumber\\
& &\left.\times\exp[-i\omega\tau+i\vec{k}^{\prime *}(\vec{r}^{\,\prime}-\vec{r})]\exp(\omega z \texttt{Im}\varepsilon)\right.\nonumber\\
& &\left.\times w_{1}(\vec{r},\vec{\theta},\, E,\,
t+T)w_{2}(\vec{r},\vec{\theta},\, E,\,
0|\vec{r}^{\,\prime},\vec{\theta}^{\prime}, E^{\prime},
\tau^{\prime})\right\}\,,
\end{eqnarray}
%%%%%%%%%%%%%%%%%%55
where $\xi$ is the set of coordinates ($\vec{r}$, $\vec{\theta}$,
$E$); $w_{1}(\vec{r}, \vec{\theta},\, E,\, t)$ is the probability
of  finding  the particle coordinates ($\vec{r}$, $\vec{\theta}$,
$E$) at time $t$; $w_{2}(\vec{r}, \vec{\theta},\, E,\,
t|\vec{r}^{\,\prime},\vec{\theta}^{\prime}, E^{\prime},\,
t^{\prime})$ is the conditional probability of finding
 the particle coordinates
($\vec{r}^{\,\prime},\vec{\theta}^{\prime}, E^{\prime} $)
 at time $t^{\prime}$ if at moment $t$ the particle
coordinates were ($\vec{r}, \vec{\theta}, E$);
$F(\theta)=1-(\theta_{x}\cos\vartheta_{x}+\theta_{y}\cos\vartheta_{y})\vartheta^{-2}$;
${\vartheta_{x}, \,\vartheta_{y},\, \vartheta_{z}\equiv\vartheta}$
are the direction angles of vector $\vec{k}(\vec{n})$.

%%%%%%%%%%%%%%%%%%%%%5
The expression for spectral--angular distribution of the intensity
of  photons whose polarization vector is perpendicular to the exit
plane  is derived from (\ref{16.15}) by substitution of
$\vartheta^{-2}$ for $\vartheta^{2}$ and
${B(\theta)=\theta_{x}\cos\vartheta_{y}-\theta_{y}\cos\vartheta_{x}}$
for $F(\theta)$.

Pay attention to the fact that some integrals in (\ref{16.15})
contain the probability densities $w_2$ which depend on the
instants of time corresponding to particle motion both in the
medium and outside it. However, it is more convenient to deal with
the densities which depend on the instants of time referring to
particle  motion in the medium or outside it alone.  With this aim
in view, make use of the following general property of the
distribution functions:
\begin{equation}
\label{16.16} w_{2}(\xi, t| \xi^{\prime}, t^{\prime})=\int
w_{2}(\xi, t| \xi^{\prime\prime},
t^{\prime\prime})w_{2}(\xi^{\prime\prime},
t^{\prime\prime}|\xi^{\prime}, t^{\prime})d\xi^{\prime\prime}\,.
\end{equation}

%%%%%%%%%%%%%%%%%%%%%%5
Substituting (\ref{16.16}) into (\ref{16.15}) and choosing the
instant of time corresponding to the moment of particle exit from
matter, i.e., $t^{\prime\prime}=0$, we obtain the expression for
$W_{s\vec{n}\omega}$ which only depends on the distribution
functions describing the particle motion in the medium or outside
it.

The probabilities $w_{1}$ and $w_{2}$ satisfy the kinetic equation
which, for example, in  a chaotic medium has the form
\begin{equation}
\label{16.17} \frac{\partial w}{\partial
t}+\frac{\vec{p}}{E}\frac{\partial
w}{\partial\vec{r}}=\left(\frac{\partial w}{\partial
t}\right)_{col}.
\end{equation}
(The case of a crystal is discussed in Chapter (\ref{ch:10}).

%%%%%%%%%%%%%%%%55
In our case the change of the collisional term
$\left(\frac{\partial w}{\partial t}\right)_{col}$ in time is due
to  scattering and  radiation processes, and it may be described
by the equations of the form:
\begin{equation}
\label{16.18} \left(\frac{\partial w^{(n)}}{\partial
t}\right)_{\mathrm{col}}=
-\sum_{n^{\prime}}g_{n^{\prime}n}w^{(n)}+\sum_{n^{\prime}}g_{nn^{\prime}}w^{(n^{\prime})}\,,
\end{equation}
where $g_{nn^{\prime}}$ is the probability of the system
transition from  state $n$ (in our case of the electron in  state
$n$) to  state $n^{\prime}$ per unit time. The probabilities
$g_{nn^{\prime}}$ may be found by conventional rules \cite{19}.

%%%%%%%%%%%%%%%%55
As a result, for example, in a chaotic medium, taking account of
the change in $w^{(n)}$, which is only due to multiple scattering
and bremsstrahlung, gives
\begin{eqnarray}
\label{16.19}
& &\left(\frac{\partial w(\vec{p},t)}{\partial t}\right)_{\mathrm{col}}=-N\sigma_{\mathrm{tot}}w(\vec{p},t)\nonumber\\
&
&+N\int\frac{d^{3}p^{\prime}}{(2\pi)^{2}}\delta(E_{p}-E_{p^{\,\prime}})
\frac{|M_{\vec{p}^{\,\prime}\vec{p}}|^{2}}{4E_{p}^{2}}w(\vec{p},t)\nonumber\\
&
&+N\int\frac{d^{3}p^{\prime}d^{3}k}{(2\pi)^{5}}\delta(E_{p}-E_{p^{\,\prime}}-k)
\frac{|M_{\vec{p}^{\,\prime}\vec{k},\vec{p}}|^{2}}{8E_{p}E_{p^{\,\prime}}k}w(\vec{p}^{\,\prime},t)\,,
\end{eqnarray}
where $N$ is the number of scatterers (nuclei) per unit volume;
the amplitude $M_{\vec{p}^{\,\prime}\vec{p}}$ describes electron
scattering in the nuclear Coulomb field, the amplitude
$M_{\vec{p}^{\,\prime}\vec{k},\vec{p}}$  describes the emission of
$\gamma$-quanta by the electron in the nuclear field;
$\sigma_{\mathrm{tot}}$ is the total cross section of all the
processes.
%%%%%%%%%%%%%%%%%%%%%%55

Give a more detailed treatment of the case when the emission of
$\gamma$-quanta may be described by the  Bethe-Heitler expression
at complete screening of the nuclear field
\cite{19,102}.\footnote{The influence of multiple scattering (the
Landau-Pomeranchuk effect)and the effect of the medium
polarization on the bremsstrahlung cross-section may be taken into
account, using the method described by Ter-Mikaelian in
\cite{63,64}.}

Turning from the probability describing the electron distribution
in the momenta to the probabilities describing particle
distribution in energies and scattering angles and taking the
appropriate transformations of (\ref{16.19}), we obtain the
following equation
\begin{eqnarray}
\label{16.20}
\left(\frac{\partial w(\vec{\theta},E, t)}{\partial t}\right)_{col}=q(E)\Delta_{\theta}w(\vec{\theta}, E, t)\nonumber\\
+K(E)w(\vec{\theta}, E, t),
\end{eqnarray}
where
$\Delta_{\theta}=\frac{\partial^{2}}{\partial\theta^{2}_{x}}+\frac{\partial^{2}}{\partial\theta^{2}_{y}}$;
$q(E)=\delta E^{-2}$; $\delta=\frac{1}{4}E^{2}_{s}$; $K(E)$ is the
integral operator of the form \footnote{It is interesting to note
that the equation obtained can be derived from  well known
equations of the shower theory \cite{103} with the terms referring
to the pair formation being dropped.}
\begin{eqnarray}
\label{16.21}
K(E)w(\vec{\theta}, E, t)=\int_{E}^{\infty}\frac{u^{2}+E^{2}-\frac{2}{3}uE}{u^{2}(u-E)}w(\vec{\theta}, u, t)du\nonumber\\
-\int^{E}_{0}\frac{u^{2}+E^{2}-\frac{2}{3}u
E}{E^{2}(E-u)}w(\vec{\theta}, E, t)du.
\end{eqnarray}
%%%%%%%%%%%%%%%%%%%%%%%%%%5

The initial conditions for the distribution functions $w_{1}$ and
$w_{2}$ have the form
\begin{eqnarray*}
& &w_{1}(t=-T)=\delta(\vec{r}-\vec{r}_{0})\delta(\theta)\delta(E-E_{0})\,,\\
&
&w_{2}(t=t^{\prime})=\delta(\vec{r}-\vec{r}^{\,\prime})\delta(\vec{\theta}-\vec{\theta}^{\prime})\delta(E-E^{\prime})\,,
\end{eqnarray*}
where $E_{0}$ is the initial energy of the particle;
\[
\frac{\vec{p}}{E}=\vec{v}\,, \vec{v} \quad \mbox{has the
components}\quad \theta_{x},\, \theta_{y},\,
1-\frac{\theta_{x}^{2}+\theta_{y}^{2}}{2}\,.
\]

Using kinetic equation (\ref{16.20}), it is possible to find the
time--dependence of the mean--square angle
$\langle\theta^{2}(E,\,t)\rangle$ of  multiple scattering of the
electron of energy $E$ within the interval $dE$. For this purpose,
we shall multiply (\ref{16.20}) by
$\theta^{2}=\theta_{x}^{2}+\theta_{y}^{2}$ and integrate it over
$\vec{r}$  and $\vec{\theta}$. As a result, we have
\begin{equation}
\label{16.22} \frac{\partial\langle\theta^{2}(E,
t)\rangle}{\partial t}=4q(E)w(E,t)+K(E)\langle\theta^{2}(E,
t)\rangle\,,
\end{equation}
where $w(E, t)=\int w(\vec{r}, \vec{\theta}, E,
t)d^{3}rd^{2}\theta$ is the probability of finding an electron
with the energy $E$ at  time $t$ if at $t=0$ its energy is
$E_{0}$.

%%%%%%%%%%%%%%%%%%%%%%5
Solving (\ref{16.22}) using the Mellin transform, find
\begin{equation}
\label{16.23} \langle\theta^{2}(E,
t)\rangle=4\delta\int\limits_{0}^{t}dt^{\prime}\int\limits_{E}^{E_{0}}\frac{dE^{\prime}}{E^{\prime}}
 w(E_{0}|E^{\prime},\, t^{\prime})w(E^{\prime}|E,
t^{\prime}-t)\,.
\end{equation}
For particular calculation of (\ref{16.23}), make use of the
approximate expression, derived by Bethe and Heitler \cite{102}:
\begin{equation}
\label{16.24} w(E_{0}, E,
t)=\frac{1}{E_{0}}\frac{\left(\ln\frac{E_{0}}{E}\right)^{\frac{t}{\ln
2}-1}}{\Gamma\left(\frac{t}{\ln 2}\right)}\,,
\end{equation}
where $t$ is measured in radiation units;
$\Gamma\left(\frac{t}{\ln 2}\right)$ is the gamma function.
Substitution of (\ref{16.24}) into (\ref{16.23}) gives
\begin{equation}
\label{16.25} \langle\theta^{2}(E,
t)\rangle=\frac{4q(E)\left(\ln\frac{E_{0}}{E}\right)^{\frac{t}{\ln
2}-1}}{E_{0}\Gamma\left(\frac{t}{\ln 2}\right)}
\int\limits_{0}^{t}\Phi\left(\frac{\tau}{\ln 2}, \frac{t}{\ln 2},
2 \ln\frac{E_{0}}{E}\right)d\tau\,,
\end{equation}
where $\Phi$ is the degenerate hypergeometric function. Expression
(\ref{16.25}) differs considerably from a simple exponential
dependence, obtained through substitution of the equality $\langle
E\rangle=E_{0}e^{-\vec{t}/L}$ into
$\langle\theta^{2}(E,t)\rangle=4qt$.

%%%%%%%%%%%%%%%5
Like in the case without losses, further analysis is convenient to
perform,  studying the equations for the functions of the type
given below, which appear in (\ref{16.15})
\begin{equation}
\label{16.26} u_{0}(\vec{\theta}, E, t +T)=\int d^{3}r
w_{1}(\vec{r},\vec{\theta},\,E,\, t +T)e^{\omega z
\texttt{Im}\varepsilon}\,
\end{equation}
\begin{equation}
\label{16.27}  u_{2}(\vec{\theta},\, \vec{\theta}^{\prime},\, E,\,
E^{\prime},\, \vec{\tau})  =\int d^{3}\rho w_{2}(\vec{\rho},\,
\vec{\theta},\, \vec{\theta}^{\prime},\, E,\, E^{\prime},
\vec{\tau})e^{-i(\omega\vec{\tau}-\vec{k}^{\prime *}
\vec{\rho})}\,.
\end{equation}

%%%%%%%%%%%%%%%%%5
In view of (\ref{16.26}) and (\ref{16.27}), multiplication of
(\ref{16.20}) by the corresponding multipliers gives gives the
following equation for $u$  in a chaotic medium
\begin{equation}
\label{16.28} \frac{\partial u}{\partial t}+A(\theta,
E)u=q(E)\Delta_{\theta}u+K(E)u\,,
\end{equation}
where in the case of $u_{0}$
\begin{equation}
\label{16.29} A(\theta, E)u\equiv
A_{0}(\theta)=-\omega\beta\left(1-\frac{1}{2}\theta^{2}\right)\texttt{Im}\varepsilon\,,
\end{equation}
in the case of $u_{2}$
\begin{eqnarray}
\label{16.30} & &A(\theta, E)\equiv A_{2}(\theta)\\
&
&=i\omega\left[1-\beta(\theta_{x}\cos\vartheta_{x}+\theta_{y}\cos\vartheta_{y})
-\beta\left(1-\frac{1}{2}\theta^{2}\right)
\left(1-\frac{1}{2}\theta^{2}+\frac{1}{2}\delta\varepsilon^{*}\right)\right]\nonumber\,,
\end{eqnarray}
where $\beta=v/c$; $c=1$.

%%%%%%%%%%%%%%%%%%%%%%%%
Thus, to find the radiation spectrum one should solve equation
(\ref{16.28}). Similar equations can be analyzed, using the
methods developed in the cascade theory \cite{103}, though the
formal solution obtain thus obtained is sophisticated in form.

Let us give a more detailed treatment of the case of radiation in
an amorphous medium when the energy losses can be neglected (the
plate thickness is much smaller than the radiation length). The
problem of photon radiation in the X-ray and optical regions by
particles passing through the matter-vacuum boundary was discussed
in many publications. However, angular, spectral and polarization
properties of the radiation produced in the presence of multiple
scattering were analyzed regardless photon absorption in the
medium, and for this reason they are not suitable for the study
of, e.g., generation of resonance photons (optical, X-ray, or
M\"{o}ssbauer ones). Below the results obtained in \cite{98,99}
are presented.

%%%%%%%%%%%%%%%%%%%%%%%%%%%%%%%%%%%  Section 17 %%%%%%%%%%%%%%%%%%%%%%

\section{Spectral-Angular Distribution in the Absence of the Energy Loss}
\label{sec:5.17}

First, consider the problem of the relation between
bremsstrahlung, transition and Cherenkov radiations in the range
of high energy $\gamma$-quanta.

It is worth mentioning that the analysis of the role of transition
radiation in the  X-ray spectral range used the expression of the
form below for dielectric permittivity of the the medium:
\begin{equation}
\label{17.1} \varepsilon=1-\frac{\omega^{2}_{L}}{\omega^{2}}\,,
\end{equation}
where $\omega^{2}_{L}=4\pi e^{2}zN/m$; $\omega_{L}$ is the
Langmuir frequency; $\omega$ is the frequency of the emitted
quantum.

 Equation (\ref{17.1}) is only valid for the energy ranges where the Compton scattering is the major mechanisms of scattering of
 $\gamma$-quanta. In the  range of high energies of $\gamma$-quanta we are concerned with, the significant contribution to
 $\varepsilon$ also comes from the pair production processes, and  $\texttt{Re}(\varepsilon-1)<\texttt{Im}\,\varepsilon$.

1. To clarify the relationship between the transition radiation
and bremsstrahlung in the high energy regions, it is necessary to
find the intensity $W_{n\omega}$ of radiation emerging in a vacuum
in the direction of motion of a particle passing through a layer
of matter. Assume for  simplicity that the layer thickness is much
larger than the absorption length of $\gamma$-quanta. In this case
in a similar manner as when solving the problem of optical
radiation of a particle entering the matter \cite{104}, one may
obtain the expression for $W_{\vec{n}\omega}$ coinciding with the
that derived by Pafomov [see \cite{104}, formulae
(27.44)-(27.49)], upon substituting in the latter
$\beta\rightarrow-\beta$ and
$\varepsilon\rightarrow\varepsilon^{*}$ in all the functions
except $\eta=[4\omega\beta q
\texttt{Im}\sqrt{\varepsilon-\sin^{2}\vartheta}]^{1/2}$. Taking
account of the fact that at high energies the angular distribution
of radiation is concentrated within a very narrow angle relative
to  the direction of particle motion, and the dielectric
permittivity of matter is close to unity, allows us to simplify
formulas (27.44)-(27.49) in \cite{104}. They take the simplest
form if the following condition ifs fulfilled
\begin{equation}
\label{17.2}
\langle\theta^{2}\rangle\ll\frac{\omega}{c}(\texttt{Im}\varepsilon)^{2}),
\end{equation}
where $\langle\theta^{2}\rangle$ is the mean-square angle of
multiple scattering per unit length. The condition (\ref{17.2})
may be recast as follows:
\begin{equation}
\label{17.3} \langle\theta^{2}\rangle L_{c}\ll\vartheta^{2}_{c},
\end{equation}
where $L_{c}=\frac{c}{\omega \texttt{Im}\varepsilon}$ is the
absorption depth of $\gamma$-quanta;
$\vartheta^{2}_{c}=\frac{c}{\omega L_{c}}$ is the squared
effective angle of quantum emission.

According to (\ref{17.3}) the formulas in question simplify, if
the mean-square  angle of multiple scattering of an electron over
the  absorption length of $\gamma$-quantum is smaller than the
squared effective angle of radiation. Then the expression for
angular and spectral distributions of the radiation intensity
$W_{\vec{n}\omega}$ takes the form
\begin{eqnarray}
\label{17.4} & &
W_{\vec{n}\omega}=\frac{e^{2}\vartheta^{2}}{4\pi^{2}c}\frac{|\delta\varepsilon|^{2}|1-\beta^{2}-
\beta-\frac{1}{2}\delta\varepsilon+\frac{1}{2}\vartheta^{2}|^{2}}
{(1-\beta^{2}+\vartheta^{2})^{2}|1-\beta-\frac{1}{2}\delta\varepsilon+\frac{1}{2}\vartheta^{2}|^{2}}\nonumber\\
&
&+\frac{e^{2}\langle\theta^{2}\rangle}{\pi^{2}\omega}\frac{\vartheta^{2}}{(1-\beta^{2}+\vartheta^{2})}
\texttt{Im}\frac{1-\beta-\frac{1}{2}\delta\varepsilon}{(1-\beta-\frac{1}{2}\delta\varepsilon+\frac{1}{2}\vartheta^{2})^{4}}\nonumber\\
&
&+\frac{e^{2}\langle\theta^{2}\rangle}{4\pi^{2}\omega\texttt{Im}\varepsilon|1-\beta-\frac{1}{2}
\delta\varepsilon+\frac{1}{2}\vartheta^{2}|}\,,
\end{eqnarray}
where $\vartheta$ is the radiation angle;
$\delta\varepsilon=\varepsilon-1$. The first term in (\ref{17.4})
describes transition radiation, the second and third ones are
non-zero only allowing for the particle scattering in a medium and
describe the interference of the transition radiation and
bremsstrahlung, and bremsstrahlung itself.

Using the condition (\ref{17.2}), one may notice that the second
and third terms are smaller than the first term, i.e., the
intensity of transition radiation is greater than that of
bremsstrahlung. To make sure, use the following common expression
for $\langle\theta^{2}\rangle_1$ \cite{63}
\begin{equation}
\label{17.5} \langle\theta^{2}\rangle_1
=4E_{s}^{2}/L_{\mathrm{r}}E^{2}\,,
\end{equation}
where $L_{\mathrm{r}}$ is the radiation length; $E_{s}=21$\,MeV;
$E$ is the electron energy. As in the energy range under
consideration $L_{c}\simeq L_{\mathrm{r}}$,  the condition
(\ref{17.2}) may be recast as
\begin{equation}
\label{17.6} E\gg E_{s}\sqrt{\frac{\omega}{c}L_{\mathrm{r}}}\,.
\end{equation}
Fulfilment of (\ref{17.6}) entails satisfying the condition
\begin{equation}
\label{17.7}
1-\beta\simeq\frac{(m_{0}c^{2})^{2}}{2E^{2}}\ll\frac{E^{2}_{s}}{E^{2}}\ll\frac{c}{\omega
L_{c}}=\texttt{Im}\varepsilon\,.
\end{equation}
Integration of (\ref{17.4}) with respect to the angles using
(\ref{17.7}), gives the following expression for the first term
describing  the intensity of transition radiation:
\begin{equation}
\label{17.8} W_{\mathrm{tr}}\simeq\frac{2e^{2}}{\pi
c}\ln\frac{E\sqrt{|\delta\varepsilon|}}{m_{0}c^{2}}\,,
\end{equation}
for the second and third terms, we obtain the estimate coinciding
with that for the bremsstrahlung \cite{104}:
\[
W_{\mathrm{br}}\simeq\frac{e^{2}\langle\theta^{2}\rangle }{\pi
\omega}\frac{1}{(\texttt{Im}\varepsilon)^{2}}
\]
or, taking into account (\ref{17.2}),
\begin{equation}
\label{17.9} W_{\mathrm{br}}\ll\frac{e^{2}}{\pi c}\,.
\end{equation}

%%%%%%%%%%%%%%%%
Thus, in the energy range, where the condition (\ref{17.2}) holds,
the intensity of bremsstrahlung is much less than that of
transition radiation of $\gamma$-quanta. If a less stringent
condition
$\langle\theta^{2}\rangle\sim\omega/c(\texttt{Im}\varepsilon)^{2}$
is fulfilled, the intensities $W_{\mathrm{tr}}$ and
$W_{\mathrm{br}}$ (and their interference) become comparable in
magnitude, so  the separate consideration of transition radiation
and bremsstrahlung is also impossible for high--energy
$\gamma$-quanta. In view of (\ref{17.6}) such a situation arises
at the electron energy
\[
E\sim E_{s}\sqrt{\frac{\omega}{c}}L\,,
\]
i.e., at $E\sim 10^{14}$\,eV for $\gamma$-quanta in the range of
several gigaelectronvolts, and substances for which $L\sim 1$\,cm
(for example, for copper $L=1.29$ cm, for lead $L=0.46$\,cm).

Interestingly enough, in the ranges of high--energy
$\gamma$-quanta the contribution of the pair production processes
to the real part of dielectric permittivity is positive, in
contrast to the negative contribution from the Compton scattering
(see (\ref{17.1}). According to \cite{105}, allowing for pair
production
\begin{equation}
\label{17.10} \texttt{Re}\,\varepsilon=1+\frac{4\pi
Nc^{2}}{\omega^{2}}\left[-\frac{z
e^{2}}{mc^{2}}+\frac{7}{18}z(z+1)\alpha^{3}a\right]\,,
\end{equation}
where $a$ is the shielding radius; $\alpha$ is the fine structure
constant, $z$ is the atomic number of the nucleus.

%%%%%%%%%%%
From (\ref{17.10}) follows that under standard conditions the
contribution of pair production processes to
$\texttt{Re}\varepsilon$ is by the order of magnitude less than
that of the Compton scattering even for heavy substances
\cite{105}. As a consequence, $\texttt{Re}\,\varepsilon<1$, and
the Vavilov--Cherenkov effect is impossible. Nevertheless,
according to (\ref{17.10}), to increase the contribution of pair
production processes to $\texttt{Re}\,\varepsilon$  is possible by
increasing the shielding radius which is attainable in plasma.
Thus, in hydrogen plasma the contribution of pairs to
$\texttt{Re}\,\varepsilon$ becomes greater than the Compton
contribution for the shielding radii $a>2\cdot 10^{-6}$\,cm. As a
result, $\texttt{Re}\,\varepsilon>1$, and Cherenkov radiation is
possible even in the high energy range in the isotropic
homogeneous medium.

3. Formula (\ref{17.4}) also enables analyzing transition and
Cherenkov radiations of resonance $\gamma$-quanta and the effect
that multiple scattering of electrons in matter produce on the
stated processes.

Let us first dwell upon the role of bremsstrahlung. To avoid the
influence of multiple scattering on transition radiation, the
condition (\ref{17.2}) should to be fulfilled. Using (\ref{17.5})
recast (\ref{17.2}) as follows
\begin{equation}
\label{17.11} E\gg
2E_{s}\sqrt{\frac{\omega}{c}\frac{L_{c}^{2}(\omega)}{L}}\,,
\end{equation}
where
\[
L_{c}(\omega)=\frac{c}{\omega \texttt{Im}\,\varepsilon(\omega)}
\]
is the absorption depth of a quantum of frequency $\omega$. From
(\ref{17.11}) follows that, for example, for $^{119}$Sn
($\hbar\omega= 24$\,keV, $L_{c}(\omega)$ in the resonance is
$2.8\cdot 10^{-4}$\,cm) multiple scattering may be neglected at
the electron energies $E\gg 400$\,MeV. Assuming that the condition
(\ref{17.11}) is fulfilled, upon integrating (\ref{17.4}) with
respect to the angles, we obtain the following expression for
spectral intensity of transition (and, if possible, Cherenkov)
radiation \cite{106}:
\begin{eqnarray}
\label{17.12} W_{\omega}& &=\frac{e^{2}}{2\pi
c}\left[1-\frac{4(1-\beta)\texttt{Re}\,\delta\varepsilon}{|\delta\varepsilon|^{2}}\right]
\ln\frac{[2(1-\beta)-\texttt{Re}\,\delta\varepsilon]^{2}+(\texttt{Im}\,\delta\varepsilon)^{2}}{4(1-\beta)^{2}}\nonumber\\
& &+\frac{e^{2}}{\pi
c}\left[\frac{\texttt{Re}\,\delta\varepsilon}{\texttt{Im}\,\delta\varepsilon}+\frac{2(1-\beta)}{\texttt{Im}\,\delta\varepsilon}
\frac{(\texttt{Im}\,\delta\varepsilon)^{2}-(\texttt{Re}\,\delta\varepsilon)^{2}}{|\delta\varepsilon|^{2}}\right]\nonumber\\
&
&\times\left[\frac{\pi}{2}-\arctan\frac{2(1-\beta)-\texttt{Re}\,\delta\varepsilon}{\texttt{Im}\,\delta\varepsilon}\right]
-\frac{e^{2}}{\pi c}\,.
\end{eqnarray}

Using the values of $\delta\varepsilon$ given in \cite{107}, one
may obtain the below estimation of the number of $\gamma$-quanta
in the center of the M\"{o}ssbauer line in the energy region
$\Delta\omega$ of the order of the level width $\Gamma$, which are
emitted by an electron with the energy of the order of 1\,GeV:
\begin{equation}
\label{17.13}
N_{\gamma}=\frac{W_{\omega}}{\hbar\omega}\frac{\Gamma}{\hbar}\simeq
10^{-12}\,.
\end{equation}

%%%%%%%%

The estimate (\ref{17.13}) was obtained in \cite{108} by
numerical solution. This estimate is also valid  for the case when
the difference $2(1-\beta)-\texttt{Re}\delta\varepsilon$ may
vanish, i.e., in the presence of the Cherenkov radiation
mechanism. Note that for the electron energy, at which
$1-\beta\ll|\delta\varepsilon|$, the radiation intensity
$W_{\omega}$ is determined by formula (\ref{17.8}), demonstrating
weak dependence on the type (form) of $\varepsilon$, unlike the
case of a thin target. Thus, to detect in the emission spectrum
the anomalies associated with the resonance level, it is necessary
that the electron energy should not be very high (for
$\delta\varepsilon\sim 10^{-5}$ the term in (\ref{17.12}) similar
to that in (\ref{17.8}) exceeds tenfold the terms  depending on
$\delta\varepsilon$ if the electron energy $E\sim 100$ GeV).

%%%%%%%%%%%%%%%%%%
Now turn to quantitative analysis of the radiation spectra in the
absence of the energy losses.

Thus, assume that in (\ref{16.28}) $K(E)=0$, i.e., neglect the
bremsstrahlung loss. Following the similar lines as given by
Pafomov \cite{104}, one may find explicit solutions of equations
(\ref{16.28}) for $u$. Substitution of thus derived expressions
for $u$ into (\ref{16.15}) and integration with respect to the
angles of electron scattering, give the following expressions for
the radiation intensity $W_{s\vec{n}\omega}$
\begin{eqnarray}
\label{17.14} &
&W_{\perp\vec{n}\omega}=\frac{e^{2}\omega^{2}}{2\pi^{2}c}\left\{\int_{0}^{\infty}dt
\int_{0}^{\infty}dt^{\prime}\frac{T}{p^{2}_{0}}\exp\left[i\omega(t-t^{\prime})(1-\beta+\frac{1}{2}\vartheta^{2})
\right.\right.\nonumber\\
&
&\left.\left.+\frac{\vartheta^{2}\eta_{0}^{4}T(t-t^{\prime})^{2}}{4p_{0}q}
\right]+2\texttt{Re}\int_{0}^{T}dt\int_{0}^{\infty}dt^{\prime}\frac{T-t}{H^{2}\cosh^{2}\eta t}\right.\nonumber\\
& &\left.\times\exp\left[-i\omega t^{\prime}
(1-\beta+\frac{1}{2}\vartheta^{2})-i\omega t(1-\beta+\frac{1}{2}\vartheta^{2}-\frac{1}{2}\delta\varepsilon)\right.\right.\nonumber\\
& &\left.\left.+\frac{\eta\vartheta^{2}}{4q(1-\vartheta^{2}+\delta\varepsilon)}\left(\eta t-\frac{\tanh\eta t}{H}\right)+S\right]\right.\nonumber\\
& &\left.+2 \texttt{Re}\int^{T}_{0}dt\int^{t}_{0}d\tau\frac{\eta_{1}}{p^{2}\cosh^{2}\eta_{2}\tau\sinh\eta_{1}(T-t)}\right.\nonumber\\
&
&\left.\times\exp\left[-i\omega\vec{\tau}(1-\beta+\frac{1}{2}\vartheta^{2}-\frac{1}{2}\delta\varepsilon^{*})
-\frac{\eta_{1}^{2}}{2q}+\frac{\eta_{2}\vartheta^{2}}{4q(1-\vartheta^{2}+\delta\varepsilon^{*})}\right.\right.\nonumber\\
&
&\left.\left.\times\left(\eta_{2}\tau-\frac{\eta_{1}\coth\eta_{1}(T-t)\tanh\eta_{2}\tau}{p}\right)\right]\right\}\,,
\end{eqnarray}
%%%%
\begin{eqnarray}
\label{17.15} &
&W_{\parallel\vec{n}\omega}=\frac{e^{2}\omega^{2}\vartheta^{2}}{4\pi^{2}c}
\left\{\int_{0}^{\infty}dt\int_{0}^{\infty}dt^{\prime}\frac{1}{p^{2}_{0}}\left(\frac{1}{p^{2}_{0}}
+2qT\vartheta^{-2}\right)\right.\nonumber\\
&
&\left.\times\exp\left[i\omega(t-t^{\prime})(1-\beta+\frac{1}{2}\vartheta^{2})
+\frac{\vartheta^{2}\eta_{0}^{4}T(t-t^{\prime})^{2}}{4p_{0}q}
\right]\right.\nonumber\\
&
&\left.+2\texttt{Re}\int_{0}^{T}dt\int_{0}^{\infty}dt^{\prime}\frac{1}{H\cosh\eta
t}
\left[1+R+\frac{2q(T-t)\vartheta^{-2}}{H\cosh\eta t}\right]\right.\nonumber\\
& &\left.\times\exp\left[-i\omega
t^{\prime}(1-\beta+\frac{1}{2}\vartheta^{2})
-i\omega t(1-\beta+\frac{1}{2}\vartheta^{2}-\frac{1}{2}\delta\varepsilon)\right.\right.\nonumber\\
& &\left.\left.+\frac{\eta\vartheta^{2}}{4q(1-\vartheta^{2}+\delta\varepsilon)}\left(\eta t-\frac{\tanh\eta t}{H}\right)+S\right]\right.\nonumber\\
& &\left.+2
\texttt{Re}\int^{T}_{0}dt\int^{t}_{0}d\tau\frac{\eta_{1}^{2}\cosh\eta_{1}(T-t)}{p^{2}\cosh^{2}
\eta_{2}\tau\sinh^{2}\eta_{1}(T-t)}\right.\nonumber\\
&
&\left.\times\left[\frac{\eta_{1}\coth\eta_{1}(T-t)}{p}+\frac{2q\vartheta^{-2}
\tanh\eta_{1}(T-t)}{\eta_{1}}\right]\right.\nonumber\\
& &\left.\times\exp\left[-i\omega\tau(1-\beta+\frac{1}{2}\vartheta^{2}-\frac{1}{2}\delta\varepsilon^{*})\right.\right.\\
&
&\left.\left.-\frac{\eta_{1}^{2}}{2q}+\frac{\eta_{2}\vartheta^{2}}{4q(1-\vartheta^{2}+\delta\varepsilon^{*})}
\left(\eta_{2}\tau-\frac{\eta_{1}\coth\eta_{1}(T-t)\tanh\eta_{2}\tau}{p}\right)\right]\right\}\,,\nonumber
\end{eqnarray}
%%%
where
\[
\delta\varepsilon=\varepsilon-1;\quad \eta_{0}=\sqrt{2i\omega\beta
q(1-\frac{1}{2}\vartheta^{2})}\,,
\]
\[
\eta=\sqrt{2i\omega\beta
q(1-\frac{1}{2}\vartheta^{2}+\frac{1}{2}\delta\varepsilon)}\,,\quad\eta_{1}=\sqrt{2\omega\beta
q \texttt{Im}\varepsilon}\,,
\]
\[
\eta_{2}=\sqrt{2i\omega\beta
q(1-\frac{1}{2}\vartheta^{2}+\frac{1}{2}\delta\varepsilon^{*})}\,,
\]
\[
p_{0}=1-\eta_{0}^{2}T(t-t^{\prime})\,,\,\,p=[\eta_{2}\tanh\eta_{2}\tau+\eta_{1}\coth\eta_{1}(T-t)]\,,
\]
\[
H=\left[1+\eta(T-t)\tanh\eta
t+\frac{\eta_{0}^{2}t^{\prime}}{\eta}(\eta(T-t)+\tanh\eta
t)\right]\,,
\]
\begin{eqnarray}
& &S=\frac{\vartheta^{2}\eta(T-t)\tanh\eta
t}{2qH(1-\vartheta^{2}+\delta\varepsilon)}
\times\left\{(\eta^{2}-\eta_{0}^{2})t^{\prime}\left[\frac{1}{2}+\frac{\cosh\eta t-1}{\eta(T-t)\sinh\eta t}\right]+\frac{\eta^{2}t^{\prime}}{2}\right.\nonumber\\
& &\times\left.\left[1+\eta t^{\prime}\coth\eta
t+\frac{t^{\prime}}{T-t}\right]\right\}\,,\nonumber
\end{eqnarray}
\begin{eqnarray}
& &R=\frac{1}{H}\left\{-\tanh\eta t\left[1-\frac{\eta(T-t)(\cosh\eta t-1)}{H\sinh\eta t\cosh\eta t}\right]\right.\\
& &\left.\times\left[2\eta(T-t)+\frac{[1+\eta_{0}^{2}t^{\prime}(T-t)](\cosh\eta t-1)}{\sinh\eta t}\right.\right.\\
& &\left.\left.+\eta t^{\prime}\left(1+\frac{\eta(T-t)(\cosh\eta
t+1)}{\sinh\eta t}\right)\right]
+\frac{\eta^{3}(T-t)^{2}t^{\prime}(\cosh\eta t-1)}{H\sinh\eta t\cosh^{2}\eta t}\right.\nonumber\\
& &\left.+\eta(T-t)\tanh\eta t\left(\eta t^{\prime}+\frac{\cosh\eta t-1}{\sinh\eta t}\right)\right.\\
& &\left.\times\left[\eta t^{\prime}\frac{1+\eta(T-t)\coth\eta
t}{H\cosh\eta t}+\frac{\cosh \eta t-1}{\eta(T-t)\sinh\eta
t}\right]\right\}\,.\nonumber
\end{eqnarray}

Formula (\ref{17.14}) describes the emission of photons polarized
perpendicular to the their exit plane. The origin of this
radiation is closely connected with scattering in a media (at
$q\rightarrow 0$ the radiation disappears), so it may be related
to bremsstrahlung.

Formula (\ref{17.15}), referring to the emission of photons
polarized in the exit plane, contains contributions associated not
only with bremsstrahlung but also with the transitional mechanism
of radiation as well as with their mutual interference. The first
term in (\ref{17.15}) describes radiation in a vacuum through
particle emission from matter into vacuum, the second term - the
interference of radiation at emission, and radiation produced on
the part of the particle trajectory in matter. The third term
contains contributions to radiation, which are caused by
scattering in matter as well as uniform motion to the point of
exit from matter.

If we are concerned with the radiation spectrum of a particle
passing through the plate of thickness $L=vT$ rather than that of
a particle passing through the matter-vacuum boundary, then as
seen from the appropriate calculations, the formulae for angular
and spectral distributions of photons polarized perpendicular to
the exit plane remain the same, but the terms of the form as given
below are to be added to expression (\ref{17.15}) %%
\begin{eqnarray}
\label{17.16} &
&W^{\prime}_{\parallel\vec{n}\omega}=\frac{e^{2}\omega^{2}\vartheta^{2}}
{4\pi^{2}c}\left\{\frac{e^{-\omega\beta T
\texttt{Im}\varepsilon}}{\omega^{2}(1-\beta+\frac{1}{2}\vartheta^{2})^{2}}
+\frac{2e^{-\omega\beta T
\texttt{Im}\varepsilon}}{\omega(1-\beta+\frac{1}{2}\vartheta^{2})}\texttt{Im}
\int_{0}^{T}\frac{dt}{\cosh^{2}\eta_{2}t}\right.\nonumber\\
& &\left.\times\exp\left[-i\omega
t(1-\beta+\frac{1}{2}\vartheta^{2}-\frac{1}{2}\delta\varepsilon^{*})
+\frac{\eta_{2}\vartheta^{2}}{4q(1-\vartheta^{2}+\delta\varepsilon^{*})}(\eta_{2}t-\tanh\eta_{2} t)\right]\right.\nonumber\\
&
&\left.+\frac{2}{(1-\beta+\frac{1}{2}\vartheta^{2})}\texttt{Im}\int_{0}^{\infty}\frac{dt^{\prime}}{p_{1}\cosh\eta
T}
\left[1-\frac{\tanh\eta T}{p_{1}}\left(\eta t^{\prime}+\frac{\cosh\eta T-1}{\sinh\eta T}\right)\right]\right.\nonumber\\
& &\left.\times\exp\left[-i\omega
t^{\prime}(1-\beta+\frac{1}{2}\vartheta^{2})
-i\omega T(1-\beta+\frac{1}{2}\vartheta^{2}-\frac{1}{2}\delta\varepsilon)\right.\right.\nonumber\\
& &+\left.\left.\frac{\eta\vartheta^{2}}{4q(1-\vartheta^{2}+\delta\varepsilon)}\right.\right.\\
& &\left.\left.\times\left(\eta T-\frac{\tanh\eta T}{p_{1}}
+\frac{(\eta t^{\prime})^{2}\sinh\eta T+2\eta t^{\prime}
(\cosh\eta
T-1)\left(\frac{\eta_{0}^{2}}{\eta^{2}}-1\right)}{p_{1}\cosh\eta
T}\right)\right]\right\}\,,\nonumber
\end{eqnarray}
%%%
where
\[
p_{1}=1+\frac{\eta_{0}^{2}t^{\prime}}{\eta}\tanh\eta T\,.
\]

The first term in (\ref{17.16}) describes radiation at stopping,
the second term, the interference between the radiation produced
on the particle trajectory in  vacuum (before it enters  the
matter) and the radiation in the matter. The third term is the
interference between the radiations appearing on the particle
trajectory in  vacuum before the particle enters the plate and
after it leaves it.

With increasing $L$, the contribution of
$W_{\parallel\vec{n}\omega}^{\prime}$ disappears leading to
conversion of the formula describing the intensity of radiation
produced in the plate into the one for the intensity of photons
produced by a particle passing from matter into vacuum. This is
understandable, as the radiation originating from the first
vacuum--matter boundary is completely absorbed in the target if
the plate thickness is much greater than the absorption depth of
$\gamma$-quanta.

Consider some limiting cases for the derived general formulas
(\ref{17.14})--(\ref{17.16}).

1. Absorption of  $\gamma$-quanta in the plate may be ignored. The
plate thickness is much smaller than the photon absorption depth.
Assuming that in (\ref{17.17})-(\ref{17.16})
$\texttt{Im}\,\varepsilon=0$, one may obtain \footnote{Formulae
(\ref{17.17}), (\ref{17.18}) coincide with the expressions derived
by V.Ye. Pafomov when analyzing the process of radiation in a
plate (see \cite{104},$\S 26$, formulae (26.13)-(26.22)). Note,
however, that the second and third summands  in (26.15) as well as
the fifth and the sixth ones in (26.16) contain errata. Formulae
(\ref{17.17}), (\ref{17.18}) were obtained by Garibyan and Yan
independently of us \cite{109}.} %%
\begin{eqnarray}
\label{17.17} &
&W_{\perp\vec{n}\omega}=\frac{e^{2}\omega^{2}q}{2\pi^{2}c}
\left\{\int_{0}^{\infty}dt\int_{0}^{\infty}dt^{\prime}\frac{T}{p^{2}_{0}}
\exp\left[i\omega (t-t^{\prime})(1-\beta+\frac{1}{2}\vartheta^{2})\right.\right.\nonumber\\
&
&\left.\left.+\frac{\vartheta^{2}\eta_{0}^{4}T(t-t^{\prime})^{2}}{4p_{0}q}\right]+2\texttt{Re}
\int_{0}^{T}dt\int_{0}^{\infty}dt^{\prime}\frac{T-t}{\tilde{H}^{2}\cosh^{2}\eta_{0}t}\right.\\
& &\left.\times\exp\left[-i\omega
t^{\prime}(1-\beta+\frac{1}{2}\vartheta^{2})
-i\omega t(1-\beta+\frac{1}{2}\vartheta^{2}-\frac{1}{2}\delta\varepsilon^{\prime})\right.\right.\nonumber\\
&
&\left.\left.+\frac{\eta_{0}\vartheta^{2}}{4q}\left(\eta_{0}t-\frac{\tanh\eta_{0}t}{\tilde{H}}+\tilde{S}\right)\right]
+2\texttt{Re}\int_{0}^{T}dt\int_{0}^{t}d\tau\frac{T-t}{\tilde{p}^{2}\cosh^{2}\eta_{0}\tau}\right.\nonumber\\
&
&\left.\times\exp\left[-i\omega\tau(1-\beta+\frac{1}{2}\vartheta^{2}-\frac{1}{2}\delta\varepsilon^{\prime})
+\frac{\eta_{0}\vartheta^{2}}{4q}\left(\eta_{0}\tau-\frac{\tanh\eta_{0}\tau}{\tilde{p}}\right)\right]\right\}\,,\nonumber
\end{eqnarray}
%%%%%%%%%
\begin{eqnarray}
\label{17.18} &
&W_{\parallel\vec{n}\omega}=\frac{e^{2}\omega^{2}\vartheta^{2}}{4\pi^{2}c}
\left\{\int_{0}^{\infty}dt\int_{0}^{\infty}dt^{\prime}\frac{1}{p^{2}_{0}}
\left(\frac{1}{p_{0}}+2qT\vartheta^{-2}\right)\right.\nonumber\\
& &\left.\times\exp\left[i\omega
(t-t^{\prime})(1-\beta+\frac{1}{2}\vartheta^{2})
+\frac{\vartheta^{2}\eta_{0}^{4}T(t-t^{\prime})^{2}}{4p_{0}q}\right]\right.\nonumber\\
& &\left.+2\texttt{Re}
\int_{0}^{T}dt\int_{0}^{\infty}dt^{\prime}\frac{1}{\tilde{H}^{2}\cosh^{2}\eta_{0}t}
\left[\frac{1+\eta_{0}t^{\prime}\tanh\eta_{0}t}{\tilde{H}}+2q(T-t)\vartheta^{-2}\right]\right.\nonumber\\
& &\left.\times\exp\left[-i\omega
t^{\prime}(1-\beta+\frac{1}{2}\vartheta^{2})
-i\omega t(1-\beta+\frac{1}{2}\vartheta^{2}-\frac{1}{2}\delta\varepsilon^{\prime})\right.\right.\nonumber\\
&
&\left.\left.+\frac{\eta_{0}\vartheta^{2}}{4q}\left(\eta_{0}t-\frac{\tanh\eta_{0}t}{\tilde{H}}
+\tilde{S}\right)\right]\right.\nonumber\\
& &\left.+2
\texttt{Re}\int_{0}^{T}dt\int_{0}^{t}d\tau\frac{1}{\tilde{p}^{2}\cosh^{2}\eta_{0}\tau}
\left[\frac{1}{\tilde{p}}+2q(T-t)\vartheta^{-2}\right]\right.\nonumber\\
&
&\left.\times\exp\left[-i\omega\tau(1-\beta+\frac{1}{2}\vartheta^{2}-\frac{1}{2}\delta\varepsilon^{\prime})
+\frac{\eta_{0}\vartheta^{2}}{4q}\left(\eta_{0}\tau-\frac{\tanh\eta_{0}\tau}{\tilde{p}}\right)\right]\right.\nonumber\\
&
&\left.+\frac{1}{\omega^{2}(1-\beta+\frac{1}{2}\vartheta^{2})^{2}}
+\frac{2}{\omega(1-\beta+\frac{1}{2}\vartheta^{2})} \texttt{Im}\int_{0}^{T}\frac{dt}{\cosh^{2}\eta_{0}t}\right.\nonumber\\
& &\left.\times\exp\left[-i\omega
t(1-\beta+\frac{1}{2}\vartheta^{2}-\frac{1}{2}\delta\varepsilon^{\prime})
+\frac{\eta_{0}\vartheta^{2}}{4q}(\eta_{0}t-\tanh\eta_{0}t)\right]\right.\nonumber\\
& &\left.+\frac{2}{\omega(1-\beta+\frac{1}{2}\vartheta^{2})}
\texttt{Im}\int^{\infty}_{0}dt^{\prime}\frac{1}{\tilde{p}^{2}_{1}\cosh^{2}\eta_{0}T}\right.\nonumber\\
& &\left.\times\exp\left[-i\omega
t^{\prime}(1-\beta+\frac{1}{2}\vartheta^{2})
-i\omega T(1-\beta+\frac{1}{2}\vartheta^{2}-\frac{1}{2}\delta\varepsilon^{\prime})\right.\right.\nonumber\\
&
&\left.\left.+\frac{\eta_{0}\vartheta^{2}}{4q}\left(\eta_{0}T-\frac{\tanh\eta_{0}T}{\tilde{p}_{1}}\right)\right]\right\}\,,
\end{eqnarray}
%%%%%%%%%5
where
\[
\eta_{0}=\sqrt{2i\omega\beta q(1-\frac{1}{2}\vartheta^{2})};\quad
p_{0}=1-\eta_{0}^{2}T(T-t^{\prime})\,,
\]
\[
\tilde{p}_{1}=1+\eta_{0}t^{\prime}\tanh\eta_{0}T;\,\,
\tilde{p}=1+\eta_{0}(T-t)\tanh\eta_{0}\tau\,,
\]
\[
\tilde{H}=1+\eta_{0}^{2}(T-t)t^{\prime}+\eta_{0}(T-t+t^{\prime})\tanh\eta_{0}t\,,
\]
\[
\tilde{S}=\frac{\vartheta^{2}\eta_{0}^{2}(T-t)t^{\prime}\tanh\eta_{0}t}{4\tilde{H}q}
\left[1+\eta_{0}t^{\prime}\coth\eta_{0}t+\frac{t^{\prime}}{T-t}\right]\,.
\]

%%%%%%%%%%%%%%%%55
2. In the case of sufficiently high particle energies when the
conditions $q\ll \omega (\texttt{Im}\,\varepsilon)^{2}$ are
fulfilled, i.e., $qL_{c}\ll\vartheta^{2}_{\gamma}$, where
$L_{c}=\frac{1}{\omega\texttt{Im}\varepsilon}$ is the absorption
depth of the $\gamma$-quantum; $\vartheta^{2}_{\gamma}=1/\omega
L_{c}$ is the squared effective angle of $\gamma$-quantum
radiation, all the functions appearing in
(\ref{17.14})-(\ref{17.16}) may be expanded in terms of a small
argument $q$.  The stated condition is satisfied, for example, for
electrons with the energy $E\sim 1$ GeV when studying the
Mossbauer radiation spectrum, or for the electrons whose energy
$E> 10^{14}$ eV, when studying the spectrum of $\gamma$quanta with
$\omega\sim 1$ GeV.
%%%%%%%%5
%%%%%%%%%%%%%%%%5
The resulting formulas have the form
\begin{eqnarray}
\label{17.19} & &W_{\perp\vec{n}\omega}=\frac{e^{2}}{8\pi^{2}}
\frac{|\varepsilon-1|^{2}q T}{(1-\beta+\frac{1}{2}\vartheta^{2})^{2}|1-\beta+\frac{1}{2}\vartheta^{2}-\frac{1}{2}\delta\varepsilon|^{2}}\nonumber\\
& &+\frac{e^{2}q}{2\pi^{2}\omega}
\frac{\texttt{Im}\varepsilon(1-\beta+\frac{1}{2}\vartheta^{2}-\delta\varepsilon^{\prime})}{(1-\beta+\frac{1}{2}\vartheta^{2})
|1-\beta+\frac{1}{2}\vartheta^{2}-\frac{1}{2}\delta\varepsilon|^{4}}\nonumber\\
&
&+\frac{e^{2}}{2\pi^{2}\omega}\frac{q}{\texttt{Im}\varepsilon|1-\beta+\frac{1}{2}\vartheta^{2}-\frac{1}{2}\delta\varepsilon|^{2}}\,,
\end{eqnarray}
\[
W_{\parallel\vec{n}\omega}=W_{\mathrm{tr}}+ \quad
(\mbox{terms}\quad \sim q)\,,
\]
where $W_{\mathrm{tr}}$ is the contribution of transition
radiation; $\delta\varepsilon^{\prime}=
\texttt{Re}(\varepsilon-1)$.

According to (\ref{17.19}), at $q\rightarrow 0$ the expression for
radiation intensity includes a $q$-independent  term (coinciding
with the expression for transition radiation of a particle
uniformly moving  perpendicular to the boundary), and the terms
proportional to $q$. The total intensity has a similar structure
\[
W_{\vec{n}\omega}=W_{\parallel\vec{n}\omega}+W_{\perp\vec{n}\omega}\,.
\]
In this approximation ($q\rightarrow 0$), the contribution to
$W_{\parallel n\omega}$ of the terms proportional to $q$, is much
smaller than the transition radiation. Here the term describing
the transition radiation is maximum for the photon exit angles
$\vartheta\sim m/E$. At the same time, the terms proportional to
$q$ and associated with the contribution to $W_{sn\omega}$ of the
trajectories  passing through matter (the last term in
(\ref{17.19}) lead to a broader angular distribution with the
effective emission angle
\[
\vartheta_{\gamma}=\frac{1}{\sqrt{\omega L_{c}}}\,.
\]

%%%%%%%%%%5
3. The case of small plate thicknesses.

If $L\ll L_{c}$,  (\ref{17.14})--(\ref{17.15}) take the most
simple form. Developing (\ref{17.14})--(\ref{17.15}) as a series
in powers of $T$, gives \cite{98,99}
\[
W_{\perp\vec{n}\omega}=\frac{e^{2}}{2\pi^{2}c}\frac{qT}{(1-\beta+\frac{1}{2}\vartheta^{2})^{2}}\,,
\]
\[
W_{\parallel\vec{n}\omega}=\frac{e^{2}\vartheta^{2}}{16\pi^{2}c}\frac{|\delta\varepsilon|^{2}\omega^{2}T^{2}}
{(1-\beta+\frac{1}{2}\vartheta^{2})^{2}}
+\frac{e^{2}qT}{2\pi^{2}c}\frac{(1-\beta-\frac{1}{2}\vartheta^{2})^{2}}{(1-\beta+\frac{1}{2}\vartheta^{2})^{4}}\,.
\]

%55555555555555
Note that considerable fluctuations of the energy losses due to
bremsstrahlung, generally speaking, cause changes in the spectra
of transition radiation and bremsstrahlung even in a thin plate
because the contribution to radiation at the frequency $\omega$
will also come from electrons with abruptly changed energy (due to
radiation of a  hard  quantum). Thus, according to \cite{98} in a
thin plate with due account of radiation losses
\begin{eqnarray*}
&
&W_{\parallel\vec{n}\omega}=\frac{e^{2}\vartheta^{2}|\varepsilon-1|^{2}(\omega
T)^{2}}{16\pi^{2}(1-\beta_{0}+\frac{1}{2}\vartheta^{2})^{2}}
+\frac{e^{2}qT(1-\beta_{0}-\frac{1}{2}\vartheta^{2})^{2}}{2\pi^{2}(1-\beta_{0}
+\frac{1}{2}\vartheta^{2})^{4}}\\
&
&+\frac{e^{2}\vartheta^{2}}{4\pi^{2}}\frac{T}{(1-\beta_{0}-\frac{1}{2}\vartheta^{2})^{2}}
\int_{E_{1}}^{E_{0}}\frac{(\beta_{0}-\beta)^{2}}{(1-\beta+\frac{1}{2}\vartheta^{2})^{2}}\sigma(E_{0}|E)dE\,,
\end{eqnarray*}
%%%%%%%%%5
where $\beta_{0}=\sqrt{1-\frac{m}{E_{0}}}$;
$\sigma(E_{0}|E)dE=N\sigma_{\gamma}(E_{0}|E)$;
$\sigma_{\gamma}(E_{0}|E)$ is the bremsstrahlung cross-section per
unit interval of energies $E$ of the electron with the initial
energy $E_{0}$; $E_{1}$ is the limiting energy (on choosing it,
see \cite{92,100}.

In the general case, the analysis of the formulas for
$W_{s\vec{n}\omega}$ even in the absence of the energy losses is
only possible when using numerical methods. Below we present the
results of such an analysis  at generating of resonance photons by
an electron passing through the plate containing $^{138}W$ nuclei
($\omega=46.5$\,keV). The stated process is of great interest in
connection with the possibility of creating the sources of
resonance radiation with the help the beams of relativistic
electrons  \cite{107}.

\begin{figure}[htp]
\centering
\epsfxsize = 12 cm \centerline{\epsfbox{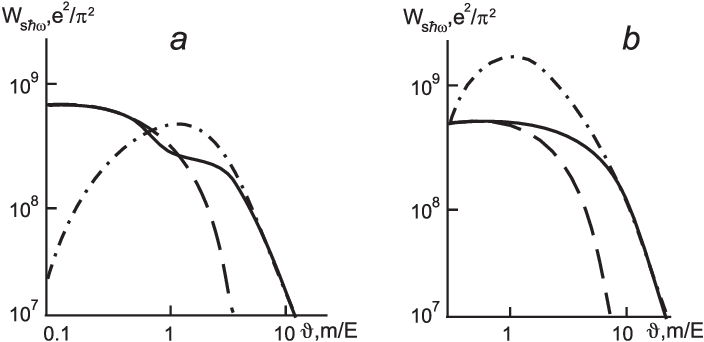}}
\caption{Angular distribution for $\gamma$-quanta produced by the
electron ($E=40$\,GeV) in the plates of thickness $L_{c}$ (a), and
$10L_{c}$ (b) the values of $L_{c}$ and $10L_{c}$  are borrowed
from \cite{58}. Solid lines --- the density of the radiation
energy $W_{\parallel{\vec{n}}\omega}$; Dashed--dot lines---
transition radiation produced by the electron passing through the
plate at a constant velocity directed normal to its surface;
Dashed lines --- the angular distribution of
$W_{\perp{\vec{n}}\omega}$} \label{ChannelingFigure7}
\end{figure}

Particle scattering in a medium appreciably affects
$W_{\parallel\vec{n}\omega}$, leading to the fact that at the
angles $\vartheta\leq m/E$ the radiation intensity of the waves
with the polarization parallel to the exit plane of
$\gamma$-quanta differs from the intensity of transition radiation
Figure (\ref{ChannelingFigure7}).

%%%%%%%%%%%%%%%%%%%%%55

We also pay attention to the  fact that at the angles
$\vartheta\leq m/E$, the major contribution to
$W_{\parallel\vec{n}\omega}$ is made by the term  associated with
electron scattering in a medium and equal to
$W_{\perp\vec{n}\omega}$. Numerical analysis of the formulas shows
that in the range of electron energies ($E\geq 1$\,GeV) we have
discussed, for the values of the radiation angles of
$\gamma$-quanta up to
$\vartheta\sim\sqrt{\texttt{Im}\varepsilon}$, the predominating
contribution to  $W_{\perp\vec{n}\omega}$ comes from the term,
caused by appearance of the fan of trajectories of electron motion
behind the plate due to multiple scattering in a medium. This
contribution increases with the growth of the electron energy at
fixed frequency of $\gamma$-quanta. For the angles $\vartheta>
m/E$, the radiation intensity $W_{\parallel\vec{n}\omega}$
coincides with the intensity of transition radiation at normal
transmission through the plate (see Figure
(\ref{ChannelingFigure7})). Moreover, the increase in the plate
thickness leads to broadening of angular distribution of the
radiation energy density of $\gamma$-quanta (see Figure
(\ref{ChannelingFigure7}(b))).

%%%%%
We have also calculated the radiation intensity
$W_{s\vec{n}\omega}$  for electrons with $E=1$ GeV in a plate of
thickness $L_{c}$. The angular distributions obtained are similar
to those given in Figure (\ref{ChannelingFigure7}) for electrons
with the energy of 40 GeV. Numerical integration of formulae for
non-resonance $\gamma$-quanta with the energy of 40 and 200 MeV,
and electrons with the energy of 40 and 200 GeV, respectively has
also been carried out. Calculations were made for tungsten plates
of thicknesses $0.05 \, L$ and $0.1 \, L$. In this case a
substantial contribution to the radiation intensity
$W_{\perp\vec{n}\omega}$ proportional to $q$ is made by the fan of
vacuum trajectories.

Thus, the fan of vacuum trajectories appreciably changes the
pattern of of angular and spectral distributions of the radiation
intensity for a particle  passing through matter-vacuum boundary.

%%%%%%%%%%%%%%%%%%%%%%%%%%%%%%%%%%%%5

%%%%%%%%%%%%%%%%%%%%%%%%%%%%%%%%%%%  Chapter 6 %%%%%%%%%%%%%%%%%%%%%%

\chapter[Scattering and Radiation in Crystals Exposed to Variable Fields]{Scattering and Radiation in Crystals Exposed to Variable Fields}
\label{ch:6}

%%%%%%%%%%%%%%%%%%%%%%%%%%%%%%%%%%%  Section 18 %%%%%%%%%%%%%%%%%%%%%%

\section[Generation of $\gamma$-quanta by Channeled Particles in the Presence of Variable Fields]{Generation of $\gamma$-quanta by Channeled Particles in the Presence of Variable Fields}
\label{sec:6.18}

Now let a crystal in which fast particle move be affected by a
variable external field (electromagnetic or sound). The latter can
influence the process of photon emission by a particle in two
ways. On the one hand, it acts directly on the particle, causing
forced vibrations in its channel, on the other hand, the field
makes the nuclei  swing. As a result, the channel where the
particle moves starts bending, thus causing the appearance of a
variable force which sets the particle into vibration.

As was repeatedly pointed out, electromagnetic radiation produced
by spontaneous radiation transitions may be considered as the
radiation of a certain oscillator (atom). Causing vibrations of
the crystal nucleus, the external field leads to oscillations of
the point of the  equilibrium position of the oscillator. Suppose
that the oscillation frequency of the equilibrium point under the
external field is much  less than that of the oscillator (i.e.,
the particle vibration frequency in the channel). In this case the
oscillator follows the oscillations of the equilibrium point
adiabatically. High-frequency vibrations of the charge, i.e.,
particle vibrations about the equilibrium position result in
spontaneous radiation, discussed above; low-frequency
oscillations, which are due to oscillations of the equilibrium
point, lead to additional electromagnetic radiation. Consider the
features of this radiation, following \cite{66,67}. (This
radiation was also discussed in \cite{110} for a particular case
of a standing acoustic wave ).

Let a crystal be affected by an external wave. As a result, the
centers of mass of the atoms execute forced oscillations
$\vec{R}_{i}^{(s)}=\vec{r}_{0}^{(s)}
\cos(\kappa\vec{R}_{i}-\Omega_{s}t)$, where $\vec{r}_{0}^{(s)}$
is the amplitude of forced oscillations; $\vec{R}_{i}$ - is the
coordinate of the equilibrium point of the i-th nucleus;
$\vec{\kappa}$ is the wave vector of the external wave;
$\Omega_{s}$ is the external wave frequency. When a particle moves
in the channel oriented along the z-axis, vibrations of nuclei
cause oscillations of the equilibrium position  of the oscillator
corresponding to the particle in the $xy$ plane  as
$\vec{r}_{\perp}^{s}=\vec{r}_{0\perp}^{s}\cos\Omega^{\prime}t$.
Here $\Omega^{\prime}=\kappa_{z}c-\Omega_{s}$ is the oscillation
frequency of the equilibrium position, $c$ is the velocity of
light. In the adiabatic case the particle trajectory in a
transverse plane  $\vec{r}_{\perp}(t)$ is determined by the sum
$\vec{r}_{\perp}(t)=\vec{r}^{s}_{\perp}(t)+\vec{r}^{c}_{\perp}(t)$,
where $\vec{r}^{c}_{\perp}(t)$ is the trajectory of the charged
particle in the channel.

Let us give a more detailed consideration of particle motion in
the channel exposed to, for example an ultrasonic wave. Assume
that the wave moves in a crystal along the z-axis. The channel
bends caused by the undulator may result in dechanneling. The
particle will not leave the channel if the
 minimal radius of curvature of the channel $\rho_{\mathrm{min}}$ satisfies the inequality following from the equilibrium conditions
\begin{equation}
\label{18.1} \frac{m\gamma
v^{2}}{\rho_{\mathrm{min}}}\simeq\frac{m\gamma
c^{2}}{\rho_{\mathrm{min}}}\leq|\nabla_{\perp}u(\vec{r}_{\perp})|_{\mathrm{max}}\,,
\end{equation}
where $m$ is the particle rest mass;
$\rho_{\mathrm{min}}=(r_{0\perp}^{s}\kappa^{2})^{-1}$;
$u(\vec{r}_{\perp})$ is the potential energy of particle
interaction with the crystal plane.

%%%%%%%%%%%55
Suppose that we consider the particles with the amplitude of free
vibrations in the channel $r_{0\perp}^{c}<d/2$ ($d$ is the channel
width). In this case the amplitude of ultrasonic vibrations should
satisfy the condition
$r_{0\perp}^{s}<4u_{\mathrm{max}}/E\kappa^{2}$, $E$ is the
particle energy. For example, in silicon $r_{0\perp}^{s}<10^{-4}$
cm for $E\sim 1$ GeV  and $\Omega_{s}\sim 2\pi\cdot 10^{7}$
s$^{-1}$. Spectral distribution of radiation induced by
oscillations of the center of equilibrium of the oscillator may be
written in the form
\begin{equation}
\label{18.2} \frac{dN}{d\omega}=\frac{e^{2}L}{4\hbar
c^{4}}(r_{0\perp}^{s}\Omega^{\prime})^{2}
\left[1-2\frac{\omega}{\omega^{\prime}_{m}}+2(\frac{\omega}{\omega^{\prime}_{m}})^{2}\right]\,,
\end{equation}
where $\omega^{\prime}_{m}=2\Omega^{\prime}\gamma^{2}$; $L$ is the
crystal thickness, $L\gg 1/\kappa$. If the potential
$u(r_{\perp})$ is harmonic, the formula for the spectrum of
radiation produced by free vibrations of a particle in the channel
is similar to (\ref{18.2}). As a result, the relation of the
intensity of radiation induced by the external wave to the
intensity of spontaneous radiation can be written as follows

%%%%%%%%%%%%%5
\begin{equation}
\label{18.3}
B\simeq\left(\frac{r_{0\perp}^{s}\Omega^{\prime}}{r_{0\perp}^{c}\Omega}\right)^{2}=\frac{4W_{s}c^{2}}{\rho
v_{s}^{3}\Omega^{2}(r_{0\perp}^{c})^{2}},
\end{equation}
where $\Omega$ and $r_{0\perp}^{c}$ are the frequency and the
amplitude of particle vibrations in the channel;
$W_{s}=\frac{1}{4}\rho\kappa^{-1}\Omega_{s}^{3}(r_{0\perp}^{s})^{2}$
is the power of the ultrasonic wave in W/cm$^{2}$, $\rho$ is the
density of the medium. Study  radiation in  the spectral range
$\omega\sim 2\Omega^{\prime}\gamma^{2}$. According to the
condition (\ref{18.1}),
$$
B_{max}\sim\left(\frac{\Omega v_{s}}{\Omega_{s}c}\right)\gg 1.
$$
So, for positrons with the energy $B\sim 1$ GeV at the ultrasonic
wave power $W_{s}\sim 10^{3}$ W/cm$^{2}$ and frequency
$\Omega_{s}\sim 2\pi 10^{7}$ s$^{-1}$, the vibration amplitude
$r_{0\perp}^{s}\sim 10^{-5}$ and the relation $B\simeq 10$ for
$r_{0\perp}^{c}\sim d/2$.

%%%%%%%%%%%%%%%

In the soft part of spectrum, it is important to take into account
refraction and absorption of photons. he appropriate expression
for the spectrum is analogous to that obtained above for
spontaneous radiation, and it reads
\begin{equation}
\label{18.4}
\frac{dN}{d\omega}=\frac{e^{2}(r_{0\perp}^{s}\Omega^{\prime})^{2}}{4\hbar
c^{4}} \left[1-2\frac{\omega}{\omega^{\prime}_{m}}+
2(\frac{\omega}{\omega^{\prime}_{m}})^{2}\right]l_{\mathrm{abs}}\left(1-e^{-\frac{L}{l_{\mathrm{abs}}}}\right)\,,
\end{equation}
where $l_{\mathrm{abs}}$ is the absorption length of the photon
with frequency $\omega$,
$\Omega^{\prime}=\kappa_{z}v_{z}-\Omega_{s}$. Note that the wave
number is related to frequency as $\kappa=\kappa(\Omega_{s})$. For
the acoustic branch in the long wave limit
\[
\kappa=\frac{\Omega_{s}}{v_{s}}\,,
 \]
 $v_{s}$ is the velocity of sound. For light
\[
\kappa=\frac{\Omega_{s}}{c}n(\Omega_{s})\,,
\]
where $n(\Omega_{s})$ is the refractive index.

%%%%
When a channeled particle moves in a crystal in the presence of a
light wave, it undergoes forced vibrations caused by the direct
effect of the force from the wave. In this case the amplitude of
forced vibrations is
\begin{equation}
\label{18.5}
x_{s}=\frac{eE_{x}}{2m\gamma}(\Omega^{2}-\Omega^{\prime
2}-i\Omega^{\prime}\Gamma)^{-1}\,,
\end{equation}
where $E_{x}=E_{0}(1-\beta n_{0}\vec{n}_{\gamma}\vec{n}_{e})$ is
the $x$-component of the external field strength in the medium
(the angle of the wave vector $\vec{\kappa}$ with the particle
momentum $\vec{p}$ is assumed to be much less than unity or close
to $\pi$).

The frequency of the emitted photon
\[
\omega=\Omega^{\prime}\left(1-\frac{v_{z}}{c}n(\omega)\cos\vartheta\right)^{-1}\,.
 \]
As the amplitude of particle forced vibrations cannot exceed the
channel width (and the radiation intensity is only determined by
the vibration amplitude and frequency), the intensity of the
radiation due to the electromagnetic wave cannot exceed the
intensity of spontaneous radiation of a channeled particle,
vibrating with the same frequency and amplitude.

%%%%%%%%%%%%%%%%%%%%%%%%%%%%%%%%%%%  Section 19 %%%%%%%%%%%%%%%%%%%%%%

\section[Coherent Scattering of Photons by a Beam of Channeled Particles. The Effect of Super-radiation]
{Coherent Scattering of Photons by a Beam of Channeled Particles.
The Effect of Super-radiation} \label{sec:6.19}

We have already pointed out in (\ref{sec:2.5}) that a channeled
particle may be considered as a fast atom. This allows us to state
that under appropriate conditions for such particles it is
possible to observe numerous effects known in atomic physics.
Moreover, the similarity of the properties of a channeled particle
and a fast atom enables using the results of the photon-atom
interaction theory rather than carrying out new calculations in
order to find any process (scattering, photon radiation or
absorption). For this purpose suffice it in the beginning to
consider the process in the coordinate system, where the initial
longitudinal  momentum of a particle is zero. In this system we
deal with a resting atom, the cross-section (amplitude) of photon
scattering by which is well known. Further, it is necessary to
convert the scattering cross-section (amplitude) to the laboratory
coordinate system according to simple rules (see, for example,
\cite{111}, p.86-97) with due account of the fact that the atom
corresponding to the channeled particle has a one-dimensional
(axial channeling) or two-dimensional (planar channeling)
momentum. For example, the amplitude of elastic coherent forward
scattering of a photon by a channeled particle
\begin{equation}
\label{19.1}
f(\omega)=\frac{\sqrt{1-\beta^{2}}}{|1-\beta_{z}n(\omega)\cos\vartheta|}f(\omega^{\prime}),
\end{equation}
where $f(\omega^{\prime})$ is the scattering amplitude in the rest
system ($v_{z}=0$)of the channeled particle;
$\omega^{\prime}=(1-\beta_{z}n(\omega)\cos\vartheta)\gamma\omega$
is the photon frequency in the system. The amplitude
$f(\omega^{\prime})$ has a usual Breit-Wigner form.

Using the optical theorem, we also immediately find the total
cross-section of  photon scattering by a channeled particle
$\sigma=\frac{4\pi c}{\omega} \texttt{Im}f(\omega)$.

Now estimate the addition $\delta\varepsilon$ to the  dielectric
permittivity of a crystal caused by photon interaction with a beam
of particles. According  to \cite{14,112}
$\delta\varepsilon=\frac{4\pi\rho Ac^{2}}{\omega^{2}}f(\omega)$,
where $\rho$ is the beam density; the difference of the constant
$A$ from unity is due to the difference between the mean field in
the medium and the local one acting on a moving atom. For the
media with $\varepsilon$ close to unity, $A\simeq 1$. The effect
of refraction by the beam is appreciable, when
$\frac{1}{2}\frac{\omega}{c}\delta\varepsilon L\geq 1$. For a
particle moving in a harmonic potential,
$$
f(\omega^{\prime})=-\frac{r_{0}}{2}\frac{\omega^{\prime}}{\omega^{\prime}-\Omega^{\prime}-i\frac{\Gamma^{\prime}}{2}},
$$
where $r_{0}$ is the classical electron radius; $\Omega^{\prime}$
and $\Gamma^{\prime}$ are the oscillation frequency of the
oscillator and the width of the transition in the particle rest
system, respectively. As a result, under the resonance conditions
refraction is great, if $\frac{2\pi\rho cr_{0}}{\gamma\Gamma}\geq
1$; $\Gamma=\Gamma^{\prime}\gamma^{-1}$ is the line width in the
laboratory system. If the particle bunch thrown onto a crystal is
accelerated as a whole, $\rho\gamma^{-1}$  is the bunch density
before the acceleration $\rho_{0}$. As a result,
$\rho_{0}\geq\frac{\Gamma}{2\pi cr_{0}L}$, i.e., $\rho_{0}\geq
10^{13}$ for $\Gamma=10^{12}$ s$^{-1}$ and $L=1$ cm. It should be
noted that according to (\ref{19.1}) the anomalous and complex
Doppler effects lead to the fact that in a resonance with a moving
oscillator an emitted hard photon appears along with a soft one.
Therefore hard photons are also effectively refracted by a beam.

Since the amplitude of photon scattering by a channeled particle
depends on the photon polarization, the channeled beam is an
optically anisotropic medium. For example, in the case of planar
channeling  a beam is a birefringent medium. Under the stated
conditions, the particle interaction through the  field of photons
is considerable. As a consequence, the formation of exciton
polaritons in such a beam is possible, and when the beam is
affected by a light pulse, the oscillators corresponding  to
channeled particles may be driven into the super-radiant state. In
their system a boson avalanche may evolve, and eventually
generation of ultrashort radiation pulses may occur \cite{20,113}
(compare with similar phenomena in atomic physics \cite{114,115}).

The presence of spin in electrons leads to spin-orbital level
splitting at axial channeling, i.e., to the appearance of fine
structure of the levels. That is why the effect of a circularly
polarized electromagnetic wave on such particles will result in
spin polarization of the electron beam (compare with the effects
of electron polarization through photoionization of atoms
\cite{116}). Note that in the transition of channeled electrons
between the states with different orbital moments, the efficiency
of their interaction with a crystal changes sharply. s-electrons
with enhanced density at nuclei dechannel faster than, for
example, $p$-electrons. If channeled unpolarized electrons are,
for example, in  $p$-state, then a circularly polarized wave,
causing the transition between one of the components of the fine
structure of this state and $s$-state, will transfer a fraction of
the electrons to  $s$-state, from which they dechannel rapidly. As
a result, the passing beam will appear to be partially polarized.
If the crystal is thin enough to neglect dechanelling processes,
in order to select polarized electrons, one can make use of the
fact that the angular distributions of electrons passing through
the crystal depend on the state in which they were in the crystal.
A circularly polarized wave, transferring electrons from
$s$-state to one of the states of the fine structure, will cause,
for example, non-zero degree of electron polarization in the
directions of the $p$-electron escape.

We also point out that, due to equally probable occupation of the
state with different projection of the orbital moment on the axis,
the degree of circular polarization of photons produced by a
particle through axial channeling is practically zero. A
circularly polarized wave, changing the occupation of different
projections of the orbital moment, causes the emission of
circularly polarized photons.

The probabilities of the induced processes discussed here are easy
to find, according to the well known rules (see, for example,
\cite{19}, $\S 44$, using the probability of a spontaneous process
for polarized particles, whose explicit form is given in
(\ref{sec:3.7}).

An intense light wave, causing the transitions between different
states of a channeled particle will lead to the fact that the
particle beam leaving the crystal will turn out to be spatially
modulated (compare with the effect of modulation of a beam passing
through a dielectric plate \cite{117,118}, and the modulation
effect at electron diffraction in a single crystal \cite{119}).
This new modulation mechanism exhibits high efficiency, and, due
to fine splitting of levels, causes spatial modulation of the
degree of polarization of the initially unpolarized particle beam
(provided the crystal is illuminated (irradiated) by a circularly
polarized wave).

As mentioned above, coherent occupation of the levels of
transverse motion at the particle entering the crystal brings
about beatings in the radiation intensity, depending on the target
thickness. The degeneracy of the levels, likewise in atomic
physics will give the opportunity to observe the burst in the
intensity of radiation produced by channeled particles when the
level crossing is stimulated, for example, by  means of crystal
bending.

%%%%%%%%%%%%%%%%%%%%%%%%%%%%%%%%%%%  Section 20 %%%%%%%%%%%%%%%%%%%%%%

\section[Induced Scattering and Radiation under Diffraction Conditions]{Induced Scattering and Radiation under Diffraction Conditions}
\label{sec:6.20}

It is common knowledge (also see above) that, due to the periodic
arrangement of atoms (nuclei) the energy spectrum of particles
($\gamma$-quanta) moving in a crystal exhibits the energy-band
structure $E_{f\vec{k}}$ ($f$ is the band number, $\vec{k}$ is the
reduced quasi-momentum). For a particle with spin the spectrum
also depends on the spin state of the incident beam \cite{14}. The
energy-band structure of the spectrum causes spontaneous
transitions between the bands accompanied by the emission of
photons, phonons, plasmons, etc., and, as a result, simulated
transitions. Simulated transitions between the bands bring about
resonance repolarization, modulation of a neutron beam, the change
in the rate of nuclear reactions in crystals, and polarization of
particles \cite{14,73,120}. As an example consider simulated
neutron transitions between the bands under the action of phonons
, i.e., under ultrasonic pumping. By the action of ultrasound on a
crystal nuclei in the equilibrium position start executing forced
vibrations according to
\begin{equation}
\label{20.1}
\delta\vec{R}_{i}(t)=\vec{a}\sin(\vec{\kappa}\vec{R}_{i}-\Omega(\kappa)t+\delta),
\end{equation}
where $\vec{a}$ is the amplitude of forced vibrations of the
nucleus; $\vec{\kappa}$ is the wave vector of phonons;
$\Omega(\kappa)$ is the phonon frequency; $\delta$ is the initial
vibration phase; $\vec{R}_{i}$ is the equilibrium coordinate of
the nucleus in the absence of phonons.

The Schrodinger equation describing  diffraction of neutrons by a
vibrating crystal has the form
\begin{equation}
\label{20.2}
i\hbar\frac{\partial\psi}{\partial t}=\left\{-\frac{\hbar^{2}}{2m}\Delta_{r}\right.\nonumber\\
\left.+\sum_{i}V[\vec{r}-\vec{R}_{i}-\vec{a}\sin(\vec{\kappa}\vec{R}_{i}-\Omega
t+\delta)]\right\}\psi,
\end{equation}
where $V$ is the coherent potential of interaction of neutrons
with nuclei.

In the expression for the potential (\ref{20.2}) perform the
summation over positions of nuclei. With this aim in view
introduce  the Fourier transform of the potential $V(q)$:
\begin{eqnarray}
\label{20.3}
u(t)\equiv\sum_{i}V(\vec{r}-\vec{R}_{i}-\delta\vec{R}_{i}(t))=\frac{1}{(2\pi)^{3}}\sum_{i}\int d^{3}qV(\vec{q})\nonumber\\
\times\exp\left\{i\vec{q}(\vec{r}-\vec{R}_{i}-\vec{a}\sin(\vec{\kappa}\vec{R}_{i}-\Omega t+\delta))\right\}\nonumber\\
=\frac{1}{(2\pi)^{3}}\sum_{i}\int d^{3}qV(\vec{q})e^{i\vec{q}(\vec{r}-\vec{R}_{i})}\left\{J_{0}(\vec{q}\vec{a})\right.\nonumber\\
\left.-2i\sum_{n=0}^{\infty}J_{2n+1}(\vec{q}\vec{a})\sin[(2n+1)(\vec{\kappa}\vec{R}_{i}-\Omega t+\delta)]\right.\nonumber\\
\left.+2\sum_{n=0}^{\infty}J_{2n}(\vec{q}\vec{a})\cos[2n(\vec{\kappa}\vec{R}_{i}-\Omega
t+\delta)]\right\}.
\end{eqnarray}

To perform the summation over $i$ in (\ref{20.3}), note that
\begin{equation}
\label{20.4}
\sum_{i}e^{-i(\vec{q}-n\kappa)\vec{R}_{i}}=\frac{(2\pi)^{3}}{v_{0}}\sum_{\tau}\delta(\vec{q}-n\vec{\kappa}-2\pi\vec{\tau}),
\end{equation}
where $2\pi\vec{\tau}$ is the reciprocal lattice vector; $v_{0}$
is the volume of the unit cell (it is assumed to be simple) in the
crystal.

Further we shall consider diffraction by a set of planes
characterized by such a vector $2\pi\vec{\tau}$ that
$2\pi\vec{\tau}a\ll 1$ ($2\pi\tau\sim10^{8}\div 10^{9}$ cm$^{-1}$,
$a\sim 10^{-10}$ cm). In this case in (\ref{20.3}) it takes only
to retain the terms containing $J_{0}(\vec{q}\vec{a})\simeq 1$ and
$J_{1}(\vec{q}\vec{a})\simeq \frac{1}{2}\vec{q}\vec{a}$. As a
result we have
\begin{eqnarray}
\label{20.5}
u(t)=\frac{1}{v_{0}}\sum_{\tau}V(2\pi\vec{\tau})e^{i2\pi\vec{\tau}\vec{r}}+\delta V_{(+)}(\vec{r})e^{i\Omega t}-\delta V_{(-)}(\vec{r})e^{-i\Omega t};\nonumber\\
\delta
V_{(\pm)}(\vec{r})=\frac{1}{2v_{0}}\sum_{\tau}V(2\pi\vec{\tau})(2\pi\vec{\tau}\vec{a})e^{i(2\pi\tau\mp\vec{\kappa})\vec{r}}e^{\pm
i\delta}.
\end{eqnarray}
In view of (\ref{20.5}) the the potential of neutron interaction
with the crystal lattice moving under ultrasonic wave may be
represented as a sum of two summands. The first one describes
particle diffraction in a static grating, the second one - the
time-periodic perturbation, which is the superposition of plane
wave traveling in the crystal. Diffraction by these waves is
possible as well as diffraction by static ones produced by the
first summand, which, unlike diffraction in the static case, is
accompanied by the change in the particle energy by the amount
divisible into $\hbar\Omega$. As a consequence, the perturbation
described by the second summand may cause resonant transitions
between the energy band.

To find the probability of the interband transition per unit time
under periodic perturbation, one should know stationary states of
an unperturbed problem. (In the case in question the stationary
wave functions describing the diffraction process in a crystal are
well known \cite{14}). If a crystal is a plate of thickness $l$,
inside the crystal the wave function of the initial state in the
two-wave Laue case has the form
\begin{equation}
\label{20.6}
\psi_{k}^{(+)}=\frac{1}{L^{3}}\sum_{\alpha=1}^{4}A_{\alpha}e^{i\vec{k}_{\alpha}\vec{r}},
\end{equation}
where $L^{3}$ is the normalization volume;
$\vec{k}_{1}=(\vec{k}_{\perp},\vec{k}_{z1})$;
$\vec{k}_{2}=(\vec{k}_{\perp}, \vec{k}_{z2})$;
$\vec{k}_{3}=\vec{k}_{1}+2\pi\vec{\tau}$;
$\vec{k}_{4}=\vec{k}_{2}+2\pi\vec{\tau}$;
$k_{z1}=k_{z}n_{1}(k_{z})$; $k_{z2}=k_{z}n_{2}(k_{z})$; $k_{z}$ is
the z-th component of the wave vector of neutrons in a vacuum; the
expressions for refractive indices under diffraction conditions
are given in \cite{14}; $\vec{k}_{\perp}$ is the component of the
wave vector of the incident wave, perpendicular to the z-axis
(parallel to the crystal surface); $2\pi\vec{\tau}$ is the
reciprocal lattice vector characterizing the family of diffracting
planes.

In the final state one should use the wave function of the type
$\psi_{k}^{(-)}$, under diffraction conditions having the form
\begin{equation}
\label{20.7}
\psi_{k^{\prime}}^{(-)}(\vec{r})=\frac{1}{\sqrt{L^{3}}}\sum_{\alpha=1}^{4}A_{\alpha}^{\prime
*}\exp(i\vec{k}^{\prime
*}_{\alpha}\vec{r})\exp\left(-i\frac{k^{\prime}}{\gamma_{0}^{\prime}}\varepsilon_{\alpha}^{\prime
*}l\right),
\end{equation}
where $\varepsilon^{\prime}_{3}=\varepsilon^{\prime}_{1}$;
$\varepsilon^{\prime}_{4}=\varepsilon^{\prime}_{2}$; $k^{\prime}$
is the wave number in a vacuum of the neutron which has undergone
the transition; $\varepsilon_{\alpha}$ and $\gamma_{0}$
determining the refractive indices  $n_{1}$ and $n_{2}$ are given
in \cite{14} (also see  (\ref{sec:4.12}).

The probability of transition per unit time that is of interest to
us is
\begin{equation}
\label{20.8}
W_{\vec{k}^{\prime}\vec{k}}=\frac{2\pi}{\hbar}|<\psi_{\vec{k}^{\prime}}^{(-)}|\delta
V_{(\pm)}|\psi_{\vec{k}}^{(+)}>|^{2}\delta(E_{\vec{k}^{\prime}}\pm\hbar\Omega-E_{k})\frac{L^{3}d^{3}k^{\prime}}{(2\pi)^{3}}
\end{equation}
where the plus (minus) sign refers to the transitions with the
energy loss (acquisition); $E_{\vec{k}}=\hbar^{2}k^{2}/2m$;
$E_{\vec{k}^{\prime}}=\hbar^{2}k^{\prime 2}/2m$.

Consider the matrix element appearing in (\ref{20.8}). Integrated
in it is performed with respect to the volume of the crystal
plate. As the wave functions and perturbation $\delta V_{(\pm)}$
are the superpositions of plane waves, then integration over the
crystal surface leads to appearance in the matrix element of
two-dimensional $\delta$-functions oft the form
$\delta(\vec{k}^{\prime}_{\perp}-2\pi\vec{\tau}^{\prime\prime}_{\perp}\pm\vec{\kappa}_{\perp}-\vec{k}_{\perp})$,
fixing the component of the momentum parallel to the plate
surface.

The stated $\delta$-functions together with the $\delta$-function
with respect to energy included into (\ref{20.8}) allow in
(\ref{20.8}) integration over $d^{3}k^{\prime}$. As a result,
one-dimensional integrals of the type as follows will remain in
(\ref{20.8})
\begin{eqnarray}
\label{20.9}
\frac{1}{L}\int_{0}^{l}\exp\left\{-i(k^{\prime *}_{\alpha^{\prime}z}\pm\kappa_{z}-k_{\alpha z})z\right\}dz\nonumber\\
=\frac{i}{L}\frac{\exp\left\{-i(k^{\prime
*}_{\alpha^{\prime}z}\pm\kappa_{z}-k_{\alpha
z})l\right\}-1}{k^{\prime
*}_{\alpha^{\prime}z}\pm\kappa_{z}-k_{\alpha z}},
\end{eqnarray}
which are  the sharp functions of the difference of the
z-projections of the wave numbers. Such integrals take on their
maximum values when the difference of the real parts of the wave
numbers vanishes. In other words, the process of neutron
interaction with a vibrating crystal is governed by
energy-momentum conservation law of the form
\begin{eqnarray}
\label{20.10}
E_{k^{\prime}}=E_{k}\mp\hbar\omega;\nonumber\\
\vec{k}^{\prime}_{\perp}=\vec{k}_{\perp}\mp\vec{\kappa}_{\perp}+2\pi\vec{\tau}^{\prime\prime}_{\perp}\nonumber\\
or\nonumber\\
\vec{k}^{\prime}_{\perp}=\vec{k}_{\perp}\mp\vec{\kappa}_{\perp}+2\pi\vec{\tau}^{\prime\prime}_{\perp}+2\pi\vec{\tau}_{\perp}\,\mbox{and so on},\nonumber\\
\end{eqnarray}
$$
\texttt{Re}(k^{\prime}_{\alpha^{\prime}z}=k_{\alpha
z}\mp\kappa_{z}+2\pi\tau_{z})
$$
or
$\texttt{Re}\,k^{\prime}_{\alpha^{\prime}z}=\texttt{Re}\,k_{\alpha
z}\mp\kappa_{z}+2\pi\tau_{z}^{\prime\prime}+2\pi_{z}$ and so on,
where $\mp\vec{\kappa}+2\pi\vec{\tau}^{\prime\prime}$ is the
momentum due to $\delta V$ interaction; $\vec{k},
\vec{k}+2\pi\vec{\tau}$ is the primary momentum due to the
particle.

If equalities (\ref{20.10}) are fulfilled, then in the case when
the crystal depth is less than the particle absorption depth, the
integral in (\ref{20.9}) equals $lL^{-1}$. Substitution of
(\ref{20.9}) into (\ref{20.8}) demonstrates that the probability
of transition per unit time integrated over $d^{3}k^{\prime}$
proves to be an oscillating function of the crystal thickness $l$
with the spatial oscillation periods determined by the difference
of the refractive indices of the plane waves involved in
diffraction. The maximum value of the probability of transition is
attained when (\ref{20.10}) is fulfilled, being equal, for
example, for neutrons exiting behind the crystal in the positive
direction of the z-axis, in the case of the exact fulfillment of
the Bragg conditions to
\begin{equation}
\label{20.11} W=\int
dW_{\vec{k}^{\prime}\vec{k}}\simeq\frac{l^{2}}{\hbar^{2}v_{z}L}\left|\frac{V(2\pi\tau^{\prime\prime})(2\pi\vec{\tau}^{\prime\prime}\vec{a})}{2v_{0}}\right|^{2}
\end{equation}
where $v_{z}$ is the z-th  component of the particle velocity.

The experimentally observable quantity is the  cross-section of
the process $\sigma=L^{3}W/v_{z}$, or the fraction of particles
that have undergone a transition, per one incident particle:
$\delta N=\sigma/L^{2}$. From (\ref{20.11}) follows that
\begin{equation}
\label{20.12} \delta N\simeq\frac{l^{2}}{\hbar^{2}v_{z}^{2}}
\left|\frac{V(2\pi\tau^{\prime\prime})(2\pi\vec{\tau}^{\prime\prime}\vec{a})}{2v_{0}}\right|^{2}.
\end{equation}
According to \cite{14} $V(2\pi\vec{\tau})$ may be expressed in
terms of the amplitude of neutron scattering by a nucleus:
\begin{equation}
\label{20.13} V(2\pi\tau)=-\frac{2\pi\hbar^{2}}{m}fe^{-W(\tau)},
\end{equation}
where $f$ is the amplitude of coherent scattering of the neutron
by the nucleus; $e^{-W(\tau)}$is the Debye-Waller factor.

In view of (\ref{20.12}) the value of  $\delta N$ is maximum when
the nuclei vibrate along  the direction of  $\tau$, e.i.,
perpendicular to the planes by which the particle is diffracted.
Note that the analogous result is also obtained in the case when
the change in the particle energy through diffraction by a
vibrating grating is ignored (static approximation, see
\cite{121}).

Estimate the magnitude of the effect. The scattering amplitudes
are of the order of $10^{-12}\div 10^{-13}$ cm, the vibration
amplitudes - $a\simeq 10^{-10}$ cm. Hence, for thermal neutrons
($v\simeq 10^{5}$ cm/sec) the fraction of particles that have
undergone a transition is  $\delta N\simeq 10^{2}l^{2}$. From this
follows that at crystal thicknesses as small as $l\simeq 10^{-1}$
cm all the particles undergo a transition with a change in energy.
At large thicknesses the perturbation theory is not applicable. In
this case it is helpful to consider the problem in terms of
effective refractive indices in a rotating coordinate system
\cite{14,117}, or to usie the conception of quasi-energy.

Vibrations of nuclei in a crystal may be caused by either an
ultrasonic or an electromagnetic wave. Under diffraction
(channeling) of charged particles in a crystal the wave affects
not only the nuclei but also the particle itself, bringing about
the additional mechanism of interband transitions.  Interband
transitions of electrons induced by ultrasound (electromagnetic
field) are accompanied by simulated radiation. Naturally,
spontaneous interband transitions also exist.

Radiation through diffraction in the case of optical transitions
between the neighboring bands in an infinite crystal without
reference to spin structure of the bands was discussed in
\cite{69,70,71}. In view of the above analysis, taking into
account a finite crystal thickness results in appearing of
oscillations of the radiation intensity, depending on $l$ and
electron energies. The dependence of the band structure of
electrons diffracting in a crystal on their spin will lead to the
dependence of the intensity and polarization properties of
radiation on the beam polarization state, as well as polarization
of a non-polarized beam ((\ref{sec:6.19})). Resonant interband
transitions of electrons will cause the appearance of a spatially
modulated beam behind the crystal. The beam modulation period will
depend on the spin orientation. As a result, the initially
non-polarized beam behind the crystal will prove to be spatially
polarized in some regions of space. Of course, the aforesaid also
refers to the electrons which moved in the channeling regime.

Let now $\gamma$-quanta be diffracted in a crystal (light in a
liquid crystal or in some other periodic structure (array)). In
this case  even in a non-magnetic crystal in a wide energy range
(from several kiloelectron-volts to tens and hundreds of
gigaelectron-volts) there is band splitting, depending on the
photon polarization state. Diffraction of Mossbauer
$\gamma$-radiation in polarized crystals is considered in
\cite{14}. At two-wave diffraction the wave functions of
$\gamma$-quanta are analogous to the functions in
(\ref{12.2})-(\ref{12.5}). For this reason the structure of the
matrix element describing the transition of a $\gamma$-quantum
from one band state to another is also similar to the structure of
the matrix element appearing in (\ref{20.8}). Consequently,
ultrasonic (electromagnetic field) induces  resonant
repolarization of the diffracting beam of $\gamma$-quanta (light
passing through a liquid crystal and etc.) under the condition
determined by the conservation laws (\ref{20.10}). The process of
the interband transition of X-rays by the action of ultrasound
without reference to the change in their frequency and
polarization through the transition was treated in \cite{121}. In
the case of Mossbauer $\gamma$-quanta it is crucial that the
change in the $\gamma$-quantum frequency through transition
described by the conservation law should be taken into account.

Due to the close connection between the phenomena of diffraction
and mirror reflection under diffraction conditions \cite{14},
analogous effects will manifest themselves for mirror reflected
waves (neutrons, $\gamma$-quanta, light) too. In fact, the process
of the interband transition of X-rays and $\gamma$-quanta under
the electromagnetic wave causing vibrations of the crystal nuclei
can be treated the process of coherent coalescence (splitting) of
a $\gamma$-quantum and an optical photon.

%%%%%%%%%%%%%%%%%%%%%%%%%%%%%%%%%%%  Section 21 %%%%%%%%%%%%%%%%%%%%%%

\section[Optical Anisotropy in a Rotating Coordinate System]{Optical Anisotropy in a Rotating Coordinate System}
\label{sec:6.21}

It has been shown above that  spectral-angular distribution of
photons produced by particles passing through a crystal, depend
considerably on the refracting properties of the medium. If a
crystal is placed in an external variable field, its refracting
properties change sharply. In particular, the effects caused by
optical anisotropy of crystals in the $\gamma$-range acquire
qualitatively new features, when the material is placed in a
time-dependent external field (electromagnetic, sound).

To consider the essence of the arising phenomena, let us begin
with a simple example of neutron refraction in a constant magnetic
field on which a time-dependent transverse variable field is
imposed \cite{120}. The Schrodinger equation describing the stated
process has the form:
\begin{equation}
\label{21.1} i\hbar\frac{\partial\psi}{\partial
t}=\left\{-\frac{\hbar^{2}}{2m}\Delta_{r}-\vec{\mu}\vec{H}(\vec{r},t)\right\}\psi,
\end{equation}
where $m$ is the neutron mass; $\vec{\mu}=\mu\vec{\sigma}$ is its
magnetic moment; $\vec{\sigma}$ is the vector made up of the Pauli
matrices $\sigma_{x}, \sigma_{y}, \sigma_{z}$;
$\vec{H}(\vec{r},t)$ is the magnetic field acting on the neutron
at point $\vec{r}$ at moment $t$ with the components
$H_{x}=H_{\perp}\cos\omega t$, $H_{y}=H_{\perp}\sin\omega t$,
$H_{z}$ is time-independent; $\omega$ is the rotation frequency of
the transverse magnetic field.

Using the explicit form of $\vec{\sigma}$, one may obtain the
following system of equations for the components $\psi_{1}$ and
$\psi_{2}$ of the spinor wave function $\psi=\left(
                 \begin{array}{c}
                   \psi_{1} \\
                   \psi_{2} \\
                 \end{array}
               \right):
$

\begin{eqnarray}
\label{21.2}
i\hbar\frac{\partial\psi_{1}}{\partial t}=-\frac{\hbar^{2}}{2m}\Delta_{r}\psi_{1}-\mu H_{z}\psi_{1}-\mu H_{\perp}e^{-i\omega t}\psi_{2};\nonumber\\
i\hbar\frac{\partial\psi_{2}}{\partial
t}=-\frac{\hbar^{2}}{2m}\Delta_{r}\psi_{2}+\mu H_{z}\psi_{2}-\mu
H_{\perp}e^{i\omega t}\psi_{1};
\end{eqnarray}
Introduce new functions  $\psi_{1}$ and $\psi_{2}$, using the
following  transformation
\begin{equation}
\label{21.3}
\psi_{1}=\varphi_{1}\exp\left(-i\frac{\omega}{2}t\right);\,\,\,
\psi_{2}=\varphi_{2}\exp\left(i\frac{\omega}{2}t\right).
\end{equation}
The transformation  (\ref{21.3}) is equivalent to that performing
the conversion to the coordinate system rotating about the z-axis
at the frequency $\omega$ \cite{122}. As a result, (\ref{21.2})
goes over to the following system:
\begin{eqnarray}
\label{21.4}
i\hbar\frac{\partial\varphi_{1}}{\partial t}+\frac{\hbar\omega}{2}\varphi_{1}=-\frac{\hbar^{2}}{2m}\Delta_{r}\varphi_{1}-\mu H_{z}\psi_{1}-\mu H_{\perp}\varphi_{2};\nonumber\\
i\hbar\frac{\partial\varphi_{2}}{\partial
t}-\frac{\hbar\omega}{2}\varphi_{2}=-\frac{\hbar^{2}}{2m}\Delta_{r}\varphi_{2}+\mu
H_{z}\psi_{2}-\mu H_{\perp}\varphi_{1}.
\end{eqnarray}
Introduction of the spinor function $\varphi=\left(
                 \begin{array}{c}
                  \varphi_{1} \\
                  \varphi_{2} \\
                 \end{array}
               \right),
               $
enables us to write (\ref{21.4}) as follows
\begin{equation}
\label{21.5} i\hbar\frac{\partial\varphi}{\partial
t}=-\frac{\hbar^{2}}{2m}\Delta_{r}\varphi-\vec{\mu}\vec{H}(\omega)\varphi,
\end{equation}
where $\vec{H}(\omega)$ has the components
$H_{x}(\omega)=H_{\perp}$, $H_{y}(\omega)=0$,
$H_{z}(\omega)=H_{z}-\frac{\hbar\omega}{2\mu}$, and at the initial
instant of time the function $\varphi$ is
\begin{eqnarray*}
\varphi(t_{0})=\left\{\begin{array}{c}
                 \psi_{1}(t_{0})\left(i\frac{\omega}{2}t_{0}\right) \\
                 \psi_{2}(t_{0})\left(-i\frac{\omega}{2}t_{0}\right)
               \end{array}\right\}.
\end{eqnarray*}

Thus, the problem of refraction of a neutron wave in a
time-dependent magnetic field has reduced to the problem of wave
refraction in a constant effective magnetic field
$\vec{H}(\omega)$ depending on frequency $\omega$.

Due to the complete equivalence of equation (\ref{21.5}) and the
equations describing neutron motion in a time-independent magnetic
field $\vec{H}(\vec{r})$, all the conclusions concerning the laws
of refraction and mirror reflection in it hold true, however, with
a considerable difference that both the refractive index and the
amplitude of the reflected neutron wave now become dependent on
the external field frequency $\omega$.

The situation when $H_{z}\gg H_{\perp}$ seems to be of particular
interest. In this case at the frequency $\omega=\frac{2\mu
H_{z}}{\hbar}$ the component of the effective field $H_{z}$
vanishes, and the effective field  $H(\omega)$ equals $H_{\perp}$,
which  is much less than the value of the magnetic field in the
absence without resonance. Hence, the refractive index
(coefficient of mirror reflection) will appear to be smaller. For
instance, if without a rotating field, the magnetic field was so
great that the neutrons experienced total  mirror reflection from
it, under the resonance conditions, the neutrons will pass through
the area occupied by the magnetic field. Similarly, the
polarization state of the neutron beam will prove to be strongly
dependent on the frequency of a variable field.

Now let neutrons (electrons and so on) be incident onto a single
crystal with polarized electrons (nuclei). Then the area occupied
by the crystal may be described in terms of a spatially periodic
effective magnetic field $\vec{B}(\vec{r})$. If the crystal is
placed in the external rotating magnetic field (or excite a
circular sound wave in it ), then  a spatially periodic
$\omega$-dependent field $\vec{B}(\vec{r},\omega)$ emerges in a
rotating system. Mathematical formulation of the particle beam
propagation in the periodic field $\vec{B}(\vec{r},\omega)$ is
completely equivalent to that describing the phenomena of
refraction, diffraction and mirror reflection of particles in
single crystals in the absence of a variable field \cite{14,120}.
Therefore the formulae for the refractive indices of a crystal
placed in a variable field under diffraction conditions are
similar \cite{14,120}.

It is common knowledge that under diffraction of particles in
crystals the effect of anomalous transmission (anomalous
suppression of inelastic processes, nuclear reactions) arises
\cite{85,123}. In the case under consideration, due to the
frequency dependence of the periodic field
$\vec{B}(\vec{r},\omega)$, a new phenomenon appears: the effect of
anomalous transmission of particles ($\gamma$-quanta) through
crystals, which depends on the frequency of the external field
(electromagnetic, sound). (The probability of inelastic processes
and nuclear reactions also depends significantly on the frequency
of the external field). It is important to emphasize that effect
of anomalous transmission and reaction suppression depending on
the frequency of the external field occurs for both instantaneous
particle ($\gamma$-quantum) intensity and the intensity averaged
over the alteration period of the  external variable field.

Note that, as shown in \cite{9,125}, even in non-magnetic
unpolarized crystals placed in an external magnetic field, one may
observe multi-frequency precession of neutron spin  and
$H$-dependent effect of suppression of nuclear reactions. When the
crystal is exposed to an external variable field (magnetic, sound)
the effects depending on the field frequency emerge: anomalous
suppression of nuclear reactions (analogous to that considered
above) and multi-frequency precession of a neutron spin.

Thus, an external variable field sharply changes refractive
properties of a crystal under diffraction, which eventually
manifests directly in the  process of radiation. By way of
example, consider  diffraction of neutrons in a constant magnetic
field.

According to \cite{14} the system of equations for the neutron
wave function $\psi(\vec{r})$ describing the dynamic diffraction
in an arbitrary magnetically-ordered crystal with polarized nuclei
has the form
\begin{eqnarray}
\label{21.6}
\left(\frac{k^{2}}{k_{0}^{2}}-1\right)\varphi(\vec{k})-\sum_{\tau}\hat{g}(\vec{\tau})\varphi(\vec{k}-2\pi\vec{\tau})=0;\nonumber\\
\varphi(\vec{r})=\sum_{\tau}\varphi(\vec{k}+2\pi\vec{\tau})\exp[i(\vec{k}+2\pi\tau)\vec{r}];
\end{eqnarray}
\begin{eqnarray}
\label{21.7}
\hat{g}(\vec{\tau})=\hat{g}_{nuc}(\vec{\tau})+\hat{g}_{mag}(\vec{\tau})\nonumber\\
=\frac{4\pi}{\Omega_{0}k_{0}^{2}}\sum_{j}(\hat{f}_{j_{nuc}}(\vec{\tau})+\hat{f}_{j_{mag}}(\vec{\tau}))e^{-2\pi\tau\vec{r}_{j}},
\end{eqnarray}
where $\hat{g}(\tau)$ is the structure amplitude;
$\hat{f}_{j}(\tau)$  is the amplitude of coherent scattering by
the j-th center included in the unit cell; $\vec{r}_{j}$ is the
coordinate of the j-th center; the summation is performed over all
the scatterers constituting the unit cell; $\Omega_{0}$ is the
unit cell volume; $\vec{k}_{0}$ is the wave vector of the neutron
incident on a crystal.

At $\vec{\tau}\neq 0$ the amplitude of coherent magnetic
scattering is defined by the expression \cite{14}
\begin{equation}
\label{21.8} \hat{f}_{j\,
mag}(\vec{\tau})=-4\pi\mu_{n}\left[\frac{(\vec{\sigma}\vec{\tau})(\vec{\tau}\vec{\mu}_{j})}{\tau^{2}}-\vec{\sigma}\vec{\mu}_{j}\right]
F_{j}(\vec{\tau})e^{-W_{j}(\tau)}.
\end{equation}
At $\vec{\tau}=0$ the magnetic contribution to $\hat{g}(0)$ has
the form
\begin{equation}
\label{21.9}
\hat{g}_{mag}(0)=\frac{2m\mu_{n}}{\hbar^{2}k_{0}^{2}}\vec{\sigma}\vec{B},
\end{equation}
where $\vec{B}$ is the macroscopic magnetic field of the target;
$m$ is the particle mass.

In the case of a non-magnetic unpolarized crystal placed in a
constant magnetic field of strength $\vec{H}$, the structure
amplitudes of (\ref{21.7}) can be written as follows:
\begin{equation}
\label{21.10}
\hat{g}(0)=\frac{4\pi}{\Omega_{0}k_{0}^{2}}\sum_{j}f_{j_{nuc}}(0)+\frac{2m\mu_{n}}{\hbar^{2}k_{0}^{2}}\vec{\sigma}\vec{H},
\end{equation}
\begin{equation}
\label{21.11} \hat{g}_{\tau\neq
0}(\vec{\tau})=g(\vec{\tau})=\frac{4\pi}{\Omega_{0}k_{0}^{2}}\sum_{j}f_{j_{nuc}}(\vec{\tau})e^{-i2\pi\tau
r_{j}}.
\end{equation}

Choose the quantization axis parallel to the direction of the
field $\vec{H}$. As a result, the operator system (\ref{21.6})
will reduce to two independent systems of equations for either
neutron spin component, parallel $\varphi_{+}$ and antiparall
$\varphi_{-}$ to the quantization axis:
\begin{eqnarray}
\label{21.12}
\left(\frac{k^{2}}{k_{0}^{2}}-1\right)\varphi_{\pm}(\vec{k})-\sum_{\tau}g_{\pm}(\vec{\tau})\varphi_{\pm}(\vec{k}-2\pi\vec{\tau})=0,\nonumber\\
g_{\pm}(0)=g_{nuc}(0)\pm\frac{2m\mu_{n}}{\hbar^{2}k_{0}^{2}}H,\,\,
g_{\pm}(\vec{\tau})=g_{nuc}(\vec{\tau}).
\end{eqnarray}

The system of equations (\ref{20.12}) has a standard form for the
dynamical diffraction theory. This enables us to immediately write
the expression for the wave function of a neutron that has passed
through the crystal plate of thickness $l$ \cite{14}:
$$
\psi(\vec{r})\left(
              \begin{array}{c}
                c_{+}\psi_{+}(\vec{r}) \\
                c_{-}\psi_{-}(\vec{r}) \\
              \end{array}
            \right),
$$
where $\left(
         \begin{array}{c}
           c_{+} \\
          c_{-} \\
         \end{array}
       \right)
$
 is the spin wave function of the neutrons incident of the plate; the z-axis is directed along the quantization axis;
\begin{eqnarray}
\label{21.13} \psi_{\sigma}(\vec{r})=
\frac{(2\varepsilon_{2}^{\sigma}-g_{\sigma}(0))\exp\left(ik_{0}\varepsilon_{1}^{\sigma}\frac{l}{\gamma_{0}}\right)-
(2\varepsilon_{1}^{\sigma}-g_{\sigma}(0))\exp\left(ik_{0}\varepsilon_{2}^{\sigma}\frac{l}{\gamma_{0}}\right)}
{2(\varepsilon_{2}^{\sigma}-\varepsilon_{1}^{\sigma})}e^{-i\vec{k}_{0}\vec{r}}\nonumber\\
-\frac{\beta
g(\vec{\tau})}{2(\varepsilon_{2}^{\sigma}-\varepsilon_{1}^{\sigma})}
\left[\exp\left(ik_{0}\varepsilon_{1}^{\sigma}\frac{l}{\gamma_{0}}\right)-\exp\left(ik_{0}\varepsilon_{2}^{\sigma}\frac{l}{\gamma_{0}}\right)\right]e^{i(\vec{k}_{0}+2\pi\vec{\tau})\vec{r}},\nonumber\\
\varepsilon_{1(2)}^{\sigma}=\frac{1}{4}\left\{(1+\beta)g_{\sigma}(0)-\beta\alpha\right.\nonumber\\
\left.\pm\sqrt{[\beta\alpha+g_{\sigma}(0)(1-\beta)]^{2}+4\beta g(\tau)g(-\tau)}\right\};\nonumber\\
\end{eqnarray}
$\sigma=\pm$ ($(+)$ corresponds to the neutrons with spin parallel
to $\vec{H}$, $(-)$ corresponds to the neutrons with the opposite
spin direction); $\gamma_{0}=\vec{k}_{0}\vec{n}/k_{0}$; $\vec{n}$
is the normal to the crystal surface;
$\alpha=2\pi\vec{\tau}(2\pi\vec{\tau}+2\vec{k}_{0})/k_{0}^{2}$ is
the quantity characterizing deviation from the exact Bragg
conditions; $\beta=\gamma_{0}/\gamma_{1}$;
$\gamma_{1}=(\vec{k}_{0}+2\pi\vec{\tau})\vec{n}/|\vec{k}_{0}+2\pi\vec{\tau}|$.
In the case of the symmetric Laue diffraction
$\gamma_{0}=\gamma_{1}$, $\beta=1$ and
\begin{equation}
\label{21.14}
\varepsilon^{\sigma}_{1(2)}=\frac{1}{4}\left\{2g_{0}(0)-\alpha\pm\sqrt{\alpha^{2}+4g(\vec{\tau})g(-\vec{\tau})}\right\}.
\end{equation}
First consider how the magnetic field influences the diffracted
neutrons. Using the expression for the wave function (\ref{21.3}),
write the expression for the intensity of the diffracted wave:
\begin{eqnarray}
\label{21.15}
I_{d}=|c_{+}\psi_{+}|^{2}+|c_{-}\psi_{-}|^{2}=|c_{+}|^{2}\beta^{2}\left|\frac{g(\tau)}{A_{+}}\right|^{2}\nonumber\\
\times\left\{\exp\left(-k_{0}\texttt{Im}2\varepsilon_{2}^{+}\frac{l}{\gamma_{0}}\right)+
\exp\left(-k_{0}\texttt{Im}2\varepsilon_{1}^{+}\frac{l}{\gamma_{0}}\right)\right.\nonumber\\
\left.-2\cos\left(k_{0}\texttt{Re}\frac{A_{+}}{2}\frac{l}{\gamma_{0}}\right)
\exp\left[-k_{0}\texttt{Im}(\varepsilon_{1}^{+}+\varepsilon_{2}^{+})\frac{l}{\gamma_{0}}\right]\right\}\nonumber\\
+|c_{-}|^{2}\beta^{2}\left|\frac{g(\tau)}{A_{-}}\right|^{2}\left\{\exp\left(-k_{0}\texttt{Im}2\varepsilon_{2}^{-}\frac{l}{\gamma_{0}}\right)\right.\nonumber\\
\left.+\exp\left(-k_{0}\texttt{Im}2\varepsilon_{1}^{-}\frac{l}{\gamma_{0}}\right)
-2\cos\left(k_{0}\texttt{Re}\frac{A_{-}}{2}\frac{l}{\gamma_{0}}\right)\right.\nonumber\\
\left.\times\exp\left[-k_{0}\texttt{Im}(\varepsilon_{1}^{-}+\varepsilon_{2}^{-})\frac{l}{\gamma_{0}}\right]\right\},
\end{eqnarray}
where
\begin{equation}
\label{21.16}
A_{\pm}=\left\{[g_{\pm}(0)(1-\beta)+\beta\alpha]^{2}+4\beta
g(\vec{\tau})g(-\vec{\tau})\right\}^{1/2}.
\end{equation}

First of all, note that in the presence of the external magnetic
field, oscillations of the intensity of the diffracted wave (the
pendulum effect), unlike those in the case when $H=0$, occur at
two spatial frequencies
\begin{equation}
\label{21.17} \kappa_{1}=k_{0}\texttt{Re}\frac{A_{+}}{2};\,\,
\kappa_{2}=k_{0}\texttt{Re}\frac{A_{-}}{2}.
\end{equation}

In symmetric diffraction the frequencies $\kappa_{1}$ and
$\kappa_{2}$ coincide and do not depend on the value of the
magnetic field:
\begin{equation}
\label{21.18}
\kappa_{1}=\kappa_{2}=\kappa=k_{0}\texttt{Re}\frac{1}{2}\sqrt{\alpha^{2}+4g(\vec{\tau})g(-\vec{\tau})}
\end{equation}
If the neutron beam is polarized parallel to the magnetic field
($c_{+}=1$, $c_{-}=0$), then $I_{d}$ oscillates at the frequency
$\kappa_{1}$, if antiparallel,  $I_{d}$ oscillates at the
frequency $\kappa_{2}\neq\kappa_{1}$.

Let an unpolarized neutron beam fall upon a crystal. In this case
$I_{d}$ is given by expression (\ref{21.15}), where
$|c_{+}|^{2}=|c_{-}|^{2}=1/2$. As $\kappa_{1}\neq\kappa_{2}$, at
certain values of the magnetic field $H$ the situation is
possible, when the contribution to $I_{d}$ coming from one of the
neutron spin components appears to be zero. Hence, at such values
of  $H$ the diffracted beam will be fully polarized.

The intense neutron beam fully polarized along the magnetic field
will be obtained at the exit from the crystal plate provided that
one of the summands in (\ref{21.15}) takes on its maximum value,
with the second summand taking on its minimum value. For
simplicity, assume that the crystal is non-absorptive, and the
exponential factors in (\ref{21.15}) are equal to unity.

When in (\ref{21.15}) the cosine in the augend equals $-1$, and in
the addend - $+1$, we obtain a beam fully polarized along the
field. The neutron beam fully polarized opposite the field is
obtained when the cosine in the augend becomes $+1$, and in the
addend it quals $-1$. In the general case this condition may be
written as follows with due account of the explicit form  for
frequencies $\kappa_{1}$ and $\kappa_{2}$
\begin{eqnarray}
\label{21.19} \mbox{for}\, p_{z}=-1\left\{\begin{array}{c}
              \texttt{Re}k_{0}\frac{1}{2}A_{+}\frac{l}{\gamma_{0}}=2\pi N, \\
              \texttt{Re}k_{0}\frac{1}{2}A_{-}\frac{l}{\gamma_{0}}=2\pi N^{\prime}+\pi;
            \end{array}\right.
\end{eqnarray}
\begin{eqnarray}
\label{21.20} \mbox{for}\, p_{z}=1\left\{\begin{array}{c}
              \texttt{Re}k_{0}\frac{1}{2}A_{+}\frac{l}{\gamma_{0}}=2\pi N+\pi, \\
              \texttt{Re}k_{0}\frac{1}{2}A_{-}\frac{l}{\gamma_{0}}=2\pi N^{\prime};
            \end{array}\right.
\end{eqnarray}
where $N$ and $N^{\prime}$ are integral numbers.

Find the phase difference $\eta$ between the components of the
wave function of neutrons corresponding to the parallel and
anti-parallel spin states when passing through the plate of
thickness $l$, the magnetic field strength $H$ being equal to
\begin{equation}
\label{21.21}
\eta=\frac{1}{2}k_{0}\frac{l}{\gamma_{0}}\texttt{Re}(A_{+}-A_{-}).
\end{equation}

Estimation of the expression (\ref{21.21}) shows that in the
asymmetric diffraction case at $(1-\beta)\simeq 10^{-1}$, $l=0.1$
cm, $\lambda\sim 1$\,{\AA} and $g(0)\simeq 10^{-6}$, the phase
difference is $\pi$ at the magnetic field strength of the order of
1000 Gs. Hence, at such strength of the magnetic field one may
obtain fully polarized neutron beams with the possibility to
specify the polarization direction by changing the direction of
the magnetic field.

By varying the value of the magnetic field strength, it is also
possible to modulate the intensity of the diffracted beam. The
degree of modulation can be close to $100\%$. Indeed, choose the
plate thickness $l$ so that in the absence of the magnetic field
the intensity of the diffracted beam would be zero. Then, as
follows from (\ref{21.15}), with the external magnetic field
imposed, the intensity becomes non-zero. With the strength of the
external magnetic field vanishing, the frequencies $\kappa_{1}$
and $\kappa_{2}$ become equal to each other, and the intensity of
the diffracted beam starts oscillating at one frequency only,
which is defined by equality (\ref{21.18}), so we have an ordinary
pendulum effect.

Now consider the influence of the magnetic field on the beam
absorption in a crystal. From (\ref{21.13}) follows that at
$g_{+}(0)\neq g_{-}(0)$  the imaginary parts of
$\varepsilon^{+}_{1(2)}$ and $\varepsilon^{-}_{1(2)}$ will be
different, and what is more, they will reveal different dependence
on the value of the magnetic field:
\begin{eqnarray}
\label{21.22}
\texttt{Im}\varepsilon^{\sigma}_{1(2)}=\frac{1}{4}(1+\beta)\texttt{Im}g_{\sigma}(0)\pm\left[(\texttt{Re}A_{\sigma})^{2}\right.\nonumber\\
\left.+(\texttt{Im}A_{\sigma})^{2}\right]^{1/2}\sin\left(\frac{1}{2}\arctan\frac{\texttt{Im}A_{\sigma}}{\texttt{Re}A_{\sigma}}\right),
\end{eqnarray}
where $A_{\sigma}$ are given by equality (\ref{21.16}).

Note that if $\texttt{Im}A_{\sigma}=0$, then
\begin{equation}
\label{21.23}
\texttt{Im}\varepsilon^{\sigma}_{1}=\texttt{Im}\varepsilon^{\sigma}_{2}.
\end{equation}
This is attained at the value of the magnetic field
\begin{eqnarray}
\label{21.24}
H=& &-(\sigma)\nonumber\\
& &\times\frac{2(1-\beta)\beta\alpha\texttt{Im}g(0)+(1-\beta)^{2}\texttt{Im}g^{2}(0)+4\beta g(\tau)g(-\tau)}{2(1-\beta)^{2}\texttt{Im}g(0)}\nonumber\\
& &\times\frac{\hbar^{2}k_{0}^{2}}{2m\mu_{n}},
\end{eqnarray}
where $(\sigma)$ indicates the neutron spin state for which
(\ref{21.24}) holds.

As a result, in the case of the asymmetric Laue diffraction under
consideration, the effect of the anomalous transmission of
particles through a crystal, and, hence, the yield of nuclear
reactions will depend on the strength of the external magnetic
field.

Analyze polarization characteristics of the diffracted neutron
beam in more detail.

To fix the idea, we shall assume that the polarization vector of
neutrons incident on a crystal $\vec{p}_{0}$ is directed
perpendicular to the quantization axis, i.e. to the z-axis (the
z-axis is directed parallel to $\vec{H}$). The direction of
$\vec{p}_{0}$ is chosen as the x-axis so that
$c_{+}=c_{-}=1\sqrt{2}$. Using (\ref{21.13}), immediately find the
components $p_{x}$ and $p_{y}$ of the neutron polarization vector
in the diffracted wave:
\begin{eqnarray}
\label{21.25} p_{x}=&
&\frac{\beta^{2}}{4}\frac{|g(\tau)|^{2}}{|(\varepsilon^{+}_{2}-\varepsilon^{+}_{1})(\varepsilon^{-}_{2}-\varepsilon^{-}_{1})|}
\left\{\cos\left[k_{0} \texttt{Re}(\varepsilon^{+}_{1}-\varepsilon^{-}_{1})\right.\right.\nonumber\\
& &\left.\left.\times\frac{l}{\gamma_{0}}+\delta\right]\exp\left[-k_{0} \texttt{Im}(\varepsilon^{+}_{1}+\varepsilon^{-}_{1})\frac{l}{\gamma_{0}}\right]\right.\nonumber\\
& &\left.-\cos\left[k_{0} \texttt{Re}(\varepsilon^{+}_{1}-\varepsilon^{-}_{2})\frac{l}{\gamma_{0}}+\delta\right]\right.\nonumber\\
& &\left.\times\exp\left[-k_{0} \texttt{Im}(\varepsilon^{+}_{1}+\varepsilon^{-}_{2})\frac{l}{\gamma_{0}}\right]\right.\nonumber\\
& &\left.-\cos\left[k_{0} \texttt{Re}(\varepsilon^{+}_{2}-\varepsilon^{-}_{1})\frac{l}{\gamma_{0}}+\delta\right]\right.\nonumber\\
& &\left.\times\exp\left[-k_{0} \texttt{Im}(\varepsilon^{+}_{2}+\varepsilon^{-}_{1})\frac{l}{\gamma_{0}}\right]\right.\nonumber\\
& &\left.+\cos\left[k_{0} \texttt{Re}(\varepsilon^{+}_{2}-\varepsilon^{-}_{2})\frac{l}{\gamma_{0}}+\delta\right]\right.\nonumber\\
& &\left.\times\exp\left[-k_{0}
\texttt{Im}(\varepsilon^{+}_{2}+\varepsilon^{-}_{2})\frac{l}{\gamma_{0}}\right]\right\},
\end{eqnarray}
where $\delta=\delta_{+}-\delta_{-}$ and the following notation is
used
$$
\frac{g(\tau)}{2(\varepsilon_{2}^{\pm}-\varepsilon_{1}^{\pm})}=\left|\frac{g(\tau)}{2(\varepsilon_{2}^{\pm}-\varepsilon_{1}^{\pm})}\right|e^{i\delta_{\pm}}.
$$
The component $p_{y}$ is obtained from $p_{x}$ by replacing
$\cos$ with $-\sin$.

Using (\ref{21.12}) and (\ref{21.13}) the differences of the
values of $\varepsilon$ appearing in (\ref{21.15}) are written as
follows:
\begin{eqnarray}
\label{21.26}
\varepsilon^{+}_{1}-\varepsilon^{-}_{1}=G+\frac{1}{4}(A_{+}-A_{-});\nonumber\\
\varepsilon^{+}_{1}-\varepsilon^{-}_{2}=G+\frac{1}{4}(A_{+}+A_{-});\nonumber\\
\varepsilon^{+}_{2}-\varepsilon^{-}_{1}=G-\frac{1}{4}(A_{+}+A_{-});\nonumber\\
\varepsilon^{+}_{2}-\varepsilon^{-}_{2}=G-\frac{1}{4}(A_{+}-A_{-});\nonumber\\
G=\frac{m\mu_{n}}{\hbar^{2}k^{2}_{0}}H(1+\beta)
\end{eqnarray}

In the case of the symmetric Laue diffraction when $\beta=1$,
$A_{+}$ and  $A_{-}$ are equal. And the neutron polarization
vector undergoes beatings with changes in $H$ at one frequency,
determined by the Larmour spin precession frequency in a magnetic
field.

At the asymmetric Laue diffraction ($\beta\neq 1$) the situation
changes drastically. $A_{+}\neq A_{-}$, and with the changes in
$H$, the neutron polarization vector undergoes beating at four
different frequencies determined by the differences (\ref{21.26}):
\begin{eqnarray}
\label{21.27} \omega_{1}=\frac{\hbar
k_{0}^{2}}{m}\texttt{Re}(\varepsilon^{+}_{1}-\varepsilon^{-}_{1});\,\,
\omega_{2}=\frac{\hbar k_{0}^{2}}{m}\texttt{Re}(\varepsilon^{+}_{1}-\varepsilon^{-}_{2});\nonumber\\
\omega_{3}=\frac{\hbar
k_{0}^{2}}{m}\texttt{Re}(\varepsilon^{+}_{2}-\varepsilon^{-}_{1});\,\,
\omega_{4}=\frac{\hbar
k_{0}^{2}}{m}\texttt{Re}(\varepsilon^{+}_{2}-\varepsilon^{-}_{2}).
\end{eqnarray}

From (\ref{21.25}) and (\ref{21.26}) follows that at relatively
small magnetic fields ($H=10^{3}\div 10^{4}$ Gs) the effect of
multi-frequency precession in a crystal should be clearly observed
even at $l\sim 10^{2}\div 10^{-1}$ cm.

Using the expression for the wave function (\ref{21.13}), we also
write the expression for the $p_{z}$ component of the polarization
vector of a diffracted wave:
\begin{eqnarray}
\label{21.28}
p_{z}=|\psi_{+}|^{2}-|\psi_{-}|^{2}=\beta^{2}\left|\frac{g(\tau)}{A_{+}}\right|^{2}\left\{\exp\left[-k_{0} \texttt{Im}2\varepsilon_{2}^{+}\frac{l}{\gamma_{0}}\right]\right.\nonumber\\
\left.+\exp\left[-k_{0}
\texttt{Im}2\varepsilon_{1}^{+}\frac{l}{\gamma_{0}}\right]
-2\cos\left(k_{0}\texttt{Re}\frac{A_{+}}{2}\frac{l}{\gamma_{0}}\right)\right.\nonumber\\
\left.\times\exp\left[-k_{0}\texttt{Im}(\varepsilon_{2}^{+}+\varepsilon_{1}^{+})\frac{l}{\gamma_{0}}\right]\right\}
-\beta^{2}\left|\frac{g(\tau)}{A_{-}}\right|^{2}\nonumber\\
\times\left\{\exp\left[-k_{0}\texttt{Im}2\varepsilon_{2}^{-}\frac{l}{\gamma_{0}}\right]+
\exp\left[-k_{0}\texttt{Im}2\varepsilon_{1}^{-}\frac{l}{\gamma_{0}}\right]\right.\nonumber\\
\left.-2\cos\left(k_{0}\texttt{Re}\frac{A_{-}}{2}\frac{l}{\gamma_{0}}\right)
\exp\left[-k_{0}\texttt{Im}(\varepsilon_{2}^{-}+\varepsilon_{1}^{+})\frac{l}{\gamma_{0}}\right]\right\}.
\end{eqnarray}

Comparison of (\ref{21.15}) and (\ref{21.28}) shows that the
longitudinal component of the polarization vector of the
diffracted beam oscillates at the same two frequencies
$\kappa_{1}$ and $\kappa_{2}$ as the intensity of the diffracted
beam does.

Further consider the expressions for the components of the
polarization vector and the intensity of the diffracted wave when
absorption can be ignored (the crystal thickness is much smaller
than the absorption depth, but of the order or greater than the
spatial precession period; this requirement can always be met for
neutrons as $\texttt{Re}g\gg \texttt{Im}g$, if only the neutron
energy does not lie in the resonance region). Then equating in
(\ref{21.15}) the exponential factors to unity, gives the
following expressions for the components of the polarization
vector:
\begin{eqnarray}
\label{21.29}
p_{x}=2\beta^{2}\frac{|g(\tau)|^{2}}{|A_{+}A_{-}|}\cos\left(k_{0}G\frac{l}{\gamma_{0}}+\delta\right)\left\{\cos k_{0}\texttt{Re}\left[G+\frac{1}{4}\right.\right.\nonumber\\
\left.\left.\times(A_{+}-A_{-})\right]\frac{l}{\gamma_{0}}-\cos
k_{0}\texttt{Re}\left[G+\frac{1}{4}(A_{+}+A_{-})\right]\frac{l}{\gamma_{0}}\right\};
\end{eqnarray}
$p_{y}$ is obtained from $p_{x}$ by substituting
$\cos\left(k_{0}G\frac{l}{\gamma_{0}}+\delta\right)$ for
$-\sin\left(k_{0}G\frac{l}{\gamma_{0}}+\delta\right)$. From this
\begin{eqnarray}
\label{21.30}
p_{x}+ip_{y}=2\beta^{2}\frac{|g(\tau)|^{2}}{|A_{+}A_{-}|}\exp\left[-i\left(k_{0}G\frac{l}{\gamma_{0}}+\delta\right)\right]\nonumber\\
\times\left\{\cos k_{0}\texttt{Re}\left[G+\frac{1}{4}(A_{+}-A_{-})\right]\frac{l}{\gamma_{0}}\right.\nonumber\\
\left.-\cos
k_{0}\texttt{Re}\left[G+\frac{1}{4}(A_{+}+A_{-})\right]\frac{l}{\gamma_{0}}\right\}.
\end{eqnarray}
According to (\ref{21.29}) and (\ref{21.30}) the longitudinal
component of the neutron polarization vector rotates at the
frequency $\Omega=(1+\beta)\mu_{n}H/\hbar$ in the plane $xy$,
i.e., in the plane normal to the magnetic field. In the case of
asymmetric diffraction $\beta\neq 1$ the rotation frequency of the
neutron polarization vector $\Omega$ does not coincide with the
Larmour frequency $2\mu_{n}H/\hbar$. Thus, under diffraction
conditions the spin rotation frequency depends not only on the
value of the magnetic field but also on the angle of incidence on
the crystal and the orientation of the crystal surface with
respect to crystallographic axes.

As follows from (\ref{21.30}) rotation is accompanied by the
oscillation of the magnitude of the longitudinal component of the
polarization vector at the frequencies $\omega_{1}$ and
$\omega_{2}$.

With the help of (\ref{21.28}) and (\ref{21.15})  we obtain the
following expressions for the longitudinal component of the
polarization vector $p_{z}$ and the intensity of the diffracted
wave $I_{d}$ for a thin plate:
\begin{eqnarray}
\label{21.31}
p_{z}=2\beta^{2}\left|\frac{g(\tau)}{A_{+}}\right|^{2}\left[1-\cos \left(k_{0}\texttt{Re}\frac{1}{2}A_{+}\frac{l}{\gamma_{0}}\right)\right]\nonumber\\
-2\beta^{2}\left|\frac{g(\tau)}{A_{-}}\right|^{2}\left[1-\cos
\left(k_{0}\texttt{Re}\frac{1}{2}A_{-}\frac{l}{\gamma_{0}}\right)\right];
\end{eqnarray}
\begin{eqnarray}
\label{21.32}
I_{d}=2\beta^{2}\left|\frac{g(\tau)}{A_{+}}\right|^{2}\left[1-\cos \left(k_{0}\texttt{Re}\frac{1}{2}A_{+}\frac{l}{\gamma_{0}}\right)\right]\nonumber\\
+2\beta^{2}\left|\frac{g(\tau)}{A_{-}}\right|^{2}\left[1-\cos
\left(k_{0}\texttt{Re}\frac{1}{2}A_{-}\frac{l}{\gamma_{0}}\right)\right];
\end{eqnarray}
It is clear from (\ref{21.31}) and (\ref{21.32}) that density
oscillations of the components of the wave function, which
correspond to the states with spin parallel and antiparallel to
the magnetic field will occur at two different frequencies
$\kappa_{1}$ and $\kappa_{2}$.

Now we shall consider thoroughly the oscillations of the
transverse component of the polarization vector. With this aim in
view recast (\ref{21.30}) as follows:
\begin{eqnarray}
\label{21.33}
p_{x}+ip_{y}=2\beta^{2}\frac{|g(\tau)|^{2}}{|A_{+}A_{-}|}\exp\left[-i\left(k_{0}G\frac{l}{\gamma_{0}}+\delta\right)\right]\nonumber\\
\times
2\sin\left(k_{0}\texttt{Re}\frac{1}{4}A_{-}\frac{l}{\gamma_{0}}\right)\sin\left[k_{0}\texttt{Re}\left(G+\frac{1}{4}A_{+}\right)\frac{l}{\gamma_{0}}\right].
\end{eqnarray}
Analyzing expressions (\ref{21.31})-(\ref{21.33}), one may see
that the oscillation frequencies of the transverse component
$p_{x}+ip_{y}$ of the polarization vector in the presence of the
external magnetic field do not coincide with those of the
longitudinal component of the polarization vector $p_{z}$. Hence,
it is always possible to select such a value of the magnetic field
(at given $g(\tau)$, $l$, $\alpha$ and $\beta$) that the
transverse component will vanish, while the longitudinal component
will be non-zero. A coherent neutron beam, initially fully
polarized along the x-axis, under diffraction conditions may
become partially or fully polarized along the z-axis.

How does absorption affect rotation of the neutron polarization
vector? Let a crystal thickness be larger than the absorption
depth of a rapidly damped wave, but smaller than the absorption
depth of an anomalously transmitted wave. In this case $p_{x}$
and $p_{y}$ may be written in the form
\begin{equation}
\label{21.34} p_{x}=\beta^{2}\frac{|g(\tau)|^{2}}{|A_{+}A_{-}|}
\cos\left\{k_{0}\texttt{Re}\left[G-\frac{1}{4}(A_{+}-A_{-})\right]\frac{l}{\gamma_{0}}+\delta\right\},
\end{equation}
$p_{y}$ is obtained by  replacing $\cos$ with $-\sin$. Write the
explicit form of the spin rotation frequency $\omega_{4}$:
\begin{eqnarray}
\label{21.35}
\omega_{4}=\frac{\hbar k_{0}^{2}}{m}\texttt{Re}\left\{\frac{m\mu_{n}}{\hbar^{2}k_{0}^{2}}H(1+\beta)\right.\nonumber\\
\left.-\frac{1}{4}\sqrt{[g_{+}(0)(1-\beta)+\beta\alpha]^{2}+4\beta g(\tau)g(-\tau)}\right.\nonumber\\
\left.+\frac{1}{4}\sqrt{[g_{-}(0)(1-\beta)+\beta\alpha]^{2}+4\beta
g(\tau)g(-\tau)}\right\}.
\end{eqnarray}
Pay attention to the fact that in this case the spin rotation
frequency no longer demonstrates linear dependence on the value of
the magnetic field, which now becomes more complicated, as
described by expression (\ref{21.35}).

The spin phenomena investigated also occur in the case of the
symmetric Laue diffraction, if the field boundary is not parallel
to the crystal surface. Now consider the case when a diffracted
wave exits through the same crystal surface on which the initial
beam falls, i.e., consider diffraction reflection of neutrons from
a non-magnetic crystal placed in a constant homogeneous magnetic
field.

As the set of dynamic equations (\ref{21.12}) in question is
perfectly analogous in form to that describing diffraction in a
crystal in the absence of a magnetic field, we can immediately
write down the coefficient of diffraction reflection for each spin
component of the neutron wave \cite{14}
\begin{eqnarray}
\label{21.36}
R_{\pm}=& &\beta^{2}|g(\tau)|^{2}\nonumber\\
&
&\times\left|\frac{1-\exp\left[i(\varepsilon_{2}^{\pm}-\varepsilon_{1}^{\pm})k_{0}\frac{l}{\gamma_{0}}\right]}
{[2\varepsilon_{1}^{\pm}-g_{\pm}(0)]-[2\varepsilon_{2}^{\pm}-g_{\pm}(0)]
\exp\left[i(\varepsilon_{2}^{\pm}-\varepsilon_{1}^{\pm})k_{0}\frac{l}{\gamma_{0}}\right]}\right|^{2},
\end{eqnarray}
where $\varepsilon^{\pm}_{1(2)}$ are specified by equality
(\ref{21.13}).

For simplicity, consider the symmetric Bragg case, when
$\beta=-1$. Then, according to (\ref{21.14}),
\begin{equation}
\label{21.37}
\varepsilon^{\pm}_{1(2)}=\frac{1}{4}[\alpha\pm\sqrt{[2g_{\pm}(0)-\alpha]^{2}-4g(\tau)g(-\tau)}].
\end{equation}

From (\ref{21.36}) follows that when the following conditions are
fulfilled
\begin{eqnarray}
\label{21.38}
\left(\alpha-2g(0)-\frac{4m\mu_{n}}{\hbar^{2}k_{0}^{2}}H\right)^{2}<4g(\tau)g(-\tau);\nonumber\\
\left(\alpha-2g(0)+\frac{4m\mu_{n}}{\hbar^{2}k_{0}^{2}}H\right)^{2}>4g(\tau)g(-\tau)
\end{eqnarray}
the reflection coefficient $R_{+}=1$, while $R_{-}\gg 1$. And vice
versa, if the conditions
\begin{eqnarray}
\label{21.39}
\left(\alpha-2g(0)-\frac{4m\mu_{n}}{\hbar^{2}k_{0}^{2}}H\right)^{2}>4g(\tau)g(-\tau);\nonumber\\
\left(\alpha-2g(0)+\frac{4m\mu_{n}}{\hbar^{2}k_{0}^{2}}H\right)^{2}<4g(\tau)g(-\tau)
\end{eqnarray}
are satisfied, then $R_{-}=1$, and $R_{+}\ll 1$.

The phenomena considered above also occur in a wave passing
through a crystal in the incident direction of the initial
(primary) beam. Outside the diffraction conditions the intensity
of the diffracted wave diminishes rapidly. At the same time the
refractive index of the wave propagating in the initial direction
contains the admixture owing to the existence of diffraction, for
example,
\begin{equation}
\label{21.40}
\varepsilon^{\sigma}_{1}=\frac{1}{2}g_{\sigma}(0)+\frac{\beta
g(\tau)g(-\tau)}{2(\beta\alpha+g_{\sigma}(0)(1-\beta))}+...
\end{equation}
Therefore  even away from diffraction, the neutron spin rotates at
the frequency different from the Larmour one. The contribution to
the  refractive index of a particle ($\gamma$-quantum) passing
through a crystal due to the summand of the type as considered in
(\ref{21.40}) affects the optical anisotropy of crystals in a hard
spectrum, and depends, in particular, on variable fields acting on
the crystal. (Under the diffraction conditions these fields
considerably modify $\varepsilon^{\sigma}_{1(2)}$ \cite{120}).

%%%%%%%%%%%%%%%%%%%%%%%%%%%%%%%%%%%%%%%%%%%%%%

%%%%%%%%%%%%%%%%%%%%%%%%%%%%%%%%%%%  Chapter 7 %%%%%%%%%%%%%%%%%%%%%%

\chapter[Interference of Independently Generated Beams of $\gamma$-quanta]{Interference of Independently Generated Beams of $\gamma$-quanta}
\label{ch:7}

%%%%%%%%%%%%%%%%%%%%%%%%%%%%%%%%%%%  Section 22 %%%%%%%%%%%%%%%%%%%%%%

\section[Interference of Independently Generated Photons]{Interference of Independently Generated Photons}
\label{sec:7.22}

The interference phenomena in beams of light generated by
independently emitting sources have been widely debated in
literature. The study of such phenomena in the X-ray band would
make it possible to carry out direct measurements of the phases of
scattering amplitudes and structure amplitudes in crystals.
However, as shown in \cite{126} it is impossible to perform such
measurements with conventional sources of X-ray and $\gamma$
radiation. Nevertheless, according to \cite{127}, high intensity
and  pointed directivity of radiation produced be relativistic
particles in crystals give hope for experimental detection of the
interference phenomenon of independently generated beams of
$\gamma$-quanta.

%%%%%%%%%%%%%%%%

In the beginning consider the nature of the phenomenon. Let us
have two excited atoms with energies $E_{a}$ and $E_{b}$ located
at points $\vec{r}_{a}$ and $\vec{r}_{b}$, and two atoms (two
elementary (simple) counters) located at points $\vec{r}_{c}$ and
$\vec{r}_{d}$. One and the same final state of the system (photon
registration by the counters at specified instants of time $t$
and $\vec{\tau}$), due to the identity of incident particles, is
achieved by two possible ways: ($a\rightarrow c$, $b\rightarrow
d$) and ($b\rightarrow c$, $a\rightarrow d$).
Observation cannot distinguish these two regimes. Therefore the
probability $P_{ab}(\vec{r}_{c}\,,\, t\,;\, \vec{r}_{d}\,,\,\tau)$
that at time $t$ radiation will interact  with the atom located at
point $\vec{r}_{c}$, and at time $\tau$, with the atom located at
point $\vec{r}_{d}$, contains the interference term and can be
represented in the form
\begin{eqnarray}
\label{22.1}
& &P_{ab}(\vec{r}_{c}\,,\, t\,;\, \vec{r}_{d}\,,\,\vec{\tau})=A\left|\frac{\exp[i(k_{a}r_{ac}-\omega_{a}t+\delta_{a})]}{r_{ac}}\right.\nonumber\\
&
&\left.\times\frac{\exp[i(k_{b}r_{bd}-\omega_{b}\tau+\delta_{b})]}{r_{bd}}+
\frac{\exp[i(k_{a}r_{ad}-\omega_{a}\tau+\delta_{a})]}{r_{ad}}\right.\nonumber\\
&
&\left.\times\frac{\exp[i(k_{b}r_{bc}-\omega_{b}t+\delta_{b})]}{r_{bc}}\right|^{2}\,,
\end{eqnarray}
where $A$ is the constant insignificant for the case in question.

%%%%%%%%%%%%%%%%%5
The expression of the type
$r^{-1}_{ac}\,\exp[i(k_{a}r_{ac}-\omega_{a}t+\delta_{a})]$  is the
wave function of a photon with the wave number
$k_{a}=\omega_{a}/c$, which is emitted at point $\vec{r}_{c}$,
where $r_{ac}=|\vec{r}_{a}-\vec{r}_{c}|$. For simplicity, it is
assumed that the atoms emit monochromatic radiation of the same
polarization. According to (\ref{22.1}), the probability
$P_{ab}(\vec{r}_{c}\,,\, t\,;\, \vec{r}_{d}\,,\,\tau)$  is
independent of random phases $\delta_{a}$ and $\delta_{b}$. If the
distance between the atoms of either pair is assumed to be much
shorter than the distance $R$ between the pairs, then (\ref{22.1})
can be written in the form
\begin{eqnarray}
\label{22.2}
P_{ab}(\vec{r}_{c}\,,\, t\,;\, \vec{r}_{d}\,,\,\vec{\tau})\approx \frac{2A}{R^{4}}\left\{1+\cos\left[k_{a}(r_{ac}-r_{ad})\right.\right.\nonumber\\
\left.\left.+k_{b}(r_{bd}-r_{bc})-(\omega_{a}-\omega_{b})(t-\tau)\right]\right\}\,,
\end{eqnarray}
where $r_{ac}=|\vec{r}_{a}-\vec{r}_{c}|$ etc. But for the particle
identity,  the first term in the braces in (\ref{22.2}) would
describe the probability of joint registration of the two photons;
the second term describes the change of this probability due to
the identity. As one may see, taking into account the particle
identity, leads to the fact that the probability
$P_{ab}(\vec{r}_{c}, t; \vec{r}_{d},\vec{\tau})$ is the
oscillating function of the coordinates that undergoes time
beatings at the difference of the frequencies of the emitted
photons.
%%%%%%%%%%%%%%%%%%%%%%
Since real sources and detectors contain many pairs of atoms,
(\ref{22.2}) should be summed over these pairs. In this case, the
interference term, which is of interest to us, does not vanish
unless the cosine appearing in (\ref{22.2})  undergoes
oscillations with the change in the positions of atoms  within the
volumes of the sources $S$ and detectors $D$. Hence, the following
inequality should hold for any pair of atoms within the limits of
$S$ and $D$
\begin{equation}
\label{22.3} k_{a}(r_{ac}-r_{ad})+k_{b}(r_{bd}-r_{bc})\ll 1\,.
\end{equation}

%%%%%%%%%%%%%%%%%
The condition (\ref{22.3}) under which the interference of
independent beams does not disappear, may as well be written as
follows
\begin{equation}
\label{22.4} \vartheta_{S}\vartheta_{D}\ll \lambda/R\,,
\end{equation}
\begin{equation}
\label{22.5} l_{S}l_{D}\ll\lambda R\,,
\end{equation}
where $\vartheta_{S(D)}=l_{S(D)}/R$ is the angle at which the
source (detector) with the lateral dimension $l_{s}(l_{d})$ is
visible from the detector (source) located at a distance $R$ from
$S(D)$; $\lambda$ is the radiation (emission) wave length (it is
supposed that $\Delta\lambda/\lambda\ll 1$).

%%%%%%%%%%%%%55

If  inequalities (\ref{22.3})--(\ref{22.5}) are fulfilled, and
$t=\tau$ (simultaneous registration of coincidences), then the
second term in (\ref{22.2}) equals the first term. Consequently,
the coincidence count  probability for identical particles is
different from the result obtained by classical count by not more
than a factor of two. From this also follows  that the possibility
to observe the interference is determined by the possibility to
register coincidences in the classical situation. If the intensity
of the sources is such that random coincidences of particles
without reference to the identity occur in the given experiment,
the interference phenomena are observed.

%%%%%%%%%%%%%%%%%%%%
Since the number of coincidences obtained during the time $T$
fluctuates, the intensity of the sources and the time $T$ should
be such that the average number of coincidences $\langle N\rangle$
during the stated time would be greater than the magnitude of
fluctuations in the number of coincidences ${\delta
N=\sqrt{\langle N^{2}\rangle-\langle
N\rangle^{2}}\sim\sqrt{\langle
N\rangle}\sim\sqrt{n_{1}n_{2}\tau_{c}T}}$ (${n_{1}(n_{2})}$ is the
number of particles  registered by counter 1(2) per unit time;
$\tau_{c}$ is the resolution time of the coincidence
circuit).\footnote{To be more specific, consider the case of small
counting rate of the coincidence circuit. Otherwise, we may talk
of the correlation function rather than of the number of
coincidences.} Hence, the inequality
\[
\frac{\langle N\rangle}{\delta
N}\simeq\sqrt{n_{1}n_{2}\tau_{c}T}>1
\]
should hold. If $\rho$ is the surface intensity of the source and
$\eta$ is the efficiency of the counters, then $n_{1,2}\simeq\rho
l_{S}^{2}(l_{D}^{2}/R^{2})\eta$. When the condition (\ref{22.5})
is fulfilled, we obtain $n_{1,2}\leq\rho \lambda^{2}\eta$. Thus,
finally we have
\begin{equation}
\label{22.6} \langle N\rangle/\delta
N\simeq\eta\rho\lambda^{2}\sqrt{\tau_{c}T}> 1\,,
\end{equation}
which coincides with the expression in \cite{128} for the case
when the length of the train of the  incident waves is comparable
with the time $\tau_{c}$.
%%%%%%%%%%%%%%%%%

Expressions (\ref{22.1}), (\ref{22.2}) are derived under the
assumption that the atoms of the source are fixed and undisturbed.
 However, generally speaking, in real
conditions it is not the case. Thermal motion and collisions of
atoms in a source  leads to the frequency modulation of the
emitted photons. In this general case, the photon wave function
may be represented as follows:
\begin{equation}
\label{22.7}
\Psi_{a}(t)\sim\exp\left\{-i[\omega_{a}t+\Omega_{a}(t)]\right\}\,,
\end{equation}
where $\Omega_{a}(t)$ is the phase change of the photon due to
thermal motion and collisions of the emitting atom $a$.

%%%

If the dimensions of the sources and detectors  satisfy
inequalities (\ref{22.3})--(\ref{22.5}), then using the wave
functions of the type (\ref{22.7}), one can write the following
expression for the probability {$P_{ab}(\vec{r}_{c}\,,\, t\,;\,
\vec{r}_{d}\,,\,\tau)$} averaged over the states of atoms $a$ and
$b$ in the source:
\begin{eqnarray}
\label{22.8}
& &P_{ab}(\vec{r}_{c}\,,\, t\,;\, \vec{r}_{d}\,,\, \tau)=\mbox{const}\,\langle |\exp\left\{-i[\omega_{a}t+\Omega_{a}(t)]\right\}\nonumber\\
& &\times\exp\left\{-i[\omega_{b}\tau+\Omega_{b}(\tau)]\right\}+\exp\left\{-i[\omega_{a}\tau+\Omega_{a}(\tau)]\right\}\nonumber\\
&
&\times\exp\left\{-i[\omega_{b}t+\Omega_{b}(t)]\right\}|^{2}\rangle\,,
\end{eqnarray}
here angle brackets mean averaging.
%%%%%%%%%%%%%%%%%%%%%%55555

Equality (\ref{22.8}) includes the quantity
\[
G_{ab}(t,
\tau)=\langle\exp\left\{-i[\Omega_{a}(t)-\Omega_{a}(\tau)+\Omega_{b}(\tau)-\Omega_{b}(t)]\right\}\rangle
\]
characterizing the kinetic processes in the source. As is seen
from (\ref{22.8}), when studying photon correlations, the
probability of registration of delayed coincidences by two
counters only depends on mutual correlations between atoms $a$ and
$b$. If we measured triple  or higher fold coincidence events, the
corresponding probabilities would only depend on mutual
correlations between three and more atoms. This is slightly
different from the situation arising in studying correlations in
the radiation scattered by a certain target for the case when the
energy spectrum of the scattered radiation being measured  depends
also on time correlations of the state of one atom. For
simplicity, let us further assume  that the correlations between
atoms $a$ and $b$ may be neglected (for example, investigating the
radiation of a gaseous source). Then
\begin{eqnarray*}
G_{ab}(t,\tau)&=&G_{a}(t,\tau)G_{b}^{*}(t,\tau)\,,\\
G_{a(b)}(t,\tau)&=&\langle\exp\left\{-i[\Omega_{a(b)}(t)-\Omega_{a,(b)}(\tau)]\right\}\rangle\,.
\end{eqnarray*}
For homogeneous systems, $G_{a}(t,\tau)=G_{b}(t,\tau)=G(t,\tau)$.

%%%%%%%%%%%%%%%%%%%%%%
In most cases of practical interest $G(t,\tau)$ may be represented
as follows:
\begin{equation}
\label{22.9} G(t-\tau)=\exp\left\{-w^{2}(t-\tau)/4\right\}\,,
\end{equation}
where
$w^{2}(t-\tau)/2=\langle[\Omega_{a}(t)-\Omega_{a}(\tau)]^{2}\rangle$.

Using (\ref{22.9}), from (\ref{22.8}) we obtain the equality
\begin{eqnarray}
\label{22.10}
& &P_{ab}(t-\tau)= \mbox{const}\,\left\{1+\exp\left\{-\texttt{Re}[w^{2}(t-\tau)]/2\right\}\right.\nonumber\\
& &\left.\times\cos[(\omega_{a}-\omega_{b})(t-\tau)]\right\}\,.
\end{eqnarray}
%%%%%%%%%%%%%%%%%%%%%%%%

Sum (\ref{22.10}) over all the pairs of atoms in the
source\footnote{Such summation can be made if the photon density
in the source is such that the simulated emission of atoms can be
neglected.} and the detector, assuming that the source emits with
equal probability the photons of only two frequencies $\omega_{1}$
and $\omega_{2}$. As a result, we obtain the  below expression for
probability $P(t-\tau)$ that one photon will be registered at
moment $t$, and the other one, at moment $\tau$:
\begin{eqnarray}
\label{22.11}
& &P(t-\tau)=\mbox{const}\,\left\{1+\exp\left[-\frac{\texttt{Re}w^{2}(t-\tau)}{2}\right]\right.\nonumber\\
&
&\left.\times\cos^{2}\left[\frac{\omega_{1}-\omega_{2}}{2}(t-\tau)\right]\right\}\,.
\end{eqnarray}

%%%%%%%%%%%%%%%%%%%%5
Thus, the curve of delayed coincidences undergoes modulated
beatings, depending on the the delay time $\theta=|t-\tau|$ at the
frequency equal to the difference of frequencies $\omega_{1}$ and
$\omega_{2}$. The frequency of beatings can, in principle, be
controlled by means of various external influences, e.g., by
placing the source to the external magnetic field.

Now determine the total number of coincidences $N$ in the given
experiment, i.e., determine the area under the curve of delayed
coincidences if the maximum delay time used in the experiment is
$\theta_{m}$.
For this we integrate (\ref{22.11}) over $\theta=t-\tau$ within
the interval $[0,\theta_{m}]$, which gives the expression of the
form
\begin{eqnarray}
\label{22.12}
N=\mbox{const}\,\left\{\theta_{m}+\int\limits_{0}^{\theta_{m}}\exp\left[-\frac{\texttt{Re}w^{2}(\theta)}{2}\right]
\cos^{2}\left[\frac{\omega_{1}-\omega_{2}}{2}\theta\right]d\theta\right\}\,.\nonumber\\
\end{eqnarray}

The first term proportional to $\theta_{m}$ would correspond to
the number of coincidences if the photons could be distinguished;
the second one gives the addition to $N$, appearing due to the
particle identity. Hence, taking into account the identity leads
to the fact that the area under the curve of delayed coincidences
is not proportional to $\theta_{m}$, as it would be for
distinguishable particles.

%%%%%%%%%%%%%%%%%%%%%%%5

Consider  (\ref{22.12}) for two limiting cases: in the first one
the perfect gas  with temperature $Q$ acts as a source, in the
second one the source is such that  the major role in modulating
the radiation frequency is played by collisions, whose influence
will be taken into account in the collision approximation. In the
former case,
$\texttt{Re}[w^{2}(\theta)]=k^{2}\bar{v}^{2}\theta^{2}$, in the
latter case
$\texttt{Re}[w^{2}(\theta)]=4\rho\sqrt{\bar{v}^{2}}\sigma\theta$;
here $\bar{v}^{2}$ is the mean--square thermal velocity of atoms,
 $\rho$ is the atomic density in the gas, and  $\sigma$ is the
collision cross section.

%%%%%%%%%%%%%%%%%%%%%%%%%%5

Substitution of the stated expression  for $w^{2}(\theta)$ into
(\ref{22.12}) gives the following expressions for the two cases:
\begin{eqnarray}
\label{22.13}
 N& &=\mbox{const}\left\{\theta_{m}+\int\limits_{0}^{\theta_{m}}\exp-
 \left(-\frac{k^{2}\bar{v}^{2}}{2}\theta^{2}\right)\right.\nonumber\\
&
&\left.\times\cos^{2}\left[\frac{\omega_{1}-\omega_{2}}{2}\theta\right]d\theta\right\}\,,
\end{eqnarray}
\begin{eqnarray}
\label{22.14}
& &N^{\prime}=\mbox{const}\left\{\theta_{m}+\frac{1}{2\Gamma}+\frac{\Gamma}{2[(\omega_{1}-\omega_{2})^{2}+\Gamma^{2}]}\right.\nonumber\\
&
&\left.-\frac{\exp(-\Gamma\theta_{m})}{2\Gamma}\left(1+\frac{a\Gamma\cos[(\omega_{1}-\omega_{2})\theta_{m}+\varphi]}
{(\omega_{1}-\omega_{2})^{2}+\Gamma^{2}}\right)\right\}\,.
\end{eqnarray}
where $ae^{i\varphi}=\Gamma+i(\omega_{1}-\omega_{2})$;
$\Gamma=2\rho\sqrt{\bar{v}^{2}}\sigma$ is the impact width of the
level\footnote{When the impact and the Doppler width of the level
can be neglected,  (\ref{22.14}), where $\Gamma$ is the natural
width of the level, holds true for the area of the delayed
coincidence  curve.}.

%%%%%%%%%%%%%55
If $\theta_{m}(\bar{v}^{2}k^{2})^{1/2}\gg 1$ and
$\Gamma\theta_{m}\gg 1$, (\ref{22.13}) and (\ref{22.14}) may be
recast as follows
\begin{eqnarray}
\label{22.15}
N\simeq \mbox{const}\,\theta_{m}\left\{1+\sqrt{\frac{\pi}{\bar{v}^{2}\theta_{m}^{2}k^{2}}}\right.\nonumber\\
\left.+\sqrt{\frac{\pi}{\bar{v}^{2}\theta_{m}^{2}k^{2}}}
\exp\left[-\frac{(\omega_{1}-\omega_{2})^{2}}{{\bar{v}^{2}}k^{2}}\right]\right\}\,,
\end{eqnarray}
\begin{equation}
\label{22.16} N\approx
\mbox{const}\,\theta_{m}\left\{1+-\frac{1}{2\Gamma\theta_{m}}+\frac{1}{2\Gamma\theta_{m}}\frac{\Gamma^{2}}{(\omega_{1}-\omega_{2})^{2}
+\Gamma^{2}}\right\}\,.
\end{equation}
%%%%%%%%%%%%%%%%5555

Thus, the area under the delayed-coincidence curve depends on the
difference $\omega_{1}-\omega_{2}$  and on the mechanism of the
radiation frequency modulation.
When $\omega_{1}=\omega_{2}$, expressions (\ref{22.15}) and
(\ref{22.16}) differ from the result obtained for classical
particles by the magnitudes
$2\sqrt{\pi/\bar{v}^{2}\theta_{m}^{2}k^{2}}$ and
$1/\Gamma\theta_{m}$, respectively.
If $(\omega_{1}-\omega_{2})^{2}/\bar{v}^{2}k^{2}\gg 1$ or
$|\omega_{1}-\omega_{2}|\gg\Gamma$, the number of coincidences
exceeds the classical result by
$\sqrt{\pi/\bar{v}^{2}\theta_{m}^{2}k^{2}}$ and
$1/2\Gamma\theta_{m}$, respectively.
This means that when the stated inequalities are fulfilled, the
photons of frequency $\omega_{1}$ and photons of frequency
$\omega_{2}$ may be considered non-identical.
At the same time, the photons of the same frequency (either
$\omega_{1}$ or $\omega_{2}$), of course remain identical to one
another, which is manifested in the fact that the magnitudes of
$N$ and $N^{\prime}$ differ from those predicted for classical
particles.

%%%%%%%%%%%%%%
Let us note in conclusion that it would be tempting to carry out
such experiments not only for optical photons but also for, e.g.,
M\"{o}ssbauer $\gamma$-quanta. However, due to the short
wavelength of $\gamma$-quanta,  it is practically impossible
nowadays to realize the conditions (\ref{22.5}) and (\ref{22.6})
using conventional sources.

%%%%%%%%%%%%%%%%
Indeed, from (\ref{22.5}) we obtain that if for light at
$\lambda\simeq 10^{-5}$\,cm and $l_{S}\sim l_{D}\sim 10^{-1}$\,cm,
it should be $R\geq 10^{3}$\,cm, then for $\gamma$-quanta
($\lambda\sim 10^{-8}$ cm) at the same dimensions of the source
and the detector, it should be $R\geq 10^{6}$ cm.

A more detailed treatment shows that this problem might be
avoided, using some artificial procedures. But even stricter
requirements are imposed by inequality (\ref{22.6}). From it
follows that with other conditions being equal, the observation
time $T_{\gamma}$ in the X-ray spectrum should be by several
orders of magnitude greater than the corresponding time $T_{c}$ in
the optical spectrum
($T_{\gamma}\approx(\lambda_{c}/\lambda_{\gamma})^{4}T_{c}$, i.e.,
$T_{\gamma}\approx 10^{12} T_{c}$). The aforesaid also refers  to
the case when a scattering target is placed between the sources
and detectors, as it may be treated just as a source of scattered
waves. Serious problems also exist for other types of radiation
(electrons and neutrons).

%%%%%%%%%%%%%%%%%%%%%%%%%%%%%%%%%%%  Section 23 %%%%%%%%%%%%%%%%%%%%%%

\section[Interference of $\gamma$-quanta Generated by the Beams of Relativistic Particles]
{Interference of $\gamma$-quanta Generated by the Beams of
Relativistic Particles} \label{sec:7.23}

Quite a different situation arises when radiation produced by
relativistic particles is used as a source\cite{127}. There are
two possible kinds of experiment: (a) radiation is produced when a
relativistic particle passes through a crystal, (b) synchrotron
radiation is diffracted in the Mossbauer crystal.

Recall (see(\ref{ch:4}) that radiation is a crystal is formed
through two mechanisms: parametric one, and radiative transitions
between the levels (regions) of transverse motion. The emerging
$\gamma$-quanta move within a narrow angle along the direction of
the particle motion and along the direction determined by the
reciprocal lattice vector $2\pi\vec{\tau}$. The number of
resonance $\gamma$-quanta, produced by one electron in a crystal,
due to the parametric effect,  for the forward direction equals
($\hbar=c=1$)
\begin{equation}
\label{23.1} N_{\gamma}^{(\tau)}=e^{2}\frac{\Gamma}{\omega_{p}},
\end{equation}
for radiation in the direction of  diffraction
\begin{equation}
\label{23.2}
N_{\gamma}^{(\tau)}=e^{2}\frac{g_{00}^{\prime\prime}(\omega_{p})}{\sqrt{|\Delta|}}\sqrt{g_{00}^{\prime\prime}(\omega_{p})}\frac{\Gamma}{|2\pi\tau|};\,\,
\frac{m^{2}}{E^{2}}<|g_{00}(\omega_{p})|,
\end{equation}
where $\Delta=g_{00}g_{11}-g_{01}g_{10}$.

 Angular divergence of the
quanta produced (see (\ref{sec:4.15})) for the forward direction
has the magnitude of the order of
\begin{equation}
\label{23.3} \theta^{(0)}\leq
\sqrt{\left(\frac{m}{E}\right)^{2}+|g_{00}|}.
\end{equation}
Angular divergence of the photons emitted along the direction of
diffraction is much less
\begin{equation}
\label{23.4} \theta^{(\tau)}\leq \sqrt{|\Delta|}.
\end{equation}
In the case in question the linear dimensions of the source are
defined by the width $d$ of the electron beam incident on the
crystal ($l_{S}=d$). Linear dimensions of the input window of the
detector, where the photons produced by the particle get equal
\begin{equation}
\label{23.5} l_{D}=d+R\theta^{(0,\tau)}.
\end{equation}
As a result, the condition (\ref{22.5}) may be written in the form
\begin{equation}
\label{23.6} d^{2}+dR\theta^{(0,\tau)}\leq \lambda R.
\end{equation}
In observation of the interference phenomena in radiation
propagating along the direction of particle motion, the condition
(\ref{23.6}) is difficult to fulfil. For example, for a crystal of
$^{57}Fe$  and the electron energy $E\sim1$ GeV, the width of the
electron beam should be  less than $10^{-4}$ cm. At the same time,
due to the fact that $\theta^{(\tau)}\ll\theta^{(0)}$, when
observing interference in the direction of diffraction under the
same conditions $d\sim 0.1$ cm.

Further we shall consider the possibility of observation of
interference in the direction of diffraction, assuming that the
condition (\ref{23.6}) is fulfilled. Let the resolution time of
the coincidence circuit $\tau_{c}$ is less than the length of the
train of the obtained $\gamma$-quantum, i.e., of the order of
magnitude $\tau_{c}\leq 1/\Gamma$. Taking into account  that the
number of particles passing through a crystal in one second is
$I/e$ ($I$ is the current strength), we have for the number of
quanta produced in the crystal in one second:
\begin{equation}
\label{23.7} n=\frac{I}{e}N^{(\tau)}_{\gamma}.
\end{equation}
Due to a small angular divergence of the radiation produced in the
crystal, all the photons get into the detector. Therefore when the
condition (\ref{23.6}) is fulfilled using the parametric effect,
we  find the following estimate for the observation time of the
interference pattern $T$:
\begin{equation}
\label{23.8}
T>\frac{|2\pi\tau|^{2}|\Delta|}{\eta^{2}\Gamma(g_{00}^{\prime\prime})^{3}I^{2}e^{2}}.
\end{equation}
When $I=10\mu A=10^{-5}\,A$, $\eta\simeq 1$, $E=1$ GeV,  from
(\ref{23.8}) follows the estimate $T\geq 10^{4}$ s for $^{57}Fe$:
$T\geq 10^{2}$ s for $^{137}W$.

Now consider the possibility of observation of the independently
generated photons, using diffraction of synchrotron radiation in a
crystal containing resonance nuclei. Applying the conditions
(\ref{22.5}), (\ref{22.6}) and the expression for the intensity of
the synchrotron radiation (see, for instance, \cite{31}), one may
obtain the expression for the observation time
\begin{equation}
\label{23.9}
T\geq\frac{3}{16\pi^{2}e^{4}}\frac{S^{2}\omega_{p}^{4}}{\Gamma
n^{2}_{e}},
\end{equation}
where $S$ is the area of the electron beam cross section; $n_{e}$
is the number of electrons in the accelerator. When $S=10^{-3}$cm,
$r=10^{3}$ cm, $n_{e}=10^{12}$ (or $I=0.1\,A$), the estimate is
$T\geq 10^{3}$ s.

The stated time may be appreciably reduced ($T\sim 1$ s), using
radiation of the powerful storage rings like those discussed in
\cite{130}, which are to be constructed. Such times are also
achieved with the help of the parametric effect at  the electron
current of the order of $10^{-4}$\, A. The phenomena being
analyzed may  be applied for direct phase analysis by introducing
M\"{o}ssbauer nuclei into the structure in question. If this
method is hampered, the M\"{o}ssbauer crystal may be used as the
source of radiation, which is subsequently diffracted by the
examined substance.

%%%%%%%%%%%%%%%%%%%%%%%%%%%%%%%%%%%  Chapter 8 %%%%%%%%%%%%%%%%%%%%%%

\chapter[Theory of Measurement of Nuclear Reaction Times Using Shadow Effect ...]{Theory of Measurement of Nuclear Reaction Times Using Shadow Effect. Yield of Reactions Induced by High-energy Particles in Crystals }
\label{ch:8}

%%%%%%%%%%%%%%%%%%%%%%%%%%%%%%%%%%%  Section 24 %%%%%%%%%%%%%%%%%%%%%%

\section[Quantum Theory of Reactions Induced by Channeled Particles]{Quantum Theory of Reactions Induced by Channeled Particles}
\label{sec:8.24}

Particle motion in a single crystal is accompanied by numerous
inelastic processes and reactions. The investigation of these
processes and reactions provides important information about
crystal structure and the properties of nuclei. In particular, the
shadow effect is widely used to explore the  nuclear reaction
times $\tau$ in the range  $\tau\leq 10^{-16}$ s \cite{131}. When
interpreting the results obtained, it is supposed that what is
measured in the experiments under discussion is the nucleus
lifetime.

Although, analyzing the fluctuations of effective cross sections
of reactions, Lyuboshitz and Podgoretky \cite{132,133,134} showed
that in strong overlap of the levels the law of the compound
nucleus decay becomes appreciably nonexponential. It was also
stated that in this case the process of inelastic scattering can
be divided into instantaneous diffraction scattering and
fluctuation scattering, associated with the decay of the compound
systems.

According to \cite{132,133,134} the characteristic time duration
of the fluctuating part of the reaction is determined by the mean
interlevel distance rather than by the level width of a compound
nucleus. (The whole analysis in \cite{132,133,134} was carried out
by using packets.)

%%%%%%%%%%%

%%%%%%%%%%%%%%%%

Within the framework of a stationary quantum mechanical theory of
scattering, we have demonstrated that, by applying monochromatic
states, it is also possible to define the nuclear decay law
\cite{135,136}. Moreover, angular distribution of secondary
particles, studied in the experiments on shadow effect \cite{131}
turned out to be determined by the correlation function of the
reaction amplitudes, which in the case of strong overlap of the
levels coincides with the function introduced in
\cite{132,133,134}.

In this regard it is worthy of mention that the possibility of
using the formulae employed in the experiments on the shadow
effect for establishing the relationship between the angular
distribution of secondary particles  and the law of the compound
nucleus decay in the case under consideration requires additional
analysis. This circumstance is attributed to the fact that until
now the theory describing the method for measuring nuclear
reaction times has been practically completely based on using
classical models involving a number of uncertain parameters
(variables) (such as the chain cutoff radius; for more details,
see \cite{131}). Whereas an essential element of quantum
consideration of the effect given in \cite{137} is the assumption
that the motion of finite particles is classical.

Below is presented a quantum mechanical theory of scattering which
enables deriving formulae relating the angular distribution of
particles - the reaction products - to  the position  distribution
function of the compound nucleus without using under-substantiated
model approximations. It is also shown that in excitation of a
group of levels, the position distribution function of the
compound nucleus undergoes spatial beating with the period
determined by the interlevel distance. This enables using shadow
effect for investigation not only the level width but also the
interlevel distance. Applicability to the shadow effect of the
hypothesis of the rapidly established statistical equilibrium in
the transverse plane of the phase space of the particle leaving
the crystal, widely used within the framework of the classical
approach is validated \cite{131}. Using the formal theory of
reactions makes it possible to directly apply the obtained results
to electron-nuclear reactions induced by relativistic particles
(e.g. electrons and positrons) too, if by the particle mass we
mean its relativistic mass.

%%%%%
So, let a particle $a$ be incident on a crystal, causing a nuclear
reaction $A(a,b)$ in it. Consider the angular distribution of
particles $b$. According to the general theory of reactions (see,
for example, \cite{111}) the  cross section of this process may be
written in the form:

%%%%%%%%%%%%
\begin{eqnarray}
\label{24.1}
\frac{d\sigma_{ab}}{d\Omega_{b}}&=&\frac{m_{a}m_{b}}{(2\pi\hbar^{2})^{2}k_{a}}\nonumber\\
&\times&\sum_{B}\int
k_{b}dE_{b}\delta(E_{bB}-E_{aA})|\langle\varphi_{bB}^{(-)}|{\cal{F}}|\varphi_{aA}^{(+)}\rangle|^{2}\,,
\end{eqnarray}
where $m_{a(b)}$ is the mass of  the particle $a(b)$; $k_{a(b)}$
is the wave number of particle $a(b)$; $E_{aA}$  is the energy of
the initial state; $E_{bB}$ is the energy  of the final state;
${\cal{F}}$ is the scattering operator; $\varphi^{(+)}$ is the
wave function of the initial state taking account of  the
interaction of particles $a$ and $A$ with the crystal, having at
infinity the asymptotics which contains diverging waves;
$\varphi^{(-)}$ is the same for the final state with the
asymptotics containing converging waves.

The shadow effect is applied to investigation of the duration time
of nuclear reaction  $\tau\leq10^{-16}$ s \cite{131}. As in this
case the width of nuclear levels which are involved in the
reaction (and the energy of particles) is much greater than the
characteristic vibration frequencies of nuclei in the crystal, in
order to find the operator ${\cal{F}}$,  the impulse approximation
may be used. According to this approximation \cite{111}, it it is
assumed that  ${\cal{F}}$  coincides with the scattering operator
describing the reaction $A(a,b)B$ with free particles, i.e.,
particles that do not interact  with the crystal:
\begin{eqnarray}
\label{24.2}
& &\langle\vec{k}_{b},\vec{k}_{B}|{\cal{F}}|\vec{k}_{a},\vec{k}_{A}\rangle\nonumber\\
&
&=(2\pi)^{3}\delta(\vec{k}_{b}+\vec{k}_{B}-\vec{k}_{a}-\vec{k}_{A})
\langle\vec{k}_{b},\vec{k}_{B}|T|\vec{k}_{a},\vec{k}_{A}\rangle\,,
\end{eqnarray}

%%%%%%%%%%%555
where the amplitude
$\langle\vec{k}_{b},\vec{k}_{B}|T|\vec{k}_{a},\vec{k}_{A}\rangle=T_{ba}(\vec{\kappa}_{f},\vec{\kappa},\varepsilon)$
depends of the relative momenta of the initial $(\vec{\kappa})$
and final $(\vec{\kappa}_{1})$ states , and the energy
$\varepsilon=\hbar^{2}\kappa^{2}/2\mu$ of the relative motion in
the initial state ($\mu=m_{a}M_{A}/(m_{a}+M_{A})$ is the reduced
mass in the initial state). As a result, (\ref{24.1}) may be
written as follows
\begin{eqnarray}
\label{24.3}
\frac{d\sigma_{ab}}{d\Omega_{b}}&=&\frac{m_{a}m_{b}}{(2\pi\hbar^{2})^{2}(2\pi)^{12}k_{a}}
\int\ldots\int k_{b}dE_{b}d^{3}k_{B}d^{3}k_{b}^{\prime}d^{3}k_{a}^{\prime}\nonumber\\
&\times&
d^{3}k_{A}\delta(E_{bB}-E_{aA})|\langle\varphi_{\vec{k}_{b}}^{(-)}|\vec{k}_{b}^{\prime}\rangle
\delta(\vec{k}_{b}^{\prime}+\vec{k}_{B}-\vec{k}_{a}^{\prime}-\vec{k}_{A})\nonumber\\
&\times&
\langle\vec{k}_{b}^{\prime},\vec{k}_{B}|T|\vec{k}_{a}^{\prime},\vec{k}_{A}\rangle
\langle\vec{k}_{a}^{\prime}|\varphi^{(+)}_{\vec{k}_{a}}\rangle\langle\vec{k}_{A}|\Phi_{A}\rangle|^{2}\,,
\end{eqnarray}
where $|\varphi^{(+)}_{\vec{k}_{a}}\rangle$ is the wave function
of particle $a$ incident on the crystal;
$|\varphi_{\vec{k}_{b}}^{(-)}\rangle$ is the wave function of
outcoming particle $b$; $|\Phi_{A}\rangle$ is the wave function of
the nucleus.
%%%%%%%%%%%%%%%%%%%%%555555

Further we shall consider quite thin crystals so that  we could
neglect energy losses of particles $a$ and $b$ participating in
the reaction (according to \cite{23} for protons with the energies
$E=0.4$ MeV in $Si$, the thickness is $l\sim 5-7\times
10^{3}$\,{\AA}, for $\alpha$ particles with the energies $E=2$ MeV
in $Au$, the thickness is $l\sim 3\times 10^{3}$\,{\AA}). In this
case the wave functions describing the phenomenon of channeling of
particles incident on the crystal and leaving it may be found,
using the method presented in (\ref{ch:1}). For example, inside
the crystal the wave function may be given as
\begin{eqnarray}
\label{24.4}
\varphi^{(+)}_{\vec{k}_{a}}=\sum_{n}c_{n\vec{k}_{a}}\Psi_{n\vec{\kappa}_{a}}(\vec{\rho})\exp(ik_{azn}z),\nonumber\\
c_{n\vec{k}_{a}}=\frac{(2\pi)^{2}}{\Omega_{0}}\int_{S}\exp(i\vec{k}_{a\perp}\vec\rho^{\prime})
\Psi_{n\vec{\kappa}_{a}}^{*}(\vec{\rho}^{\prime})d^{2}\rho^{\prime},
\end{eqnarray}
where the $z$-axis of the coordinate system is directed parallel
to the family of axes (planes) along which the particle is
channeled; $\Psi_{n\vec{\kappa}_{a}}(\vec{\rho})$ is the  Bloch
function; the coordinate $\vec{\rho}=(x,\,y)$; $n$ is the index of
the energy zone of the particle's transverse motion;
$\vec{k}_{a\perp}$ is the component of vector $\vec{k}_{a}$
perpendicular  to the $z$-axis; $\vec{\kappa}_{a}$ is the reduced
wave vector, corresponding to $\vec{k}_{a\perp}$;
\[
k_{azn}=\sqrt{k_{a}^{2}-\frac{2m_{a}}{\hbar^{2}}\varepsilon_{n}(\vec{\kappa}_{a})}\,,
\]
$\varepsilon_{n}(\vec{\kappa}_{a})$ is the energy of the
transverse motion of the particle in zone  $n$; $\Omega_{0}$ is
the volume of the two--dimensional unit cell of the crystal in
plane $\rho$. Integration with respect to $d^{2}\rho^{\prime}$ is
performed over the two--dimensional unit cell in plane $\rho$.

%%%%%%%%%%%%%%%%%%%%%%%555

The state $\varphi^{(-)}$ is found, using the relation
$\varphi^{(-)}_{\vec{k}}(\vec{r})=\varphi^{(+)*}_{-\vec{k}}(\vec{r})$.
%%%%%%%%%5
From the form of the wave functions
$\varphi^{(+)}_{\vec{k}_{a}}(\varphi^{(-)}_{\vec{k}_{b}})$ follows
that the matrix elements
$\langle\vec{k}_{a}^{\prime}|\varphi^{(+)}_{\vec{k}_{a}}\rangle$
and
$\langle\varphi_{\vec{k}_{b}}^{(-)}|\vec{k}_{b}^{\prime}\rangle$
make the intermediate momenta $\vec{k}_{a}^{\prime}$ and
$\vec{k}_{b}^{\prime}$  close to $\vec{k}_{a}$ and $\vec{k}_{b}$
with the accuracy of the order of $1/l$ for
$k_{a\parallel}^{\prime}$ and $k_{b\parallel}^{\prime}$ ($l$ is
the crystal thickness, the symbol $\parallel$ denotes the
components of the momentum parallel to the axis (plane) along
which the particle is channeled), which is much smaller than
$k_{a(b)}\vartheta_{L}$, where $\vartheta_{L}$ is the Lindhard
angle. The  uncertainty of momentum $k_{A}$ is of the order of
$1/r_{0}$, where $r_{0}$ is the  vibration amplitude of the
nucleus.

Pay attention to the fact that the reaction amplitude $T_{ba}$ can
be presented in the form
\begin{equation}
\label{24.5}
T_{ba}(\vec{\kappa}_{f},\vec{\kappa},\varepsilon)=T_{D}(\vec{\kappa}_{f},\vec{\kappa},\varepsilon)+
\sum_{\lambda}\frac{\gamma_{\lambda
b}(\vec{\kappa}_{f})\gamma_{\lambda
a}(\vec{\kappa})}{\varepsilon(\kappa)-E_{\lambda}
+\frac{1}{2}i\Gamma_{\lambda}}\,,
\end{equation}
where $T_{D}$ is the amplitude of the direct reaction;
$\gamma_{\lambda b}(\gamma_{\lambda a})$ in the general case, are
complex quantities, which, at weak overlap of the levels, coincide
with partial widths for the transitions $\lambda\rightarrow b$,
$\lambda\rightarrow a$; $E_{\lambda}$ is the energy of the
resonance $\lambda$; $\Gamma_{\lambda}$ is its width. The
dependence of  $T_{D}$ and $\gamma_{\lambda a(b)}$ on the momenta
is determined by the spatial domain  of the order of the nuclear
dimension in size. For this reason, the momentum uncertainty of
the incident and outcoming particles, which is caused by their
interaction with the crystal ($\leq 10^{10}$\,cm$^{-1}$), can be
ignored in $T_{D}$ and $\gamma_{\lambda a(b)}$. As a consequence,
$\vec{\kappa}_{f}$ and $\vec{\kappa}$ in them may be equated to
the vacuum values of the momentum of particles
($b$~---~$\vec{k}_{b}$) and ($a$~---~$\vec{k}_{a}$), respectively.

%%%%%%%%%%%%%%%%%%%%%%5
Note also that in integration with respect to the component of the
relative momentum, which is parallel to the incident direction of
the primary particle, the contribution from the resonant
denominator in (\ref{24.5}) will be determined by the  residues at
points
\begin{eqnarray*}
\kappa_{\parallel}=\left(\frac{2\mu}{\hbar^{2}}E_{\lambda}-\kappa^{2}_{\perp}-i\frac{\mu\Gamma_{\lambda}}{\hbar^{2}}\right)^{1/2}\\
\simeq\left(\frac{2\mu}{\hbar^{2}}E_{\lambda}\right)^{1/2}
\left(1-\frac{\hbar^{2}\kappa^{2}_{\perp}}{4\mu
E_{\lambda}}-i\frac{\Gamma_{\lambda}}{4E_{\lambda}}\right)\,.
\end{eqnarray*}
As the transverse relative momentum
\[
\vec{\kappa}_{\perp}=\frac{M_{A}}{m_{a}+M_{A}}\vec{k}_{a\perp}^{\prime}-\frac{m_{a}}{m_{a}+M_{A}}\vec{k}_{a\perp}
\]
is limited by the matrix elements, the contribution of
$\kappa_{\perp}$ to the real part of the pole appears to be small
and should be ignored.

Thus, one may consider that in integration with respect to the
intermediate momentum, the amplitude $T_{ba}$ depends only on the
component of the relative momentum  $\kappa_{\parallel}$ that is
parallel to the momentum of the incident particle:
\[
\vec{\kappa}_{\parallel}=\frac{M_{A}}{m_{a}+M_{A}}k_{a\parallel}^{\prime}-\frac{m_{a}}{m_{a}+M_{A}}k_{a\parallel}\,.
\]

%%%%%%%%%%%%%%%%%%%%%%%%%%%%%%55

As a result, we obtain the following expression for the
differential cross section:
\begin{equation}
\label{24.6} \frac{d\sigma_{ab}}{d\Omega_{b}}=D\int
d^{3}r_{b}|\varphi_{\vec{k}_{b}}^{(-)}(\vec{r}_{b})|^{2}Q(\vec{r}_{b})\,,
\end{equation}
where the constant
\begin{eqnarray*}
& &D=\frac{2\pi m_{a}m_{b}}{(2\pi\hbar^{2})^{2}k_{a}}\,,\\
& &Q(\vec{r}_{b})=\left|\int dr_{A\parallel}\left\{\int d\kappa_{\parallel}\frac{m_{a}+M_{A}}{m_{a}}T_{ba}(\kappa_{\parallel})\right.\right.\\
& &\left.\times\exp\left[-i\frac{m_{a}+M_{A}}{m_{a}}\kappa_{\parallel}(r_{b\,\parallel}-r_{A\,\parallel})\right]\right\}\\
& &\times\varphi_{\vec{k}_{a}}^{(+)}\left(\vec{r}_{b\perp};\frac{m_{a}+M_{A}}{m_{a}}r_{b\, \parallel}-\frac{M_{A}}{m_{a}}r_{A\,\parallel}\right)\\
& &\left.\times\Phi_{A}(\vec{r}_{b\perp}-\vec{R}_{i\perp};\,
r_{A\, \parallel}-z_{i})\right|^{2}\,.
\end{eqnarray*}

%%%%%%%%%%%%%%%%%%%%%%%
Expression (\ref{24.6}) should  be averaged over the coordinates
of the equilibrium positions of the excited nuclei $\vec{R}_{i}$,
which are assumed to be uniformly distributed over the crystal
volume. Upon such averaging equality (\ref{24.6}) may be written
as follows
\begin{equation}
\label{24.7} \frac{d\sigma_{ab}}{d\Omega_{b}}=D\int
d^{3}r_{b}n_{\vec{k}_{b}}(\vec{r}_{b})\tilde{Q}(\vec{r}_{b})\,,
\end{equation}
\begin{eqnarray}
\label{24.8}
\tilde{Q}(\vec{r}_{b})=\sum_{f}n_{\vec{k}_{a}f}(\vec{r}_{b\perp})\Phi_{A}^{2}(\vec{r}_{b\perp})\nonumber\\
\times\left|\int dr_{A\, \parallel}\left\{\int d\kappa_{\parallel}\frac{m_{a}+M_{A}}{m_{a}}T_{ba}(\kappa_{\parallel})\right.\right.\nonumber\\
\left.\left.\times\exp\left[-i\frac{m_{a}+M_{A}}{m_{a}}\kappa_{\parallel}(r_{b\,\parallel}-
r_{A\,\parallel})\right]\right\}\right.\nonumber\\
\left.\times\exp\left(-ik_{\parallel\,af}\frac{M_{A}}{m_{a}}r_{A\,\parallel}\right)\Phi_{A}(r_{A\,
\parallel})\right|^{2}\,,
\end{eqnarray}
where
\begin{equation}
\label{24.9}
n_{\vec{k}_{b}}(\vec{r}_{b})=\sum_{n}|c_{n_{\vec{k}_{b}}}|^{2}|\psi_{n\vec{\kappa_{b}}}(\vec{r}_{b})|^{2}\,,
\end{equation}
\begin{equation}
\label{24.10}
n_{\vec{k}_{a}f}(\vec{r}_{b\perp})=|c_{f_{\vec{k}_{a}}}|^{2}|\psi_{f\vec{\kappa_{a}}}(\vec{r}_{b\perp})|^{2}\,,
\end{equation}
%%%%%%%%%%%%%%%%%%5555
$n_{\vec{k}_{b}}(\vec{r}_{b})$ only depends on the components of
vectors $\vec{r}_{b}$ and $\vec{k}_{b}$ perpendicular  to the
direction of the axes (planes) along which particle $b$ is
channeled and has the meaning of density distribution of particles
$b$ in the transverse plane relative to the stated axes (planes);
$n_{\vec{k}_{a}f}(\vec{r}_{b\perp})$ only depends on the
projection of vector $\vec{r}_{b\perp}$ which  is perpendicular to
the direction of the axes along which the incident particle $a$
moves and has the meaning of density distribution of particles $a$
occupying the transverse energy level $f$ in the plane
perpendicular to the channeling axes (planes).

Discuss the derived expression in more detail. The quantity
$d\sigma_{ab}/d\Omega_{b}$ determines up to a constant the angular
distribution of the flow $J$ of  particles $b$  produced through
the reaction. On the other hand, the flow  $J$ can be found by
solving the Schr\"odinger equation of the form
\begin{equation}
\label{24.11}
(\Delta_{r}+k^{2}-u(\vec{r}))\psi(\vec{r})=q(\vec{r})\,,
\end{equation}
where $q(\vec{r})$ is the current distribution amplitude of the
source of particles;
\[
u(\vec{r})=\frac{2m}{\hbar^{2}}V(\vec{r})\,,
\]

$V(\vec{r})$ is the potential in which the emitted particle moves.

The solution of (\ref{24.11}) has the form
\begin{equation}
\label{24.12} \psi(\vec{r})=\int
G(\vec{r},\vec{r}^{\,\prime})q(\vec{r}^{\,\prime})d^{3}r^{\prime}\,,
\end{equation}
where $G(\vec{r},\vec{r}^{\,\prime})$ is the Green function of
(\ref{24.11}).

%%%%%%%%%%%%%%%%5
According to \cite{14}, in moving in an arbitrary potential
\begin{equation}
\label{chan_24.13}
\lim_{r\rightarrow\infty}G(\vec{r},\vec{r}^{\,\prime})=-\frac{1}{4\pi}\frac{\exp(ikr)}{r}\psi_{\vec{k}}^{(-)*}(\vec{r}^{\,\prime})\,.
\end{equation}
As a result,
\begin{equation}
\label{24.14}
\psi(\vec{r})=-\frac{1}{4\pi}\frac{\exp(ikr)}{r}\int\psi_{\vec{k}}^{(-)*}(\vec{r}^{\,\prime})q(\vec{r}^{\,\prime})d^{3}r^{\,\prime}\,.
\end{equation}
From (\ref{24.12}) follows that the flow of particles produced by
the source is:
\begin{equation}
\label{24.15}
J=\frac{\mbox{const}}{r^{2}}\left|\int\psi_{\vec{k}}^{(-)*}(\vec{r}^{\,\prime})q(\vec{r}^{\,\prime})d^{3}r^{\prime}\right|^{2}\,.
\end{equation}
Let us average (\ref{24.15}) over the distribution of currents in
the source and assume that the source is  spatially incoherent,
i.e.,
\[
\langle
q(\vec{r}^{\,\prime})q(\vec{r}^{\,\prime\prime})\rangle=Q(\vec{r}^{\,\prime})\delta(\vec{r}^{\,\prime}-\vec{r}^{\,\prime\prime})\,.
\]
As a consequence,
\begin{equation}
\label{24.16}
J=\frac{\mbox{const}}{r^{2}}\int\left|\psi_{\vec{k}}^{(-)}(\vec{r}^{\,\prime})\right|^{2}Q(\vec{r}^{\,\prime})d^{3}r^{\prime}\,.
\end{equation}
Comparison of (\ref{24.16}) with (\ref{24.6}) and (\ref{24.7})
gives that $\tilde{Q}(\vec{r}_{b})$ has a meaning of distribution
density of the emitting points of the source, which in our case
are the nuclei produced via coalescence of particles $a$ and $A$.
%%%%%%%%%%5

Note that the quantity $n_{\vec{k}_{b}}(\vec{r}_{b})$ that
appeared upon averaging of the cross-section over the positions of
nucleus $A$ is, in fact, the diagonal element of the density
matrix of particles $b$, which determines their distribution in
the transverse plane of the channel. In terms of classical theory,
this  means that the angular distribution of the emitted particles
is defined by the statistical equilibrium density in the phase
space of the transverse particle motion (in \cite{138} direct
calculation showed that within the classical limit, the density
$n_{\vec{k}_{b}}(\vec{r}_{b})$ coincides with the classical
equilibrium density). Thus, the assumption about the fast
established statistical equilibrium in the transverse plane of the
phase space, regarded as a hypothesis in the classical derivation
of angular distributions, is quite substantiated from the quantum
viewpoint.

Substitute expression (\ref{24.5}) for the reaction amplitude into
formula (\ref{24.8}) defining the  distribution of the emitting
points $\tilde{Q}(\vec{r}_{b})$:
\begin{eqnarray}
\label{24.17}
\tilde{Q}(\vec{r}_{b})&=&\sum_{f}n_{\vec{k}_{a}f}(\vec{r}_{b\perp})\Phi^{2}_{A}(\vec{r}_{b\perp})
\left|T_{D}\exp\left(-ik_{a\parallel f}\frac{M_{A}}{m_{a}}r_{b\parallel}\right)\right.\nonumber\\
& &\left.\times\Phi_{A}(r_{b\parallel})-i\pi\sum_{\lambda}\frac{2M_{A}\gamma_{\lambda b}\gamma_{\lambda a}}{\hbar^{2}\kappa^{\prime}_{\lambda}}\right.\nonumber\\
&
&\left.\times\int\exp\left[-i(\kappa_{\lambda}^{\prime}-i\kappa_{\lambda}^{\prime\prime})\frac{m_{a}+
M_{A}}{m_{a}}|r_{b\parallel}-r_{A\parallel}|\right.\right.\nonumber\\
&
&\left.\left.-ik_{af\parallel}\frac{M_{A}}{m_{a}}r_{A\parallel}\right]\Phi_{A}(r_{A\parallel})dr_{A\parallel}\right|^{2}\,,
\end{eqnarray}
where
\[
\kappa_{\lambda}^{\prime}=\sqrt{\frac{2\mu}{\hbar^{2}}E_{\lambda}}\,,\quad
\kappa_{\lambda}^{\prime\prime}=\sqrt{\frac{2\mu}{\hbar^{2}}}\frac{\Gamma_{\lambda}}{4\sqrt{E_{\lambda}}}\,,\quad
\kappa_{\lambda}=\kappa_{\lambda}^{\prime}-i\kappa_{\lambda}^{\prime\prime}\,.
\]

%%%%%%%%%%%%%%%5

According to (\ref{24.17}), the distribution of the emitting point
is formed by the superposition of damped waves and stretches in
the incident direction of the primary particle (integration in
(\ref{24.17}) is, in fact performed over all $r_{A\parallel}\leq
r_{b\parallel}$; at $r_{A\parallel}> r_{b\parallel}$ the integrand
oscillates extremely rapidly and the stated domain of integration
can be discarded).

As would be expected, the rate of wave damping is determined by
the lifetime of the compound nucleus at level $\lambda$ and by its
velocity. Indeed, the index of power of the damped exponent is:
\[
\delta=k_{\lambda}^{\prime\prime}\frac{m_a+M_A}{m_a}=
\frac{1}{2}\frac{\mu}{\hbar^{2}\kappa_{\lambda}^{\prime}}\Gamma_{\lambda}\frac{m_{a}+M_{A}}{m_{a}}\,.
\]
Integration  over $dr_{A\parallel}$ leads to the fact that
\[
\kappa_{\lambda}^{\prime}=k_{a\parallel}\frac{\mu}{m_{a}}
\]
with the accuracy up to the momentum associated with the thermal
vibrations of nucleus $A$ in the lattice. Hence, one can write
\[
\delta=\frac{1}{2}\frac{\Gamma_{\lambda}}{\hbar}\frac{m_{a}+M_{A}}{\hbar
k_{a\parallel}}\,, \quad \mbox{i.e.},\quad
\delta=\frac{1}{2}\frac{1}{v_{c}\tau_{\lambda}}\,,
\]
where
\[
\frac{1}{\tau_{\lambda}}=\frac{\Gamma_{\lambda}}{\hbar}\] and
 \[v_{c}=\frac{\hbar k_{a\parallel}}{m_{a}+M_{A}}
 \]
  is the velocity of the compound nucleus.

Note that according to (\ref{24.17}), the interference of the
direct and resonance scattering channels has an appreciable
influence on the shape of $\tilde{Q}(\vec{r}_{b})$ at short
lifetimes of the compound nucleus, when the magnitude of its
spatial displacement is of the order of the vibration amplitude of
the nucleus in a crystal. In this case, the first and second terms
in (\ref{24.17}) overlap most strongly, causing a significant
deviation from a conventionally used exponential distribution law
even when only one level is excited
\[
\tilde{Q}(\vec{r}_{b})\sim\exp\left(-\frac{r_{b\parallel}-r_{A\parallel}}{v_{c}\tau}\right)\,.
\]
If a group of levels is excited, the superposition of waves
$\lambda$ entering into (\ref{24.17}) brings about spatial
oscillations of the distribution $\tilde{Q}(\vec{r}_{b})$. The
oscillation period $l_{\lambda\lambda^{\prime}}$ is defined by the
energy difference of the excited resonances:
\[
l_{\lambda\lambda^{\prime}}=2\pi
v_{c}\hbar/E_{\lambda}-E_{\lambda^{\prime}}\,.
\]

%%%%%%%%%%%%%%%%%%%%%%%%%%%%5
Thus, the shadow effect is applicable for determining the lifetime
of a compound nucleus and the distance between the levels (at
$E_{\lambda}-E_{\lambda^{\prime}}\geq\Gamma_{\lambda}$,
$\Gamma_{\lambda^{\prime}}$) even  in the case when a
monochromatic particle beam is incident on the crystal. Note that
the quantity $\Delta E\sim\hbar v_{c}/r_{0}$ acts as  effective
non-monochromaticity, $r_{0}$ is the vibration amplitude of the
nuclei in the lattice.

Now assume that a certain group of resonance levels is excited by
a beam of particles which have the energy spread considerably
exceeding the maximum distance between these levels. Then the
cross-section, and  hence, (\ref{24.17}) should be averaged over
the stated spread, which comes to integration of (\ref{24.17})
with respect to $k_{a\parallel\gamma}$. As a result, (\ref{24.17})
simplifies, taking the form
\begin{eqnarray}
\label{24.18}
\tilde{Q}(\vec{r}_{b})&=&\sum_{f}n_{\vec{k}_{a}f}(\vec{r}_{b\perp})
\Phi_{A}^{2}(\vec{r}_{b\perp})\left\{|T_{D}|^{2}\Phi_{A}^{2}(r_{b\parallel})\right.\nonumber\\
& &\left.-\frac{4\pi^{2}}{\Delta
k_{\parallel}}\frac{m_{a}}{M_{A}}|T_{D}|
\sum_{\lambda}\frac{2M_{A}\gamma_{\lambda b}\gamma_{\lambda a}}
{\hbar^{2}\kappa^{\prime}_{\lambda}}\sin\delta_{D}\Phi_{A}^{2}(r_{b\parallel})\right.\nonumber\\
& &\left.+\frac{2\pi^{3}}{\Delta
k_{\parallel}}\frac{m_{a}}{M_{A}}\int_{-\infty}^{r_{b\parallel}}\left|\sum_{\lambda}
\frac{2M_{A}\gamma_{\lambda b}\gamma_{\lambda a}}{\hbar^{2}\kappa^{\prime}_{\lambda}}\right.\right.\nonumber\\
&
&\left.\left.\times\exp\left[-\frac{i}{\hbar}\left(E_{\lambda}-i\frac{1}{2}\Gamma_{\lambda}\right)
\frac{r_{b\parallel}-r_{A\parallel}}{v_{c}}\right]\right|^{2}\right.\nonumber\\
&
&\left.\times\Phi_{A}^{2}(r_{A\parallel})dr_{A\parallel}\right\}\,,
\end{eqnarray}
where $T_{D}=|T_{D}|\exp(i\delta_{D})$; $\Delta k_{\parallel}$ is
the domain of averaging. The oscillations appearing in
(\ref{24.18}) are the time--to--space conversion of a well-known
phenomenon of time oscillations in the radiation intensity, which
arise through level excitation by a non-monochromatic packet. In
the experiments on the shadow effect they permit studying not only
the lifetime of the levels but also the distance between them.

%%%%%%%%%%%55
If a large number of neighboring levels are excited in the
reaction, in practice to explicitly find the sums involved in
(\ref{24.18}) is a complicated task. However, if assumed that the
level are randomly distributed over the excitation region, the
expression for $\tilde{Q}(\vec{r}_{b})$ may be derived by
averaging (\ref{24.8}) over the distribution of these levels
(according to \cite{132,133,134}, averaging of the cross-section
over the energy spread in the beam and over the level
distribution leads to one and the same result). As a consequence,
the average value of $\langle
T^{*}_{ba}(\kappa_{\parallel})T_{ba}(\kappa_{\parallel}^{\prime})\rangle$
will enter into the reaction cross section in (\ref{24.7}).

Suppose that the levels are statistically independent, as well as
the energy spread in the beam is much greater than the average
interlevel distance and the width of the levels, but it does not
exceed the interval over which the levels are concentrated.  In
this case the average value of $\langle
T^{*}_{ba}(\kappa_{\parallel})T_{ba}(\kappa_{\parallel}^{\prime})\rangle$
is the function of the difference of its arguments
\cite{132,133,134}:
\begin{equation}
\label{24.19} \langle
T^{*}_{ba}(\kappa_{\parallel})T_{ba}(\kappa_{\parallel}^{\prime})\rangle=g_{ba}
(\kappa_{\parallel}-\kappa_{\parallel}^{\prime})\,.
\end{equation}

Note that in \cite{132,133,134} the amplitude correlation function
is presented as the function of energies, rather than the function
of the wave numbers of relative motion. Integration in
(\ref{24.7}) and (\ref{24.8}) with respect to $\kappa_{\parallel}$
and $\kappa_{\parallel}^{\prime}$ gives
\begin{equation}
\label{24.20} \frac{d\sigma_{ab}}{d\Omega_{b}}=(2\pi)^{2}D\int
d^{3}r_{b}n_{\vec{k}_{b}}(\vec{r}_{b})\langle
Q(\vec{r}_{b})\rangle\,,
\end{equation}
%%%%%%%%%%%%%%555
\begin{eqnarray}
\label{24.21} \langle Q(\vec{r}_{b})\rangle &=&
n_{\vec{k}_{a}}(\vec{r}_{b\perp})\Phi_{A}^{2}(\vec{r}_{b\perp})
\int_{-\infty}^{r_{b\parallel}}g_{ba}(r_{b\parallel}-r_{A\parallel})\nonumber\\
&\times&\Phi_{A}^{2}(r_{A\parallel})dr_{A\parallel}\,,
\end{eqnarray}
\begin{eqnarray}
\label{24.22}
& & g_{ba}(r_{b\parallel}-r_{A\parallel})=\frac{1}{2\pi}\int g_{ba}(\kappa)\nonumber\\
&
&\times\exp\left[-i\kappa\frac{m_{a}+M_{A}}{m_{a}}(r_{b\parallel}-r_{A\parallel})\right]
\frac{m_{a}+M_{A}}{m_{a}}d\kappa\,,
\end{eqnarray}
\begin{equation}
\label{24.23}
n_{\vec{k}_{a}}(\vec{r}_{b\perp})=\sum_{f}n_{\vec{k}_{a}f}(\vec{r}_{b\perp})\,.
\end{equation}

When the crystal is illuminated by particles under the conditions
when channeling phenomenon for them is absent, the density
$n_{\vec{k}_{a}}(\vec{r}_{b\perp})$ is independent of the
coordinate, being a constant. If the time  in the delay-time
distribution function, obtained in \cite{132,133,134} is expressed
in terms of the travel distance, i.e.
$t=(r_{b\parallel}-r_{A\parallel})/v_{c}$, then the stated
function coincides with the function
$g_{ba}(r_{b\parallel}-r_{A\parallel})$, appearing in
(\ref{24.21}). From this follows that
\[
g_{ba}(r_{b\parallel}-r_{A\parallel})=g_{ba}\left(\frac{r_{b\parallel}-r_{A\parallel}}{v_{c}}\right).
\]

%%%%%%%%%%%%%%%%%%%%%%55
Recall now that the density $n_{\vec{k}_{b}}(\vec{r}_{b})$ only
depends on the component $\vec{\rho}$ of vector $\vec{r}_{b}$,
which lies in the plane perpendicular to the axes (planes) along
which particle $b$ is channeled. Therefore, if we introduce the
coordinate system with the $x$-, $y$-axes lying in this plane,
(\ref{24.20}) may be written as follows:
\begin{equation}
\label{24.24} \frac{d\sigma_{ab}}{d\Omega_{b}}=(2\pi)^{2}D\int
d^{2}\rho_{b}n_{\vec{k}_{b}}(\vec{\rho}_{b})\langle\widetilde{Q(\vec{\rho}_{b})}\rangle\,,
\end{equation}
where $\langle\widetilde{Q(\vec{\rho}_{b})}\rangle=\int
dz_{b}\langle Q(\vec{\rho}_{b})\rangle$ describes the distribution
of the emitting points in the transverse plane $\vec{\rho}_{b}$.

%%%%%%%%%%%%%%%555

 The integral in (\ref{24.24}) only by the specific expression $\langle\widetilde{Q(\rho_{b})}\rangle$  differs from the integral, determining averaged over the crystal thickness yield of nuclear reaction $I(\vec{k}_{b\perp})$ which is excited upon entering the crystal of particle $b$ with the  momentum $\vec{k}_{b\perp}$ transverse with respect to the axes along which the identical (akin) particle emitted by a nucleus is channeled (see, for example, [\cite{23}, formula (4.12)], where the quantity
$I(\vec{k}_{b\perp})$ includes the squared absolute value of the
wave function of the excited nucleus instead of
$\langle\widetilde{Q(\rho_{b})}\rangle$). In the case when nuclear
reaction times are too short for the compound nucleus to get
displaced over the distance larger than the vibration amplitude
$r_{0}$ of particle $A$, the quantity
$\langle\widetilde{Q(\rho_{b})}\rangle$ is "smeared" over the
region with spatial dimensions of the order of $r_{0}$.  As a
consequence, the angular distribution of the number of particles
which have left  the crystal coincide in form with the angular
dependence of the yield $I(\vec{k}_{b\perp})$ of nuclear
reactions. According to \cite{23}, the distribution of
$I(\vec{k}_{b\perp})$ is minimal for entrance angles close to zero
and grows with the entrance angles approaching the Lindhard angle,
forming a breastwork due to the contribution from the over-barrier
states. With further increase in the entrance angle it drops,
approaching the magnitude characteristic of a disordered medium.
Hence, the angular distribution of the particles leaving the
crystal will also be the same as described above, which agrees
with the experimentally observed pattern \cite{131}.

%%%%%%%%%%%%%%%%%%%%%%%%%%%55
With the increase in the reaction time, the function
$\langle\widetilde{Q(\rho_{b})}\rangle$ becomes more and more
smeared and the shadow depth decreases (with the increase in the
vibration amplitude of the nuclei, the depth of the minimum in the
nuclear reaction yield $I(\vec{k}_{b\perp})$ diminishes
\cite{23}).

Thus, the expressions derived above in the general case solve the
problem of the relationship between the angular distributions of
the particles which have left the crystal and the function
$g_{ba}$.

%%%%%%%%%%%%%%%%%55

Expression (\ref{24.24}) may be further particularized by
substituting a quasiclassical expression for the particle
distribution density $n_{\vec{k_{b}}}(\rho_{b})$. It should be
pointed out here that though the  motion of a heavy particle or,
for example, a relativistic positron is quasiclassical, in
analyzing the reaction yield (the intensity of the particles
produced), one should use quantum mechanical expressions
(\ref{24.24}) rather than classical formulas. This can be
explained by fact that the function $Q(\rho)$ is generally nonzero
in a classically inaccessible for positively charged particles
range of the potential of  particle interaction with the plane (or
axis). It immediately follows from the representation form of
(\ref{24.24}) that the reaction yield depends on the sign of the
particle charge. In the case of positrons, the maximum of the
density $n_{\vec{k}_{b}}(\rho_{b})$ does not coincide with that of
function $Q(\rho_{b})(|\Phi_{A}(\rho_{b})|^{2})$. For electrons,
the overlap of functions $n_{\vec{k}}(\rho)$ and
$Q(\rho)(|\Phi_{A}(\rho)|^{2})$ is  most complete. As a
consequence, in the case of electrons, the integral of the form
(\ref{24.24}) is maximal, i.e., the yield of nuclear reactions is
maximal.

%%%%%%%%%%%%%%%%%%555
Now discuss in more detail the features of the electron and
positron distribution over the levels. With this aim in view, pay
attention to the fact that in the quasi-classical limit, passing
from summation to integration over the entering points and vice
versa allows giving a simple geometric interpretation of the
particle distribution over the levels. Indeed, consider the
distribution of incident particles in a unit cell. In the planar
case this is the distribution over the domain of length $a$. The
probability to find a particle within the interval $dx$ of this
domain is $dx/a$ or
\[
\frac{1}{a}dx=\frac{1}{a}\frac{dx}{d\varepsilon_{n}}\frac{d\varepsilon_{n}}{dn}dn=\frac{2\pi}{aT_{n}|V^{\prime}|}dn\,,
\]
where it is taken into account that
 \[
 \varepsilon_{n}=\frac{p_{x}^{2}}{2m}+V(x)\,, \quad \frac{d\varepsilon_{n}}{dn}=\frac{2\pi}{T_{n}}\,,\quad V^{\prime}=\frac{dV}{dx}\,.
  \]
In other words, we may consider the equation
\[
\varepsilon_{n}=\frac{p_{x}^{2}}{2m}+V(x)
 \]
 as the one performing the conversion from variables $x(n)$ to variables $n(x)$ (similarly, in the axial case).
  From this follows (also see (\ref{sec:1.3}) that, for example, at zero entrance angle in the the case of positrons,
 the largest fraction of particles is concentrated at the bottom of the well. For electrons in a planar potential,
  the largest fraction of particles is outside the narrow part of the well, and, consequently,
  the largest fraction of electrons is concentrated at the levels near the top of the well.
  Analogous estimations are also possible for the axial case.

%%%%%%%%%%%%%%%%%%%%%%%%%%%%%%%%%%%  Chapter 9 %%%%%%%%%%%%%%%%%%%%%%

\chapter[Spin Rotation  and Radiative Self-Polarization of Particles in Bent Crystals]{Spin Rotation  and Radiative Self-Polarization of Particles Moving in Bent Crystals}
\label{ch:9}

%%%%%%%%%%%%%%%%%%%%%%%%%%%%%%%%%%%  Section 25 %%%%%%%%%%%%%%%%%%%%%%

\section{Spin Rotation of Relativistic Particles Passing Through a Crystal}
\label{sec:9.25} With the growth in energy of particles their spin
precession frequency in external fields diminishes, in the
ultra-relativistic case being determined by the anomalous magnetic
moment \cite{19}. As a result, for example, in a magnetic field
$H$ of strength $10^{4}$ Gs the electron (proton) spin precession
frequency $\omega=2\mu^{\prime}H/\hbar$ ($\mu^{\prime}$ is the
anomalous part of the magnetic moment) is $10^{8}$ s$^{-1}$, and
the spin rotation angle over  one centimeter path length $l$ is
just $\vartheta=\omega\frac{l}{c}\approx 10^{-2}$ rad. It turns
out, however, that at particle channeling in a crystal there
appears precession leading to the spin rotation angle of the order
of hundreds of radians over one centimeter path length
\cite{9,10}.

%%%%%%%%%%%%%%%%%5

If a crystal is nonmagnetic, then the equation for the spin
polarization vector $\vec{\zeta}$ may be written in the form (see,
for example, \cite{19}, $\S 41$)
\begin{equation}
\label{25.1}
\frac{d\vec{\zeta}}{dt}=\frac{2\mu^{\prime}}{\hbar}[\vec{E}(\vec{\zeta}\vec{n})-\vec{n}(\vec{\zeta}\vec{E})]\,,
\end{equation}
where $\vec{E}$ is the electric field at the point of particle
location; $\vec{n}=\vec{v}/c$; $\vec{v}$ is the particle velocity.

Intracrystalline fields $\vec{E}$ are large, reaching the values
of $10^{7}$ CGSE and even greater. Therefore  from (\ref{25.1})
follows  that for constant intracrystalline fields, the spin
precession frequency could reach $10^{11}$\,s$^{-1}$ and the angle
$\vartheta$ could be of the order of $10$\,rad/\,cm.

%%%%%%%%%%%%%%5

However, when a particle moves through a crystal in arbitrary
direction, the field $\vec{E}$, likewise in an amorphous medium,
takes on random values at the particle location point. As a
consequence, such a field causes spin depolarization.

Under channeling conditions the situation is basically different.
If the crystal bending radius is  $\rho_{0}$, then the  beam of
protons with the energy of $1-10^{2}$ GeV will change its
direction following the crystal bend up to the radii of curvature
of $\rho_{0}\sim 1$ cm \cite{8a,8b,139,140}, i.e., the particle
will move along a curved path. The stated motion is due to a
constant mean electric field acting on a particle in a bent
crystal \cite{8a,8b}. The magnitude of the field reaches $10^{9}$
SGSE.

%%%%%%%%%%%%%%5

Equation (\ref{25.1}) for a particle moving in a crystal, for
example, in a planar channel, bent to a radius of curvature
$\rho_{0}$ around the $y$-axis, has a form ($v_{y}=0$, $E_{y}=0,$
the trajectory lies in the $x$, $z$ plane)
\begin{equation}
\label{25.2}
\frac{d\zeta_{x(z)}}{dt}=\pm\frac{2\mu^{\prime}}{\hbar}(E_{x}n_{z}-E_{z}n_{x})\zeta_{x(z)}\,.
\end{equation}
The position vector $\vec{\rho}=(x,z)$ of a particle in such a
channel rotates about the $y$-axis with the frequency
$\Omega=c/\rho_{0}$. Its magnitude oscillates about the particle
equilibrium position $\rho_{0}^{\prime}$ in the channel  with the
frequency $\Omega_{k}$, amplitude $\alpha$, and initial phase
$\delta$. In the explicit form $x=\rho(t)\cos\Omega t$,
$z=\rho(t)\sin\Omega t$,
$\rho(t)=\rho_{0}^{\prime}+\alpha\cos(\Omega_{k}t+\delta)$. We
point out that, due to the presence of centrifugal forces in a
bent crystal, the equilibrium point $\rho_{0}^{\prime}$ does not
coincide with the position $\rho_{0}$ of the minimum of the
electrostatic  potential $\varphi(\rho)$ of the channel, as it
occurs in a straight channel. For example, when moving in a
harmonic well
\[
\varphi=-\frac{k}{e}\frac{(\rho-\rho_{0})^{2}}{2}\,,\quad
\rho_{0}^{\prime}-\rho_{0}=-\frac{2E}{k\rho_{0}}\,,
\]
 $E$ is the particle energy.

%%%%%%%%%%%55
Integration of (\ref{25.2}) in the polar coordinate system gives
($|\vec{\zeta}|=1$, $\vec{E}=-\vec{\nabla}\varphi$)
\begin{equation}
\label{25.3} \zeta_{z(x)}=\begin{array}{c}
               \cos \\
               \sin
             \end{array}
\left\{\frac{2\mu^{\prime}\Omega}{\hbar
c}\int^{t}_{0}\rho\frac{d\varphi}{d\rho}dt^{\prime}+\arctan\frac{\zeta_{x}(0)}{\zeta_{z}(0)}\right\}\,.
\end{equation}

For a harmonic well,  (\ref{25.3}) accurate up to the terms of the
order $(\rho_{0}^{\prime}-\rho_{0})/\rho_{0}$ and
$\alpha\rho_{0}^{-1}\ll 1$ can be written in the form
\begin{equation}
\label{25.4} \zeta_{z(x)}(t)=\begin{array}{c}
               \cos \\
               \sin
             \end{array}
\left\{\omega
t+\beta[\sin(\Omega_{k}t+\delta)-\sin\delta]+\arctan\frac{\zeta_{x}(0)}{\zeta_{z}(0)}\right\}\,,
\end{equation}
where
\[
\omega=\frac{2\mu^{\prime}}{\hbar}E(\rho_{0}^{\prime}) \]
 and
\[
E(\rho_{0}^{\prime})=-\frac{k}{e}(\rho_{0}^{\prime}-\rho_{0})
 \]
 is the electric field at the location point of the particle center of equilibrium in a bent crystal;
 \[
 \beta=-\frac{2\mu^{\prime}k a}{\hbar e\Omega_{k}}\,.
 \]

%%%%%%%%%%%%%%%%%5
The coefficient $\beta$ in (\ref{25.4}) is small (for Si the
coefficient $k=4\cdot 10^{17}$\,eV/\,cm$^{2}$, $\Omega_{k}\simeq
10^{13}$\,s$^{-1}$ for protons with $E\sim 100$\,GeV, as a result,
$\beta=\simeq 10^{-2}$). Neglecting the term containing $\beta$,
we obtain that the spin rotates with frequency $\omega$ (with
growing energy $\Omega_{k}\sim 1/\sqrt{E}$, the coefficient
$\beta\sim \sqrt{E}$ increases, and the spin rotation turns into
oscillations at frequency $\omega$ and  the frequencies multiple
of $\Omega_{k}$).  Due to a large magnitude of the field
$E(\rho_{0}^{\prime})$ curving the particle trajectory
($E(\rho_{0}^{\prime})\sim 10^{7}-10^{9}$ CGSE), the frequency
$\omega\simeq 10^{11}-10^{13}$\,s$^{-1}$ and the rotation angle
$\vartheta\sim 10-10^{3}$\,rad/\,cm.

If the radius of curvature $\rho_{0}\rightarrow \infty$ (a
straight channel), then only spin oscillations due to the term
containing $\beta$ remain. In this case a significant spin
rotation   occurs only at high energies (at low energies it is
absent).

%%%%%%%%%%%%%%%%%%%%%%%%%%%%%%%%%%%  Section 26 %%%%%%%%%%%%%%%%%%%%%%

\section{Spin Rotation at Deflection of a Charged Relativistic Particle in the Electric Field}
\label{sec:9.26}

For relativistic particles moving in an arbitrary electric field,
there is a simple relation between the spin precession angle and
the change in the direction of particle momentum \cite{141}. In
the case of planar channeling, this relationship enables one to
determine the spin rotation angle in the  effect considered in
(\ref{sec:9.25}) without turning to particular models describing
the distribution of the intracrystalline field. Below when
considering this problem, we shall follow the line of reasoning
given by Lyuboshitz in \cite{141}.

We shall proceed from the Bargmann-Michel-Telegdi equation
\cite{19} describing the spin behavior of a relativistic particle
moving quasiclassically in an external electric field. Let $m$ be
the particle mass; $e$ its charge, $\zeta$ the spin polarization
vector referred to an "instantaneous" rest system; $\gamma$  the
Lorentz factor; $\vec{l}$ the unit vector in the velocity
direction; $g$  the gyromagnetic ratio (by definition, the
magnetic moment $\mu=\frac{eg}{2mc}\hbar s$, where $s$ is the
particle spin). According to \cite{19},
\[
\frac{d\vec{\zeta}}{dt}=[\vec{\Omega}\vec{\zeta}],
\]
where $t$ is the time in the lab reference frame,
\begin{equation}
\label{26.1} \vec{\Omega}=
-\frac{e}{2mc}g\left\{\vec{H}-\frac{\gamma-1}{\gamma}\vec{l}(\vec{H}\vec{l})+\left[\vec{E}\frac{\vec{v}}{c}\right]\right\}
-(\gamma-1)\left[\vec{l}\,\frac{d\vec{l}}{dt}\right]\,,
\end{equation}
$\vec{E}$ and $\vec{H}$ are the strengths of the electric and
magnetic fields at the particle location point.
%%%%%%%%%%%%%%%%%%%%%%%%%%%5
The first term in (\ref{26.1}) for the angular velocity of
precession $\vec{\Omega}$ may be written as
\[
-\frac{e}{2mc\gamma}g\vec{H}^{*}\,,
\]
 where $\vec{H}^{*}$ is the
magnetic field strength in the intrinsic frame of reference; the
term
\[
-(\gamma-1)\left[\vec{l}\,\frac{d\vec{l}}{dt}\right]\,,
\]
corresponds to the Thomas spin precession \cite{142}.

From the equation of motion
\begin{equation}
\label{26.2}
\frac{d\vec{p}}{dt}=e\vec{E}+\frac{1}{c}[\vec{v}\vec{H}]
\end{equation}
follows that the instantaneous angular velocity of rotation of a
particle momentum is defined by formula
\begin{equation}
\label{26.3}
\vec{\Omega}_{0}=\left[\vec{l}\,\frac{dl}{dt}\right]=-\frac{e}{mc\gamma}\left\{\vec{H}-\vec{l}(\vec{H}\vec{l})+
\frac{\gamma^{2}}{\gamma^{2}-1}\left[\vec{E}\frac{\vec{v}}{c}\right]\right\}\,.
\end{equation}
Comparison of (\ref{26.1}) and (\ref{26.3}) shows that in the
absence of a magnetic field vectors $\vec{\Omega}$ and
$\vec{\Omega}_{0}$ are parallel (or antiparallel) to one another
and are related as
\begin{equation}
\label{26.4}
\vec{\Omega}=\left[(g-2)\frac{\gamma^{2}-1}{2\gamma}+\frac{\gamma-1}{\gamma}\right]\vec{\Omega}_{0}
\end{equation}
or
\begin{equation}
\label{26.5}
\vec{\Omega}=\left[\frac{1}{2}(g-2)\gamma+\frac{\gamma}{\gamma+1}\right]\frac{v^{2}}{c^{2}}\vec{\Omega}_{0}\,.
\end{equation}
%%%%%%%%%%%%%%%%%%%%%%%%%%%%%%%%%%%%%%%%%%%555

It is clear that if the trajectory of a charged particle in the
electric field is a plane curve, vectors $\vec{\Omega}(t)$ and
$\vec{\Omega}_{0}(t)$ have constant direction along the normal
$\vec{n}$ to the plane of motion
($\vec{\Omega}_{0}(t)=\Omega_{0}(t)\vec{n}$,
$\vec{\Omega}(t)=\Omega(t)\vec{n}$). In this case the angle of the
polarization vector  precession around the normal $\vec{n}$ is
\begin{equation}
\label{26.6} \theta(t)=\int_{0}^{t} \left[(g-2)
\frac{\gamma^{2}(t^{\prime})-1}{2\gamma(t^{\prime})}+\frac{\gamma(t^{\prime})-1}{\gamma(t^{\prime})}\right]
\frac{d\theta_{0}(t^{\prime})}{dt^{\prime}}dt^{\prime}\,,
\end{equation}
where
$\theta_{0}(t)=\int_{0}^{t}\Omega_{0}(t^{\prime})dt^{\prime}$ is
the angle between the particle initial momentum and its  momentum
at time $t$.

If the kinetic energy of a particle moving along the trajectory
practically does not change, the relation between the angles of
spin and momentum rotation is defined by formula
\begin{equation}
\label{26.7}
\theta=\left[(g-2)\frac{\gamma^{2}-1}{2\gamma}+\frac{\gamma-1}{\gamma}\right]\theta_{0}\,.
\end{equation}
In the nonrelativistic case
\begin{equation}
\label{26.8}
\theta=\frac{1}{2}(g-1)\frac{v^{2}}{c^{2}}\theta_{0}\,.
\end{equation}

%%%%%%%%%%%%%%%%%%%%%%%%%%%%%
Note that for sufficiently small sections of the trajectory, the
relation (\ref{26.7}) also holds true even when the direction of
vectors $\vec{\Omega}_{0}$ and $\vec{\Omega}$ changes with time.
In this case the axis of spin rotation through the angle $\theta$
is perpendicular to the plane containing the initial and final
momenta of the particle. \footnote{For a spin wave function, the
equation of  precession in the electric field has a form
\[
i\frac{\partial\Psi(t)}{\partial
t}=(\vec{\Omega}(t)\hat{s})\Psi(t)\,,
\]
where $\vec{\Omega}(t)$ is defined according to
(\ref{26.4}--(\ref{26.5}), $\hat{s}$ is the spin operator. In
non-planar motion, the operators $\vec{\Omega}(t)\hat{s}$ taken at
different instants of time do not commute with one another, and
the symbolic representation of the solution is as follows
\[
\Psi(t)=\hat{T}\exp\left(-i\int_{0}^{t}\hat{s}\vec{\Omega}(t^{\prime})dt^{\prime}\right)\Psi(0)\,,
\]
where $\hat{T}$ is the chronological operator \cite{23}. In the
first approximation of the perturbation theory
\[
\Psi(t)=\left(1-i\hat{s}\int_{0}^{t}\vec{\Omega}(t^{\prime})dt^{\prime}\right)\Psi(0)\,.
\]
For the polarization vector this corresponds to the equality
\[
\vec{\xi}(t)=\vec{\xi}(0)+\left[\int_{0}^{t}\vec{\Omega}(t^{\prime})dt^{\prime}\vec{\xi}(0)\right]\,.
\].}

%%%%%%%%%%%%%%%%%%%%%%55

It is essential that allowing for radiative damping practically
does not  change the relations derived. Indeed, radiative
deceleration comes to the appearance of an additional electric
field in the intrinsic reference frame of a charged particle. This
field is unlikely to affect the magnetic moment, so formula
(\ref{26.1}) for the angular velocity of precession does not
change. On the other hand, the retardation force in the lab
reference frame, which is to be introduced into the right-hand
side of equation (\ref{26.2}) at $\vec{H}=0$ has the form
\cite{68}
\[
\vec{f}=\frac{2e^{4}}{3m^{2}c^{5}}\gamma^{2}v\vec{l}\left\{\vec{E}^{2}-\frac{v^{2}}{c^{2}}(\vec{E}\vec{l})^{2}\right\}
\]
(here the terms negligibly small in comparison with the Lorentz
force are discarded). As the retardation force $\vec{f}$ is
directed opposite to the velocity, it makes zero contribution to
angular velocity

\[
\vec{\Omega}_{0}=\left[\vec{l}\frac{d\vec{l}}{dt}\right]\,.
\]
 Hence, vectors $\vec{\Omega}$ and $\vec{\Omega}_{0}$ still
satisfy relations (\ref{26.4})-(\ref{26.5}).

%%%%%%%%%%%%%%%%%%%%%%%%5
In the presence of an external magnetic field the parallelism of
vectors $\vec{\Omega}$ and $\vec{\Omega}_{0}$ is, generally
speaking, violated, except for the case of motion in a transverse
magnetic field at $\vec{E}=0$ when
\[
\vec{\Omega}=\left(\frac{g-2}{2}\gamma+1\right)\vec{\Omega}_{0}.
\]
It is easy to see that in the ultra-relativistic limit ($\gamma\gg
1$) at arbitrary fields $\vec{E}$ and $\vec{H}$, the following
approximate equality holds accurate up to the terms
$eH/mc\gamma^{2}$
\begin{equation}
\label{26.9}
\vec{\Omega}=\vec{\Omega}_{0}\left(\frac{g-2}{2}\gamma+1\right)+\frac{ge}{2mc\gamma}\vec{l}(\vec{H}\vec{l})\,.
\end{equation}
%%%%%%%%%%%%%%%%%%55

%%%%%%%%%%%%%%%%%5
Consider some particular applications of formulae
(\ref{26.4})-(\ref{26.7}).

\textit{Motion in a homogeneous electric field.} Let at $t=0$ a
particle be at the origin or coordinates, the initial momentum
$p=mv\gamma$ be directed  along the y-axis, and the electric field
strength - along the x-axis. Then the calculation from formula
(\ref{26.6}) gives the following expression for the angle of spin
rotation about the z-axis:
\begin{equation}
\label{26.10} \theta=\frac{e E
y(t)}{2mc^{2}}(g-2)+\arccos\frac{\gamma+\cosh\frac{eEy(t)}{pc}}{\gamma\cosh\frac{eEy(t)}{pc}+1},
\end{equation}
where
\[
y(t)=\frac{pc}{eE}arcsinh\frac{eE(t)}{mc\gamma}
\]
The deflection angle of the particle in the electric field is
\begin{equation}
\label{26.11} \theta_{0}=\arctan\frac{eEt}{p}.
\end{equation}
If the particle kinetic energy varies insignificantly, then
$\theta_{0}\simeq \frac{eEt}{p}\simeq\frac{eEy}{pv}\ll 1$, and
formula (\ref{26.10}) goes over to (\ref{26.7}).

\textit{Planar channeling in bent crystals.} Curving the
trajectory of a charged particle  moving  along  the bent channel
is due to the existence of the perpendicular to the momentum mean
electric field, whose magnitude can reach $10^{7}-10^{8}$ SGSE. In
\cite{9,10} is shown that, due to this fact, when
ultra-relativistic particles are channeled in bent crystals the
rotation angle of the polarization vector takes on large values
(see (\ref{sec:9.25})). It is interesting that this angle may be
found from formulae (\ref{26.7}) or (\ref{26.6}), without turning
to particular models describing the distribution of the
intracrystalline field. Indeed, suppose that the momentum of a
channeled particle is parallel to the bending plane. Then the
particle deflection angle $\theta_{0}$ coincides with the crystal
bending angle, and the spin rotation axis is perpendicular to the
plane of bending. For a proton the radiation energy losses are
vanishingly small, and  relation (\ref{26.7}) holds true. At
$\gamma\gg 1$, find $\left(\frac{g-2}{g}=1.79\right)$:
\begin{equation}
\label{26.12}
\theta=\left(1+\frac{g-2}{2}\gamma\right)\theta_{0}=(1+1.91\varepsilon)\theta_{0},
\end{equation}
where $\varepsilon$ is the proton energy, GeV. According to
(\ref{26.12}) the proton spin rotates in the same direction as the
momentum does. At $\varepsilon=10$ GeV the spin precession angle
is 20 times as large as the momentum deflection angle.

Note that at the given radius of curvature $R$ the maximum energy
of particles, which are also "captured" into the channeling regime
in a bent crystal, is $eE_{max}R$, where $E_{max}$ is the maximum
strength of the electric field. As $\theta_{0}=y/R$, where $y$ is
the length of the trajectory, the spin rotation angle of the
proton, corresponding to the maximum energy is only determined by
the values of $E_{max}$ and $y$:
\[
\theta_{max}=\frac{g-2}{2}\frac{eE_{max}y}{mc^{2}}\simeq 5.79\cdot
10^{-7} E_{max}y.
\]
If $E_{max}=10^{7}$ SGSE, $y=1$ cm, $R=100$ cm, then
$\varepsilon\sim 300$ GeV and $\theta_{max}\sim 6$ rad. This value
agrees with the estimates given in \cite{9,10} and
(\ref{sec:9.25}).

In the case of channeling of positrons,  radiation losses at
achievable energies can be significant, and searching for the spin
rotation angle one should use relation (\ref{22.6}), which takes
account of the change in the kinetic energy in motion. The
corresponding design equation takes the form
$\left(\frac{g-2}{2}\simeq 1.16\cdot 10^{-3}\right)$
\begin{equation}
\label{26.13} \theta=(1+2.27\bar{\varepsilon})\theta_{0},
\end{equation}
where
$\bar{\varepsilon}=\frac{c}{l}\int_{0}^{l/c}\varepsilon(t)dt$ is
the mean value of energy, GeV.

\textit{Scattering by the electrostatic (Coulomb) potential.} At
quasi-classical scattering of a charged particle at the angle
$\theta_{0}$ in a static field of the system of charges, rotation
of the polarization vector is defined by formula (\ref{26.6}).
Integration in (\ref{26.6}) is made along the unclosed trajectory,
uniquely  determined by the scattering angle and plane.  The
rotation axis of the polarization vector is probably perpendicular
to the scattering plane. When speaking about scattering at small
angles, within the region of particle motion the potential energy
is small in comparison with the kinetic one, and thus, the
connection between the spin rotation angle and the scattering
angle is specified by relation (\ref{26.7}).

For the Coulomb scattering the latter statement also holds true
beyond pure classical description of a particle motion in the
electric field, which have been used until now. In this case the
main contribution to the amplitude of scattering at the angles
$\theta_{0}\ll 1$ comes from the region of high impact parameters
$\rho\gg\hbar/p$, where the particle potential energy is much
smaller than its kinetic energy. Therefore we shall apply the
eikonal approach, enabling representation of  the amplitude of
scattering at small angles as follows \cite{16}
\begin{equation}
\label{26.14} a(\theta_{0})=-\frac{ik}{2\pi}\int_{0}^{\infty}\rho
d\rho\left\{(e^{iS(\rho)\hbar}-1)\int_{0}^{2\pi}e^{-ik\theta_{0}\rho\cos\psi}d\psi\right\}
\end{equation}
Here $k=p/\hbar$; $S(\rho)=\frac{1}{v}\int_{+\infty}^{-\infty}
u(\rho_{1}z)dz$ is the difference of the classical action
integrals for a straight trajectory with the impact parameter
$\rho$ with and without interaction; $\psi$ is the angle of vector
$\rho$, perpendicular to the particle momentum with the scattering
plane. Formula (\ref{26.14}) is valid for both non-relativistic
and relativistic energies. To take into account spin precession in
the electric field, let us multiply the function
$\exp(iS(\rho)/\hbar)$  in (\ref{26.14}) by the rotation matrix
\begin{equation}
\label{26.15}
\hat{R}(\theta(\rho))=\exp\left(-i\hat{s}\frac{[\vec{k}\vec{\rho}]}{k\rho}\theta\rho\right),
\end{equation}
where $\theta(\rho)$ is the spin rotation angle corresponding to
the motion of the charged particle along the classical (close to
straight) trajectory with the impact parameter $\rho$; $\hat{s}$
is the spin operator. It has been shown above that the angle
$\theta(\rho)$  is connected with the angle of the momentum
deflection $\theta_{0}(\rho)$  for the same trajectory by relation
(\ref{26.7}). On the other hand, the angle $\theta_{0}(\rho)$ is
determined in terms of the action function:
\begin{equation}
\label{26.16} \theta_{0}(\rho)=\frac{1}{\hbar
k}\frac{d}{d\rho}S(\rho).
\end{equation}
Thus, provided that the angles $\theta(\rho)$ and
$\theta_{0}(\rho)$ are small in (\ref{26.14})
$\exp(iS(\rho)/\hbar)$ should be replaced by
\begin{eqnarray}
\label{26.17}
\exp(i S(\rho)/\hbar)\hat{R}(\theta(\rho))\simeq\exp(i S(\rho)/\hbar)\left[1-ib\frac{d S(\rho)}{d\rho}\right.\nonumber\\
\left.\times\frac{\cos\psi}{\hbar k}\hat{s}_{z}+ib\frac{d
S(\rho)}{d\rho}\frac{\sin\psi}{\hbar k}\hat{s}_{y}\right],
\end{eqnarray}
where
\begin{equation}
\label{26.18}
b=\left(\frac{g-2}{2}\frac{\gamma^{2}-1}{\gamma}+\frac{\gamma-1}{\gamma}\right)
\end{equation}

(it is assumed that the z-axis is directed parallel to the normal
to the scattering plane, the x-axis - along the initial momentum
of a particle). Upon integration with respect to the angle $\psi$,
the formula for the scattering amplitude takes the form
\begin{eqnarray}
\label{26.19}
\hat{A}(\theta_{0})=a(\theta_{0})+b\hat{s}_{z}\int_{0}^{\infty}J_{1}(k\theta_{0}\rho)
\left[\frac{d}{d\rho}\exp(i S(\rho)/\hbar)\right]\rho d\rho;\nonumber\\
a(\theta_{0})=-ik\int_{0}^{\infty}J_{0}(k\theta_{0}\rho)
\left[\exp(i S(\rho)/\hbar)-1\right]\rho d\rho.
\end{eqnarray}
Using well known relations for the Bessel function
\[
\frac{d}{d\rho}(\rho J_{1}(k\theta_{0}\rho))=k\theta_{0}\rho
J_{0}(k\theta_{0}\rho);\,
\int_{0}^{\infty}J_{0}(k\theta_{0}\rho)\rho
d\rho=\frac{2\delta(\theta_{0}^{2})}{k^{2}},
\]
we obtain with the accuracy up to the terms of the order of
$\theta_{0}^{2}$
\begin{equation}
\label{26.20}
\hat{A}(\theta_{0})=a(\theta_{0})(1-i\hat{s}_{z}b\theta_{0}).
\end{equation}
From this the angle of spin rotation about the z-axis is
$b\theta_{0}$.

It may be argued that this result within the range of angles
$\theta_{0}\ll 1$,  $\theta=b\theta_{0}\ll 1$ is not bound by any
additional conditions. Within the quasi-classical limit the
requirement of smallness is only imposed on the scattering angle
$\theta_{0}$, while the spin rotation angle at ultra-relativistic
energies  may take on any values.

Such consideration has nothing to do with the use of relativistic
equations, being applicable to particles with arbitrary spin and
gyromagnetic ratio. In the case of  scattering of electrons with
not very high energies ($\gamma\simeq 10^{3}$) in a Coulomb  field
of a nucleus of charge $ze$, the anomalous magnetic moment of the
electron may be neglected, which  according to (\ref{26.19}) and
(\ref{26.20}) gives
\begin{eqnarray}
\label{26.21}
A(\theta_{0})\simeq& & \frac{2z e^{2}}{p v\theta_{0}^{2}}\frac{\Gamma\left(1-i\frac{z e^{2}}{\hbar v}\right)}{\Gamma\left(1+i\frac{z e^{2}}{\hbar v}\right)}\exp\left(2i\frac{z e^{2}}{\hbar v}\ln\frac{\theta_{0}}{2}\right)\nonumber\\
&
&\times\left[1-i\hat{\sigma}_{z}\frac{\gamma-1}{2\gamma}\theta_{0}+\theta(\theta_{0}^{2})\right],
\end{eqnarray}
where $\hat{\sigma}_{z}$ is the Pauli matrix. And the spin
rotation angle is
\begin{equation}
\label{26.22} \theta=\frac{\gamma-1}{\gamma}\theta_{0}.
\end{equation}
Relations (\ref{26.21}) and (\ref{26.22}) may be obtained
independently on the basis of the solution of the Dirac equation
in the limit (in extreme case) $\theta_{0}\ll 1$ (see\cite{144}).
At non-relativistic energies
$\theta=\frac{v^{2}}{2c^{2}}\theta_{0}$, and at ultra-relativistic
energies $\theta=\theta_{0}$, which corresponds to helicity
conservation.

In conclusion we shall point out an  interesting consequence of
relation (\ref{26.4}): if the gyromagnetic ratio satisfies the
condition
\begin{equation}
\label{26.23} 1<g<2,
\end{equation}
then at the energy $\varepsilon_{0}=\frac{g}{g-2}mc^{2}$, the
electric field does not influence particle spin at all (the
angular velocity of precession vanishes, though the magnetic
moment is nonzero). This is a purely relativistic effect caused by
cancellation between the "dynamic" and  Thomas precessions. At
$\varepsilon<\varepsilon_{0}$,  spin rotates in the same direction
as  the momentum does, at $\varepsilon>\varepsilon_{0}$, it
rotates in the opposite direction. For example, a deuteron with
$g=1.72$, as well as some nuclei (e. g. $^{6}Li$ with $g=1.64$),
satisfies the condition (\ref{26.23}). \footnote{For nuclei the
quantity $g$ is associated with the so-called nuclear gyromagnetic
ratio by formula $g=g_{\mathrm{nuc}}A/z$ ($A$ is the number of
nucleons in a nucleus, $z$ is the atomic number).} For a deuteron
$\varepsilon_{0}=11.5$ Ge\,V. Analogous phenomenon also occurs in
a transverse magnetic field, providing that $0<g<2$. The energy at
which the polarization vector preserves constant direction in this
case equals
\[
\tilde{\varepsilon}_{0}=\frac{2}{2-g}mc^{2}
\]
(for a deuteron, for example,
$\tilde{\varepsilon}_{0}=13.4$\,GeV).

%%%%%%%%%%%%%%%%%%%%%%%%%%%%%%%%%%%  Section 27 %%%%%%%%%%%%%%%%%%%%%%

\section{Depolarization of Fast Particles Moving in Matter}
\label{sec:9.27}

As it has been shown (see (\ref{sec:9.26}), at small deflection of
charged particles from the initial direction in the magnetic
field, the spin polarization vector $\vec{\zeta}$ rotates around
the normal to the plane passing through the initial and final
momenta $\vec{p}_{0}$ and  $\vec{p}_{1}$ through the angle
\begin{equation}
\label{27.1}
\theta=\left[(g-2)\frac{\gamma^{2}-1}{2\gamma}+\frac{\gamma-1}{\gamma}\right]\theta_{0}.
\end{equation}
Here $\theta_{0}$ is the angle of momentum $\vec{p}_{0}$ with
momentum $\vec{p}_{1}$; $\gamma$ is the Lorentz factor; $g$ is the
gyromagnetic ratio (by definition the magnetic moment
$\mu=\frac{e\hbar}{2mc}gs$, where $e$ is the charge, $m$ is the
mass, $s$ is the spin of the particle). At small change in the
kinetic energy, providing that $\theta\ll 1$, $\theta_{0}\ll 1$,
formula (\ref{27.1}) holds true irrespective of the character of
the  intermediate motion of the particle in question.\footnote{In
the particular case of quasi-classical motion along the plane
trajectory the angles $\theta$ and $\theta_{0}$ in (\ref{27.1})
may take on any value (see \cite{141}).} It is easy to see that
when $\theta\ll 1$, the angle of deflection of  the polarization
vector from the initial direction $\vec{\zeta}_{0}$ is
\begin{equation}
\label{27.2} \theta=\theta\sin\psi,
\end{equation}
where $\psi$ is the angle of $\vec{\zeta}_{0}$ with vector
$[\vec{p}_{0}\vec{p}_{1}]$. Relations  (\ref{27.1}), (\ref{27.2})
allow calculating the degree of depolarization of the charged fast
particle moving in a macroscopic medium \cite{145}. Below we shall
follow the same line of reasoning as given in \cite{145}.

Consider the case of longitudinal polarization ($\psi=\pi/2$). It
is clear that at multiple scattering of a particle in the Coulomb
field of nuclei and electrons the mean values of the transverse
components of the momentum and polarization vector are zero. Thus,
vector $\langle\vec{\zeta}_{\parallel}\rangle$ preserves its
direction. Despite the fact that
$\langle\vec{\zeta}_{\perp}\rangle=0$, the quantity
$\langle\tilde{\theta}^{2}\rangle=\langle\theta^{2}\rangle=\langle\vec{\zeta}^{2}_{\perp}\rangle/\zeta^{2}_{\parallel}$
is nonzero. As a result, the particle undergoes depolarization
(the value of $|\vec{\zeta}_{\parallel}|$ decreases).

According to (\ref{27.1}) the mean-square  angle of deflection of
the polarization vector $\vec{\zeta}_{\parallel}$ from the initial
direction when the particle is passing through a thin layer of
matter $\Delta l$ is related to the mean-square angle of the
multiple Coulomb scattering in this layer as
\begin{equation}
\label{27.3}
\langle\tilde{\theta}^{2}\rangle=\langle\theta^{2}\rangle=
\left[\frac{g-2}{2}\frac{\gamma^{2}-1}{\gamma}+\frac{\gamma-1}{\gamma}\right]^{2}\langle\theta_{0}^{2}\rangle
\end{equation}
It is known that $\langle\theta_{0}^{2}\rangle$ is described with
good accuracy by the expression \cite{146,147}
\begin{equation}
\label{27.4}
\langle\theta_{0}^{2}\rangle=z^{2}\left(\frac{E_{s}}{m}\right)^{2}
\frac{\gamma^{2}}{(\gamma^{2}-1)^{2}}\frac{\Delta l}{L_{rad}},
\end{equation}
where $z=e/e_{0}$ is the ratio of the particle charge to the
electron charge; $m$  is the particle mass; $E_{s}=21$ MeV;
$L_{rad}$ is the radiation length for an electron. Substituting
(\ref{27.4}) into (\ref{27.3}) and taking into account that at
small $\langle\tilde{\theta}^{2}\rangle$ the degree of
polarization is
\begin{equation}
\label{27.5}
\eta=1-\langle\cos\tilde{\theta}\rangle\simeq\frac{1}{2}\langle\tilde{\theta}^{2}\rangle,
\end{equation}
we come to the formula describing depolarization of longitudinally
polarized particles:
\begin{equation}
\label{27.6}
\eta_{\parallel}=\frac{1}{2}z^{2}\left(\frac{E_{s}}{m}\right)^{2}
\left[\frac{g-2}{2}+\frac{1}{\gamma+1}\right]^{2}\frac{\Delta
l}{L_{rad}},
\end{equation}

If a particle is polarized in the direction perpendicular to the
momentum, then at fixed angle $\psi$ of the polarization vector
$\vec{\zeta}_{\perp}$  with the normal to the scattering plane
$\vec{\eta}$, in view of (\ref{27.2}),
$\eta_{\perp}=\eta_{\parallel}\sin^{2}\psi$. At averaging over the
azimuth angle, a factor $1/2$ appears.

Thus, when polarized particles pass through the layer of matter,
their depolarization in the transverse direction is half as much
as depolarization in the longitudinal direction
$\eta_{\perp}=\frac{1}{2}\eta_{\parallel}$. This leads to the fact
that in the general case the initial angle $\Phi$ of the
polarization vector with the momentum increases by
\begin{equation}
\label{27.7} \Delta\Phi=\frac{1}{4}\eta_{\parallel}\sin 2\Phi
\end{equation}
($\eta_{\parallel}$ is determined from formula (\ref{27.6})).
$\Delta\Phi$ reaches its maximum value at $\Phi=\pi/4$ and
vanishes at $\Phi=0$ and $\Phi=\pi/2$. And the degree of
depolarization
\begin{equation}
\label{27.8}
\eta=\left(1-\frac{1}{2}\sin^{2}\Phi\right)\eta_{\parallel}=(2-\sin^{2}\Phi)\eta_{\perp}.
\end{equation}
According to (\ref{27.6}) and (\ref{27.8}), at non-relativistic
energies
\begin{equation}
\label{27.9}
\eta=\frac{1}{8}z^{2}\left(1-\frac{1}{2}\sin^{2}\Phi\right)
\left(\frac{E_{s}}{m}\right)^{2}(g-1)^{2}\frac{\Delta l}{L_{rad}},
\end{equation}
whereas at ultra-relativistic energies
\begin{equation}
\label{27.10}
\eta=\frac{1}{8}z^{2}\left(1-\frac{1}{2}\sin^{2}\Phi\right)\left(\frac{E_{s}}{m}\right)^{2}(g-2)^{2}\frac{\Delta
l}{L_{rad}},
\end{equation}
We point out that the basic formula (\ref{27.6}) holds for a layer
of fixed thickness, passing through which a particle loses a small
fraction of its energy, and the condition $\eta\ll 1$ should also
be satisfied. With the latter condition preserved, it is easy to
take into account the energy losses by substituting expression
(\ref{27.6}) into the integral
\begin{equation}
\label{27.11}
\eta_{\parallel}=\frac{1}{2}z^{2}\left(\frac{E_{s}}{m}\right)^{2}\int_{0}^{\Delta
l}
\left[\frac{g-2}{2}+\frac{1}{\gamma(x)+1}\right]^{2}\frac{dx}{L_{rad}},
\end{equation}
where $\gamma(x)$ is the Lorentz factor of the particle at the
distance $x$ from the front boundary of matter. Here relations
(\ref{27.7}) and (\ref{27.8}) remain valid, as well as expression
(\ref{27.9}) for non-relativistic energies.

From (\ref{27.11}) follows that the degree of depolarization of
the longitudinally polarized protons or antiprotons ($|z|=1,\,
\frac{g-2}{2}=1.79$, $m=938$ MeV) is described by expression
\begin{equation}
\label{27.12} \eta_{p\parallel}=0.8\cdot 10^{-3}\int_{0}^{\Delta
l} \left(1+\frac{0.28}{1+0.53 T(x)}\right)^{2}\frac{dx}{L_{rad}},
\end{equation}
where $T(x)=E(x)-m_{p}c^{2}$ is the kinetic energy, GeV. It is
easy to see that on the nuclear collision length in lead
$\left(\frac{\Delta l}{L_{rad}}\sim 20\right)$  protons with the
energy $T\geq 1$ GeV proton get depolarized by $1.5-2 \%$.

In a similar manner one can estimate the degree of depolarization
of a passing beam of neutral particles with the magnetic moment
$\mu=\frac{e_{0}\hbar}{2m_{p}c}g_{a}s_{a}$ (for a neutron
$g_{n}=-3.82$, and for a $\Lambda$-particle $g_{\Lambda}=-1.2$).
Indeed, in the first approximation  the neutral particle moves in
the same electric field as the charged particle deflected through
small angles. In the given electric field the spin rotation angles
of the particle in question are related to those of the proton as
(see \cite{141}).
\begin{equation}
\label{27.13}
b(\gamma)=g_{a}/\left(g_{p}-\frac{2\gamma}{\gamma+1}\right).
\end{equation}
In view of (\ref{27.13}), (\ref{27.8}) and (\ref{27.11}) the
degree of depolarization of arbitrary polarized neutral particles
is energy-independent and described by the expression
\begin{eqnarray}
\label{27.14}
\eta_{a}=\left(1-\frac{1}{2}\sin^{2}\Phi\right)\int_{0}^{\Delta l}b^{2}(\gamma)\frac{\partial\eta_{p\parallel}}{\partial z}dx\nonumber\\
=\frac{1}{8}\left(1-\frac{1}{2}\sin^{2}\Phi\right)\left(\frac{E_{s}}{m_{p}}\right)^{2}g^{2}_{a}\frac{\Delta
l}{L_{rad}}.
\end{eqnarray}

This result may also be obtained in a different way, considering
the change in polarization at Schwinger scattering of a neutral
particle with a nonzero magnetic moment in the Coulomb nuclear
field. In the unit event of Schwinger scattering the polarization
vector of scattered particles
$\vec{\zeta}=-\vec{\zeta}_{0}+2\vec{n}(\vec{\zeta}_{0}\vec{n})$,
where $\vec{n}$ is the unit vector along the normal to the
scattering plane \cite{19}. Upon averaging over the azimuth angle
we have $\vec{\zeta}_{\parallel}=-\vec{\zeta}_{0\parallel}$,
$\vec{\zeta}_{\perp}=0$. From this follows that in the layer of
thickness $\Delta l$
\[
\eta_{a\parallel}=2\eta_{a\perp}=2N(\int\sigma_{Sch}(\theta_{0})d\Omega)\Delta
l.
\]
Here $N$ is the number of nuclei per unit volume;
\[
\sigma_{Sch}(\theta_{0})=\frac{1}{4}\left(\frac{ze^{2}}{m_{p}c^{2}\theta_{0}}
\right)^{2}g_{a}^{2}F^{2}(m_{a}\sqrt{\gamma^{2}-1}\theta_{0});
\]
$F$ is the form factor including electron screening of the nuclear
field  and the influence of the finite size of a nucleus.
Integration with respect to the solid angle gives (compare with
similar calculations in \cite{146})
$$
2N\int \sigma_{Sch}(\theta_{0})d\Omega\Delta l\simeq\frac{1}{8}
\left(\frac{E_{s}}{m}\right)^{2}g^{2}\frac{\Delta l}{L_{rad}}.
$$
From (\ref{27.14}) follows, in particular, that the degree of
depolarization of longitudinally polarized neutrons on the
radiation length $\eta_{n\parallel}=9\cdot 10^{-4}$; for
$\Lambda$-hyperons $\eta_{\Lambda\parallel}=0.9\cdot 10^{-4}
\Delta l/L_{rad}$.

In view of the smallness of factor $g-2$ the energy-dependence of
depolarization of $\mu$-mesons and electrons is more appreciable
than that of the protons. For longitudinally polarized
$\mu$-mesons  ($m_{\mu}=105.6$ MeV, $(g-2)/2=1.16\cdot 10^{-3}$),
formula (\ref{27.11}) has the form
\begin{equation}
\label{27.15} \eta_{\mu\parallel}=5\cdot 10^{-3}\int_{0}^{\Delta
l}\left(2.32\cdot10^{-3}+\frac{1}{1+4.7T(x)}\right)^{2}\frac{dx}{L_{rad}},
\end{equation}
where $T(x)$ is the kinetic energy of the $\mu$-meson, GeV. The
degree of depolarization of $\mu$-mesons passing through the layer
of lead can reach $10\%$, while for  media with a small atomic
numbers it is as low as a fraction of a percent. As for electrons,
the approach developed here is only applicable provided that
$\Delta l\ll L_{rad}$, $T\gg 15\sqrt{\frac{\Delta l}{L_{rad}}}$
MeV. And
\begin{equation}
\label{27.16}
\eta_{e\parallel}=220\left(2.32\cdot10^{-3}+\frac{1}{T}\right)^{2}\frac{\Delta
l}{L_{rad}},
\end{equation}
At the energies $T<10\sqrt{\frac{\Delta l}{L_{rad}}}$ electrons
become completely depolarized.

%%%%%%%%%%%%%%%%%%%%%%%%%%%%%%%%%%%  Section 28 %%%%%%%%%%%%%%%%%%%%%%

\section{Oscillations of Polarization of a  Fast Channeled Particle Caused by its Quadrupole Moment}
\label{sec:9.28}

In (\ref{sec:9.25}) we considered the effect of spin rotation of a
relativistic particle channeled in a non-magnetic bent crystal,
which is  caused by the action of the crystal electric field (also
responsible for the rotation of a particle momentum) on tits
dipole magnetic moment. It turns out that for particles (nuclei,
ions) with spin $I\geq 1$ the presence of multipole moments, first
of all, the quadrupole one, results in spin rotation even at
motion in a straight channel \cite{148}.

%%%%%%%%%%%%%%%%%%

First consider a nonrelativistic particle  with the quadrupole
moment $Q$. In view of the quasi-classical character of its motion
in the channel, we may write the equation of motion for its moment
as follows:
\begin{equation}
\label{28.1}
\frac{d\hat{I}_{i}}{dt}=\frac{e}{3\hbar}\varepsilon_{ikl}\varphi_{kn}\hat{Q}_{ln}\,,
\end{equation}
where $\hat{I}_{i}$ is the operator of the particle spin
projection;
\[
\hat{Q}_{ln}=\frac{3Q}{2I(2I-1)}
\left\{\hat{I}_{ln}-\frac{2}{3}I(I+1)\delta_{ln}\right\}
\]
is the operator of its quadrupole moment,
$\hat{I}_{ln}=\hat{I}_{l}\hat{I}_{n}+\hat{I}_{n}\hat{I}_{l}$;

\[\varphi_{ln}=\frac{\partial^{2}\varphi}{\partial x_{l}\partial
x_{n}}
\]
is the the second-derivative of the electrostatic potential of the
channel at the point of particle location; $\varepsilon_{ikl}$  is
the totally antisymmetric unit tensor. It is essential that due to
the Lorentz factor compensation through relativistic
transformation of $\varphi_{ik}$ and $t$,  equations (\ref{28.1})
are applicable for a relativistic channeled particle as well. The
change in polarization over the unit length of the particle flight
is energy-independent.
%%%%%%%%%%%%%%%%%5555
In the case of a particle moving near the channel center, it is
possible to employ the harmonic approximation for $\varphi$. Here
the quantities $\varphi_{ik}$ do not depend on the coordinates,
and the solution of (\ref{28.1}) simplifies considerably. So, for
a particle with spin $I=1$ moving in the direction of the $z$-axis
($x$, $y$  are the principal axes of the tensor $\varphi_{ik}$),
we get
\begin{eqnarray}
\label{28.2}
\hat{I}_{x}(t)&=&\hat{I}_{x}(0)\cos\alpha\omega t+\hat{I}_{yz}(0)\sin\alpha\omega t\,,\nonumber\\
\hat{I}_{y}(t)&=&\hat{I}_{y}(0)\cos\omega t-\hat{I}_{zx}(0)\sin\omega t\,,\nonumber\\
\hat{I}_{z}(t)&=&\hat{I}_{z}(0)\cos(1-\alpha)\omega
t+\hat{I}_{xy}(0)\sin(1-\alpha)\omega t\,,
\end{eqnarray}
\begin{eqnarray}
\label{28.3}
\hat{I}_{xx}(t)&=&\hat{I}_{xx}(0),\,\hat{I}_{yy}(t)=\hat{I}_{yy}(0),\,\hat{I}_{zz}(t)=\hat{I}_{zz}(0),\nonumber\\
\hat{I}_{xy}(t)&=&\hat{I}_{xy}(0)\cos(1-\alpha)\omega t-\hat{I}_{z}(0)\sin(1-\alpha)\omega t\,,\nonumber\\
\hat{I}_{yz}(t)&=&\hat{I}_{yz}(0)\cos\alpha\omega t-\hat{I}_{x}(0)\sin\alpha\omega t\,,\nonumber\\
\hat{I}_{zx}(t)&=&\hat{I}_{zx}(0)\cos\omega
t-\hat{I}_{y}(0)\sin\omega t\,,
\end{eqnarray}
where $\omega=\frac{eQ}{2\hbar}\varphi_{xx}$;
$\alpha=\frac{\varphi_{yy}}{\varphi_{xx}}$. From relations
(\ref{28.2}), (\ref{28.3}) follows that when a fully polarized
particle with  spin directed along the z-axis enters a crystal,
the mean values of the following projections of spin  and
quadrupolarization  will change with time
\begin{equation}
\label{28.4} \langle\hat{I}_{z}(t)\rangle=\cos(1-\alpha)\omega
t\,,\quad
\langle\hat{Q}_{xy}(t)\rangle=-\frac{3}{2}Q\sin(1-\alpha)\omega
t\,.
\end{equation}
Similarly, at the initial polarization:
%%%%%%%%%%%%%%%%%%%%%%%%%%%%%%%%%%%%%%%%%%%55
(a) in the $x$-direction
\begin{equation}
\label{28.5} \langle\hat{I}_{x}(t)\rangle=\cos\alpha\omega
t\,,\quad
\langle\hat{Q}_{yz}(t)\rangle=-\frac{3}{2}Q\sin\alpha\omega t\,,
\end{equation}

(b) in the $y$-direction
\begin{equation}
\label{28.6} \langle\hat{I}_{y}(t)\rangle=\cos\omega
t\,,\quad\langle\hat{Q}_{zx}(t)\rangle=\frac{3}{2}Q\sin\omega t\,.
\end{equation}

Thus, polarization of the particle with $Q\neq 0$ moving in a
straight channel undergoes oscillations as the particle advances
into the target. Here in the case of the transverse initial
polarization of the particle at $\alpha=1$ linear oscillations
occur, and  spin rotation only appears at $\alpha\neq 1$, i.e., at
the asymmetry of the channel field.
%%%%%%%%%%%%%%%%%%%%5
Estimate the magnitude of the effect. In axial channeling of a
positively charged particle, the values of field inhomogeneity can
be as large as of the order of $10^{18}$\,V/cm$^{2}$. In this
case, for a  bare nucleus ($Q\sim 10^{-24}$\,cm$^{2}$) $\omega\sim
10^{9}$\,s$^{-1}$, i.e., the polarization can change by $1\%$ over
the path length  of about 1 cm. For an ion  passing through a
crystal, due to antishielding, the effective field on the nucleus
may increase by several orders of magnitude. As a result, the
change in the polarization can  increase by several orders of
magnitude.

A negatively charged elementary particle, for example, an
$\Omega$-hyperon in  the case of channeling will move inside the
atomic layer or in the region of the nuclear tube along  the
crystallographic axis. Here the electric fields (and  their
inhomogeneities) are considerably higher than in the interplanar
channel, so the appreciable rotation of spin may occur even  at
quite small values of $Q$. For example, for a nuclear tube in lead
$\varphi_{xx}\sim 10^{20}$\,V/cm$^{2}$ and the value of
$\omega\sim 10^{9}$\,s$^{-1}$ is attained at $Q\sim
10^{-26}$\,cm$^{2}$. The measurement of the polarization of
$\Omega^{-}$ under such conditions may provide unique information
about the hyperon quadrupole moment.

Note also that in a bent channel, for a positively charged
particle the magnitude of the spin rotation due to  the quadrupole
moment, generally speaking, should grow at the cost of trajectory
displacement closer to the atomic plane. However, in this case
spin rotation due to magnetic moment should be simultaneously
taken into account.

%%%%%%%%%%%%%%%%%%%%%%%%%%%%%%%%%%%  Section 29 %%%%%%%%%%%%%%%%%%%%%%

\section{Radiative Self-Polarization of Spin of Fast Particles in Crystals}
\label{sec:9.29}

Let a particle move in a channel bent with the radius of curvature
$R$ around the z-axis. The particle motion along the curved path
in such a channel means that here the particle is affected by the
electric field $\varepsilon$ perpendicular to the particle
momentum. Therefore  the particle  in its rest frame is affected
by the magnetic filed $H=\gamma\varepsilon$ directed along the
z-axis, where $\gamma$ is the particle Lorentz factor. In the
magnetic field spin undergoes radiative transitions between the
states with different spin projections on the the field direction.
These spontaneous transitions lead to accumulation of particles at
a lower energy level, i.e. to the beam polarization along the
$z$-axis, if it has not been polarized when entering the crystal
\cite{9}.

%%%%%%%%%%%%%%%%%%%%%%%%%%%%

%%%%%%%%%%%%%%%%%%%%%%%%%%%%%%55
A detailed description of the self-polarization effect can be
given, using the equation for spin motion in an external
electromagnetic field with due account of radiative damping
\cite{11}. Suppose that the crystal is non-magnetic. In this case
the spin polarization vector of a particle satisfies the equation
of the form (compare with \cite{149}, p. 204)
\begin{equation}
\label{29.1} \frac{d\vec{\zeta}}{dt} =
\frac{e}{m}\left(\frac{\mu^{\prime}}{\mu_{0}}+\frac{1}{1+\gamma}\right)[\vec{\zeta}[\vec{v}\vec{E}]]
 -T^{-1}
\left(\vec{\zeta}-\frac{2}{9}\vec{v}(\vec{v}\vec{\zeta})+\frac{8}{5\sqrt{3}}\frac{[\vec{v}\vec{w}]}{|\vec{w}|}\right)\,,
\end{equation}
where $\mu^{\prime}$ is the anomalous part of the magnetic moment
(it depends on the particle energy); $\mu_{0}$ is the Bohr
magneton;
$T^{-1}=5\sqrt{3}\alpha\hbar^{2}\gamma^{5}|\vec{w}|^{3}/8m^{2}$ is
the damping constant; $\alpha=1/137$; $c=1$ is the velocity of
light; $\vec{v}$ is the particle velocity; $m$ is its mass;
$\vec{w}$ is the acceleration. The augend in (\ref{29.1})
describes the effect of spin rotation in a bent crystal (see
(\ref{sec:9.25})). The addend leads to the effect of radiative
polarization of the beam.

Consider the projection of the polarization vector $\vec{\zeta}$
on the $z$-axis, around which the crystal is bent. If the particle
undergoes planar channeling around the $z$-axis in $xy$-plane,
then the first term on the right--hand side of (\ref{29.1})) has a
zero projection onto the $z$-axis, i.e.,
\begin{equation}
\label{29.2}
\frac{d\zeta_{z}}{dt}=-T^{-1}\zeta_{z}-8(5\sqrt{3}T)^{-1}[\vec{v}\times\vec{w}]_{z}|\vec{w}|^{-1}\,.
\end{equation}
The solution of this equation has the form
\begin{eqnarray}
\label{29.3}
\zeta_{z}(t)&=&\zeta_{z}(0)\exp\left(-\int_{0}^{t}T^{-1}(t^{\prime})dt^{\prime}\right)-8(5\sqrt{3})^{-1}\nonumber\\
&
&\times\exp\left(-\int_{0}^{t}T^{-1}(t^{\prime})dt^{\prime}\right)\int_{0}^{t}dt^{\prime}T^{-1}(t^{\prime})
\frac{[\vec{v}\times\vec{w}]_{z}}{|\vec{w}|}\nonumber\\
&
&\times\exp\left(\int_{0}^{t}T^{-1}(t^{\prime\prime})dt^{\prime\prime}\right)\,.
\end{eqnarray}
%%%%%%%%%%%%%%%%%%%%

Generally speaking, the particle trajectory in the channel is
known. For instance, in a bent planar channel
$x=\rho(t)\cos\Omega_{b}t$, $y=\rho(t)\sin\Omega_{b}t$. When the
potential is harmonic $k(\rho-R)^{2}/2$,
$\rho(t)=\rho_{0}+\alpha_{1}\cos(\Omega t+\delta)$, where
$\rho(t)$ is the radius of the particle orbit;
$\Omega_{b}=c/\rho_{0}$ is the rotation frequency in a bent
crystal; $\Omega$ is the oscillation frequency in the channel;
$\alpha_{1}$ is the oscillation amplitude; $\delta$ is the initial
phase; $\rho_{0}$ is the radius of curvature of the particle
equilibrium trajectory in the channel. The value of displacement
of the  particle equilibrium trajectory from the channel center
$\Delta=m\gamma/kR$ is limited by the channel width $\Delta<d/2$.

%%%%%%%%%%%%%%%%5
In the case when $\alpha_{1}/\Delta\ll 1$, the acceleration equals
$1/R$ with high precision. As a result,
\begin{equation}
\label{29.4}
\zeta_{z}(t)=\zeta_{z}(0)e^{-t/T_{0}}+8(5\sqrt{3})^{-1}(1-e^{-t/T_{0}})\,,
\end{equation}
where
$T_{0}^{-1}=(5\sqrt{3}/8)\alpha(\hbar\gamma/m)^{2}(\gamma/R)^{3}$.
From (\ref{29.4}) follows that at times $t>T_{0}$, the value of
$\zeta_{z}=8(5\sqrt{3})^{-1}\simeq 0.924$ irrespective of the
value of the initial polarization (i.e., the beam appears to be
polarized along the crystal bending axis $z$). For example, at
channeling of positrons with the energy of 100\,GeV and $R\sim
12$\,cm, the polarization length $T_{0}$  in $(110)$ channel of a
single crystal of  tungsten  is approximately 1 cm. The estimates
show that in the case of axial channeling of electrons with the
energy of 50\,GeV and $R\sim 10$\,cm the same polarization length
may be attained even in single crystals of relatively light
elements (e.g. silicon). The  length of self-polarization
decreases rapidly with the growth of particle energy. Note also
that  over the length $T_{0}$  the particle emits  one photon,
i.e., the intensity of this type of radiation is very high.

%%%%%%%%%%%%%%%5

At $\alpha_{1}/\Delta>1$ at the exit from the crystal the
magnitude of the projection of the polarization vector
$\zeta_{z}(t)$ as a function of crystal thickness is the sum of a
non-oscillating and oscillating with frequency $\Omega$ terms.
Upon averaging over the initial state of the beam, only
non-oscillating part remains, which vanishes with the increase in
$\alpha_{1}/\Delta$. The aforesaid means that even in the limiting
(extreme) case of an undeformed crystal ($R\rightarrow \infty,
\,\Delta\rightarrow 0$ for the given particle trajectory there
appears a nonzero projection of the polarization vector
$\zeta_{z}(t)$ oscillating with frequency, which vanishes after
averaging over all the initial points of particle entrance into
the crystal. However, the intensity of electromagnetic radiation
accompanying radiation polarization of channeled particles will be
high in this case too.

As was mentioned above, the anomalous magnetic moment  depends on
the particle energy. In the case of channeling of charged
particles, the parameter $\chi=e\hbar\varepsilon\gamma/m^{2}$ (see
\cite{149}) may be of the order of unity and greater. Therefore
the  phenomenon of spin precession of charged particles described
above opens up possibilities for experimental investigation of the
dependence of  radiative corrections on particle energy.

It should be emphasized that with $\chi$ approaching unity in the
spin-flip process, a very hard quantum is emitted. Therefore if in
the experiment the electrons are selected by energy as well, then
the degree of polarization of the beam will turn out to be higher.
In this case the theory based on equation (\ref{29.1}) is not
suitable. The process may be studied, for example, using the
density matrix formalism.

%%%%%%%%%%%%%%%%%%%%%%%%%%%%%%%%%%%%%%%%%%%%%%%%555

%

%%%%%%%%%%%%%%%%%%%%%%%%%%%%%%%%%%%  Chapter 10 %%%%%%%%%%%%%%%%%%%%%%

\chapter[The Influence of Radiative Transitions on Particles Channeling in Crystals]{The Influence of Radiative Transitions on Channeling of Charged Particles in Crystals}
\label{ch:10}

%%%%%%%%%%%%%%%%%%%%%%%%%%%%%%%%%%%  Section 30 %%%%%%%%%%%%%%%%%%%%%%

\section[Particle Lifetime at the Transverse Motion Level]{Particle Lifetime at the Transverse Motion Level}
\label{sec:10.30}

Radiative transition of a channeled particle from one level to
another is accompanied by the change in its energy and momentum.
Therefore one should expect that such transitions  may affect the
character of particle motion in a crystal. In particular, the
redistribution of the initial population of transverse motion
levels, which will influence the beam divergence at the crystal
exit \cite{44,150a,150b}.

Classical theory of the influence of electromagnetic radiation on
the motion of channeled particles was first given by
Bonch-Osmolovsky and Podgoretsky in \cite{151,152}, quantum theory
- by Grubich and the author in \cite{153,154}. For example, as
shown in \cite{151,152,153,154}, development of electromagnetic
cascade  in a crystal is possible in the length considerably
smaller than the radiation length. A similar conclusion was later
made in \cite{155}.

The possibility in principle to change the angular divergence of a
beam under channeling conditions is due to the fact that different
quasi-classical transverse momentum corresponds to different
levels of transverse motion.

To estimate the rate of the process in question, let us first find
the radiation width of excited levels in the model of a
rectangular well \cite{44,150a,150b}. Let us pass to the
coordinate system with a zero longitudinal particle momentum. In
this case the particle moves between two barriers of height
$u^{\prime}-\gamma u$ ($\gamma=E/m$, $u$ is the height of the
potential). To determine the radiation length, apply the dipole
approximation:
\begin{equation}
\label{30.1}
\Gamma_{n}=\sum_{nn^{\prime}}\frac{4e^{2}\omega_{nn^{\prime}}^{3}}{3}|x_{nn^{\prime}}|^{2},
\end{equation}
where
$\omega_{nn^{\prime}}=\frac{\pi}{2ma^{2}}(n^{2}-n^{\prime^{2}})$
is the transition frequency; $x_{nn^{\prime}}$ is the matrix
element of the coordinate for the transition between the states
$n$ and $n^{\prime}$. For a rectangular potential well of the
channel the matrix element has the form
\begin{equation}
\label{30.2}
x_{nn^{\prime}}=-\frac{2a}{\pi^{2}}\frac{4nn^{\prime}}{(n^{2}-n^{\prime^{2}})^{2}}.
\end{equation}
As a result, the expression for the radiation width of the level
$n$, and correspondingly, for the lifetime at the level $\tau_{n}$
in the lab system may be written as follows:
\begin{equation}
\label{30.3}
\frac{1}{\tau_{n}}=\frac{A}{\gamma}\sum_{n^{\prime}<n}\frac{n^{\prime^{2}}n^{2}}{n^{2}-n^{\prime^{2}}},
\end{equation}
where $A=\frac{32}{3}\frac{e^{2}\pi^{2}}{m^{3}a^{4}}$.

Taking into account that in the high-energy range there are many
levels $\left(n_{max}\sim\left(\frac{2ma^{2}}{\pi^{2}}\gamma
u\right)^{1/2}\right)$ in a well, to obtain accurate enough
estimate, in equation (\ref{30.3}) one may substitute summation
for integration, which yields the expression
\begin{equation}
\label{30.4} \frac{1}{\tau_{n}}\simeq\frac{A}{2\gamma}n^{3}\ln 2n.
\end{equation}
From (\ref{30.4}) follows that the lifetime for a particle with
maximum probability of residing at level $n_{max}$ (this
corresponds to the particle incident on a crystal at the Lindhard
angle $\vartheta_{L}$) is
\begin{equation}
\label{30.5}
\frac{1}{\tau_{max}}\simeq\frac{e^{2}}{a\gamma}\left(\frac{\gamma
u}{m}\right)^{3/2}\ln ma^{2}u\gamma.
\end{equation}

Thus, from (\ref{30.5}) follows that, e.g., for a positron of
energy $E\sim 1$ GeV, the length over which  the level population
will drop by a factor of $e$ is $l_{n\, max}\sim n\sim
10^{-3}-10^{-2}$ cm. As also seen from (\ref{30.4}) the length
$l_{n}$ should grow with the decrease in  $n$. For instance, for
particles entering the crystal at the angle one-tenth as large as
the Lindhard angle, $n=n_{max}/10$ and $l_{n}=l_{max}\cdot
10^{3}\sim1$ - 10 cm.

Let the angular divergence of the beam incident on the crystal be
$\vartheta\sim 10^{-4}$ rad, i.e., of the same order of magnitude
as the critical angle for channeling. In this case for positrons,
in fact, all the levels in the potential well of the channel are
populated. According to (\ref{30.4}), (\ref{30.5}) after a beam of
positrons of energy $E\sim 1$ GeV passes through  a single crystal
with the thickness $L\sim 0.1-1$ cm, one should expect an order of
magnitude decrease in the angular divergence of the beam.

With the influence of multiple scattering on the beam evolution in
the channel ignored, the particle distribution over the levels can
be analyzed relatively simply. Let us assume that the initial
population over the levels is equally probable. The calculation
(Fig. 8) will be carried out using the kinetic equation of the
form \cite{156}
$$
\frac{\partial\rho_{n}}{\partial t}
=-\sum_{n^{\prime}<n}W_{nn^{\prime}}\rho_{n}+\sum_{n^{\prime}>n}W_{nn^{\prime}}\rho_{n^{\prime}}
$$

\bigskip

\textit{Figure 8. The change in the population of the transverse
motion levels for a particle passing 0.1-cm-thick crystal target.}

\bigskip

The expression for $W_{nn^{\prime}}$ is obtained from formula
(\ref{30.3}) if summation over $n^{\prime}$ is ignored, i.e.,
$$
W_{n^{\prime}n}=\frac{A}{2\gamma}\frac{n^{2}n^{\prime^{2}}}{n^{2}-n^{\prime^{2}}}.
$$

Interestingly enough, to obtain the above estimates, the
difference between the real potential of the channel and the
harmonic one appears to be of principal importance. In a harmonic
well the lifetime at the excited level is easy to find from the
classical formula for radiative damping
$\tau=\left(\frac{2}{3}\frac{e^{2}\Omega^{2}}{m}\right)^{-1}(\hbar=c=1)$.
According it for the length over (in) which the level population
will reduce by a factor of $e$, we get the estimate $l\sim 10$ cm,
i.e., in the case of harmonic potential the phenomenon of
radiative cooling is practically absent \cite{156}.

It should be noted, however, that the obtained estimates of the
the radiative cooling rate do not take into account the processes
leading to the increase in the magnitude of the transverse
momentum of the channeled particle, such as, for example, multiple
scattering in the channel. Allowing for multiple scattering can
appreciably affect the features of the motion of a channeled
particle. Consider this process in more detail.\footnote{The
results presented in (\ref{sec:10.31}) and (\ref{sec:10.32}) were
obtained together with A.O.Grubich.}

%%%%%%%%%%%%%%%%%%%%%%%%%%%%%%%%%%%  Section 31 %%%%%%%%%%%%%%%%%%%%%%

\section[Classical Theory of Channeling of Charged Particles with Due Account of Radiation Energy Losses]
{Classical Theory of Channeling of Charged Particles with Due
Account of Radiation Energy Losses} \label{sec:10.31}

As the number of the transverse energy levels of a channeled
ultra-relativistic charged particle moving in a potential well
formed by the crystal axes (planes) is great, we shall use the
classical theory as the first step towards the description of the
particle motion.  In classical thermodynamics  the equation of
motion of a charged particle  in an external field with the
account of radiation slowdown has the form \cite{68}
\begin{equation}
\label{31.1} \frac{d\vec{\rho}}{dt}=\vec{F}_{L}+\vec{F}_{rad},
\end{equation}
where $\vec{\rho}=m\gamma\vec{v}$ is the particle momentum;
$F_{L}=e\left(\vec{\varepsilon}+\frac{1}{c}[\vec{v}\vec{H}]\right)$
is the Lorentz force;
\begin{equation}
\label{31.2}
F_{rad}=\frac{2e^{2}\gamma^{2}}{2c^{3}}\left[\vec{w}+\frac{\vec{v}(\vec{v}\vec{w})}{c^{2}}\gamma^{2}+
\frac{3\vec{w}(\vec{v}\vec{w})}{c^{2}}\gamma^{2}+\frac{\vec{v}(\vec{v}\vec{w})^{2}}{c^{4}}\gamma^{4}\right]
\end{equation}
is the radiative friction force; $\vec{w}=d\vec{v}/dt;
\vec{w}=d\vec{w}/dt$; $\gamma$ is the Lorentz factor. Using a well
known formula of relativistic dynamics \cite{68}
\begin{equation}
\label{31.3}
m\gamma\vec{w}=\vec{F}-\frac{1}{c^{2}}(\vec{v}\vec{F})\vec{v};\,
\vec{F}=\vec{F}_{L}+\vec{F}_{rad},
\end{equation}
enables one to write equation (\ref{31.1}) in the form convenient
for further analysis $(c=1)$:
\begin{equation}
\label{31.4}
\vec{w}-\frac{\vec{F}_{L}-\vec{v}(\vec{v}\vec{F}_{L})}{m\gamma}=
\frac{2}{3}r_{e}\gamma[\vec{w}+3\gamma^{2}\vec{w}(\vec{v}\vec{w})];
\end{equation}
\begin{equation}
\label{31.5}
\gamma-\frac{\vec{F}_{L}-\vec{v}}{m}=\frac{2}{3}r_{e}\gamma^{4}\vec{v}[\vec{w}+3\gamma^{2}\vec{w}(\vec{v}\vec{w})];
\end{equation}
where $r_{e}=e^{2}/m$ is the classical electron radius. In a
non-magnetic crystal $\vec{F}_{L}=-\nabla u$, where $u$ is the
potential energy of particle interaction with the crystallographic
axes (planes), which is averaged over thermal vibrations of the
crystal lattice. Recall that equation (\ref{31.5}) is the
corollary to equation (\ref{31.4}) and  formula
$\gamma=(1-v^{2})^{-1/2}$ ,and it may be written as, for example,
\begin{equation}
\label{31.6} \dot{\gamma}-\gamma^{3}\vec{v}\vec{w}.
\end{equation}

It is almost impossible to solve equation (\ref{31.1}) without
using numerical methods. Therefore let us dwell on the
simplifications that are may be realized in the original
equations.

As is known, equation (\ref{31.1}) is applicable when in one of
the reference frames $F_{rad}\ll F_{L}$. Therefore instead of
equation (\ref{31.4}) an approximate equation is usually used,
which is obtained by substitution into the right-hand side of
equation (\ref{31.4}) of the particle acceleration expressed in
terms of an external electromagnetic field acting on a particle:
\begin{equation}
\label{31.7}
\vec{w}^{(0)}=[\vec{F}_{L}-\vec{v}(\vec{v}\vec{F}_{L})]/m\gamma.
\end{equation}
The condition of smallness of the radiative friction force
$\vec{F}_{rad}$ as compared with the external force affecting the
charge $\vec{F}_{L}$ has the form \cite{68}
\begin{equation}
\label{31.8} \gamma\ll\gamma_{quant}=\frac{m}{r_{e}\vec{F}_{L}}.
\end{equation}

However, classical electrodynamics becomes unsuitable due to the
production of electron-positron pairs in the external
electromagnetic field yet in the range of  energies (see
\cite{68}, p.267)
\begin{equation}
\label{31.9} \gamma\sim\gamma_{S}=\gamma_{quant}/137,
\end{equation}
at which the external field $\varepsilon^{\prime}$ acting on the
particle in the instantaneous rest frame
($\varepsilon^{\prime}=\gamma\varepsilon$) attains the value of
the Schwinger field $\varepsilon_{quant}=m^{2}/e\hbar$.

In this regard it is interesting that when the condition
\begin{equation}
\label{31.10}
\gamma\geq\gamma_{sa}=\frac{m^{2}|(\vec{v}\vec{\nabla})\vec{F}_{L}|}{2r_{e}|F^{3}_{L}|}
\approx\frac{2m^{2}v_{\perp}}{r_{e}F^{2}_{L}d}
\end{equation}
is fulfilled, in the right-hand side of equations (\ref{31.4}) and
(\ref{31.5}) in the second order perturbation theory, the terms
leading to the particle "self-acceleration" prevail. The magnitude
of the Lorentz factor $\gamma_{sa}$ appears to be of the same
order of the magnitude as $\gamma_{S}$ (implying qualitative
estimates we assumed that $|(\vec{v}\vec{\nabla})\vec{F}|\approx
v_{\perp}4F_{L}/d$, where $d$ is the channel width).\footnote{When
deriving inequality (\ref{31.10}) it was taken into account that
the velocity of the channeled particle is directed at a small
angle with the crystallographic axes (planes) forming the channel
$|v_{\perp}|\ll 1$, as well as the fact that the Lorentz force
acting on the particle in the channel is transverse.}. Indeed,
assuming that $v_{\perp}\approx (F_{L}x/m\gamma)^{1/2}$, we obtain
$$
\gamma_{sa}\approx\gamma_{S}\alpha^{-1}(r_{e}x)^{1/3}(d/2)^{-2/3},
$$
where $x$ is the amplitude of the particle vibrations in the
channel; $\alpha=1/137$ . Thus, at
$x=d/4\gamma_{sa}\approx\gamma_{S}\alpha^{-1}(r_{e}/d)^{1/3}\sim\gamma_{S}$.

One might suppose that the application of the Dirac-Lorentz
equation in the energy range $\gamma\sim\gamma_{S}$ (due to
quantum effects), though not being quite correct, nevertheless may
give a correct qualitative pattern of the motion of particles with
the energy $\gamma\sim\gamma_{S}$ channeled in the crystal.
However, according to the estimates obtained, application of the
classical equation of motion in the case of a conventionally used
approximate equation (\ref{31.4}) with the radiative force
$\vec{F}_{rad}=\vec{F}_{rad}(\vec{w}^{(0)})$ in the range of
energies $\gamma\approx \gamma_{S}$ is quite problematic.

In the range of energies $\gamma\ll \gamma_{S}$ the principal
terms in right-hand sides of equations (\ref{31.4}) and
(\ref{31.5}) are those proportional to the derivative of the
particle acceleration $\vec{w}$. As a consequence, to solve the
problem, one may use the approximate equations
\begin{equation}
\label{31.11}
\vec{w}-\vec{w}^{(0)}=\frac{2}{3}r_{e}\gamma\dot{\vec{w}};
\end{equation}
\begin{equation}
\label{31.12}
\dot{\gamma}-\frac{\vec{F}_{L}\vec{v}}{m}=2r_{e}\gamma^{6}(\vec{v}\vec{w})^{2}.
\end{equation}
When the crystal thickness is not very large and the
time-dependence of  factor $\gamma$ may be neglected, for the
simplest forms of the potential $u$ the solution of the equation
of motion (\ref{31.11}) may be found explicitly. For example, at
planar channeling in a harmonic potential $u(x)=kx^{2}/2$ the
approximate solution of equation (\ref{31.11}) for particle
transverse vibrations in the channel has the form
\begin{equation}
\label{31.13} x(t)=x_{m}\cos(\Omega t+\varphi)e^{-t/\tau},
\end{equation}
where $\Omega^{2}=k/m\gamma$; $\varphi$ is the initial phase;
$\tau=3m/r_{e}k$ .

The solution is similar for the case of axial channeling in a
two-dimensional harmonic potential $u(\vec{\rho})=k\rho^{2}/2$
($\vec{\rho}$ is the radius-vector of the particle in the plane
perpendicular to the crystallographic axes which form axial
channels).

Substitution into right-hand sides of equations (\ref{31.11}),
(\ref{31.12}) of the quantity $\vec{w}$ corresponding to the
zero-order approximation (of)(\ref{31.7}) gives approximate
equations of the form \cite{151,152}:\footnote{The equation of
motion for a longitudinal component of the radius-vector is not
presented, as we are mainly concerned with the transverse motion
in the channel.}
\begin{equation}
\label{31.14}
\vec{w}_{\perp}-\frac{\vec{F}_{L}}{m\gamma}=\frac{2}{3}\frac{r_{e}}{m}(\vec{v}\vec{\nabla})\vec{F}_{L};
\end{equation}
\begin{equation}
\label{31.15}
\dot{\vec{\gamma}}=\frac{\vec{F}_{L}\vec{v}}{m}-\frac{2}{3}\frac{r_{e}}{m^{2}}\gamma^{2}F^{2}_{L};
\end{equation}
Harmonic potential is often used when considering planar
channeling of positively charged particles. Since the real
potential may contrast sharply with the harmonic one, we shall
dwell on the quantitative comparison of the features of particle
motion in different potentials.

A thorough a review of different model potential is given in
\cite{3}. In particular, it is shown that in the case of
channeling of positively charged particles  the potentials of the
channels formed by the planes $(110)$ of a single crystal of
silicon are well described by the harmonic potential (see
\cite{3}, Fig.9).

Let us consider how a harmonic potential approximates  planar
channels of single crystals of other chemical elements (Fig. 9).

\bigskip

\textit{Figure 9. The potentials of single crystals: planar
channels (solid curves), harmonic channels (dashed curves).}

\bigskip

The potentials in Fig 9. are depicted in the space region from the
channel center to the point located at the distance equal to the
shielding radius $a$ from the equilibrium position of atoms of the
crystallographic plane, forming the channel wall. The curves are
calculated from the formulae
$$u_{p}^{L}(x)=u(x)-u(0);\,\, u(x)=u^{L}(x+d_{p}/2)+u^{L}(x-d_{p}/2),
$$
where the Lindhard potential is
$$
u^{L}(x)=2\pi
nze^{2}ad_{p}\left[\left(\frac{x^{2}}{a^{2}}+3\right)^{1/2}-\frac{x}{a}\right]
$$
($n$ is the density of atoms in the crystal; $z$ is the nucleus
charge; $a$ is the shielding radius; $d_{p}$ is the distance
between the planes). At point $x=d_{p}/2-a$ harmonic potentials
$u_{p}(x)=kx^{2}/2$ are equal to the potential $u_{p}^{L}(x)$.

Interestingly enough, the elasticity constant $k_{L}$ found from
the equality
$$
k_{L}(d_{p}/2-a)^{2}=u_{p}^{L}(d_{p}/2-a)
$$
is well described  by the quantity
\begin{equation}
\label{31.16} \tilde{k}=4\pi n_{e}e^{2},
\end{equation}
used by Bonch-Osmolvsky and Podgoretsky \cite{151,152} (here
$n_{e}$ is the electron density in the central part of the
channel). The magnitudes of the attenuation length
$\Lambda=3mc^{2}/kr_{e}$, calculated  with the help of the
elasticity coefficients $k_{L}$ and $\tilde{k}$ (electron density
$n_{e}=nz/2.72)$ are given in the Table. The accepted expression
for $n_{e}$ corresponds to the uniform distribution of the crystal
electrons in the central part of the channel with the density
which is by a factor of $e$ smaller than the mean electron density
$\bar{n}_{e}=zn$.\footnote{Henceforth the model of the harmonic
potential $u_{p}(x)$ with the elasticity constant (\ref{31.16} is
used more than once for quantitative assessments; by the channel
width $d$ we shall mean $(d_{p}-2a)$.}
\begin{table}
\label{table:} \caption{}
\begin{center}
\begin{tabular}{|c|c|c|c|}
\hline $ Crystal $ & $ Channel $ & $\lambda_{L} cm$
& $ \lambda cm $\\
  \hline
Si & (100) &  8.8 & -   \\
Si & (110) & 11.64 & 11.86 \\
Ge & (100) & 5.6   & 5.88 \\
Cu & (110) & 2.92 &3.38  \\
  \hline
\end{tabular}
\end{center}
\end{table}

The phase trajectory of the transverse motion of a positron  (Fig.
10, curve 1) channeled in $(110)$ channel of the single crystal of
silicon can be found by means of numerical solution of the system
of equations (\ref{31.14})-(\ref{31.15}) in the Moliere potential.
The phase trajectory of the particle moving in the Moliere
potential  is only slightly different from a circle which is the
phase trajectory of the harmonic motion (curve 2).

\bigskip

\textit{Figure 10. The phase trajectory of the transverse motion
of a positron: 1. The trajectory of a particle moving in the
Moliere potential; 2 - the trajectory of the harmonic motion.}

\bigskip

In the case of planar channeling of positrons, the value of
$R=d_{p}/2a$ serves as a criterion for applicability of the
harmonic potential. The smaller the value of $R$, the better the
harmonic potential approximates the channel potential in the range
$|x|\leq d_{p}/2-a$. So, for the elements given in the Table we
have: in silicon for the channel $(100)$ $R\simeq 3.5$, for the
channel $(110)$ $R\simeq 4.9$; in tungsten for the channel $(111)$
$R\simeq 4.1$, for the channel $(100)$ $R\simeq 9.9$; the harmonic
approximation in this case appears to be of little use.

Now consider the change in the total energy and the attenuation of
the transverse velocity. In the initial stage of motion the losses
of the particle total energy $\varepsilon(L)=E_{0}-E(L)$ and the
attenuation of the amplitude of the particle transverse velocity
$\theta_{m}$, are of linear character due to the presence of the
radiative friction force  (Figure 11):
\begin{equation}
\label{31.17} \varepsilon(L)=\alpha_{\varepsilon}L;
\end{equation}
\begin{equation}
\label{31.18} \theta_{m}(L)/\theta_{m}(0)=1-\alpha_{\theta}L.
\end{equation}

\bigskip

\textit{Figure 11. Radiation energy losses as a function of
thickness.}

\bigskip

Note that at the energies used  in the experiments with channeled
particles until recently ($E\leq 10$ GeV), the linear laws
(\ref{31.17}), (\ref{31.18}) are valid, for example, in diamond
and silicon targets up to the crystal thicknesses as large as
several centimeters. Indeed, in the case of particle motion in a
harmonic potential it follows from equations (\ref{31.14}),
(\ref{31.15}) that if the conditions $\gamma_{0}\theta_{0}<1$, or
$t\ll \tau/\gamma_{0}^{2}\theta_{0}^{2}$,
$>\gamma_{0}\theta_{0}>1$ are fulfilled, corresponding to the
smallness of radiation energy losses $\varepsilon/E_{0}\ll 1$, the
particle trajectory is determined by expression (\ref{31.13}), and
the character of the changes in its total energy - by expression
\begin{equation}
\label{31.19}
\gamma(t)=\frac{\gamma_{0}}{1+(1-e^{-2t/\tau})\gamma_{0}^{2}\theta_{0}^{2}/2}
\end{equation}
where $\gamma_{0}$ is the Lorentz factor corresponding to the
initial energy of a particle $E_{0}$; $\theta_{0}=x_{0}\Omega_{0}$
is the initial amplitude of the particle transverse velocity in
the channel. From (\ref{31.19}) we obtain the below equality for
the coefficient $\alpha_{\varepsilon}$ at motion in a harmonic
potential
\begin{equation}
\label{31.20}
\alpha_{\varepsilon}=\frac{m\gamma_{0}^{3}\theta_{0}^{2}}{\tau}.
\end{equation}
the coefficient $\alpha_{\theta}=\tau^{-1}$. From (\ref{31.19}) is
also seen that  in the case $\gamma_{0}\theta_{0}<1$ the
approximate solution of (\ref{31.19}) is applicable for times
$t>\tau$ too.

Joint solution of equations (\ref{31.14}), (\ref{31.15}) for the
harmonic potential $u_{p}(x)$ was obtained in \cite{152}.
Therefore the coefficients $\alpha_{\varepsilon}$,
$\alpha_{\theta}$ can certainly be found from equations (62) and
(63) of \cite{152}. The above analysis shows that the range of
energies and crystal thicknesses, where (\ref{31.17}),
(\ref{31.18}) have linear solutions is quite broad; the correct
expression for the coefficient $\alpha_{\varepsilon}$ follows just
from the solution of (\ref{31.13}) and
(\ref{31.15})\footnote{Recall that at $\gamma\gg 1$ the augend on
the right-hand side of equation (\ref{31.15}) may be neglected.}
at the initial stage of motion
($t\ll\tau/\gamma_{0}^{2}\theta_{0}^{2}$)
$\alpha_{\theta}=\tau^{-1}$ at any values of
$\gamma_{0}^{2}\theta_{0}^{2}$ (compare \cite{152}).

To avoid possible misunderstandings, note that harmonic
approximation is not suitable in the cases when even slight
nonlinearity of the potential $u$ is of importance, for example,
in the case of resonance action of electromagnetic or ultrasonic
fields on channeled particles.

Now recall the presence of multiple scattering of a channeled
particle by the fluctuating part of the  potential of interaction
with the grating. Multiple scattering can be taken into account
by introducing into the right-hand side of equation (\ref{31.1})
of a random force $\vec{F}(\vec{\rho}, \vec{\theta}, t)$
describing the events of inelastic collisions between the particle
and the atoms of the crystal lattice. \footnote{The radius-vector
$\vec{\rho}$ describes the particle transverse motion in the
($x,y$) plane. The velocity
$\vec{v}_{\perp}\equiv\vec{\theta}=\vec{\rho}$.} It is known
\cite{157} that from the stochastic equation of motion one may go
over to the Einstein-Fokker equation for the probability  density
$w(\vec{\rho},\vec{\theta}, t)$ of finding the particle at moment
$t$ in the space region $(\vec{\rho},\vec{\rho}+d\vec{\rho})$ with
the velocity $\vec{\theta}$ which is a part of the interval
$(\vec{\theta},\vec{\theta}+d\vec{\theta})$. For the equation of
motion (\ref{31.14}) with the random force $\vec{F}(\vec{\rho},
t)$. The Einstein-Fokker equation has the form
\begin{eqnarray}
\label{31.21} \frac{\partial w}{\partial
t}+\vec{\theta}\frac{\partial w}{\partial
\vec{\rho}}+\frac{\partial}{\partial \vec{\theta}}
\left\{w\left[\frac{\vec{F}_{L}}{m\gamma}+\frac{2r_{e}}{3m}(\vec{v}\vec{\nabla})\vec{F}_{L}\right]\right\}\nonumber\\
=\sum_{ik}\frac{\partial^{2}{\cal{D}}_{ik}(\rho)w}{\partial\theta_{i}\partial\theta_{k}},
\end{eqnarray}
where
${\cal{D}}_{ik}(\vec{\rho})=c_{ik}(\vec{\rho},\vec{\rho})/2$;
$c_{ik}(\vec{\rho},\vec{\rho}^{\prime})\delta(t-t^{\prime})=\langle{\cal{F}}_{i}(\vec{\rho},t)\times{\cal{F}}_{k}(\vec{\rho}^{\prime},t)\rangle$;
${\cal{F}}_{k}(\vec{\rho},t)$ are normally distributed random
fields with zero mean values
$\langle{\cal{F}}_{i}(\vec{\rho},t)\rangle=0$.

Radiation energy losses described by equation (\ref{31.15}) are of
a continuous character. Therefore they may  be taken into account
upon passing from equation (\ref{31.21}) to the equation for the
probability density $w(\vec{\rho},\vec{\theta}, E, t)$ containing
in the right-hand part a differential term $m\frac{d\gamma w}{dE}$
describing the change in the number of particles in the energy
range $(E, E+dE)$. However, for a qualitative analysis of the
problem it is possible to use directly the set of two differential
equations (\ref{31.15}), (\ref{31.21}) (see, for example,
\cite{151,152}).

In addition to the electromagnetic radiation, generated by a
particle moving in a potential well $u$, a channeled particle also
emits $\gamma$ quanta  through scattering by a fluctuating part of
the interaction potential ("ordinary" bremsstrahlung). Large
straggling of the radiation energy losses is typical of such
bremsstrahlung with the Bethe-Heitler spectrum of the form
$\omega^{-1}$ \cite{98,102}. The quantitative theory in this case
should be based on the kinetic equation with the collision
integral describing  the  bremsstrahlung processes in the
right-hand side (see \cite{98}).  For not very thick crystals (for
example, those of silicon with the thickness of about 1 cm) the
usual bremsstrahlung loss may be ignored.\footnote{In \cite{158}
usual bremsstrahlung loss was taken into account by introducing
into the right-hand side of equation of the type of (\ref{31.6})
of a term equal to the magnitude of the average bremsstrahlung
energy loss per unit time, i.e. the approximation of the
continuous losses was used. Such a method of allowing for
bremsstrahlung loss is erroneous \cite{98}.}

In \cite{152} it is shown that when a charged particle moves in a
one-dimensional harmonic potential, it is possible to obtain from
equation of  type (\ref{31.21}) the closed systems of differential
equations  for the first- and second-order moments. The given
statement, generally speaking, is a particular case of the general
theorem holding for linear systems \cite{157}.

So, for a harmonic potential, a similar system of closed
differential equations may be obtained from the equation of motion
(\ref{31.11}) with a random force $\vec{F}(t)$. Here instead of
the system of three equations, we obtain a closed system of six
differential equations for the moments
$\langle\vec{\rho}^{(k)}\vec{\rho}^{(e)}\rangle$, where
$\vec{\rho}^{(k)}=\frac{d^{(k)}}{dt{(k)}}\vec{\rho}, k,l=0, 1,2$.

Below we shall dwell on the analysis of the system of differential
equations for two-dimensional moments
$\langle\vec{\rho}^{(k)}\vec{\rho}^{(l)}\rangle$ ($k, l=0, 1$)
which follows from the equation of motion (\ref{31.14}) with a
random force $\vec{F}(t)$ and a harmonic potential $u$. Implying
the qualitative analysis of the problem, we assume here that the
random force $\vec{F}(t)$ is independent of $\vec{\rho},
\vec{\theta}$. For the potential $u=k\rho^{2}/2$ the desired
equations are obtained from equations (28) of \cite{152} by a
simple substitution of one-dimensional moments $\langle
x^{2}\rangle$, $\langle \theta^{2}\rangle$, $\langle
x\theta\rangle$ for two-dimensional ones:
\begin{eqnarray}
\label{31.22}
\frac{d}{dt}\langle \rho^{2}\rangle=2\langle\vec{\rho}\vec{\theta}\rangle;\nonumber\\
\frac{d}{dt}\langle\vec{\rho}\vec{\theta}\rangle=\langle
\theta^{2}\rangle-\frac{2}{\tau}\langle
\vec{\rho}\vec{\theta}\rangle
-\Omega^{2}\langle \rho^{2}\rangle;\nonumber\\
\frac{d}{dt}\langle
\theta^{2}\rangle=-2\Omega^{2}\langle\vec{\rho}\vec{\theta}\rangle-\frac{4}{\tau}
\langle \theta^{2}\rangle+4D.
\end{eqnarray}
Then, following the similar lines as in \cite{152}, supplement
equations (\ref{31.22}) with  averaged equation (\ref{31.15}):
\begin{equation}
\label{31.23}
\dot{\gamma}=\vec{v}\vec{F}-\frac{2}{\tau}\gamma^{3}\Omega^{2}\langle
\vec{\rho}^{2}\rangle.
\end{equation}
The first term on  the right-hand side of (\ref{31.23}) describes
the change in the particle energy caused by the work done by the
Lorentz force, and in the case $\gamma\gg 1$  it may be dropped.
The second term, proportional to $\langle \vec{\rho}^{2}\rangle$
corresponds to the energy emitted by a relativistic harmonic
oscillator per unit time.

If the incursion  of the root-mean-square angle of multiple
scattering of the channeled particle during the velocity
relaxation time of its transverse motion in the channel $\tau$  is
much greater than the angle $\theta^{2}_{0}$, then the force
$\vec{F}_{rad}$ in the equation of motion can apparently  be
neglected. In the model under consideration this condition
satisfies the following inequality
\begin{equation}
\label{31.24} 4D\tau\gg\theta_{0}^{2}
\end{equation}

It is also obvious that in the case in question the linear law of
motion is valid for not  large times $t$:
\begin{eqnarray}
\label{31.25}
\langle \theta^{2}\rangle=2D(E_{0}t+\langle \theta^{2}\rangle;\nonumber\\
\langle\rho^{2}\rangle=\langle
\theta^{2}\rangle/\Omega^{2}(E_{0}),
\end{eqnarray}
suitable providing that  $\varepsilon(t)/E_{0}\ll 1$. Then
according to (\ref{31.23}), (\ref{31.25}), we obtain
\begin{eqnarray}
\label{31.26}
\gamma(t)=\gamma_{0}\left\{1+\frac{2\gamma_{0}^{2}t}{\tau}(\langle
\theta^{2}_{0}\rangle+Dt)\right\}^{-1}
\end{eqnarray}
(in perfect agreement with formula (\ref{31.19})).

Figure 12 exemplifies  the comparison between the numerical
solution of the system of equations (\ref{31.22}), (\ref{31.23})
and approximate solution of (\ref{31.25}), (\ref{31.26}).

\bigskip

\textit{Figure 12. The root-mean-square angle of multiple
scattering  and energy losses as a function of thickness.}

\bigskip

Planar channeling of positrons in $(110)$ channel of a silicon
single crystal at $\langle \theta^{2}_{0}\rangle=0$ (in the case
of planar channeling  $D$ should be replaced by  $1/2D$ in
formulae (\ref{31.22}), (\ref{31.25}), (\ref{31.26})) is
considered. As seen from graphs, the two solutions agree well. The
diffusion coefficient $D$, as well as for protons \cite{23,24},
is taken equal to $d=kD_{chaot}$, where $k=z_{v}/z^{2}$ ($z_{v}$
is the number of valence electrons);
$D_{chaot}=E^{2}_{S}/4E^{2}L_{R}$; $E^{2}_{S}=4\pi m^{2}/\alpha$;
$\alpha=1/137$; $L_{R}$ is the radiation unit of length. Note that
in view of the aforesaid characteristics of  $k_{L}$ and $k$, the
value of the coefficient $k=1/z\cdot 2.72$ is more precise.

At $\langle\theta^{2}(0)\rangle=0$ the law of variation of
$\gamma/\gamma_{0}$ which follows from (\ref{31.26}) is
independent of the particle initial energy, as the diffusion
coefficient $D\sim \gamma_{0}^{-2}$. Therefore the corresponding
(dashed) curve in Fig. 12 is  universal for different initial
energies $E_{0}$. Dotted curve in Fig. 12 represents the
dependence of $\gamma/\gamma_{0}$ $(E_{0}=10^{3}$ GeV), used in
\cite{158}. The discrepancy with the exact solution
($\langle\theta^{2}(0)\rangle=0$)is quite large. As a result, the
calculations carried out in the stated work give the incorrect
picture of the evolution of angular distributions
$\gamma_{2}(\theta,\theta_{0}, t) (\theta_{0}=0)$\footnote{In
\cite{158} in the equation of the the type  (\ref{31.23}) the
constant equal to $d_{p}/4^{2}$ is used instead of the moment
$\langle x^{2}\rangle$, and thus obtained dependence $\gamma(t)$
is then substituted into the solution of the kinetic equation of
the type (\ref{31.21}).}

From (\ref{31.24}) and (\ref{31.26}) one can easily find the
ranges ($\gamma,t)$ where the approximation (\ref{31.25}),
(\ref{31.26}) is applicable:
\begin{eqnarray}
\label{31.27}
\gamma\ll\gamma_{1}=\tilde{\gamma}\frac{\tau}{T}\eta^{-2},\, t<T=\sqrt{\frac{\alpha L_{R}\tau}{\pi k}};\nonumber\\
\gamma_{1}\sim\gamma\ll\gamma_{rad}=4\gamma_{1}\frac{\tau}{T},\,t\ll
L_{E}(\gamma)=\tau\frac{\tilde{\gamma}}{\gamma}\eta^{-2},
\end{eqnarray}
where $\tilde{\gamma}=m/2u_{0}$ is the magnitude of the Lorentz
factor which corresponds to the particle energy $E$, at which the
critical angle
$$
\theta_{cr}=(2u_{0}/E)^{1/2}=(\gamma\tilde{\gamma})^{-1/2}
$$
equals $\gamma^{-1}$,
$\langle\theta^{2}(0)\rangle=\frac{1}{2}\theta^{2}_{0}$,
$\theta_{0}=\eta\theta_{cr}$. Note that in the case of planar
channeling of positrons $\frac{\tau}{T}=\sqrt{3{\cal{L}}_{R}}$
(here ${\cal{L}}_{R}$ is the  radiation logarithm
$({\cal{L}}_{R}\simeq\ln(191z^{-1/3})$ \cite{63,64},
$k=\tilde{k}$).

It  should also be pointed out that the range $(\gamma, t)$
determined by the relations (\ref{31.27}) is rather large. For
example, at $\eta=1/2\,
4\gamma_{1}\frac{\tau}{T}\sim10^{2}\tilde{\gamma}$.

Further make use of the derived relations for seeking the
dechanneling length $L_{D}$. Define $L_{D}$ as the pathway where
$\langle\rho^{2}(L_{D})\rangle=d^{2}/4$. As a result,
\begin{equation}
\label{31.28}
L_{D}=(1-\eta^{2})/2D\gamma\tilde{\gamma}=T\gamma(1-\eta^{2})/\gamma_{D},
\end{equation}
where $\gamma_{D}=2\tilde{\gamma}\tau/T$. It is easy to see that
the expression (\ref{31.28}) holds true at any $\eta \in[0,1]$ in
the energy range $\gamma<\gamma_{D}$. Thus, the motion of charged
particles in a wide  range of energies and crystal thicknesses is
described by the approximate solution of (\ref{31.25}),
(\ref{31.26}).

Consider the diffusion coefficient for channeled particles in more
detail. The coefficient $D_{chaot}$, corresponding to the multiple
Coulomb scattering of ultra-relativistic electrons (positrons) in
an amorphous medium is well known and equal to $1/4$ of the
root-mean-square angle of particle multiple scattering per unit
time: $D_{chaot}=\frac{E_{S}^{2}}{4E^{2}L_{R}}$ \cite{63,64}.

In the channeling regime the diffusion coefficient $D$ depends on
the particle trajectory in the channel (the charge sign and the
energy of the particle transverse motion $E_{\perp}$). For
example, at channeling of negatively charged particles moving in
the vicinity of nuclei, $D^{(-)}>D_{chaot}$ and vice versa, for
positively charged particles moving in the peripheral area of
atoms forming a channel, $D^{(+)}<D_{chaot}$. (Hereinafter the
superscripts $+(-)$ will be dropped, unless this leads to
misunderstanding). The diffusion coefficient $D$ is normally
written as the sum $D=D_{nuc}+D_{e}$, with the first term
corresponding to scattering of channeled particles by a screened
potential of nuclei, and the second one corresponding to
scattering by valence electrons (conduction electrons).

The concrete form of the coefficient $D_{e}$  is based on some
model of electron distribution in the crystal. In the simplest
case the distribution of valence electrons is considered
homogeneous. In the case of electron channeling in tubes or layers
$D_{nuc}\gg D_{e}$. We note further that $D_{chaot}$ is
proportional to the density of nuclei per unit volume of the
crystal, so it is natural to suppose that  $D^{(-)}_{nuc}$ is
approximately equal to the diffusion coefficient in an amorphous
medium, where the density of nuclei is the same as that in tubes
or layers, i.e., $D_{nuc}^{(-)}\simeq k_{nuc}D_{chaot}$, where
$k_{nuc R}=(d_{R}/a_{R})^{2};$ $k_{nuc\, p}=d_{p}/a_{p}$; $a_{k}$,
$a_{p}$ is the tube radius and the layer width, respectively;
$d_{R}$ is the distance between the axes along which the channeled
particle moves. At channeling of positrons in the central part of
the channel $D_{e}\gg D_{nuc}$, and $D_{e}$, as mentioned above,
is taken equal to $k_{e}D_{chaot}$, where $k_{e}=z_{v}/z^{2}$
\cite{3}. Hence, the approximate solution of (\ref{31.25}),
(\ref{31.26}), as well as expression (\ref{31.28}) for the
dechanneling length $L_{D}$ are also suitable for describing
electron motion in tubes or in layers.

Discuss the general pattern of channeling of light
ultra-relativistic particles in a harmonic potential we obtained.
The domain of applicability of the classical equation of motion is
determined by the following two-sided inequality:
\begin{equation}
\label{31.29}
(40\lambda/d)^{2}\gamma=\gamma_{min}<\gamma\ll\gamma_{S}=(d/2\lambda\eta)\tilde{\gamma};\,
\lambda=\hbar/mc.
\end{equation}
If the particle Lorentz factor is $\gamma<\gamma_{min}$, then
there are only a few energy levels in a potential well of height
$U_{0}$, and, as a consequence, the classical description of
motion proves to be impossible. On the other hand, in the range
$\gamma\sim\gamma_{s}$ the quantum effects gain importance
\cite{151,152}.

The energy range, determined by inequalities (\ref{31.29})
stretches approximately for four orders of magnitude: from
$\gamma\sim 10^{2}$ to $\gamma\sim 10^{6}$ ($d\sim
1$\,{\AA},$\eta\sim 1/2$). By the character of motion inside the
channel it is helpful to divide the initial energies of channeled
particles  into three intervals:

I. $\gamma_{min}<\gamma<\gamma_{D}$;

II. $\gamma_{D}<\gamma<\gamma_{rad}$;

III. $\gamma_{rad}<\gamma$.

Within interval I the the channeled particle motion is entirely
determined by its multiple inelastic scattering by the atoms of
the crystal lattice, and the change in the particle energy may be
neglected. Within interval II it is also determined by multiple
scattering, but considerably affected by the radiation energy
losses. Within interval III the character of the particle motion
becomes affected by radiation friction  (the right-hand side of
equation (\ref{31.14})).

For energies defining the limits of the stated intervals the
following relation holds:
$$
\gamma_{min}\ll\gamma_{D}\ll\gamma_{rad}\sim\gamma_{S}(d\sim 1\,
\mbox{\AA},\, \eta\sim 1/2).
$$
Therefore interval III, where the due account of the radiative
recoil is important, is practically beyond the applicability of
classical description.

At $\gamma\sim\gamma_{S}$ the characteristic frequency of emitted
$\gamma$-quanta is $\omega_{eff}\sim E$. But for multiple
scattering of particles in the channel, the radiative recoil would
also appear to be important in the energy range
$\gamma\ll\gamma_{S}$, when $\omega_{eff}\ll E$.

For particles channeled either in layers or in tubes
\cite{151,152}, $\gamma_{min}<\gamma_{S}(d\ll 1$\,{\AA}, and,
hence, in the cases mentioned above the application of classical
description is strongly restricted.

%%%%%%%%%%%%%%%%%%%%%%%%%%%%%%%%%%%  Section 32 %%%%%%%%%%%%%%%%%%%%%%

\section[Quantum Theory of Channeling Electrons and Positrons Allowing for Multiple Scattering and Radiation Energy Losses]
{Quantum Theory of Channeling Electrons and Positrons Allowing for
Multiple Scattering and Radiation Energy Losses} \label{sec:10.32}

The most consistent description of the transmission of
relativistic charged particles through crystals may be achieved by
means of quantum consideration of the process. This circumstance
is due to the fact that in the range of not very high energies (of
the order of several megaelectronvolts) there are only several
levels for a transverse electron motion in a potential well, while
at high energies, the emission of hard photons with the energy of
the order of the particle energy is possible, which makes the
account of quantum recoil crucial. If the radiation processes are
of no importance, then the kinetic equations derived by Kagan and
Kononetz \cite{23,24} my be used to describe channeling. With the
growth of energy of the particles, radiation is gaining greater
importance, and in order to describe the behavior of electrons and
positrons in crystals we have to introduce into the kinetic
equations the collisional term cause by the photon radiation
\cite{159}.

For detailed treatment of charged particles in the crystal and the
electromagnetic radiation they produce it is necessary to find the
density matrix $\rho(t)$ of the system crystal-particles-photons.
The sated density matrix satisfies the quantum Liouville equation
($\hbar=c=1$)
\begin{equation}
\label{32.1} i\frac{\partial\rho}{\partial t}=[H_{tot}, \rho]
\end{equation}
with the Hamiltonian
\begin{equation}
\label{32.2}
H_{tot}=H_{e}+H_{\gamma}+H_{c}+V_{ec}+V_{e\gamma}+V_{\gamma c},
\end{equation}
where $H_{e}$, $H_{\gamma}$ are the Hamiltonians of free particles
and photons, respectively; $H_{c}$ is the crystal Hamiltonian;
$V_{ij}$ are the operators of interaction between the subsystems
$i$ and $j$ ($i,j=e,\gamma,c$).

It is convenient  to obtain first from equation (\ref{32.1}) the
equations describing the time change of the diagonal non-diagonal
parts of the density matrix \cite{160}. The equation for the
diagonal part of the density matrix describing the time evolution
of a certain small subsystem $a$ (the incident particle, and the
$\gamma$-quanta it produced,) which interacts with a large
subsystem $\beta$,has the form
\begin{equation}
\label{32.3} \frac{\partial\rho_{\alpha}}{\partial
t}=\sum_{\alpha^{\prime}}\dot{w}_{\alpha^{\prime}\alpha}\rho_{\alpha}+
\sum_{\alpha^{\prime}}\dot{w}_{\alpha\alpha^{\prime}}\rho_{\alpha^{\prime}},
\end{equation}
where $\rho_{\alpha}=(sp_{\beta}\rho)_{\alpha\alpha}$ is the
diagonal matrix element  of the matrix $sp_{\beta}\rho$;
$sp_{\beta}$ is the trace over the states of subsystem $\beta$
from the full density matrix. The probabilities of transition per
unit time are directly connected with the scattering operator
\cite{19}. Their explicit form for the processes of photon
radiation through radiative transitions and bremsstrahlung was
obtained in previous sections. The expressions for $\dot{w}$
describing the process of pair production see in
(\ref{sec:10.33}).

Before passing to a detailed treatment of equation (\ref{32.3}),
it is useful to derive it for the case when the particle
interaction with a crystal may be described in term of the
perturbation theory. We shall neglect the influence of  usual
bremsstrahlung on the electron and positron behavior.

Equation for the density matrices of the electron
$\rho_{e}(t)=sp_{\gamma c}\rho(t)$ and photon
$\rho_{\gamma}=sp_{ec}\rho(t)$ subsystems may be found by taking
the trace  $sp_{\gamma c}$ and $sp_{ec}$ of  both parts of
equation (\ref{32.1}):
\begin{equation}
\label{32.4} i\frac{\partial}{\partial t}\rho_{e}=[H_{0},
\rho_{e}]+sp_{\gamma c}[V,\rho];
\end{equation}
\begin{equation}
\label{32.5} i\frac{\partial}{\partial
t}\rho_{\gamma}=[H_{\gamma}, \rho_{\gamma}]+sp_{e c}[V,\rho],
\end{equation}
where $H_{0}=H_{e}+\overline{V}_{ec}$; $\overline{V}_{ec}=
\sum_{c} \rho_{c(0)}^{c,c}V_{ec}^{c,c}$ is the operator $V_{ec}$
averaged over the crystal states \cite{23}; $\rho_{c(0)}$ is the
equilibrium density matrix of the crystal, diagonal in the
representation of the eigenfunctions of the Hamiltonian $H_{c}$;
$H_{c}|c\rangle=E_{c}|c\rangle$; $V\equiv W+V_{e\gamma}+V_{\gamma
c}$; $W=(V_{ec}-\overline{V}_{ec})$ and responsible for inelastic
scattering of channeled particles by the lattice atoms;
$$
[A,B]=AB-BA;\, sp_{\alpha\beta}\rho
A=\sum_{\alpha\beta\alpha^{\prime}\beta^{\prime}}\rho^{\alpha\beta,\alpha^{\prime}\beta^{\prime}}
A^{\alpha^{\prime}\beta^{\prime},\alpha\beta},|\gamma, c\rangle
$$
$$
=|\gamma\rangle|c\rangle,\, |e, c\rangle=|e\rangle|c\rangle.
$$
Using the integral representation of equation (\ref{32.1})
\begin{equation}
\label{32.6} \rho(t)=S(t)\rho(0)S^{+}(t)-i\int_{-t}^{0}d \tau
S^{+}(\tau)[V, \,\rho(t+\tau)]S(\tau),
\end{equation}
int the second order over the operator $V$ from (\ref{32.4}),
(\ref{32.5}) we obtain the system of integro-differential
equations
\begin{eqnarray}
\label{32.7}
\frac{\partial\rho_{e}}{\partial t}+& &i[H_{0}, \rho_{e}]\nonumber\\
& &=\int_{-t}^{0}d\tau sp_{\gamma
c}[S^{+}(\tau)[V,S(\tau)\rho(t)S^{+}(\tau)]S(\tau),\,V];
\end{eqnarray}
\begin{eqnarray}
\label{32.8}
\frac{\partial\rho_{\gamma}}{\partial t}+i[H_{\gamma}, \rho_{\gamma}]\nonumber\\
=\int_{-t}^{0}d\tau sp_{e
c}[S^{+}(\tau)[V,S(\tau)\rho(t)S^{+}(\tau)]S(\tau),V];
\end{eqnarray}
where $S(\tau)=e^{-iH^{\prime}\tau}$;
$H^{\prime}=H_{0}+H_{\gamma}+H_{c}$, $t_{0}=0$, where it is taken
into account that at the initial time $t=0$ of the particle
entrance the crystal, the density matrix
\begin{equation}
\label{32.9} \rho(0)=\rho_{e}(0)\rho_{\gamma}(0)\rho_{c}(0),
\end{equation}
where the crystal density matrix $\rho_{c}(0)=sp_{\gamma
e}\rho(0)$.

Neglect the interaction of $\gamma$-quanta with the crystal
($V_{\gamma c}=0$). let us also consider that the state of the
medium does not change during the particle transmission through
the crystal. As a consequence, we may write
\begin{equation}
\label{32.10} \rho(t)=\rho_{c}(0)\rho_{e\gamma}(t),
\end{equation}
where $\rho_{e\gamma}(t)$ is the density matrix of the subsystem
particles-photons. As a result
\begin{equation}
\label{32.11} \frac{\partial\rho_{e}}{\partial t}+i[H_{0},
\rho_{e}]=sp_{\gamma c}I_{1}+sp_{\gamma}I_{2};
\end{equation}
\begin{equation}
\label{32.12} \frac{\partial\rho_{\gamma}}{\partial
t}+i[H_{\gamma}, \rho_{\gamma}]=sp_{e}I_{2}.
\end{equation}
The integral
\begin{eqnarray}
\label{32.13}
I_{2}=& &\int_{-t}^{0}d\tau[e^{i(H_{0}+H_{\gamma})\tau}[V_{e\gamma},e^{-i(H_{0}+H_{\gamma})\tau}\rho_{e\gamma}(t)e^{i(H_{0}+H_{\gamma})\tau}]\nonumber\\
& &\times e^{-i(H_{0}+H_{\gamma})\tau},V_{e\gamma}].
\end{eqnarray}
The integral $I_{1}$  is obtained  from the integral in
(\ref{32.13}) upon by replacing in it the operator $H_{\gamma}$
with $H_{c}$, $V_{e\gamma}$ - with $W$ and $\rho_{e\gamma}(t)$ -
with $\rho(t)$. Note that as a result of fulfilment of equality
$sp_{\gamma}\rho_{e\gamma}(t)=\rho_{e}(t)$ the expression
$sp_{\gamma c}I_{1}$ in (\ref{32.11}), as expected is equal to the
right-hand side of equation (2.5) in \cite{23}.

As in \cite{23}, let the lower limit of integration in the
expressions of $I_{1}$ and $I_{2}$ tend to $-\infty$ . In thus
obtained integrals of the type
\begin{equation}
\label{32.14} \lim_{T\rightarrow\infty}\int_{-T}^{0}e^{\pm
xt}dt=\pm i\frac{{\cal{P}}}{x}+\pi\delta(x)
\end{equation}
we may neglect the summands with principal values of ${\cal{P}}/x$
which lead to renormalization of the energy spectrum. As a result,
the expression $sp_{\gamma}I_{2}$ in the representation of the
eigenfunctions of the Hamiltonian $H_{0}$ has the form
\begin{eqnarray}
\label{32.15} sp_{\gamma}I_{2}^{e,
e^{\prime}}=\pi\sum_{e^{\prime\prime}e^{\prime\prime\prime}_{\gamma\gamma^{\prime}\gamma^{\prime\prime}}}
\left\{\rho_{e\gamma}^{e^{\prime\prime}\gamma,
e^{\prime\prime\prime}\gamma^{\prime\prime}}(t)V_{e\gamma}^{e\gamma^{\prime},e^{\prime\prime}\gamma}
V_{e\gamma}^{e^{\prime\prime\prime}\gamma^{\prime\prime},e^{\prime}\gamma^{\prime}}\left[\delta\left( E_{e}+\right.\right.\right.\nonumber\\
\left.\left.\left.+E_{\gamma^{\prime}}-E_{e^{\prime\prime}}-E_{\gamma}\right)
+\delta(E_{e^{\prime}}+E_{\gamma^{\prime}}-E_{e^{\prime\prime\prime}}-E_{\gamma^{\prime\prime}})\right]\right.\nonumber\\
\left.-\rho_{e\gamma}^{\gamma
e^{\prime\prime\prime},e^{\prime}\gamma^{\prime\prime}}(t)V_{e\gamma}^{e\gamma^{\prime\prime},e^{\prime\prime}\gamma^{\prime}}
V_{e\gamma}^{e^{\prime\prime}\gamma^{\prime},e^{\prime\prime\prime}\gamma}\delta
(E_{e^{\prime\prime}}+E_{\gamma^{\prime}}-E_{e^{\prime\prime\prime}}-E_{\gamma})\right.\nonumber\\
-\left.\rho_{e}^{e,e^{\prime\prime\prime}}(t)V_{e\gamma}^{e^{\prime\prime\prime}\gamma^{\prime\prime},e^{\prime\prime}\gamma^{\prime}}V_{e\gamma}^{e^{\prime\prime}\gamma^{\prime},e^{\prime}\gamma}
\delta(E_{e^{\prime\prime}}+E_{\gamma^{\prime}}+E_{e^{\prime\prime\prime}}-E_{\gamma^{\prime\prime}})\right\}
\end{eqnarray}

Due to the symmetry of the integral (\ref{32.15}) about the
operators acting on vectors $|e\rangle$ and $|\gamma\rangle$, the
matrix element $sp_{e}I_{2}^{\gamma,\gamma^{\prime}}$ may be
obtained from expression (\ref{32.15}) be substitution of the
subscript $e$ into $\gamma$, and the subscript $\gamma$ into $e$.
The matrix element $sp_{\gamma c}I_{1}^{e,e^{\prime}}$ appearing
in the left-hand side of equation (\ref{32.11}) is obtained from
(\ref{32.15}) with the operator $V$ replaced by $W$, and the
subscripts $\gamma$ and $c$ and the density matrix
$\rho_{e\gamma}(t)$ - by $\rho_{c}(0)\rho_{e}(t)$ (the explicit
form of this matrix element see also in \cite{23}, formula (2.7)).

Now use factorization $\rho_{e\gamma}=\rho_{e}\rho_{\gamma}$  for
the density matrix $\rho_{e\gamma}(t)$. Averaging of the equations
obtained over times greater in comparison with the oscillation
period of non-diagonal elements $\rho_{e}^{e,e^{\prime}}(t)$,
$\rho_{\gamma}^{\gamma\gamma^{\prime}}(t)$ gives the set of
balance equations (\ref{32.3}). Attenuation of non-diagonal
elements of the density matrix $\rho_{e}(t)$ due to inelastic
scattering of channeled particles by the crystal lattice is
studied in \cite{23}.

Thus, equation (\ref{32.3}) has the form
\begin{eqnarray}
\label{32.16}
\frac{\partial}{\partial t}\rho_{e}^{e,e}(t)=\sum_{e^{\prime}}\left\{(w_{e,e}^{W}+w_{e^{\prime}e}^{S})\rho_{e}^{e^{\prime}}(t)\right.\nonumber\\
\left.-(w^{W}_{ee^{\prime}}+w_{ee^{\prime}}^{S})\rho_{e}^{e,e}(t)\right\},
\end{eqnarray}
where
\begin{equation}
\label{32.17}
w_{ee^{\prime}}^{S}=2\pi\sum_{cc^{\prime}}|w^{ec,e^{\prime}c^{\prime}}|^{2}\rho_{c}^{c,c}(0)\delta(E_{e}-E_{e^{\prime}}+E_{c}-E_{c^{\prime}})
\end{equation}
is the probability of the transition $e\rightarrow e^{\prime}$
caused by inelastic scattering of channeled particles by the atoms
of the  crystal lattice;
\begin{equation}
\label{32.18} w_{ee^{\prime}}^{S}=2\pi\sum_{\nu}|V_{e\gamma}^{e
0,e^{\prime}1\nu}(\omega_{\nu})|^{2}\delta(E_{e}-E_{e^{\prime}}-\omega)
\end{equation}
is the probability of a spontaneous radiative transition
$e\rightarrow e^{\prime}$ per unit time; the function
$\rho_{e}^{e,e}(t)$ equals the probability density of finding the
channeled particle at moment $t$ in the state $|e\rangle$.

Int the high-energy region it is possible to use impulse
approximation when calculating the probability of inelastic
scattering of a channeled particle by the atoms of the crystal
lattice $w_{ee^{\prime}}^{W}$. As a result
\begin{equation}
\label{32.19}
w_{ee^{\prime}}^{W}=2\pi\delta(E-E^{\prime})\sum_{i}(\langle|V_{i}^{e^{\prime},e}|^{2}\rangle_{c}-|\langle
V_{i}^{e^{\prime},e}\rangle_{c}|^{2}),
\end{equation}
where $\langle
A\rangle_{c}=\sum_{cc^{\prime}}\rho_{c}^{c,c^{\prime}}(0)A^{c,c^{\prime},c}$.
The potential energy of the particle interaction with the i-th
atom of the crystal lattice
\begin{equation}
\label{32.20}
V_{i}(\vec{r})=\pm\left(\frac{e^{2}z}{|\vec{r}-\vec{r}_{i}|}-\sum_{j=1}^{z}
\frac{e^{2}}{|\vec{r}-\vec{r}_{i}-\vec{r}_{ij}|}\right),
\end{equation}
where $\vec{r}_{i}=\vec{r}_{0i}+\vec{u}_{i}$ is the radius-vector
of the center of inertia of the atom; vector $\vec{r}_{0i}$
determines the equilibrium position of the lattice atom;
$\vec{r}_{ij}$ is the radius-vector of the j-th electron with
respect to the atom center of inertia.

upon corresponding calculations, we obtain the following
expression for the probability $w_{ee^{\prime}}^{W}$, for example,
in the case of planar channeling \cite{159}:
\begin{eqnarray}
\label{32.21}
w^{W}_{ee^{\prime}}=\frac{(2\pi)^{3}\delta(E-E^{\prime})}{L^{3}}4e^{4}n_{at}\sum_{\tau_{x}q_{x}}
\frac{J^{(1)}(q_{x})J^{(1)^{*}}(q_{x}^{\prime})}{(qq^{\prime})^{2}}\nonumber\\
\times\left\{z[F(\tau_{x})-FF^{\prime}]e^{-w^{\prime}(\tau_{x})}+z^{2}(1-F)(1-F^{\prime})\right.\nonumber\\
\left.\times(e^{-w(\tau_{x})}-e^{-w(q)}e^{-w(q^{\prime})})\right\},
\end{eqnarray}
where $J^{(1)}(q_{x})\equiv
J^{(1)}_{nk,n^{\prime}k^{\prime}}(q_{x})=\int
dx\psi_{nk}(x)\psi^{*}_{n^{\prime}k^{\prime}}(x)e^{iq_{x}^{x}}$;
$\vec{q}_{p}=\vec{p}_{p}^{\prime}-\vec{p}_{p}$;
$\vec{p}_{p}=(p_{y}, p_{z})$;
$\vec{q}^{\prime}=(q^{\prime}_{x},\vec{q}_{p})$;
$q^{\prime}_{x}=q_{x}+\tau_{x}$; $F(q)\equiv F$; $F^{\prime}\equiv
(q^{\prime})$; $\vec{q}=(q_{x},\vec{q}_{p})$;
$F(q)=z^{-1}\int\rho(\vec{r})e^{-i\vec{q}\vec{r}}d^{3}r$;
$e\rho(r)$ is the electron charge density in the atom;
$F(\vec{q})$ is the atomic form-factor; $L^{3}$ is the crystal
volume; $n_{at}$ is the density of atoms; $e^{-w(q)}$ is the
Debye-Waller factor.

If the functions $\varphi(x)=L^{-1/2}e^{ip_{x}x}$ are introduced
into the integrals $J^{(1)}$ instead of the Bloch functions, then
expression (\ref{32.21}) goes over into the probability of
particle scattering in a disoriented crystal
\begin{eqnarray}
\label{32.22}
w^{W}_{ee^{\prime}chaot}=(2\pi)^{3}L^{-3}\delta(E-E^{\prime})\frac{4e^{4}n_{at}}{q^{4}}\left\{z(1-F^{2})\right.\nonumber\\
\left.+z^{2}(1-F)^{2}(1-e^{-2q^{2}u^{2}})\right\},
\end{eqnarray}
where $q_{x}=p^{\prime}_{x}-p_{x}$. The augend in (\ref{32.22}),
proportional to the atomic number $z$ is equal to the probability
of inelastic scattering of a particle by electron shell of the
atoms; the addend, proportional to $z^{2}$ is the probabilities of
inelastic scattering of the particle by oscillating atoms
(photons) without changing their intrinsic state.

For single crystals with $z>10$ in (\ref{32.21}), (\ref{32.22})
scattering by photons of the order of $z^{2}$ acts the main part.
generally speaking, inelastic scattering of a particle by electron
shells of the atoms of the order of $z$ may be neglected. Cooling
of the majority of crystals does not lead to considerable
suppression of scattering by photons, with the possible exception
of the crystals with a low Debye temperature $\theta_{D}$
($\theta_{D}\leq 100$ K). The total scattering probability
$w_{e}^{W}=\sum_{e^{\prime}}w_{ee^{\prime}}^{W}$. In the case of
planar channeling the particle state in the crystal is described
by a set of quantities $(\vec{p}_{p}, k, n)$. In view of the
completeness condition of the Bloch functions
\begin{equation}
\label{32.23}
\tilde{F}(\tau_{x})=\sum_{n^{\prime}k^{\prime}}J^{(1)}(q_{x})J^{(1)^{*}}(q^{\prime}_{x})\int
dx|\psi_{nk}(x)|^{2}e^{-i\tau_{x}x}.
\end{equation}
the integral (\ref{32.23}) is, in fact, the form factor of the
channeled particle. In an axial channeling regime , we get the
following form factor instead of (\ref{32.23})
\begin{equation}
\label{32.24} \tilde{F}(\vec{\tau}_{\perp})=\int
d^{2}\rho|\psi_{e}(\vec{\rho})|^{2}e^{-i\tau_{\perp}\vec{\rho}}.
\end{equation}
Next write the total probability of inelastic scattering as a
series in terms of the reciprocal lattice vectors
\begin{equation}
\label{32.25} w_{e}^{W}=\sum_{\vec{\tau}}w_{e}^{W}(\vec{\tau}).
\end{equation}
In the planar channeling regime summation is made over the vectors
$\vec{\tau}_{x}$, while in the case of axial channeling of a
particle, vector $\vec{\tau}$ in (\ref{32.25}) equals
$\vec{\tau}_{\perp}=(\vec{\tau}_{x},\vec{\tau}_{y})$.

At $\vec{\tau}=0$ the form factor $\tilde{F}(0)=1$. As a result
the first term of series (\ref{32.25}) corresponds to the
probability of inelastic scattering of a particle in a disoriented
crystal (\ref{32.22}). The next following terms of the series
determine the correction to $w_{e\, chaot}^{W}$ which depends on
the form factor of the channeled particle. (\ref{32.23}),
(\ref{32.24}).

The probability of inelastic scattering of a particle in a
channeling regime (\ref{32.21}) differs from the scattering
probability in a disoriented crystal (\ref{32.22}) not only by the
presence of a form factor, (\ref{32.23}) (which reflects the
peculiarities of the particle transverse  motion) , but also by
the temperature dependence. Indeed, $w_{e}^{W}(\tau_{x})\sim
e^{-w(\tau_{x})}$. As seen, the terms of the series (\ref{32.25})
bear the same relationship to the crystal temperature  as the
probability of elastic scattering of a particle by the crystal,
i.e.,proportional to $e^{-2w}$ The appearance of the multiplier
$e^{-w}$ becomes obvious, if we recall that with the increase in
the crystal temperature, the probability of inelastic scattering
of electrons moving in tubes or in layers should tend to the
magnitude of inelastic scattering of a particle in a disoriented
crystal (amorphous medium) $w_{e\, chaot}^{W}=w_{e}^{W}(0)$.

Introduce the coefficient
\begin{equation}
\label{32.26} k=1+\sum_{\vec{\tau}\neq
0}w_{e}^{W}(\vec{\tau})/w_{e\, chaot}^{W}.
\end{equation}
The magnitude of the sum in (\ref{32.26}) depends on the sign of
the form factor $ \tilde{F}(\tau)$ which determines the sign of
the terms of the series $w_{e}^{W}(\vec{\tau})\neq 0$ , and on the
cutoff efficiency  of the series $\sum_{\vec{\tau}\neq
0}w_{e}^{W}(\vec{\tau})$, which is defined by the absolute value
of $|\tilde{F}(\vec{\tau})|$ and the multiplier $e^{-w}$. The sum
\begin{equation}
\label{32.27} \sum_{\tau_{x}\neq
0}\tilde{F}(\tau_{x})=2\sum_{l=1}^{\infty}\int
dx|\psi_{nk}(x)|^{2}\cos(2\pi lx/d_{p}),
\end{equation}
where $\tau_{x}=2\pi l/d_{p}$ ($l=0, \pm 1, \pm 2, ...)$.
Therefore for particles moving in the vicinity of the
crystallographic axes ( for example, electrons channeled in
layers), for which the effective distribution  width
$|\psi_{nk}(x)|^{2}$ is much smaller than the channel width, the
main contribution to the series (\ref{32.27}) will come from the
summands with $l$ from 1 to $l_{max}\gg 1$. As a result the
coefficient $k\gg 1$. In the case of channeling of positrons with
transversal energy much smaller than the height of the potential
barrier $u$, the series (\ref{32.27}) will be alternating, and the
coefficient $k$ will turn out to be less than unity.

Consider the cutoff of the series (\ref{32.25}) caused by the
temperature factor $e^{-w}$. The quantity $w$ included in the
Debye-Waller factor, equals $w(\tau)=\vec{\tau}^{2}\langle
u^{2}\rangle/6$ where $\langle u^{2}\rangle$ is the mean-square
displacement of the atoms of the crystal lattice due to
temperature oscillations. Thus, the exponent $e^{-w(\tau_{x})}$
leads to the cutoff of the  series (\ref{32.25}), starting with
$l_{max}\simeq d_{p}/\pi\sqrt{\langle u^{2}\rangle}$.

For most crystals at temperatures $T\sim 300$ K the probability of
inelastic scattering of a particle by photons (the term of the
order of $z^{2}$ in (\ref{32.21}) is of the same order of
magnitude as  the probability of scattering a particle by
stationary atoms $\sim z^{2}(1-F)^{2}$. Therefore in qualitative
consideration of channeled particle motion in the space of
transverse impulses (within the diffusion approximation) it is
possible to apply  the diffusion coefficient $D\simeq
k_{n}D_{chaot}$, where the diffusion coefficient
$D_{chaot}=\frac{1}{4}(E_{S}/E)^{2}L_{R}^{-1}$ describes the
elastic multiple scattering of electrons and positrons in
amorphous substance.

When channeling electrons in  layers $k_{n}\simeq l_{max}$ and,
hence, $D\simeq(d_{p}/\pi\sqrt{\langle u^{2}\rangle})D_{chaot}$.
The estimate of the magnitude of the diffusion coefficient $D$
found here may also be obtained with the help of the following
pictorial presentations. The diffusion coefficient $D_{chaot}$ is
proportional to the density of the nuclei per unit volume of the
crystal. Therefore it is natural to suppose that the coefficient
$D$ is approximately equal to the diffusion coefficient in
amorphous substance, where the density of the nuclei is the same
as that in layers. As a result, $k\simeq d_{p}/2\sqrt{\langle
u^{2}\rangle}$ \cite{159}. As seen, both estimates of the
magnitude of $k$ differ by a numerical factor equal to $2/\pi$. In
the case of electron channeling in tubes, the factor $e^{-w}$
leads to the cutoff of the series (\ref{32.25}). beginning with
$l_{max}\simeq[3d^{2}/2\pi\langle u^{2}\rangle]$ (for
simplicity,the crystal is assumed to have a simple cubic lattice,
with the axes $\langle 100\rangle$,  $\langle 010\rangle$,
$\langle 001\rangle$ being considered). Thus, according to
(\ref{32.26}) for electron channeled in tubes $k_{n}\simeq
10^{-1}d^{2}/\langle u^{2}\rangle$.

The number of electrons in the electron core of the crystal
lattice atom, generally speaking, equals  some $z_{cor}\leq z$.
Therefore in the general case in (\ref{32.21}) in the summand of
the order of $z$ due to inelastic scattering of a channeled
particle by stationary atoms, $z$ should be replaced by $z_{cor}$
The remained $z-z_{cor}$ electrons are distributed over the entire
volume of the unit cell of the crystal. Here the scattering
probability $w_{ee^{\prime}}^{W}$ will include the the form factor
describing the distribution of valence (free) electrons together
with the form factor of the atomic cores  9atoms forming the
crystal unit cell). If the distribution of the valence electrons
is supposed to be homogeneous over the entire volume of the unit
cell, then the coefficient $k_{e}$ corresponding to the
contribution to scattering due to elastic collisions with valence
electrons, may be assumed equal to the ration $z_{e}/z^{2}$
\cite{23,24}, where $z_{e}=(z-z_{cor})$ is the number of valence
electrons. as a result, $D=D_{n}+D_{e}$ (here
$D_{n}=k_{n}D_{chaot},\, D_{e}=k_{e}D_{chaot}$).

Obviously, the coefficient $k_{e}=(2.72\,\, z)^{-1}$ offered above
for describing multiple scattering of positrons in the central
region of the planar channel ($|x|\leq d_{p}/2-a$) corresponds to
the total concentration of crystal electrons in the center of the
channel, found ("restored") by the potential $u(x)$. Note that at
large $z$ the magnitude of the coefficient is $k_{e}=z_{e}/z^{2}$.
For example, for tungsten $z/2.72 \,\,z_{e}\simeq 14$.

Using the simplest models, let us consider the evolution of the
diagonal elements of the density matrix of a charged particle
$\rho_{e}^{e,e}(t)$ in a crystal.  Neglecting the change in the
total energy $E$, we obtain a conventional balance equation for
the probability
$p_{e}^{nn}(t)=\sum_{\vec{p}_{p}k}\rho_{e}^{e,e}(t)$ of finding
the particle at moment $t$ at the level $n$
\begin{equation}
\label{32.28} \frac{\partial\rho_{n}}{\partial
t}=\sum_{n^{\prime}}(w_{n^{\prime}
n}\rho_{n^{\prime}}-w_{nn^{\prime}}\rho_{n}),
\end{equation}

where $\rho_{n}\equiv\rho_{e}^{n,n}(t)$;
$w_{nn^{\prime}}=w_{nn^{\prime}}^{S}+w_{nn^{\prime}}^{W}$ is the
probability of transition $n\rightarrow n^{\prime}$.

First suppose that inelastic scattering is absent
($w_{nn^{\prime}}^{W}=0$). Then the evolution of the function
$\rho_{n}(t)$ in a crystal is fully determined by radiative
transitions $n\rightarrow n^{\prime}$. In the case of a harmonic
potential $u(x)$ the probability of radiative transitions in
dipole approximation
$w_{nn^{\prime}}^{S}=2\pi\vec{\tau}^{-1}\delta_{n^{\prime}, n-1}$.
From equation (\ref{32.28}) follows that the mean value of the
level number $\overline{n}(t)=\overline{n}(0)e^{-2t/\tau}$, where
$\tau=3m/kr_{e}$. As seen, the relaxation time of the quantum and
classical oscillators is the same ($E^{quan}_{\perp}\sim n\sim
e^{-2t/\tau}, E^{cl}_{\perp}\sim\theta_{0}e^{-2t/\tau}$). The
lifetime of the harmonic oscillator
$\tau_{n}=1/w_{nn^{\prime}}^{S}=(3d/2\alpha\eta)\tilde{\gamma}\sqrt{\tilde{\gamma}/\gamma}$)
at the level $n(n\gg1, \gamma\gg 1)$ is much less than the
relaxation time
$\tau=3m/kr_{e}=(3d^{2}/4r_{e})\tilde{\gamma}(\eta=n/n_{max}$,
$n_{max}=(d/4\lambda)\sqrt{\gamma/\tilde{\gamma}}$,
$\lambda=m^{-1}$, $\gamma=E/m$, $\alpha=1/137$,
$r_{e}=\lambda\alpha$, $d$ is the channel width).

Here we, certainly, discuss the evolution of only those states of
the channeled particles which correspond to the energy levels
lying inside the potential well formed by crystal planes (axes).
An interesting and complicated problem of the mutual kinetics of
sub- and over-barrier states of a channeled particle calls for
special study.

The sharper the potential of the channel is the shorter the
relaxation time is in comparison with the relaxation time of a
harmonic oscillator $\vec{\tau}$. The limiting form of a sharp
potential is a rectangular well. Therefore it seems to be of
interest to analyze the radiation kinetics of channeled particles
in both cases. A detailed treatment of particle motion in a
harmonic potential was given in (\ref{sec:10.31}). Here we shall
dwell on the analysis of motion of a  particle channeled in an
infinite high rectangular potential well.

The probability of a spontaneous radiative transition of a
particle channeled in a rectangular potential well in dipole
approximation is  (see (\ref{sec:10.30}))
\begin{equation}
\label{32.29}
w_{nn^{\prime}}^{S}=\frac{A}{\gamma}\frac{n^{2}n^{\prime
2}}{n^{2}-n^{\prime 2}},
\end{equation}
where $n-n^{\prime }=1,3,5,\ldots $ . The particle lifetime at
this level is given by (\ref{30.4}).

Comparison of expressions for the particle lifetime at the level
$n$ in a rectangular and harmonic potential wells show that the
quantities $\vec{\tau}_{n}$ appear to be of the same order of
magnitude. Nevertheless, due to the fact that in a rectangular
potential well far transitions are allowed, unlike those in ia
harmonic potential well, the relaxation time $E_{\perp}$ in a
rectangular well is $\vec{\tau}_{rec}\ll \vec{\tau}$. As an
illustration of the foregoing, consider two examples.

Substitute into the balance equation
\begin{equation}
\label{32.30} \frac{\partial\rho_{n}}{\partial
t}=\sum_{n^{\prime}}(w_{n^{\prime}n}\rho_{n^{\prime}}-
w_{nn^{\prime }}\rho_{n}),
\end{equation}
the probability (\ref{32.29}) of spontaneous radiation
$w_{un^{\prime}}^{S}$ calculated in dipole approximation.
Numerical solution of this equation (Fig. 13) was made for a model
example, where at the initial time $t=0$ only one level of the
transverse energy of motion is populated
\begin{eqnarray}
\label{32.31} \rho_{n}(0)=\left\{\begin{array}{c}
              1,\, n=49; \\
              0,\, n\neq 49.
            \end{array}\right.
\end{eqnarray}

\bigskip

\textit{Figure 13. Time change of the distribution $\rho_{n}(t)$:
a - for short times; b - for long times.}

\bigskip

The width of the well $d=1.92$\,{\AA},  $n_{max}=70$, the particle
energy $E=1.0$ GeV (the given figures "correspond" to a
rectangular potential well for a positron channeled in the channel
$(110)$ of a single crystal of silicon with the initial transverse
energy $E_{\perp}(0)=12$ eV).

At short times of the order of the lifetime at the level $n_{0}$
($t\sim 4.5\cdot 10^{-2}$ cm) the distribution function
$\rho_{n}(t)$ depends on the parity of the level number due to the
dependence of the probability of radiative transition
$w_{nn^{\prime}}^{S}$ in the model of the rectangular well on the
initial and final states (see Fig.13, a). However, with the
increase in the time $t$ the difference between the probabilities
$\rho_{n}(t)$ for the levels $n=2m$ and $n=2m+1$ is rapidly
smoothed over.

For the times $t\gg10^{-2}$ cm the function $\rho_{n}(t)$  may,
for the sake of convenience,  be approximated by a continuous
curve (solid) (Fig.14).

\bigskip

\textit{Figure 14. Distribution of $\rho_{n}(t)$ for large times.}

\bigskip

Dotted in Figs. 13 and 14 show the distribution function
$\rho_{\overline{n}}(t)$ corresponding to the mean value
\begin{equation}
\label{32.32} \overline{n}(t)=\sum_{n=1}^{n_{max}}n\rho_{n}(t).
\end{equation}

The mean value of the level number $\overline{n}$, as well as the
rms value $\overline{n}^{2}$ and the dispersion
$\sigma^{2}=\overline{n}^{2}-\overline{n}^{2}$ as the function of
time are given in Figure 15.

\bigskip

\textit{Figure 15. Time change of $\overline{n}$,
$\overline{n}^{2}$, $\sigma$}

\bigskip

A high value of dispersion $\sigma$, reaching $\overline{n}^{2}$
($\tau_{rec}$) at the depth $t\sim \tau_{rec}(c=1)$, is a
characteristic of the particle dynamics in a rectangular potential
well caused by radiative transitions between the levels of the
transverse energy of motion

Further assume that at the initial time ($t=0$) the beam incident
on the crystal leads to the uniform filling of the entire set of
the levels inside the potential well $\rho_{n}(0)=1/n_{max}$
($n\leq n_{max}$). A similar pattern of the level population is
likely to occurs when a wide beam with the angular divergence
$\theta\sim \theta_{cr}$ is incident on the crystal.

Consider non-radiative transitions. Solving the balance equation
(\ref{32.28}), we obtain the distributions  $\rho_{n}(t)$ (Figure
16, dashed curves).

\bigskip

\textit{Figure 16. Distribution in a rectangular well for
positrons ($E=1.2$ GeV (a) and 20 GeV (b)) with account of
multiple scattering (dashed curves) and ignoring  it (solid
curves): 1 - initial; 2 - for the plate of thickness $10^{-2}$ cm;
3 - for the plate of thickness $10^{-1}$ cm; 4 - for the plate of
thickness $0.5$ cm}

\bigskip

The diffusion coefficient $D=0.16 \,\,z_{v}z^{-2} D_{chaot}$ (the
magnitude of the coefficient $D$ is close to the experimental
one). In Figure 16 it is seen that multiple scattering prevails
over radiative transitions and fully determines the distribution
function $\rho_{n}(t)$. It is easy to see that the function
$\rho_{n}(t)$ are described by the solutions of an  ordinary
diffusion equation
\begin{eqnarray}
\label{32.33}
\rho_{n}(t)\simeq\rho_{n}(0)\Phi(\zeta);\, \Phi(\zeta)=\frac{2}{\sqrt{2\pi}}\int_{0}^{\zeta}e^{-x^{2}/2}dx;\nonumber\\
\zeta=\frac{n_{max}+1-n}{\sqrt{2Dt}}.
\end{eqnarray}
Note that in view of the accepted model (the particle that has
quitted the channel due to non-radiative transition $n\rightarrow
n^{\prime}$ ($n^{\prime}>n_{max}$) never returns into the
channel), the distribution $\rho_{n}(t)$ does not preserve time
normalization: $\sum_{n=1}^{n_{max}}\rho_{n}(t)\leq 1$.

With the growth of particle energy the probability of
non-radiative transitions $w_{nn^{\prime}}^{W}$ falls rapidly, and
the effect of radiative transitions  on the change in the
distribution function $\rho_{n}(t)$ is growing.

The process of radiative diminution of the transverse momentum of
channeled particles discussed above should be distinguished from
from the process of the damping of the amplitude of transverse
oscillations of a channeled particle. The rates of relaxation of
the transverse momentum and the oscillation amplitude are the same
only in a harmonic well. In a steep-walled well relaxation of the
transverse momentum is practically not accompanied by  the
reduction of the amplitude of transverse oscillations. From this
follows that  in spite of a possible decrease in the transverse
momentum, the damping of the oscillation amplitude is not
observed: in a steep-walled well it does not exist, and in a
harmonic well the relaxation time is large.

%%%%%%%%%%%%%%%%%%%%%%%%%%%%%%%%%%%  Section 33 %%%%%%%%%%%%%%%%%%%%%%

\section[Pair Production by $\gamma$-quanta in Crystals Under Channeling Conditions]{Pair Production by $\gamma$-quanta in Crystals Under Channeling Conditions}
\label{sec:10.33}

%%%%%%%%%%%%%%%%%%%%%%%%

The cross section for pair production  by $\gamma$-quanta in
crystals which takes account of possible channeling of electrons
and positrons has the form
\begin{equation}
\label{33.1}
d\sigma=2\pi\delta(\omega-E-E_{1})\overline{|M^{\prime}(\vec{p_{1}},\vec{k},\vec{p})|^{2}}
\frac{d^{3}pd^{3}p_{1}}{8(2\pi)^{6}\omega EE_{1}}\,,
\end{equation}
%%%%%%%%%%%%%%%%%%%%%%%%%%%%
where
\[
M^{\prime}(\vec{p_{1}},\vec{k},\vec{p})\equiv
e\sqrt{4\pi}M^{\prime}_{fi} =e\sqrt{4\pi}\int
d^{3}r\psi_{\vec{p}_{1}}^{(-)*}(\vec{r})\vec{\alpha}\vec{e}e^{i\vec{k}\vec{r}}\psi_{-\vec{p}}^{(+)}(\vec{r})\,;
\]
$\psi_{p_{1}}^{(-)}$ is the electron wave function;
$\psi_{-p}^{(+)}$ is the wave function with negative energy
$(-E)$. The  asymptotics of this function should have a form of a
diverging  spherical wave. The positron wave function (formed from
$\psi_{-\vec{p}}^{(+)*}$) will the asymptotics of a converging
spherical wave as required for a final particle \cite{19}.

Similarly to the pair production in a shielded Coulomb potential,
all characteristics of pair production in crystals may be found
from the expression for the cross section of the  bremsstrahlung
process, using the transform $(E, \omega,
\vec{e}^{\,*})\rightarrow (-E, -\omega, \vec{e})$. In the
expressions for $I_{1}$, $\vec{I}_{2}$, $\vec{I}_{3}$ the
transmitted momentum $\vec{q}=\vec{p}+\vec{p}_{1}-\vec{k}$. As a
result, we have
%%%%%%%%%%%%%%%%%
\begin{equation}
\label{33.2}
M^{\prime}_{fi}=2\sqrt{\frac{E}{E_{1}}}w_{1}^+\left\{(E_{1}-E)(\vec{g}\vec{e})-i\omega\vec{\sigma}
[\vec{g}\times\vec{e}]\right\}w\,.
\end{equation}
Calculation of the trace appearing in (\ref{33.1}) with due
account of polarization of all the final particle gives \cite{164}
\begin{equation}
\label{33.3} d\sigma=e^{2}\delta(\omega-E-E_{1})\textrm{Tr} \,\rho
M^{\prime
+}\rho_{1}M^{\prime}\frac{d^{3}pd^{3}p_{1}}{4(2\pi)^{4}\omega
EE_{1}}\,,
\end{equation}
where
\begin{eqnarray}
& &\textrm{Tr}\,\rho M^{\prime
+}\rho_{1}M^{\prime}=4\frac{E}{E_{1}}\left\{\frac{\omega^{2}}{2}|\vec{g}|^{2}-2EE_{1}
(1-\vec{\zeta}\vec{\zeta}_{1})|\vec{g}\vec{e}|^{2}\right.\nonumber\\
& &
\left.-\frac{\omega^{2}}{2}\texttt{Re}\left\{|\vec{g}|^{2}\vec{\zeta}\vec{\zeta}_{1}-
2(\vec{g}\vec{\zeta})(\vec{g}^{\,*}\vec{\zeta}_{1})\right\}
+\omega E_{1}\texttt{Re}\left\{\left[|\vec{g}|^{2}\vec{\zeta}\vec{e}^{\,*}\right.\right.\right.\nonumber\\
&
&\left.\left.\left.-2(\vec{g}\vec{e}^{\,*})(\vec{g}^{\,*}\vec{\zeta})\right](\vec{\zeta}\vec{e})\right\}
+\omega E\texttt{Re}\left\{\left[|\vec{g}|^{2}\vec{\zeta}_{1}\vec{e}^{\,*}\right.\right.\right.\nonumber\\
&
&\left.\left.\left.-2(\vec{g}\vec{e}^{\,*})(\vec{g}^{\,*}\vec{\zeta}_{1})\right](\vec{\zeta}\vec{e})\right\}
+\frac{\omega}{2}|\vec{g}|^{2}(E\vec{\zeta}+E_{1}\vec{\zeta}_{1})[i\vec{e}\times\vec{e}^{\,*}]\right.\nonumber\\
&
&\left.-\frac{\omega}{2}\texttt{Re}\left\{|\vec{g}|^{2}(E_{1}\vec{\zeta}
+E\vec{\zeta}_{1})[i\vec{e}\times\vec{e}^{\,*}]-2\left(\vec{g}\left(E_{1}\vec{\zeta}\right.\right.\right.\right.\nonumber\\
&
&\left.\left.\left.\left.+E\vec{\zeta}_{1}\right)\right)(\vec{g}^{\,*}[i\vec{e}\times\vec{e}^{\,*}])\right\}
-\frac{\omega}{2}\left[\omega(1-\vec{\zeta}\vec{\zeta}_{1})[i\vec{e}\times\vec{e}^{\,*}]\right.\right.\nonumber\\
&
&\left.\left.-(E-E_{1})[[\vec{\zeta}\times\vec{\zeta}_{1}]\times[i\vec{e}\times\vec{e}^{\,*}]]
+(E-E_{1})(\vec{\zeta}-\vec{\zeta}_{1})\right.\right.\nonumber\\
& &\left.\left.-2\texttt{Re}\left\{\vec{e}^{\,*}((E\vec{\zeta}
+E_{1}\vec{\zeta}_{1})\vec{e})\right\}\right][i\vec{g}\times\vec{g}^{\,*}]\right\}\,.\nonumber
\end{eqnarray}
%%%%%%%%%%%%%%%%%%%%%%%%%%%%%%%%%%%%%
Vector
\[
\vec{g}=\vec{I}_{2\perp\vec{k}}+\frac{1}{2}\vec{n}_{\vec{p}\perp\vec{k}}I_{1}
+\frac{m}{2E}\vec{n}_{\parallel}I_{1}\,,
\]
\[
\vec{I}_{2}=\frac{i}{2E}\int
e^{-i\vec{q}\vec{r}}\varphi^{(-)*}_{\vec{p}_{1}}(\vec{r})\vec{\nabla}_{r}\varphi_{-\vec{p}}^{(+)}(\vec{r})d^{3}r\,,
\]
\[
I_{1}=\int
e^{-i\vec{q}\vec{r}}\varphi^{(-)*}_{\vec{p}_{1}}(\vec{r})\varphi_{-\vec{p}}^{(+)}(\vec{r})d^{3}r\,.
\]
Due to mathematical equivalence of  (\ref{33.3}) and (\ref{6.27}),
integration of (\ref{33.3}) over the variables of one of the
particles (an electron or a positron) gives, e.g., for the number
of produced positrons, the expression coinciding in form with
(\ref{7.2}):
\begin{eqnarray}
\label{33.4} &
&\frac{d^{2}N_{e^{+}}}{dEd\Omega_{e^{+}}}=\frac{e^{2}E^{2}}{4\pi^{2}\omega}\texttt{Re}\sum_{nfj}Q_{nj}e^{i\tilde{\Omega}_{nj}L}
\left[\frac{1-\exp(iq_{zif}L)}{q_{zif}}\right]\nonumber\\
& &\times\left[\frac{1-\exp(iq_{znf}L)}{q_{znf}}\right]
\left\{\frac{\omega^{2}}{2E_{1}^{2}}\vec{g}_{nf}\vec{g}_{jf}^{\,*}-2\frac{E}{E_{1}}(1-\vec{\zeta}\vec{\zeta}_{1})\right.\nonumber\\
&
&\left.\times(\vec{g}_{nf}\vec{e})(\vec{g}_{jf}^{\,*}\vec{e})-\frac{\omega^{2}}{2E_{1}^{2}}
\texttt{Re}\left\{(\vec{g}_{nf}\vec{g}_{jf}^{\,*})(\vec{\zeta}\vec{\zeta}_{1})
-2(\vec{g}_{nf}\vec{\zeta})(\vec{g}_{jf}^{\,*}\vec{\zeta}_{1})\right\}\right.\nonumber\\
&
&\left.+\frac{\omega}{E_{1}}\texttt{Re}\left\{[(\vec{g}_{nf}\vec{g}_{jf}^{\,*})(\vec{\zeta}\vec{e}^{\,*})
-2(\vec{g}_{nf}\vec{e}^{\,*})(\vec{g}_{jf}^{\,*}\vec{\zeta})](\vec{\zeta}_{1}\vec{e})\right\}\right.\nonumber\\
& &\left.+\frac{\omega
E}{E_{1}^{2}}\texttt{Re}\left\{[(\vec{g}_{nf}\vec{g}_{jf}^{\,*})\vec{\zeta}_{1}\vec{e}^{\,*}
-2(\vec{g}_{nf}\vec{e}^{\,*})(\vec{g}_{jf}^{\,*}\vec{\zeta}_{1})](\vec{\zeta}\vec{e})\right\}\right.\nonumber\\
&
&\left.+\frac{\omega}{2E_{1}^{2}}(\vec{g}_{nf}\vec{g}_{jf}^{\,*})(E\vec{\zeta}
+E_{1}\vec{\zeta}_{1})[i\vec{e}\times\vec{e}^{\,*}]\right.\nonumber\\
&
&\left.-\frac{\omega}{2E_{1}^{2}}\texttt{Re}\left\{(\vec{g}_{nf}\vec{g}_{jf}^{*})
(E_{1}\vec{\zeta}+E\vec{\zeta}_{1})[i\vec{e}\times\vec{e}^{\,*}]
-2\left(\vec{g}_{nf}\left(E_{1}\vec{\zeta}\right.\right.\right.\right.\nonumber\\
&
&\left.\left.\left.\left.+E\vec{\zeta}_{1}\right)\right)(\vec{g}_{jf}^{\,*}[i\vec{e}\times\vec{e}^{\,*}])\right\}
-\frac{\omega}{2E_{1}^{2}}\left[\omega(1-\vec{\zeta}\vec{\zeta}_{1})[i\vec{e}\times\vec{e}^{\,*}]\right.\right.\nonumber\\
&
&\left.\left.-(E-E_{1})[[\vec{\zeta}\times\vec{\zeta}_{1}]\times[i\vec{e}\times\vec{e}^{*}]]
+(E-E_{1})(\vec{\zeta}-\vec{\zeta}_{1})\right.\right.\nonumber\\
& & \left.\left.-2\texttt{Re}\left\{\vec{e}^{\,*}((E\vec{\zeta}
+E_{1}\vec{\zeta}_{1})\vec{e})\right\}\right][i\vec{g}_{nf}\times\vec{g}_{jf}^{\,*}]\right\}\,.
\end{eqnarray}

%%%%%%%%%%%%%%%%%%%%%%%%%%%%%%%%%%%%%%%%%%%
In (\ref{33.4}) the quantity
\begin{eqnarray}
\label{chan_33.5}
\vec{g}_{nf}&\equiv&\frac{1}{2E}(\vec{I}^{\,\prime}_{2nf}+\vec{p}_{z\perp\vec{k}}I_{1nf}^{\prime}
+m\vec{n}_{\parallel}I_{1nf}^{\prime})\,,\nonumber\\
I_{1nf}^{\prime}&=&N_{\perp}\int_{S}e^{i\vec{k}\vec{\rho}}\psi^{*}_{f\vec{\kappa}_{1}}
(\vec{\rho})\psi_{n,-\vec{\kappa}}(\vec{\rho})d^{2}\rho\,,\nonumber\\
\vec{I}^{\prime}_{2nf}&=&iN_{\perp}\int_{S}e^{i\vec{k}\vec{\rho}}\psi^{*}_{f\vec{\kappa}_{1}}
(\vec{\rho})\vec{\nabla}_{\rho}\psi_{n,-\vec{\kappa}}(\vec{\rho})d^{2}\rho\,,
\end{eqnarray}
the subscript $f$ refers to electron states, subscripts $n$ and
$j$ to positron ones; $\vec{\kappa}$ is obtained by reduction of
the momentum $\vec{p}_{\perp}$ to the first Brillouin  zone;
$\vec{\kappa}_{1}$ is the same for the momentum
($\vec{k}_{\perp}-\vec{p}_{\perp}$); $E_{1}=\omega-E$;
$Q_{nj}=c_{n}(-\vec{p}_{\perp})c_{j}^{*}(-\vec{p}_{\perp})$ (see
 the definitions in Chapter III); $q_{znf}=p_{zn}+p_{1zf}-k_{z}$;
$p_{zn}=\sqrt{p^{2}-2m\varepsilon_{n\kappa}(E)}$;
$p_{1zf}=\sqrt{p_{1}^{2}-2m\varepsilon_{f\kappa_{1}}(E_{1})}$.
It should be pointed out that at pair production, as well as in
the process of photon generation (see (\ref{7.2}) the oscillations
of $d^{2}N_{e^{+}}$,  depending on the thickness $L$ which enter
into (\ref{33.4}) cannot be ignored in the general case. This is
due to the fact that there are a lot of closely spaced and even
degenerate levels in the structure of transverse levels. For
example, at axial channeling there is level degeneracy in the sign
of the projection of the orbital moment, and the over-barrier
states  are, in fact, continuous. The contribution of such states
at different energies has not been  interrogated yet. For
non-degenerate  states the features of $1/q_{zjf}$ and $1/q_{znf}$
are not the same, and at $\tilde{\Omega}_{nj}L\gg 1$ the
interference terms may be discarded.\footnote{Level degeneracy in
the axial case affects attenuation of non-diagonal elements of the
density matrix, which should be taken into account when analyzing
the bursts of nuclear reactions discussed in \cite{15,23,24}.}

%%%%%%%%%%%%%%%%%%%%

If we are not concerned about the polarization of the  particles
produced, we should assume that in (\ref{33.4}) $\vec{\zeta}$ and
$\vec{\zeta}_{1}$ are zero, and multiply the result obtained by 4.
In consequence, for example, for non-degenerate states we have
\begin{eqnarray}
\label{33.6}
\frac{d^{2}N_{e^{+}}}{dEd\Omega_{e^{+}}}=\frac{2e^{2}E^{2}L}{\pi\omega}\sum_{nf}Q_{nn}\delta(q_{znf})\left\{\frac{\omega^{2}}{2E_{1}^{2}}|\vec{g}_{nf}|^{2}\right.\nonumber\\
\left.-2\frac{E}{E_{1}}|\vec{g}_{nf}\vec{e}|^{2}-\frac{\omega^{2}}{2E_{1}^{2}}[i\vec{e}\times\vec{e}^{*}][i\vec{g}_{nf}\times\vec{g}_{nf}^{*}]\right\}.
\end{eqnarray}

According to (\ref{33.4}) and (\ref{33.6}) the cross section of a
channeled particle production depends on the photon polarization
state. In particular, the cross section of  production of a pair
undergoing planar channeling is different for photons with
polarization vector perpendicular (parallel) to the plane in
question. Hence, for such $\gamma$-quanta a crystal exhibits
dichroism. And flux of  non-polarized $\gamma$-quanta incident on
the crystal will get polarized.

The effect we have considered above will be fundamentally
different from the effect of $\gamma$-quanta polarization by
crystals discussed by Cabibbo (see, e.g., \cite{63,64}). The
effect considered by Cabbibo corresponds in a crossed channel with
coherent bremsstrahlung; the effect we have considered corresponds
with the formation of polarized photons through radiative
transitions between the levels of channeled particles. Moreover,
e.g., at zero entrance angles  with respect to a crystallographic
plane, the effect considered by Cabibbo is zero, whereas in our
case it is non-zero. From the Kramers-Kroning dispersion relations
follows that alongside with absorption, the real parts of
refractive indices will be different for different photon
polarizations. In other word, the crystal for high-energy
$\gamma$-quanta appears to be birefringent \cite{83}.

The general formulae (\ref{6.27}), (\ref{33.3}) are also suitable
for describing radiation and pair production processes  in bent
crystals. In particular, in (\ref{3.27}) the terms proportional to
$\zeta_{1}$ describe the effect of radiation self-polarization of
spin in bent crystals  if the crystal thickness is less than the
length of self-polarization, which was established in \cite{9}.
The similarity of the formulas for bremsstrahlung and pair
formation implies that in bent crystals even non-polarized
$\gamma$-quanta will produce polarized electrons and positrons.
Polarized particles, in turn, will produce polarized (having
circular polarization) $\gamma$-quanta.

%%%%%%%%%%%%%%%%%%%%%%%%%%%%%%%%%%%%%%%%%%%%

%%%%%%%%%%%%%%%%%%%%%%%%%%%%%%%%%%%  Section 34 %%%%%%%%%%%%%%%%%%%%%%

\section[Nuclear Optics of Crystals at High Energies]{Nuclear Optics of Crystals at High Energies}
\label{sec:10.34}

We studied the formulae describing photon radiation in crystals in
the Sommerfeld-Maue approximation for the wave function of
electrons (positrons) and in the two-wave approximation for the
wave function of $\gamma$-quanta produced. The existence of the
rotation effect and particle self-polarization means that in thick
crystals we must go beyond the scope of the Sommerfeld-Maue
approximation and use the wave functions, which are the solution
of the Dirac equation including the anomalous magnetic moment of
the electron. Moreover, with the increasing frequency of the
produced photon, when the wave length of a $\gamma$-quantum
appears to be much  shorter than the distance between atoms
(nuclei), for wave functions of a photon
$\vec{A}_{\vec{k}\vec{S}}(\vec{r})$ and other particles (e.g.,
neutrons) it is possible to apply the approximation similar to
that used for describing electron and positron channeling (see
$\S2$, \cite{83}). When a $\gamma$-quantum moves at a small angle
with respect to the planes (axes) of a single crystal, one may
introduce the averaged over the plane (chain of atoms) dielectric
permittivity. In this regard we may talk about the existence of
channeling of $\gamma$-quanta and any of other particles, (e.g.,
neutrons, $K^{0}$-mesons) \cite{83}.

It is worthy of mention that the crystal structure of a target is
of impact even at very high energies (e.g., gigaelectronvolt and
higher) of $\gamma$-quanta. In particular, due to channeling of
pairs produced by $\gamma$-quanta,  the intensity of transmitted
$\gamma$-quanta demonstrates  a pronounced dependence on the
rotation angle of the crystal with respect to momentum of the
incident beam (see also the previous section). Moreover, even at
such high energies the crystal proves to be optically anisotropic
(the birefringence phenomenon appears) \cite{83}. One may
theoretically describe  this effect, recalling that when a photon
moves in an external field (electric, magnetic), due to the vacuum
polarization by the field, the area occupied by the field is
characterized by the dielectric permittivity tensor of the form
\cite{149}
\[
\varepsilon_{ij}=\delta_{ij}+2g_{1}(\kappa)F_{i}^{(1)}F_{j}^{(1)}+2g_{2}(\kappa)F_{i}^{(2)}F_{j}^{(2)},
\]
where $\vec{F}^{(1)}=\vec{E}_{\perp}+[\vec{n}\vec{H}]$;
$\vec{F}^{(2)}=[\vec{n}\vec{F}_{1}]$;
$\vec{E}_{\perp}=\vec{E}-\vec{n}(\vec{n}\vec{E})$;
$\vec{n}=\vec{k}/k$ (the definition of functions $g_{1}$ and
$g_{2}$ see in \cite{149}). From this, when a $\gamma$-quantum
moves, for example, at a small angle with respect to the
crystallographic plane, the dielectric permittivity of a crystal
for photons, whose polarization is parallel to the plane differs
from that for $\gamma$-quanta, whose polarization is perpendicular
to the crystallographic plane. As a result, birefringence arises.
As seen from estimates, to transform a linearly polarized photon
into a circularly polarized one, the crystal length of the about 1
cm is sufficient. As a consequence, it is definitely impossible to
neglect the refraction effects in thick crystals even at high
energies of emitted photons. The investigation of birefringence of
$\gamma$-quanta in electric fields produced by crystallographic
planes even now allows studying the effects of vacuum polarization
by external fields.

Let us give a brief review of the theory of photon radiation in
crystals, which is not restricted by the Sommerfeld-Maue
approximation \cite{72}. To analyze the photon radiation in
crystals, make use of the general quantum theory of reactions.

Let a plane wave of momentum $\vec{p}$ and energy $E$ describing
the primary particle be incident on a crystal with the volume $V$.
It is well known that at the distances larger in comparison with
the size of the objects, alongside with a primary wave, there are
spherical waves describing secondary particles. To find the
radiation cross section (the transition probability per unit time,
the number of emitted photons) within the framework of the
time-independent theory of reactions, it is necessary that the
wave function of the primary particle exactly allowing for its
interaction with the crystal and having the asymptotic of the
diverging spherical wave $(\psi_{p}^{(+)}(\vec{r}))$ should be
taken as the initial wave function. The wave functions of final
particles have the asymptotic of a converging wave
$\psi_{p}^{(-)}(\vec{r}(A^{(-)}_{\mu\vec{k}}(\vec{r})$ for a
photon). Using the general formulae \cite{19}, we have the
following expression for the transition probability per unit time
with photon emission ($\hbar=c=1$):
\begin{equation}
\label{34.1}
dW=2\pi\delta(E-E_{1}-\omega)|M(\vec{p}_{1},\vec{k};\vec{p})^{2}|\frac{d^{3}p_{1}d^{3}k}{8(2\pi)^{6}L^{3}EE_{1}\omega},
\end{equation}
where $M(\vec{p}_{1},\vec{k};\vec{p})=\int
d^{3}r\overline{\psi_{\vec{p}_{1}}^{(-)}}(\vec{r})\gamma^{\mu}\psi_{\vec{p}}^{(+)}(A^{(-)*}_{\mu\vec{k}}(\vec{r})$;
$L^{3}$ is the normalization volume; $\psi^{\pm}$ are the exact
solutions of the Dirac equation; $(A^{(-)}_{\mu\vec{k}}$ are the
exact solutions of Maxwell's equations.

Recall that according to the rules (see \cite{19}, p. 285) the
stated wave functions do not include a factor of the type
$1/\sqrt{2EL^{3}}$. It should be noticed that in view of \cite{14}
a particle in a medium is affected by an effective potential
expresses in terms of the elastic scattering amplitude. This
amplitude is a complex value. At high energies the contribution to
the imaginary part of the amplitude comes from, e.g.,
bremsstrahlung and the pair production effect. For this reason at
large energies  the functions $\psi^{\pm}$ satisfy the Dirac
equation with a complex potential.

When considering the reactions with polarized particles, it is
helpful to represent the solutions of $\psi$  and $A$ in the form
explicitly including final polarization of particles too. With
this aim in view, write $\psi(\vec{r}=\hat{\psi}(\vec{r})u$ and
$A_{\mu}=B^{\nu}_{\mu}e_{\nu}$, where $u$ is the bispinor
characterizing the polarized state of an electron (positron) in a
plane wave outside the crystal; $e_{\nu}$ - is the photon
polarization vector in the plane wave outside the crystal. As a
result the matrix element
\begin{equation}
\label{34.2} M=\bar{u}_{\vec{p}_{1}}G^{\nu}u_{\vec{p}}e_{\nu};\,
G^{\nu}= \int
d^{3}r\hat{\overline{\psi}}^{(-)}_{\vec{p}_{1}}(\vec{r})\gamma^{\mu}\hat{\psi}_{p}^{(+)}(\vec{r})B^{\nu}_{\mu}(\vec{r}).
\end{equation}
Upon introducing polarization density matrices of the initial
$\rho$ and final $\rho_{1}$ electrons and a photon $\rho_{\mu\nu}$
the squared matrix element entering into (\ref{34.1}) is:
\begin{eqnarray}
\label{34.3}
\overline{|M|^{2}}=sp \,\rho_{1}G^{\mu}\rho \bar{G}^{\nu}\rho_{\mu\nu}=T^{\mu\nu}\rho_{\mu\nu},\nonumber\\
T^{\mu\nu}=sp\,\rho_{1}G^{\mu}\rho G^{\nu}
\end{eqnarray}
($T^{\mu\nu}$ is the linear function of the polarization vectors
of the initial $\vec{\zeta}$ and final $\vec{\zeta}_{1}$ electrons
(positrons)). Therefore the explicit form of $T$ as a function of
$\vec{\zeta}$ and  $\vec{\zeta}_{1}$ may be written as follows:
\begin{equation}
\label{34.4}
T^{\mu\nu}=\alpha^{\mu\nu}+\vec{\beta}^{\mu\nu}\vec{\zeta}+\vec{\beta}^{\mu\nu}_{1}\vec{\zeta}_{1}
+\gamma^{\mu\nu}_{il}\zeta_{i}\zeta_{1l}.
\end{equation}
Carrying out further analysis, we take into account that
integration in (\ref{34.2}) is made over the he range with linear
dimensions exceeding the linear dimensions of a crystal by only
the magnitude of the vacuum coherent length of radiation. That is
why, when considering the radiation process in a crystal whose
lateral dimensions are much larger than its thickness, in order to
find the wave functions, one may use the wave functions describing
scattering of a plane wave by a crystal plate of finite
dimensions. According to (\ref{sec:1.1}), (\ref{sec:1.2}) in this
case the time-independent wave function of a particle (photon) in
the area occupied by the crystal is defined by the superposition
of the Bloch functions. Since the Bloch functions may be
represented as a superposition of plane waves (their explicit form
see in (\ref{sec:1.2})), integration in (\ref{34.2}) with respect
to the coordinates in the plane parallel to the crystal surface
leads to the $\delta$-function describing the law of conservation
of the component of a transmitted momentum, which is parallel to
the crystal surface. The stated $\delta$-function together with
the energy $\delta$-function enables performing in (\ref{34.4})
explicit integration with respect to $\vec{p}_{1}$ (or $\vec{k}$).
As a result, we obtain a spectral-angular distribution of emitted
photons (or ejected electrons).

From (\ref{34.2})-(\ref{34.4}) follows that all the amplitudes
$\alpha,\,\beta,\,\gamma$ appearing in (\ref{34.4}) will be the
squared absolute values of the superposition of the functions:
\[
\sum
c_{nn^{\prime}}\frac{e^{-iq_{znn^{\prime}}l}-1}{q_{znn^{\prime}}},
\]
where $q_{znn^{\prime}}$ is the longitudinal momentum transmitted
to the crystal. The stated superpositions oscillate with the
crystal thickness.

According to \cite{14}, when particles and $\gamma$-quanta move in
crystals, a number of polarization phenomena arise
(multi-frequency change of  polarization characteristics of
electrons, positrons and $\gamma$-quanta, depending on thickness;
the effect of anomalous transmission depending on the external
field frequency). All these phenomena also manifest themselves in
the case under study. In other words, $T^{\mu\nu}$ and, hence, the
intensity and polarization characteristics of emitted photons
(electrons)  are the oscillating functions of the energy of
particles, crystal thickness and the frequency of the variable
external field (sound, electromagnetic) imposed on the crystal.

%%%%%%%%%%%%%%%%%%%%%%%%%%%%%%%%%%%  Section 35 %%%%%%%%%%%%%%%%%%%%%%

\section[Surface Channeling of Charged Particles]{Surface Channeling of Charged Particles}
\label{sec:10.35}

The experiments \cite{162,163,164} studying refraction of ion
beams from crystal surfaces revealed sharp anomalies in the number
of particles refracted by the crystal at rotation of the crystal
surface about the axes perpendicular to it. The quantum mechanical
explanation of this effect is given below \cite{165}.

Let a particle beam be incident on the surface of the crystal
which occupies the half-space area $z>0$ at the glancing angle
$\alpha$. Choose the x-axis perpendicular to a certain family of
crystallographic planes, and the y-axis parallel to the stated
family of planes. Suppose that particles are incident at a small
angle $\beta$ with respect to the stated planes. Then the general
view of the wave reflected from the crystal surface can be written
as follows:
\begin{equation}
\label{35.1}
\psi=e^{i\vec{k}_{0}\vec{r}}+\sum_{\tau}A_{\tau}e^{i(k_{0x}+2\pi\tau)x}e^{ik_{0y}y}e^{-ik_{z}(\tau)z},
\end{equation}
where
$k_{z}(\tau)=(k_{0}^{2}-(k_{0x}+2\pi\tau)^{2}-k_{0y})^{1/2}=(k_{0z}^{2}+k_{0x}^{2}-(k_{0x}+2\pi\tau)^{2})^{1/2}$
is found from the condition of the wave energy conservation
(preservation) at elastic scattering.

When deriving (\ref{35.1}), it was taken into account that, due to
the periodicity of the potential of the selected family of planes
along the x-axis, the parallel $x$-component  of the momentum of
the wave reflected from the crystal may only differ from the
initial value of $k_{0x}$ by the reciprocal lattice vector
$2\pi\vec{\tau}$.

The general solution of the Schrodinger equation describing the
particle motion inside the crystal in the case under study has the
form
\begin{equation}
\label{35.2}
\psi=\sum_{nk_{x}}c_{n}(k_{x})\psi_{nk_{x}}(x)e^{ik_{y}y}e^{i\kappa_{zn}(k_{x})z},
\end{equation}
where
$\kappa_{n}(k_{x})=(k_{0z}^{2}+k_{0x}^{2}-q_{n}^{2}(k_{x}))^{1/2}$;
$q_{n}^{2}(k_{x})=\frac{2m}{\hbar^{2}}E_{n}(k_{x})$;
$E_{n}(k_{x})$ is the particle energy in a periodic
one-dimensional potential of the family of planes in question in
the range $n$ as a function of the wave number $k_{x}$;
$\psi_{nk_{x}}(x)=\frac{1}{\sqrt{N_{x}}}e^{ikx}u_{nk}(x)$ is the
Bloch function. Unlike the case of mirror reflection of neutrons
and $\gamma$-quanta under diffraction conditions \cite{14}, in the
the case of diffraction of charged particles we consider the
two-wave approximation is not applicable.

Joining (ref({35.1}) and (\ref{35.2}) at the crystal boundary, we
may obtain the following expressions for the amplitudes of
refracted waves $A_{\tau}$ and coefficients $c_{n}(k_{x})$:
\begin{eqnarray}
\label{35.3}
A_{\tau=0}=\sum_{n}\frac{k_{0z}-\kappa_{n}(k_{x})}{k_{0z}+\kappa_{n}(k_{x})+\delta_{n}\left(\frac{2\pi l}{a}\right)}\nonumber\\
\times\frac{1}{\Omega}\left|W_{n}\left(\frac{2\pi
l}{a}\right)\right|^{2}g_{n}\left(\frac{2\pi l}{a}\right);
\end{eqnarray}
\begin{eqnarray}
\label{35.4}
A_{\tau\neq 0}=\sum_{n}\frac{2k_{0z}}{k_{0z}+\kappa_{n}(k_{x})+\delta_{n}\left(\frac{2\pi l}{a}\right)}\nonumber\\
\times\frac{1}{\Omega}W_{n}\left(2\pi \tau+\frac{2\pi
l}{a}\right)W_{n}^{*}\left(\frac{2\pi l}{a}\right)
g_{n}\left(\frac{2\pi l}{a}\right);
\end{eqnarray}
\begin{eqnarray}
\label{35.5} c_{n}\left(\frac{2\pi
l}{a}\right)=\frac{2\sqrt{N_{x}}k_{0z}W_{n}^{*}\left(\frac{2\pi
l}{a}\right)}{k_{0z}+\kappa_{n}(k_{x})+\delta_{n}\left(\frac{2\pi
l}{a}\right)}
g_{n}\left(\frac{2\pi l}{a}\right);\nonumber\\
c_{n}(k_{x}\neq k_{0x})=0,
\end{eqnarray}
where $g_{n}\left(\frac{2\pi l}{a}\right)$ satisfies the set of
equations of the form
\begin{eqnarray}
\label{35.6}
g_{n}\left(\frac{2\pi l}{a}\right)=1+\sum_{n^{\prime}\neq n}\left[\sum_{\tau\neq 0}(k_{0z}-k_{z}(\tau))\right.\nonumber\\
\times\frac{1}{\Omega}W_{n}^{*}\left(2\pi \tau+\frac{2\pi l}{a}\right)W_{n^{\prime}}\left(2\pi \tau+\frac{2\pi l}{a}\right)\nonumber\\
\times\frac{W_{n^{\prime}*}\left(\frac{2\pi
l}{a}\right)}{W_{n}^{*}\left(\frac{2\pi l}{a}\right)}
\frac{g_{n^{\prime}}\left(\frac{2\pi l}{a}\right)}{k_{0z}+\kappa_{n^{\prime}}(k_{x})+\delta_{n^{\prime}}\left(\frac{2\pi l}{a}\right)};\nonumber\\
\delta_{n}\left(\frac{2\pi l}{a}\right)\equiv\sum_{\tau\neq
0}(k_{z}(\tau)-k_{0z})\frac{1}{\Omega}
\left|W_{n}\left(2\pi\tau+\frac{2\pi l}{a}\right)\right|^{2};
\end{eqnarray}
Integration is made over the unit cell volume $\Omega$; $l$ is the
integer part of $\frac{k_{0x}a}{2\pi}$.;
$$
W_{n}(2\pi\tau)=\int e^{-i2\pi\tau x}u_{nk_{0x}}(x)dx.
$$
To clarify the structure of expressions (\ref{35.3}),
(\ref{35.4}), recall that the ordinary amplitude of a mirror
reflected wave has the form
\begin{equation}
\label{35.7} B=\frac{k_{0z}-k_{z}}{k_{0z}+k_{z}},
\end{equation}
where $k_{z}=k_{0z}n$; $n$ is the refractive index.

Comparison of (\ref{35.3}) and (\ref{35.7}) shows that the
amplitude of a mirror reflected wave $A_{0}$ under channeling
conditions can be represented as a superposition of the amplitudes
describing mirror reflection from the medium with the "refractive
index" $n=\kappa_{n}(k_{x})/k_{oz}$ which depends on the number of
the zone where the particle incident on the crystal is captured in
it.

Consider the dependence of the reflected wave amplitude  on the
gliding angle $\alpha$ and the azimuth angle $\beta$. First note
that with the change of the angle $\beta$, the magnitude of the
component of the particle momentum $k_{0x}$ lso changes, and,
hence, the magnitude of the number $l$. On the other hand, the
coefficients $W_{n}(2\pi l/a)$ as a function of the  number of the
range $n$ have the maximum distribution at $n=2l$, which falls
rapidly with the increase in the difference $n-2l$, with the
distribution $W_{n}(2\pi l/a)$ getting sharper at large values of
the number $l$. The features mentioned above result in the fact
that at $\beta$ exceeding a certain critical angle which has the
same order of magnitude as the Lindhard angle, $g_{n}(2\pi
l/a)\rightarrow 1$,  $\delta_{n}(2\pi l/a)\rightarrow 0$. As  a
consequence, $A_{0}\rightarrow B$, $A_{\tau\neq 0}\rightarrow 0$,
and we go over to a known mirror reflection pattern.

Let now the glancing angle of ions $\alpha$  be much greater than
the total mirror reflection angle
$\vartheta_{cr}=\sqrt{u_{mean}/E}$ ($u_{mean}$ is the mean energy
of particle-crystal interaction). In this case , if the angle
$\beta$ is greater than the Lindhard angle, the intensity of
reflected particles is small. However, at $\beta\rightarrow 0$,
one may see (see formulae (\ref{11.3}), (\ref{11.4}) and the above
mentioned features of the coefficients $W_{n}(2\pi l/a)$) that
though the amplitude $A_{0}$ remains small, the amplitudes
$A_{\tau\neq 0}$ ($A_{\tau\neq 0}\sim W_{0}(\tau)W_{0}^{*}(0)$)
become greater. As a consequence, one may observe the increase in
the intensity of reflected particles (as $2\pi\tau/k\ll 1$, for
$Ar$ with $E=30$ keV $k\sim 10^{12}\, cm^{-1}$, the particles with
$\tau\neq 0$ move practically in the plane of mirror reflection).
The situation is different when ions fall on the crystal at
gliding angle comparable with the angle of total mirror
reflection. In this case at $\beta$ greater than the Lindhard
angle the total mirror reflection of the wave is observed, i.e.,
$A_{0}\rightarrow 1$, and $A_{\tau\neq 0}\rightarrow 0$. At the
same time at $\beta=0$ the amplitude $A_{0}$ is small, as the here
low energy levels for which vector $\kappa_{n}\sim k_{0z}$ play
the leading part. The amplitude $A_{\tau\neq 0}$  also vanishes at
$k_{0z}\rightarrow 0$, due to the limitation of $\delta_{n}(2\pi
l/a)$. As a result at $\beta=0$ the minimum in the intensity of
reflected particles should be observed. The qualitative picture
given here is in good agreement with the experimental results
\cite{162}. Indeed, the total mirror reflection angle for ions
$A_{\tau}$ with $E=30$ keV and the crystal of $Cu$ is about
$4^{\circ}$. In the experiment at the gliding angle
$\alpha=10-15^{\circ}$ and the azimuth angle $\beta=0$ the maximum
intensity of scattered particles was observed, at the same time at
$\alpha=5^{\circ}$ and $\beta=0$ the minimum was observed. It
should be emphasized that the phenomenon discussed is of the
general character and it should occur for surface channeling of
light particles (electrons and positrons). It is natural that
electrons and positrons undergoing surface channeling emit photons
due to transitions between the levels (ranges) of transverse
motion.  Induced transitions caused by a polarized electromagnetic
wave result in particle polarization (compare (\ref{sec:6.19})).
Quantum modulation of a reflected beam also emerges.

\printindex


\begin{thebibliography}{165}

\bibitem[{Afanas'ev and Aginyan(1978)}]{80} Afanas'ev, A.~M. and Aginyan, M.~A. (1978).  \emph{Zh. Eksp. Teor. Fiz.} \textbf{74} p. 570 [ \emph{Sov. Phys. JETP} \textbf{47} p. 300].
 \bibitem[{Afanasiev  and  Kagan(1965)}]{123} Afanasiev, A.  and  Kagan, Yu. (1965). \emph{Zh. Eksp. Teor. Fiz.} \textbf{48}  p. 327.
 \bibitem[{Akhiezer \emph{et~al.}(1979)Akhiezer, Akhiezer and Shulga}]{60} Akhiezer, A.~I., Akhiezer, I.~A. and Shulga, N.~F. (1979).  \emph{Zh. Eksp. Teor.  Fiz.} \textbf{76} p.  1244. %% see 87
\bibitem[{Akhiezer \emph{et~al.}(1979)Akhiezer, Akhiezer and Shul'ga}]{87} Akhiezer, A.~I., Akhiezer,  I.~A.  and Shulga,  N.~F. (1979). Theory of bremsstrahlung of relativistic electrons and positrons in crystals, \emph{Sov. Phys. JETP} \textbf{49}, 4, pp. 631--635.
\bibitem[{Akhiezer and Shul'ga(1980)}]{155} Akhiezer, A.~I. and  Shulga, N.~F. (1980). On electromagnetic showers in crystalline media   \emph{JETP Lett.} \textbf{32}, 4, pp. 294--296 (\emph{Pis'ma. Zh. Eksp. Teor.  Fiz.} \textbf{32}, 4, p. 318).
 \bibitem[{Alferov  \emph{et~al.}(1977a)Alferov, Bashmakov and Bessonov}]{95a} Alferov, D.~F.,  Bashmakov, Yu.~A. and Bessonov, Ye.~G. (1977a). On classical theory of induced electromagnetic radiation of charged particles in modulators, in \emph{Preprint FIAN}, No 162 (Moscow) ; Zh. Tekh. Fiz. \textbf{48}  1592.
\bibitem[{Alferov  \emph{et~al.}(1977b)Alferov, Bashmakov and Bessonov}]{95b}  Alferov, D.~F.,  Bashmakov, Yu.~A. and Bessonov, Ye.~G. (1977b). \emph{Zh. Tekh. Fiz.} \textbf{48}  p. 1592.
\bibitem[{Alguard \emph{et ~al.}(1979)Alguard, Swent, Pantell, Berman, Bloom and Datz}]{55} Alguard, M.~J., Swent, R.~L., Pantell, R.~H., Berman, B.~L. Bloom, S.~D. and Datz, S. (1979). Observation of radiation from channeled positrons,  \emph{Phys. Rev. Lett.} \textbf{42}, 17, pp. 1148--1151.
\bibitem [{Allen and Eberly(1975)}]{114} Allen, L. and  Eberly, J.~H. (1975). \emph{Optical Resonance and Two-level Atoms} (Wiley, New York).
\bibitem[{Avakyan \emph{et~al.}(1975)Avakyan,  Aginyan,  Garibyan and Yan Shi}]{79} Avakyan, A.~L., Aginyan, M.~A., Garibyan, G. M. and Yan Shi (1975). \emph{Zh. Eksp. Teor. Fiz.} \textbf{68}   2038.
\bibitem[{Baier \emph{et~al.}(1973)Baier, Katkov and Fadin}]{149} Baier, V.~N., Katkov, V.~M.  and Fadin, V.~S. (1973). \emph{Radiation of Relativistic Electrons} (Atomizdat, Moscow) [in Russian].
\bibitem[{Baier \emph{et~al.}(1979)Baier, Katkov and Strakhovenko}]{58} Baier, V.~N., Katkov, V.~M. and Strakhovenko, V.~M. (1979). On radiation of relativistic positrons at channeling,   \emph{Phys. Lett. A} \textbf{73}, 5--6, pp. 414--416.
\bibitem[{Baryshevskii(1971)}] {74} Baryshevskii,  V.~G. (1971). Scattering of light by a flow of electrons through a crystal, \emph{Dokl. Akad. Nauk BSSR}  \textbf{15}, 4, pp. 306--308.
\bibitem [{Baryshevskii(1974)}]{97} Baryshevskii, V.~G. (1974).  Effect of energy losses on bremsstrahlung by relativistic electrons, \emph{Zh. Eksp. Teor. Fiz.} \textbf{67}  pp. 1651-1659 (\emph{Sov. Phys. JETP} \textbf{40} (1975) 821).
\bibitem [{Baryshevsky(1976)}]{14}  Baryshevsky, V.~G. \emph{Nuclear  Optics of Polarized Media} (Bel. State Univer., Minsk) (in Russian).
 \bibitem [{Baryshevsky(1979a)}]{120} Baryshevsky, V.~G. (1979a).  Optical Anisotropy of Matter in a Hard Spectrum in the Presence of Variable External Fields, in \emph{Radiation-Induced Phenomena in Condensed Media} (MIFI, Moscow).
\bibitem [{Baryshevsky(1979b)}]{136} Baryshevsky, V.~G. (1979b). Theory of measuring the duration of nuclear reactions using the blocking effect, \emph{Sov. J. Nucl. Phys.} \textbf{30}, 3, pp. 448--452 [Yad. Fizika  \textbf{30}  867].
\bibitem [{Baryshevsky(1979c)}]{9}  Baryshevsky, V.~G. (1979c). Radiative self-polarization and spin precession of particles moving in crystals, \emph{Dokl. Akad. Nauk BSSR} \textbf{23}, 5, pp. 438--439.
 \bibitem [{Baryshevsky(1979d)}]{10} Baryshevsky,  V.~G. (1979d). Spin rotation of  ultra-relativistic particles passing through a crystal, \emph{Pis'ma. Zh. Tekh. Fiz.} \textbf{5}, 3, p.  182--184.
\bibitem [{Baryshevsky(1979e)}]{84} Baryshevsky,  V.~G. (1979e). Coherent neutron--optical (optical) resonance, \emph{Dokl. Akad. Nauk BSSR } \textbf{23}, 12, pp. 1107--1109.
\bibitem [{Baryshevsky(1979f)}]{83}  Baryshevsky, V.~G. (1979f).  Coherent Processes in Crystals at High Energies, in  \emph{Proceedings of XIV Winter School LNPI} (LNPI, Leningrad) p. 158.
 \bibitem [{Baryshevsky(1980a)}]{113} Baryshevsky, V.~G. (1980a). \emph{Dokl. Akad. Nauk SSSR} \textbf{255} 331.
\bibitem [{Baryshevsky(1980b)}]{73} Baryshevsky,  V.~G.  (1980b). \emph{Izv. Akad. Nauk  BSSR} ser. fiz.-mat. \textbf{3}  p. 117.
 \bibitem [{Baryshevsky(1980c)}]{72}  Baryshevsky, V.~G. (1980c). Emission of photons by polarized electrons  passing through a single crystal,  \emph{Dokl. Akad. Nauk BSSR} \textbf{24}  pp. 510--512.
\bibitem [{Baryshevsky(1980d)}]{20} Baryshevsky,  V.~G. (1980d). Channeling, emission and reactions in crystals at high energies, in, \emph{Physics of High Energies:  Proceedings of XV Winter  School LNPI} (LNPI, Leningrad, 1980) pp.199--217.
\bibitem[{Baryshevskii(1981)}]{125}  Baryshevskii, V.~G. (1981).
Multifrequency precession of the neutron spin in a uniform
magnetic field, \emph{JETP Lett.} \textbf{33}, 1, pp. 74--77
(\emph{Pis'ma Zh. Eksp. Teor. Fiz.} \textbf{33}, 1, pp. 78--81).
\bibitem[{Baryshevsky and Chevganov(1979)}]{26} Baryshevsky, V.~G. and Chevganov, B.~A. (1979).
\emph{Preliminary Programme and Abstracts of the X Conference on
the Problems of Application of Charged Particle Beams for Studying
the Composition and Properties of Matter} (Moscow) p. 72.
\bibitem[{Baryshevskii and Dubovskaya(1976a)}]{34}  Baryshevskii, V.~G. and  Dubovskaya, I.~Ya. (1976a). Complex and
anomalous Doppler effect for channelized positrons (electrons),
\emph{Dokl.  Akad. Nauk SSSR } \textbf{231}, 6, pp. 1335--1338
(\emph{Sov. Phys. Dokl.}  \textbf{21} p. 741).
\bibitem[{Baryshevsky and Dubovskaya(1976a)}]{5} Baryshevsky,  V.~G. and  Dubovskaya, I.~Ya. (1976a). in
Proceedings of the VIII All-Union Conference on Physics of
Interaction of Charged Particles with Crystals, Moscow, 1976,
p.51, p. 276.
\bibitem[{Baryshevsky and Dubovskaya(1977a)}]{150a} Baryshevsky, V.~G. and  Dubovskaya, I.~Ya. (1977a). Radiation cooling of charged beams  \emph{Phys. Lett. A } \textbf{62}, 1,  pp. 45-46.
\bibitem [{Baryshevsky and Dubovskaya(1977b)}]{150b}Baryshevsky, V.~G. and  Dubovskaya, I.~Ya. (1977b). \emph{Pis'ma Zh. Tekh. Fiz.}  \textbf{3} p. 500; Errata, ibid p.1100.
\bibitem[{Baryshevskii and Dubovskaya(1977c)}]{165} Baryshevskii, V.~G.  and  Dubovskaya, I.~Ya. (1977c).
Surface channeling of chraged particles,  \emph{Fiz. Tverd. Tela}
\textbf{19}, 2,  pp. 597--599.
\bibitem[{Baryshevskii and Dubovskaya(1977d)}]{17}  Baryshevskii, V.~G. and  Dubovskaya, I.~Ya. (1977d).
Coherent radiation of the channelling positron (electron),
\emph{Phys. Status Solidi (b)} \textbf{82}, 1,  pp. 403--412.
 \bibitem [{Baryshevsky and Dubovskaya(1978)}]{42} Baryshevsky, V.~G. and  Dubovskaya, I.~Ya. (1978).
 \emph{Izv. Akad. Nauk  BSSR} ser. fiz.-mat. \textbf{4}  p. 78.
\bibitem[{Baryshevskii and Feranchuk(1971)}] {75}  Baryshevskii, V.~G. and  Feranchuk, I.~D. (1971).
On transition radiation of gamma-quanta in a crystal, \emph{Zh.
Eksp. Teor. Fiz.} \textbf{61} pp. 944; Errata, ibid \textbf{64}
(1973) p. 760 (\emph{Sov. Phys. JETP} \textbf{34}(1972) p. 50.;
Errata ibid \textbf{37} (1973) 605).
\bibitem[{Baryshevskii and Feranchuk(1973)}]{76} Baryshevsky,  V.~G. and Feranchuk, I.~D. (1973).
Theory of radiation from charged particles in crystals, \emph{Izv.
Akad. Nauk  BSSR }, Ser. fiz.-mat. \textbf{2}  pp. 102--108.
 \bibitem [{Baryshevsky and Feranchuk(1974)}]{90} Baryshevsky,  V.~G. and  Feranchuk, I.~D. (1974).
 \emph{Dokl. Akad. Nauk BSSR } \textbf{18}  p. 449.
\bibitem[{Baryshevskii and Feranchuk(1974)}]{61} Baryshevskii, V.~G. and  Feranchuk, I.~D. (1974).
 Quantum  theory of emission of radiation by electrons in crystals, \emph{Dokl. Akad. Nauk BSSR }\textbf{18} pp. 499--502.
\bibitem[{Baryshevskii and Feranchuk(1976)}] {77} Baryshevskii, V.~G. and  Feranchuk, I.~D. (1976).
The X-ray radiation of ultrarelativistic electrons in a crystal,
\emph{Phys. Lett A} \textbf{57}, 2, pp. 183--185.
\bibitem [{Baryshevsky and Feranchuk(1980a)}]{127}  Baryshevsky, V.~G. and   Feranchuk, I.~D. (1980a). Ultrarelativistic
particle radiation in a crystal and observation of the $\gamma$-$\gamma$ correlations, \emph{Phys. Lett. A} \textbf{76}, 5-6, pp. 452--454.
\bibitem [{Baryshevsky and Feranchuk(1980b)}]{82} Baryshevsky, V.~G. and   Feranchuk, I.~D. (1980b). in
\emph{Proceedings of the X All-Union Conference on Non-linear and Coherent Optics} (Kiev--Moscow)  p. 89.
\bibitem [{Baryshevsky and Grubich(1978)}]{153} Baryshevsky, V.~G. and Grubich, A.~O. (1978). In
\emph{Proceedings of the IX All-Union Conference on Physics of Interaction of Charged Particles with Crystals} (Moscow) p. 105.
\bibitem [{Baryshevsky and Grubich(1979a)}]{11} Baryshevsky, V.G. and Grubich, A.O.  (1979a).
Radiative self-polarization of fast particles in bent crystals, \emph{Pis'ma. Zh. Tekh. Fiz.} \textbf{5}, 24,  pp. 1527--1530.
 \bibitem [{Baryshevsky and Grubich(1979b)}]{159} Baryshevsky,  V.~G. and  Grubich, A.~O. (1979b).
 In  \emph{Proceedings of the X All-Union Conference on Physics of Interaction of  Charged Particles with Crystals} (Moscow) p. 24.
 \bibitem [{Baryshevsky and Grubich(1979c)}]{100} Baryshevsky, V.~G. and Grubich, A.~O. (1979c).
 in, \emph{Proceedings of the X All-Union Conference on Physics of Interaction of Charged Particles with Crystals} (Moscow, 1979). p. 23.
 \bibitem[{Baryshevskii and Ngo Dan Nyan(1974)}]{106}  Baryshevskii, V.~G. and Ngo Dan Nyan,
 Bremsstrahlung transition and Cherenkov radiation of high energy $\gamma$-quanta, \emph{Yad. Fizika } \textbf{20}, 6, pp. 1219--1222.
\bibitem[{Baryshevskii and Podgoretskii(1968)}]{126} Baryshevskii, V.~G. and  Podgoretskii, M.~I. (1968).
Some remarks concerning the interference of independent light
beams, \emph{Zh. Eksp. Teor. Fiz.}  \textbf{55}
(1968) pp. 312--. (\emph{Sov. Phys. JETP}) \textbf{28} (1969) p. 165). %check
 \bibitem [{Baryshevsky and Sokolsky(1980)}]{148} Baryshevsky,  V.~G. and Sokolsky, A.~A.  (1980).
 On the existence of the effect of oscillations of the polarization of a fast particle (channeling
 particle with a quadrupole moment), \emph{Pis'ma  Zh. Tekh. Fiz. } \textbf{6}, 23, pp. 1419--1421.
\bibitem [{Baryshevsky and Tkacheva(1978)}]{135} Baryshevsky, V.~G. and Tkacheva, V.~I. (1978).
Quantum theory of measuring the duration of nuclear reactions in crystals, \emph{Dokl. Akad. Nauk BSSR } \textbf{22}, 1,  pp. 29--31.
\bibitem[{Baryshevsky \emph{et~al.}(1975)}]{12} Invetion Certificate 482 834 SSSR,The method of Production of X-radiation/
Baryshevsky, V.~G., et al.  Published in 1975, N 32 BI %????????? ??? ?????? ?????? ?? ?????? !!
\bibitem[{Baryshevskii \emph{et~al.}(1976)Baryshevskii, Grubich and Ngo Dan Nyan}]{99} Baryshevsky, V.~G.,
Grubich, A.~O. and Ngo Dan Nyan  (1976). Angular, spectral, and polarizationproperties of radiation emitted by
high--energy electrons passing through a layer of matter, \emph{Vestnik BGU}, All-Russian Scientific and Technical Information Institute of Russian Academy of Sciences, deposit No 3554.
\bibitem[{Baryshevskii \emph{et~al.}(1977)Baryshevskii, Grubich and Ngo Dan Nyan}]{98} Baryshevskii, V.~G.,
Grubich, A.~O. and Ngo Dan Nyan (1977). Angular, spectral, and polarization properties of radiation emitted by
high-energy electrons passing through a layer of matter, \emph{Sov. Phys. JETP} \textbf{45}, 6, pp. 1068-1072.
\bibitem[{Baryshevsky \emph{et~al.}(1978) Baryshevsky, Dubovskaya and Grubich}]{154} Baryshevsky, V.~G.,
 Dubovskaya, I.Ya. and Grubich, A.O. (1978). \emph{Vestnik Bel. State Univ. } All-Russian Scientific and
 Technical Information Institute of Russian Academy of Sciences, deposit No 318.
 \bibitem[{Baryshevsky \emph{et~al.}Baryshevsky, Dubovskaya and Feranchuk(1978)}]{140} Baryshevsky,  V.~G., Dubovskaya, I.~Ya.
and   Feranchuk, V.~G. (1978). in  \emph{Proceedings of the IX
All-Union Conference on Physics of Interaction of Charged
Particles with Crystals} (Moscow) p.105.
\bibitem[{Baryshevsky \emph{et~ al.}(1978)Baryshevsky, Grubich and Dubovskaya}]{44} Baryshevsky, V.~G., Grubich, A.~O.
and Dubovskaya, I.Ya. (1978). Diffraction of radiation from channeled charged particles, \emph{Phys. Stat. Sol. (b)} \textbf{88}, 1, pp. 351--358.
\bibitem[{Baryshevsky \emph{et~ al.}(1979)Baryshevsky, Grubich and Dubovskaya}]{46} Baryshevsky, V.~G., Grubich, A.~O. and Dubovskaya, I.Ya. (1979). \emph{Izv. Akad. Nauk  BSSR} ser. fiz.-mat. \textbf{6} p. 72.
\bibitem[{Baryshevsky \emph{et~ al.}(1980a)Baryshevsky, Grubich and Dubovskaya}]{45} Baryshevsky, V.~G., Grubich, A.~O. and Dubovskaya, I.Ya. (1980a). On photon production by channeled electrons (positrons) \emph{Phys. Stat. Sol. (b)} \textbf{99}, 1, pp. 205--213.
\bibitem[{Baryshevsky \emph{et~al.}(1980b) Baryshevsky, Grubich and Dubovskaya}]{67} Baryshevsky, V.~G., Grubich, A.~O. and Dubovskaya, I.Ya. (1980b).  Generation of $\gamma$-quanta by channeled particles in the presence of a variable external field,  \emph{Phys. Lett. A}  \textbf{77}, 1,  pp. 61--64.
\bibitem[{Baryshevsky \emph{et~al.}(1980c)Baryshevsky, Grubich and Dubovskaya}]{47}
 Baryshevsky, V.~G.,  Grubich, A.~O. and Dubovskaya, I.~Ya. (1980c). Eletromagnetic radiation of channeled  particles in an absorptive crystal, \emph{Izv. Akad. Nauk BSSR}, \textbf{4}  pp. 81--86.
 \bibitem[{Baryshevsky \emph{et~al.}(1980d)Baryshevsky, Grubich and Dubovskaya}]{66} Baryshevsky, V.~G., Grubich, A.~O. and Dubovskaya, I.Ya. (1980d). Photon emission by channeled particles in the presence of an ultrasonic (electromagnetic) wave, \emph{Dokl. Akad. Nauk BSSR} \textbf{24}, 3,   pp. 226--229.
 \bibitem[{Baryshevsky \emph{et~al.}(1980e)}]{65}  Baryshevsky V.~G. \emph{et~al.}(1980e). Radiation of  Channeled Particles in a Single Crystal, in\emph{ Preprint INR, Acad. Sci. SSSR P-0166} (Moscow).
 \bibitem [{Bazylev \emph{et~al.}(1980)Bazylev, Glebov and Zhevago}]{57a}  Bazylev, V.~D.,  Glebov, V.~N. and  Zhevago, N.~K. (1980). \emph{Zh. Eksp. Teor.  Fiz.} \textbf{78} p. 62;  Zh. Eksp. Teor.  Fiz. \textbf{80} (1981) 608.
 \bibitem [{Bazylev \emph{et~al.}(1981)Bazylev, Glebov and Zhevago}]{57b}  Bazylev, V.~D.,  Glebov, V.~N. and  Zhevago, N.~K. (1981). \emph{Zh. Eksp. Teor.  Fiz.} \textbf{80} p. 608.
\bibitem [{Bazylev and Zhevago(1977)}]{53} Bazylev, V.~D. and  Zhevago, N.~K. (1977). \emph{Zh. Eksp. Teor.  Fiz.} \textbf{73}  p. 1697.
\bibitem [{Belenky(1948)}]{103}  Belenky, S.~Z. (1948). \emph{Shower Processes in Cosmic Rays} (Gostehizdat, Moscow).
\bibitem [{Beloshitsky and Kumakhov(1977)}]{48} Beloshitsky, V.~V. and Kumakhov, M.~A. (1977). \emph{Doklady Akad. Nauk SSSR} \textbf{237}  p. 71.
 \bibitem [{Beloshitsky and Kumakhov(1978)}]{52} Beloshitsky, V.~V. and Kumakhov, M.~A. \emph{Zh. Eksp. Teor.  Fiz.} \textbf{74}  p. 1244.
 \bibitem [{Belyakov(1975)}]{86}  Belyakov, V.~A. (1975). Diffraction of M\"{o}ssbauer gamma rays in crystals, \emph{Sov. Phys. Usp.} \textbf{18}, 4, pp. 267--291 (\emph{ Usp. Fiz. Nauk} \textbf{115}, 4, pp. 553--601).
\bibitem [{Berestetsky \emph{et~al.}(1968) Berestetsky, Lifshitz and Pitaevsky}]{19} Berestetsky, V.~B., Lifshitz, E.~M. and  Pitaevsky, L.~P. (1968). \emph{Relativistic Quantum Theory} (Moscow, in Russian).
 \bibitem [{Bessonov(1978)}]{94} Bessonov, Ye.~G. (1978). The peculiarities of studying radiation of particles in modulators of various types, in \emph{Preprint FIAN} No 35 (Moscow).
 \bibitem [{Bogdanov \emph{et~al.}(1979)Bogdanov, Nagibarova and Nagibarov}]{115} Bogdanov, Ye.~I., Nagibarova, I.~A., Nagibarov, V.~R. (1979). Quantum theory of self-induced transparency, \emph{Sov. Phys. JETP}  \textbf{50} 2, pp. 253--256 (\emph{Zh. Eksp. Teor. Fiz.} \textbf{77} p. 498).
\bibitem [{Bonch-Osmolovskii and Podgoretskii(1978)}]{151} Bonch-Osmolovskii, A.~G. and  Podgoretskii,  M.~I. (1978)\emph{JINR Reports, P-2-11250}.
\bibitem [{Bonch-Osmolovskii and Podgoretskii(1979)}]{152} Bonch-Osmolovskii, A.~G. and Podgoretskii, M.~I. Channeling of ultrarelativistic particles, \emph{Sov. J. Nucl. Phys.}  \textbf{29}, 2, pp.  216--225 (\emph{Yad. Fizika } \textbf{29} p. 432).
\bibitem [{Bricman \emph{et~al.}(1978)}]{147} Bricman, C. (1978). Review of Particles Properties: Particle data group, \emph{Phys. Lett. B} \textbf{75}, 1, pp. i--xxi;  \emph{Phys. Lett. B} \textbf{75}, 2, pp. 1--250.
\bibitem [{Callaway(1964)}]{21} Callaway, J. (1964).  \emph{Energy Band Theory} (Academic Press, New York and London).
\bibitem [{Cue \emph{et~al.}(1980)}]{56} Cue, N.,  Bonderup, E., Marsh, B.~B., Bakhru, H., Benenson, R.~E.,  Haight, R., Inglis, K., and  Williams, G.O. (1980). Transitions between bound states for axially channeled MeV electrons,  \emph{Phys. Lett. A} \textbf{80}, 1, pp. 26--28.
\bibitem [{Delone and Fedorov(1979)}]{116} Delone, N.~ B. and Fedorov, M.~ V. (1979). Polarization of photoelectrons in the ionization of unpolarized atoms, \emph{Sov. Phys. Usp.} \textbf{22} pp. 252--269  (\emph{Uspekhi Fiz. Nauk} \textbf{127} p. 651).
\bibitem [{Dubovskaya(1978)}]{156} Dubovskaya, I.~Ya.(1978).  \emph{Coherent radiation of X-ray Photons and $\gamma$-quanta by Channeled Charged Particles}, Ph.D thesis, Belarusian State University, Minsk.
 \bibitem[{Dykhne(1961)}]{27} Dykhne,  A.~M. (1961). Quasiclassical particles in a one--dimensional periodic potential field, \emph{Zh. Eksp. Teor. Fiz.} \textbf{40}  pp. 1423--1426.
%
\bibitem[{Entin (1979)}]{121} Entin, I.~R. (1979). \emph{Zh. Eksp. Teor.  Fiz.} \textbf{77}  p. 312.
%

\bibitem [{Fedorov \emph{et~al.}(1973)Fedorov, Kiryanov and Smirnov}]{70} Fedorov, V.~V.,  Kiryanov,  K.~Ye. and Smirnov,  A.~I. (1973). \emph{Zh. Eksp. Teor.  Fiz.} \textbf{64}  p. 1452.
\bibitem [{Fedorov and Smirnov(1974)}]{69} Fedorov, V.~V. and  Smirnov, A.~I. (1974). \emph{Zh. Eksp. Teor.  Fiz.} \textbf{66}  p. 566.
\bibitem [{Fedorov(1980b)}]{119}  Fedorov,   V.~V. (1980b). Influence of Pendellosung effect on the degree of optical modulation of an electron beam diffracted in a crystal,  \emph{Sov. Phys. JETP} \textbf{51}, 2,  pp. 394--396  (\emph{Zh. Eksp. Teor. Fiz.} \textbf{78} p. 782).
\bibitem [{Fedorov(1980a)}]{71} Fedorov, V.~V. (1980a). \emph{Zh. Eksp. Teor.  Fiz.} \textbf{78} p. 46.
\bibitem [{Feinberg and Khizhnyak(1957)}]{89} Feinberg,  Ya.~B. and  Khizhnyak, N.~A. (1957). \emph{Zh. Eksp. Teor.  Fiz.} \textbf{32}  p.  883.
\bibitem [{Feranchuk (1979c) }]{129} Feranchuk, V.~G. (1979c).  \emph{Zh. Tekh. Fiz.}  \textbf{49} 1552.
\bibitem [{Feranchuk(1979a)}]{31} Feranchuk,  I.~D. (1979a). On the shape of the radiation spectrum of relativistic charged particles, \emph{Zh. Tekh. Fiz.} \textbf{49} pp. 1552--1554.
 \bibitem [{Feranchuk(1979b)}]{81} Feranchuk, I.~D.  (1979b). \emph{Kristallografiya} \textbf{24}  p. 289.
\bibitem [{Fok(1948)}]{29} Fok, V.~A. (1948). Fresnel's laws of refraction  and the laws of diffraction, \emph{Sov. Phys. Usp.}  \textbf{36}, 11,  pp. 308--327.
\bibitem [{Frank(1942)}]{38} Frank, I.~M. (1942). Doppler effect in a refracting medium, \emph{Izv. Akad. Nauk SSSR} Ser. Fiz. \textbf{6} pp. 3--31.
\bibitem [{Frank(1959)}]{39} Frank, I.M. (1959). On the role of the  group velocity of light at radiation in a refracting medium, \emph{Zh. Eksp. Teor. Fiz.} \textbf{36}, 3, pp  823--831.
\bibitem [{Frank(1969)}]{40} Frank, I.~M. (1969). Peculiarities of the short--wave part of the Doppler spectrum in a medium, \emph{Preprint JINR P4-4647} (Dubna).
\bibitem [{Frank(1979)}]{41} Frank, I.~M. (1979). Einstein and optics, \emph{Sov. Phys. Usp.} \textbf{22}, 12, pp. 975--986.
 \bibitem [{Garibian and Yan(1976)}]{109} Gariban,  G.~M. and Yan Shi (1976).  X-ray emission by an ultrarelativistic charge in a plate with allowance for multiple scattering, \emph{Sov. Phys. JETP} \textbf{43}pp. 848-- [\emph{Zh. Eksp. Teor. Fiz.} \textbf{70} (1976) 1627].
\bibitem [{Garibyan and Yan Shi(1972)}]{78} Garibyan, G.~M. and  Yan Shi (1972) \emph{Zh. Eksp. Teor.  Fiz. } \textbf{63} p.  1196.
\bibitem [{Gemmell(1974)}]{3} Gemmell, D.~S.  (1974). Channeling and related effects in the motion of charged particles through crystals, \emph{Rev. Mod. Phys.} \textbf{46}, 1, pp. 129--227.
\bibitem [{Ginzburg(1940)}]{35} Ginzburg,  V.~L.  (1940). \emph{Zh. Eksp. Teor.  Fiz.} \textbf{10} p. 584.
\bibitem [{Gluckstern and Lin(1964)}]{144} Gluckstern, R.~I. and Lin, Shin-R. (1964).  Relativistic Coulomb scattering of electrons  \emph{J. Math. Phys.}  \textbf{5}  p. 1594.
\bibitem [{Goldberger and Watson(1965)}]{128} Goldberger,  M.~L.  and Watson K.~M. (1965). Fluctuations with time of scattered--particle intensities, \emph{Phys. Rev.} \textbf{137}, 5B, pp. B1396--B1409.
\bibitem [{Goldberger and Watson(1984}]{111} Goldberger  M.~L. and  Watson, R.~M. \emph{Collision Theory} (Wiley, New York).
\bibitem [{Gradstein and Ryzhik(1980)}]{30}  Gradstein, I.~S. and  Ryzhik, I.~M. (1980). \emph{Table of Integrals, Series and Products}  (Academic Press, New York).
\bibitem[{Heitler(1984)}]{102}  Heitler,  W. (1984). \emph{The Quantum Theory of Radiation} (Dover Publications, New York).
 \bibitem [{Hirsch \emph{et ~al.}(1965)Hirsch, Howie, Nicholson, Pashley and Whelan}]{13}  Hirsch, P.~B., Howie, A.,  Nicholson, R.~B., Pashley, D.~W. and Whelan,  M.~J. (1965).  \emph{Electron Microscopy of Thin Crystals} (Butterworths, London) .
 \bibitem [{Kagan and Kononets(1970)}]{15} Kagan, Yu.  and  Kononets, Yu. (1970). Theory of channeling effects. I. \emph{Zh. Eksp. Teor. Fiz.} \textbf{58} pp. 226--244 (\emph{Sov. Phys. JETP } \textbf{31}, p. 124).
\bibitem [{Kagan and Kononets(1973)}]{23} Kagan, Yu. and Kononets, Yu. (1973). Theory of the channeling effect: II. Influence of inelastic collisions, \emph{Sov. Phys. JETP}\textbf{37} p. 530 (\emph{Zh. Eksp. Teor. Fiz.}  \textbf{64} 1042).
 \bibitem [{Kagan and Kononets(1974)}]{24} Kagan, Yu. and Kononets, Yu. (1974). Theory of channeling effect. Fast particle energy--losses, \emph{Zh. Eksp. Teor. Fiz.}  \textbf{66} pp. 1693--1711.
 \bibitem [{Kalashnikov \emph{et~al.}(1972)Kalashnikov, Koptelov and Ryazanov}]{4} Kalashnikov, N.~P., Koptelov, E.~A. and Ryazanov, M.~I.  (1972). \emph{Fiz. Tverd. Tela} \textbf{14} p. 1211.
 \bibitem [{Kalashnikov \emph{et~al.}(1975) Kalashnikov, Koptelov and Strikhanov}]{33} Kalashnikov,  N.~P., Koptelov, E.~A. and  Strikhanov, M.~N. (1975) in  \emph{Proceedings of the VII All-Union Conference on Physics of Interaction of Charged Particles with Crystals} (Moscow)  p. 36.
\bibitem [{Kalashnikov \emph{et~al.}(1985)Kalashnikov, Remizovich and Ryazanov}]{25} Kalashnikov, N.~P., Remizovich,  V.~S. and  Ryazanov, M.~I. (1985). \emph{Collisions of Fast Charged Particles in Solids} (Gordon and Breach, New York).
 \bibitem [{Kalashnikov and Koptelov(1979)}]{49} Kalashnikov, N.~P. and  Koptelov, E.~A. (1979). Characteristic Bremsstrahlung of Ultra-relativistic Electrons in Single Crystals, in \emph{Preprint INR, Acad. Sci. SSSR P-0054} (Moscow).
\bibitem [{Kalashnikov and Strikhanov(1975)}]{18}  Kalashnikov, N.~P.  and Strikhanov, M.~N. (1975).  Theory of diffractional scattering of fast positive particles in a single crystal, \emph{Zh. Eksper. Teor. Fiz.} textbf{69} p. 1253--1262.
\bibitem [{Kalashnikov and Olchak(1979)}]{50} Kalashnikov,  N.~P. and   Olchak, A.~S.  (1979). \emph{Interaction of Nuclear Radiations with Single Crystals} (Moscow Phys. Eng. Inst., Moscow) (in Russian).
\bibitem [{Kalashnikov and Strikhanov(1980)}]{51} Kalashnikov, N.~P. and Strikhanov, M.N. The Theory of Electromagnetic Radiation of Ultra-relativistic Particles in a Single Crystal (1980). in\emph{ Preprint Moscow Phys. Eng. Inst.} N  88 (Moscow) (in Russian).
 \bibitem [{Kapitsa(1979)}]{130}  Kapitsa, S.~P. (1979). Seminar on large European projects, \emph{Sov. Phys. Usp.} \textbf{22} pp. 939--941 (\emph{Uspekhi Fiz. Nauk} \textbf{129}  p. 549).
 \bibitem [{Kaplin and Vorobiev(1978)}]{139} Kaplin, V.~V.  and  Vorobiev, S.~A. (1978). \emph{Pis'ma Zh. Tekh. Fiz. } \textbf{4}  p.  196.
\bibitem [{Karamyan \emph{et~al.}(1973)Karamyan, Melikov and Tulinov}]{131}  Karamyan, S.~A., Melikov, Yu.~V. and Tulinov, A.~F. (1973). Use of the blocking effect to measure nuclear reaction times, \emph{Sov. J. Particles Nucl.} \textbf{4}, 2, pp. 196--216.
\bibitem [{Kolpakov(1973)}]{43} Kolpakov, A.~V. (1973).  \emph{Yad. Fiz.} \textbf{16} p. 1003.
\bibitem [{Komarov and Pisarevsky(1965)}]{91} Komarov, L.~I. and  Pisarevsky, A.~N. (1965). \emph{Prib. Tekh. Eksp. }  \textbf{4}   p. 226.
\bibitem [{Kumakhov(1976)}]{7} Kumakhov, M.~A. (1976). On the theory of electromagnetic radiation  of charged particles in a crystal, \emph{Phys. Lett. A} \textbf{57}, 1, pp. 17--18.
\bibitem [{Kumakhov(1977)}]{36} Kumakhov, M.~A. (1977). \emph{Zh. Eksp. Teor.  Fiz.} \textbf{72}  p. 1489.
\bibitem [{Landau and Lifshitz(1967)}]{68} Landau, L.~D. and  Lifshitz, E.~M. (1967). \emph{The  Theory of Field} (Nauka, Moscow)  (in Russian)
\bibitem [{Landau and Lifshitz(1977)}]{16} Landau, L.~D. and Lifshitz, E.~M. (1977). \emph{Quantum Mechanics: Non-Relativistic Theory}, in  Landau, L.~D. and Lifshitz, E.~M. \emph{Course of Theoretical Physics} Vol. 3, 3rd edn. (Pergamon Press).
\bibitem [{Lax(1951)}]{112} Lax, M. (1951). Multiple scattering of waves, \emph{Rev. Mod. Phys.} \textbf{23}, 4, pp. 287--310.
\bibitem [{Lindhard(1965)}]{2} Lindhard, J. (1965). \emph{Math.-Fys. Medd. Dan. Vid. Selsk.} \textbf{34}, 14; \emph{Sov. Phys. Usp.} \textbf{99} (1969) p. 249.
\bibitem [{Luttinger and Kohn(1958)}]{160}  Luttinger, J.~M. and  Kohn, W. (1958). Quantum theory of electrical transport phenomena. II, \emph{Phys. Rev.} \textbf{109}, 6,  pp. 1892--1909.
\bibitem [{Lyubosihtz and Podgoretskii(1976)}]{132} Lyubosihtz, V.~L.  and Podgoretskii, M.~I. (1976). Fluctuations  of effective cross sections in a unitary theory, \emph{Sov. J. Nucl. Phys.}   \textbf{24}, 1, pp. 110--116 (\emph{Yad. Fiz.}  \textbf{24} pp. 214--226).
\bibitem [{Lyubosihtz(1978a)}]{133} Lyubosihtz, V.~L. (1978a). Duration of nuclear reactions for strongly overlapping resonance levels, \emph{Sov. J. Nucl.} \textbf{27}, 4, pp. 502--507 (\emph{Yad. Fizika}  \textbf{27} (1978) 948).
\bibitem [{Lyubosihtz(1978b)}]{134} Lyuboshitz V.~L. (1978b). Unitary sum rules and collision times in strong overlap of resonance levels, \emph{JETP Lett.} \textbf{28}, 1, pp. 30--34.
\bibitem [{Lyubosihtz(1980a)}]{141} Lyubosihtz, V.~L. (1980a). Spin rotation associated with the deflection of a relativistic charged particle in an electric field, \emph{Sov. J. Nucl. Phys.} \textbf{31}, 4, pp. 509--512.
\bibitem [{Lyubosihtz(1980b)}]{145} Lyubosihtz, V.~L. (1980b). Depolarization of fast particles travelling through matter, \emph{Sov. J. Nucl. Phys.}  \textbf{32}, 3, pp. 362--365.
\bibitem [{M\"{o}ller(1972)}]{142}  M\"{o}ller, C. (1972). \emph{The Theory of Relativity } (Oxford Univ. Press, London).
\bibitem [{Mashkova \emph{et~al.}(1970)Mashkova, Molchanov and Skripka}]{162} Mashkova, Ye.~S.,  Molchanov, V.~A. and Skripka, Yu.~G. (1970). \emph{Dokl. Akad. Nauk SSSR } \textbf{190} p.  73.
\bibitem [{Mashkova \emph{et~al.}(1971)Mashkova, Molchanov and Skripka}]{163} Mashkova, Ye.~S.,  Molchanov, V.~A. and Skripka, Yu.~G. (1970). \emph{Dokl. Akad. Nauk SSSR } \textbf{198} p.  809.
\bibitem [{Morse and Feshbach(1953)}]{101} Morse, P.~M. and Feshbach, H. (1953). \emph{Methods of Theoretical Physics }(Mc Graw Hill, New York).
\bibitem [{Nikishov(1979)}]{88a} Nikishov, A. I. (1979). Intense external fields in quantum electrodynamics, in \emph{Quantum electrodynamics of phenomena in intense fields}  (Nauka , Moscow)  (Akademiia Nauk SSSR, Fizicheskii Institut, Trudy.)  \textbf{111} pp. 152--271 [in Russian].
\bibitem [{Olsen and Maximon(1959)}]{62} Olsen, H. and  Maximon, L.~C. (1959). Photon and electron polarization in high--energy bremsstrahlung and pair production with screening, \emph{Phys. Rev.}  \textbf{114}, 3,  pp. 887--904.
\bibitem [{Pafomov(1969)}]{104}  Pafomov, V.~Ye. (1969). Radiation of a charged particle in the presence of a separating boundary, \emph{Trudy FIAN} \textbf{44} pp. 28--167 (\emph{Proc (TR) P.N. Lebedev Phys. Inst. (USSR)} \textbf{44} (1971) pp. 25--157 (Engl. Transl.)).
\bibitem [{Perelshtein and Podgoretsky(1970)}]{107}  Perelshtein, E.~A. and Podgoretsky, M.~I.(1970). Transition radiation in domain of resonance $\gamma$-quanta, \emph{Yad. Fizika}  \textbf{12} pp. 1149--1153.
\bibitem [{Pinsker(1974)}]{85} Pinsker, Z~.G.  (1974).  \emph{Dynamic Scattering of X-Rays in Perfect Crystals } (Nauka, Moscow)  (in Russian).
\bibitem [{Plotnikov \emph{et~al.}(1979)Plotnikov, Kaplin and Vorobiev}]{110} Plotnikov, S.~V.,  Kaplin, V.~V. and Vorobiev, S.~A. (1979). Preliminary Programme and Abstracts of the X Conference on the Problems of Application of Charged Particle Beams for Studying the Composition and Properties of Matter (Moscow) p. 28.
\bibitem [{Podgoretsky(1977a)}]{92} Podgoretsky, M.~I. (1977a). \emph{Report JINR P2-10986} (Dubna).
\bibitem [{Podgoretsky(1977b)}]{93} Podgoretsky, M.~I. (1977a). \emph{Report JINR P2-11140} (Dubna).
\bibitem [{Podgoretsky(1980)}]{59}  Podgoretsky, M.~I. (1980). \emph{Yad. Fiz.} \textbf{31} p. 417.
\bibitem [{Pokrovskii and  Khalatnikov(1961)}]{28} Pokrovskii, V.~L. and Khalatnikov, I.~M. (1961). On superbarrier reflection of high--energy particles,  \emph{Zh. Eksp. Teor.  Fiz.} \textbf{40} pp.1713--1719.
\bibitem [{Ritus(1970)}]{88b} Ritus, V. I. (1979). Quantum effects in the interaction of elementary particles with an intense electromagnetic field, in \emph{Quantum electrodynamics of phenomena in intense fields} (Nauka, Moscow (Akademiia Nauk SSSR, Fizicheskii Institut, Trudy.)  \textbf{111} pp. 5--151 [in Russian].
\bibitem [{Rossi and Greisen(1948)}]{146} Rossi, B. and   Greisen, K. (1948). \emph{Interaction of Cosmic Rays with Matter} (Mir, Moscow) (in Russian).
\bibitem [{Ryabov(1970)}]{32} Ryabov, V.~A. (1970). \emph{Zh. Eksp. Teor. Fiz.}  \textbf{58}  2446.
\bibitem [{Ryabov(1975)}]{138}  Ryabov, V.~A. (1975). Quantum theory of the inelastic scattering of channeled particles, \emph{Sov. Phys. JETP} \textbf{40}, 1, pp. 77--81. (\emph{Zh. Eksp. Teor. Fiz} \textbf{67} pp. 150--160).
\bibitem [{Rytov(1966)}]{157} Rytov, S.~M. (1966).  \emph{Introduction to Statistical Radiophysics} (Nauka, Moscow) (in Russian).
\bibitem [{Samsonov(1978)}]{108} Samsonov, V.~M. (1978). Cherenkov and transition radiations in the $\gamma$-resonance frequency region, \emph{Sov. Phys. JETP} \textbf{48}, 1, pp. 44--47.
\bibitem [{Skripka(1974)}]{164}  Skripka, Yu.~G. (1974). \emph{Ukr. Fiz. Zh.} \textbf{19} p.  1731.
\bibitem [{Slichter(1963)}]{122} Slichter,  Ch.~P.  (1963). \emph{Principles of Magnetic Resonance} (Harper and Row, New York).
\bibitem [{Sommerfeld and Bethe(1938)}]{22} Sommerfeld, A. and Bethe, H. (1933). \emph{Electron Theory of the Metals}, in \emph{Manual of Physics} Volume. 24--2 (Heidelberg: Springer publishing house) pp. 333--622.
\bibitem [{Swent \emph{et~ al.}(1979)Swent, Pantell, Alguard, Berman, Bloom and Datz}]{54}Swent, R.~L.,  Pantell, R.~H., Alguard, M.~J.,  Berman, B.~L. Bloom, S.~D. and Datz, S. (1979). Observation of channeling radiation from relativistic electrons \emph{Phys. Rev. Lett.} \textbf{43}, 23, pp. 1723--1726.
\bibitem [{Ter-Mikaelian(1969)}]{63} Ter-Mikaelian, M.~L. (1969). \emph{Influence of the Medium on Electromagnetic Processes at High Energies}  (Armenian Academy of Sciences, Erevan) (in Russian).
\bibitem [{Ter-Mikaelian(1972)}]{64} Ter-Mikaelian, M.~L. (1972). \emph{High Energy Electromagnmetic Processes in Condensed Media} (Interscience Tracts in Physics and Astronomy, vol. 28, Willey, New York).
\bibitem [{Thompson(1968)}]{1}  Thompson, M.~W. (1968). The channeling of particles in crystals, \emph{Contemp. Phys.} \textbf{9}, 4, pp. 375--398.
\bibitem [{Toptygin(1964)}]{105} Toptygin, I.~N. (1964). Theory of bremsstrahlung and pair production in a medium, \emph{Zh. Eksp. Teor. Fiz.} \textbf{46} pp. 851--862.
\bibitem [{Tsyganov(1976a)}]{8a}  Tsyganov, E.~N. (1976a). Some aspects of the mechanism of a charged particle penetration through a monocrystal, Tech. Rep. TM-682, Fermilab., Batavia.
\bibitem [{Tsyganov(1976b)}]{8b}  Tsyganov, E.~N. (1976b). Estimates of cooling and bending process for particle penetration through a monocrystal Tech. Rep. TM-684 Fermilab., Batavia.
\bibitem [{Varshalovich and D'yakonov(1970)}]{117}   Varshalovich, A.~D. and D'yakonov, M.~I. (1970). Concerning the effect of Schwarz and Hora, \emph{JETP Lett.} \textbf{11},12,  pp. 411--413 (\emph{Pis'ma. Zh. Eksp. Teor. Fiz.} \textbf{11} 594).
\bibitem [{Varshalovich and D'yakonov(1971)}]{118} Varshalovich, A.~D. and D'yakonov, M.~I. (1971).  \emph{Zh. Eksp. Teor. Fiz.} \textbf{60} p. 90.
\bibitem [{Vedel' and  Kumakhov(1979)}]{158}  Vedel, R. and  Kumakhov,  M.A. (1979). \emph{Pis'ma. Zh. Tekh.  Fiz.} \textbf{5}  p. 689.
\bibitem [{Vorobiev \emph{et~al.}(1975)Vorobiev, Kaplin and Vorobiev}]{6} Vorobiev, A.~A.,  Kaplin, V.~V. and Vorobiev, S.~A. (1975). Radiation of electrons transmitted through the crystal, \emph{Nucl. Instrum. Methods} \textbf{127}, 2,  pp. 265--268.
\bibitem [{Yazaki and Yoshida(1974)}]{137} Yazaki, K. and Yoshida, Sh. (1974). Wave-packet description of nuclear lifetime measurements by crystal blocking experiments, \emph{Nucl. Phys. A}  \textbf{232}  pp. 249--268.
\bibitem [{Zhevago(1978)}]{37} Zhevago, N~.K. (1978). \emph{Zh. Eksp. Teor.  Fiz.} \textbf{75}  p. 1389.



\end{thebibliography}
\end{document}